\documentclass[11pt,reqno]{amsart}

\usepackage[ruled,vlined]{algorithm2e}
\usepackage[sorting = none, style=numeric-comp,maxbibnames=99,giveninits=true]{biblatex}
\DefineBibliographyExtras{UKenglish}{}
\usepackage{lmodern}
\usepackage{appendix}

\usepackage{etoolbox}

\usepackage{geometry}
\preto\section{\ifnum\value{section}=0\addtocontents{toc}{\vskip5pt}\fi}

\makeatletter
\def\subsection{\@startsection{subsection}{1}
	\z@{.5\linespacing\@plus.7\linespacing}{-.5em}
	{\normalfont\itshape\bfseries}}
\patchcmd{\@sect}
{\@addpunct.}
{}
{}
{}
\makeatother

\usepackage[british,UKenglish]{babel}
\addto\captionsbritish{}
\addto\captionsUKenglish{}

\usepackage{amsthm}
\usepackage{dsfont}
\usepackage{latexsym}
\usepackage{amsfonts}
\usepackage{amssymb}
\usepackage{mathtools}
\usepackage{braket}
\usepackage{thmtools}
\usepackage{bbm}

\usepackage{enumerate}
\usepackage{paralist}
\usepackage{tkz-euclide}

\usepackage{tabularx}
\usepackage{array,rotating,makecell,multirow}
\usepackage{colortbl}
\usepackage{hhline}
\usepackage{graphicx}
\graphicspath{{figures/}}
\usepackage{tikz}
\usetikzlibrary{shapes.geometric, arrows, shadings, intersections, calc, plotmarks, arrows.meta, patterns, decorations.pathmorphing,decorations.pathreplacing}
\usetikzlibrary{backgrounds}
\RequirePackage[skins]{tcolorbox}
\usepackage{xfrac}
\usepackage[framemethod=tikz]{mdframed}
\usepackage{soul}
\usepackage{xspace}
\usepackage{capt-of}
\usepackage{csquotes}
\usepackage{float}
\usepackage{hyperref}
\providecommand{\theHalgocf}{\thealgocf}
\renewcommand{\theHalgocf}{main.\arabic{algocf}}
\usepackage{cleveref}

\tikzset{
	position label/.style={
		below = 3pt,
		text height = 1.5ex,
		text depth = 1ex
	},
	brace/.style={
		decoration={brace, mirror},
		decorate
	}
}

\makeatletter
\pgfdeclarepatternformonly{my north east lines}{\pgfqpoint{-0pt}{-0pt}}{\pgfqpoint{3pt}{3pt}}{\pgfqpoint{3pt}{3pt}}{
		\pgfsetlinewidth{0.4pt}
		\pgfpathmoveto{\pgfqpoint{0pt}{0pt}}
		\pgfpathlineto{\pgfqpoint{3pt}{3pt}}
		\pgfpathmoveto{\pgfqpoint{2.8pt}{-.2pt}}
		\pgfpathlineto{\pgfqpoint{3.2pt}{.2pt}}
		\pgfpathmoveto{\pgfqpoint{-.2pt}{2.8pt}}
		\pgfpathlineto{\pgfqpoint{.2pt}{3.2pt}}
		\pgfusepath{stroke}}
\makeatother

\newdimen\LineSpace
\tikzset{
	line space/.code={\LineSpace=#1},
	line space=5pt
}

\tikzstyle{box} = [rectangle, rounded corners, minimum width=3cm, minimum height=1cm, align = center, draw=black]
\tikzstyle{arrow} = [thick,->,>=stealth]

\hypersetup{colorlinks=true,linkcolor=blue!70,citecolor=blue}

\newcommand{\doi}[1]{\textsc{doi}: \href{http://dx.doi.org/#1}{\nolinkurl{#1}}}

\DeclareCiteCommand{\citetitle}
{\boolfalse{citetracker}\boolfalse{pagetracker}\usebibmacro{prenote}}
{\ifciteindex{\indexfield{indextitle}}{}\printtext[bibhyperref]{\printfield[citetitle]{labeltitle}}}
{\multicitedelim}{\usebibmacro{postnote}}

\DeclareCiteCommand{\citeauthor}[\hypersetup{citecolor=black}]
{\boolfalse{citetracker}\boolfalse{pagetracker}\usebibmacro{prenote}}
{\ifciteindex{\indexnames{labelname}}{}\printtext[bibhyperref]{\printnames{labelname}}}
{\multicitedelim}{\usebibmacro{postnote}}

\newtheoremstyle{myplain}{5pt}{5pt}{\itshape}{0pt}{\bfseries}{}{5pt plus 1pt minus 1pt}{}
\newtheoremstyle{mydefinition}{5pt}{5pt}{\normalfont}{0pt}{\bfseries}{}{5pt plus 1pt minus 1pt}{}
\newtheoremstyle{mainresult}{5pt}{5pt}{\itshape}{0pt}{\bfseries}{}{5pt plus 1pt minus 1pt}{}

\theoremstyle{myplain}
\newtheorem{thm}{Theorem}
\newtheorem*{thm*}{Theorem}

\newtheorem{cor}[thm]{Corollary}
\newtheorem{lem}[thm]{Lemma}
\newtheorem{prop}[thm]{Proposition}
\newtheorem*{prop*}{Proposition}
\newtheorem{assum}{Assumption}
\newtheorem*{assum*}{Assumption}

\renewcommand\theassumst{\arabic{assumst}*}

\theoremstyle{mainresult}
\newtheorem{result}{Result}

\theoremstyle{mydefinition}
\newtheorem{defi}[thm]{Definition}

\theoremstyle{remark}
\newtheorem{ex}{Example}

\newtheorem*{rmk*}{Remark}

\crefname{thm}{theorem}{theorems}
\Crefname{thm}{Theorem}{Theorems}
\crefname{result}{result}{results}
\Crefname{result}{Result}{Results}
\crefname{conj}{conjecture}{conjectures}
\Crefname{conj}{Conjecture}{Conjectures}
\crefname{cor}{corollary}{corollaries}
\Crefname{cor}{Corollary}{Corollaries}
\crefname{lem}{lemma}{lemmas}
\Crefname{lem}{Lemma}{Lemmas}
\crefname{prop}{proposition}{propositions}
\Crefname{prop}{Proposition}{Propositions}
\crefname{assum}{assumption}{assumptions}
\Crefname{assum}{Assumption}{Assumptions}
\crefname{assumst}{assumption}{assumptions}
\Crefname{assumst}{Assumption}{Assumptions}
\crefname{defi}{definition}{definitions}
\Crefname{defi}{Definition}{Definitions}
\crefname{ex}{example}{examples}
\Crefname{ex}{Example}{Examples}
\crefname{rmk}{remark}{remarks}
\Crefname{rmk}{Remark}{Remarks}

\newcommand{\dist}{\operatorname{dist}}
\newcommand{\diam}{\operatorname{diam}}

\newcommand*\Bell{\ensuremath{\boldsymbol\ell}}

\newcommand{\ad}{a^\dagger}

\newcommand{\C}{\mathbbm{C}}

\newcommand{\N}{\mathbbm{N}}
\newcommand{\Z}{\mathbbm{Z}}

\newcommand{\R}{\mathbbm{R}}

\newcommand{\RE}{\textrm{Re}}
\newcommand{\IM}{\textrm{Im}}

\newcommand{\cB}{\mathcal{B}}

\newcommand{\cD}{\mathcal{D}}

\newcommand{\cH}{\mathcal{H}}

\newcommand{\cL}{\mathcal{L}}
\newcommand{\cK}{\mathcal{K}}
\newcommand{\cM}{\mathcal{M}}
\newcommand{\cN}{\mathcal{N}}
\newcommand{\cO}{\mathcal{O}}
\newcommand{\cP}{\mathcal{P}}

\newcommand{\cS}{\mathcal{S}}
\newcommand{\cT}{\mathcal{T}}

\newcommand{\cV}{\mathcal{V}}
\newcommand{\cW}{\mathcal{W}}

\renewcommand{\epsilon}{\varepsilon}

\newcommand{\n}{\textbf{n}}

\newcommand{\dd}{\textbf{d}}

\DeclareMathOperator{\dom}{dom}

\xdefinecolor{tumblue}{RGB}{0,101,189}
\xdefinecolor{tumgreen}{RGB}{162,173,0}
\xdefinecolor{tumred}{RGB}{229,52,24}
\xdefinecolor{tumivory}{RGB}{218,215,203}
\xdefinecolor{tumorange}{RGB}{227,114,34}
\xdefinecolor{tumlightblue}{RGB}{152,198,234}

\xdefinecolor{tuebired}{RGB}{161,030,59}

\xdefinecolor{mygreen}{RGB}{108,187,69}
\xdefinecolor{myorange}{RGB}{255,215,0}

\newcommand{\vertiii}[1]{{\left\vert\kern-0.25ex\left\vert\kern-0.25ex\left\vert #1
	\right\vert\kern-0.25ex\right\vert\kern-0.25ex\right\vert}}

\newcommand{\abs}[1]{\left|\hspace{-0.2ex}#1\hspace{-0.2ex}\right|}

\newcommand{\ketbra}[2]{\left|#1\right\rangle\!\!\left\langle#2\right|}
\newcommand{\tr}[1][~]{\ifthenelse{\equal{#1}{~}}{\operatorname{Tr}}{\operatorname{Tr}\left[#1\right]}}

\title[Learning and simulation via finite-energy locality]{\texorpdfstring{Learning and simulating bosonic systems \\via finite-energy locality}{Learning and simulating bosonic systems via finite-energy locality}}

\author[M\"obus, Bluhm, Caro, Werner, Rouz\'e]{%
	\normalfont Tim M\"obus\textsuperscript{1,2}, Andreas Bluhm\textsuperscript{3}, Matthias C. Caro\textsuperscript{4,5},\\
	Albert H. Werner\textsuperscript{6}, Cambyse Rouz\'{e}\textsuperscript{1,2,7}
}

\newcommand{\frontmatteraffiliations}{%
	\begin{minipage}{0.96\textwidth}
		\centering\footnotesize
		\textsuperscript{1}Department of Mathematics, Technical University of Munich, 80333 Munich, Germany\\
		\textsuperscript{2}Munich Center for Quantum Science and Technology (MCQST), 80333 Munich, Germany\\
		\textsuperscript{3}Univ.\ Grenoble Alpes, CNRS, Grenoble INP, LIG, 38000 Grenoble, France\\
		\textsuperscript{4}Department of Computer Science, University of Warwick, Coventry CV4 7AL, UK\\
		\textsuperscript{5}Dahlem Center for Complex Quantum Systems, Freie Universit\"at Berlin, Berlin, Germany\\
		\textsuperscript{6}Department of Mathematical Sciences, University of Copenhagen, 2100 Copenhagen, Denmark\\
		\textsuperscript{7}Inria, T\'{e}l\'{e}com Paris -- LTCI, Institut Polytechnique de Paris, 91120 Palaiseau, France\\[0.5em]
		\texttt{moebustim@gmail.com, andreas.bluhm@univ-grenoble-alpes.fr, matthias.caro@warwick.ac.uk}\\
		\texttt{werner@math.ku.dk, cambyse.rouze@inria.fr}
	\end{minipage}
}

\makeatletter
\def\@setauthors{%
	\begingroup
	\trivlist
	\centering\@topsep30\p@\relax
	\advance\@topsep by -\baselineskip
	\item\relax
	{\normalfont\normalsize\begin{tabular}{c}\authors\end{tabular}\par}
	\vspace{0.75em}
	{\normalfont\frontmatteraffiliations\par}
	\endtrivlist
	\endgroup
}
\makeatother

\begin{document}

\maketitle

\medskip

\addtocontents{toc}{\protect\setcounter{tocdepth}{0}}

\textbf{Bosonic devices promise applications in simulation, sensing and quantum error correction, but infinite-dimensional local Hilbert spaces obstruct the locality tools that make qubit dynamics efficiently learnable and simulable. We establish a finite-energy locality principle for geometrically local bosonic open systems satisfying photon-number moment propagation. It compares unbounded generators with Galerkin cutoffs, transferring finite-dimensional Lieb--Robinson, product-formula and circuit techniques to bosons with explicit errors. For this moment-controlled class, we obtain, to our knowledge, the first model-independent weak Lieb--Robinson bounds beyond Bose--Hubbard-type dynamics, together with quantitative Trotter and simulation guarantees for polynomial bosonic GKSL generators. As a central application, coherent-state preparation and local heterodyne detection suffice to learn coefficients of a known bounded-degree polynomial Hamiltonian ansatz local on a bounded-growth interaction graph to accuracy $\varepsilon$ and failure probability $\delta$, with sample complexity and total evolution time both $\widetilde{\mathcal{O}}(\varepsilon^{-2}\log(m/\delta))$, where $m$ counts on-site and interaction terms. This matches the best known finite-dimensional locality-assisted scaling in accuracy and system size, up to polylogarithmic factors. The assumptions hold for Bose--Hubbard and quadratic dynamics without added dissipation; for more general local polynomial Hamiltonians they can be supplied natively or engineered by known multi-photon loss in stabilized bosonic architectures.}

\section*{Finite-energy locality}
	Learning and simulating many-body quantum dynamics are dual tasks. In simulation, one starts from a classical description of the generator and aims to reproduce the resulting evolution; in Hamiltonian learning, one interrogates an unknown evolution and aims to recover that description. In finite-dimensional lattice systems, both tasks are powered by the same locality principle: Lieb--Robinson bounds \cite{LiebRobinson.1972,Kliesch.2014,Haah.2022optimal,Stilckfranca.2024}. They make it possible to replace a global evolution by dynamics on a local neighbourhood, enabling efficient product-formula simulation \cite{Lloyd.1996,Haah.2021} and Hamiltonian learning protocols whose cost grows only logarithmically with the number of local parameters \cite{Stilckfranca.2024}. Related finite-dimensional work covers local-measurement, sample-efficient, structure-learning and Heisenberg-limited multi-qubit protocols \cite{Bairey.2019learning,Anshu.2021sample,Moebus.2026lindbladians,Huang.2023learning,Dutkiewicz.2023,Brahmachari.2026heisenberg}.

	Continuous-variable (CV) and bosonic systems are central to quantum communication, sensing, analogue simulation and hardware-efficient quantum error correction \cite{Braunstein.2005quantum,Aasi.2013enhanced,Flurin.2017observing,Mirrahimi.2014dynamically,Ofek.2016extending,Guillaud.2019repetition}. Yet the locality theory that is routine for qubits does not extend automatically to oscillators. Local bosonic Hamiltonians are unbounded, finite-energy states can develop large photon-number tails, and in full generality information propagation can be arbitrarily fast \cite{Eisert.2009}. This leaves a gap between the experimental relevance of bosonic platforms and the rigorous algorithmic guarantees available for their characterization.

	We close this gap for a broad and physically motivated class of geometrically local bosonic dynamics. The key condition is not a hard local Hilbert-space cutoff, but stability and propagation of photon-number moments, which can be observed in natural settings such as Gaussian Hamiltonians, the Bose--Hubbard model and stabilized oscillator arrays. This moment propagation estimate is the main technical engine of the paper: it permits a Galerkin-type finite-energy approximation of unbounded bosonic generators, through which finite-dimensional locality and approximation tools can be lifted back to the original infinite-dimensional dynamics with explicit error bounds. Under these assumptions we prove, to our knowledge, the first weak bosonic Lieb--Robinson bounds that are model-independent within the moment-propagation class and go beyond the well-understood Bose--Hubbard setting, together with quantitative Trotter product formulas for local GKSL evolutions and a Hamiltonian learning scheme achieving the best known finite-dimensional complexities up to polylogarithmic factors. The assumptions are satisfied by the Bose--Hubbard model and by quadratic Hamiltonians without additional dissipation. For more general polynomial Hamiltonians, they can be supplied natively or deliberately engineered through known multi-photon loss terms of the form used to stabilize cat-code manifolds \cite{Mirrahimi.2014dynamically,Michael.2016new,Guillaud.2019repetition,Guillaud.2023,CostaRico.2025sobolev}. In this setting, engineered or intrinsic dissipation acts as a regularizing resource.

\begin{figure}[H]
		\centering
		\includegraphics[width=\textwidth]{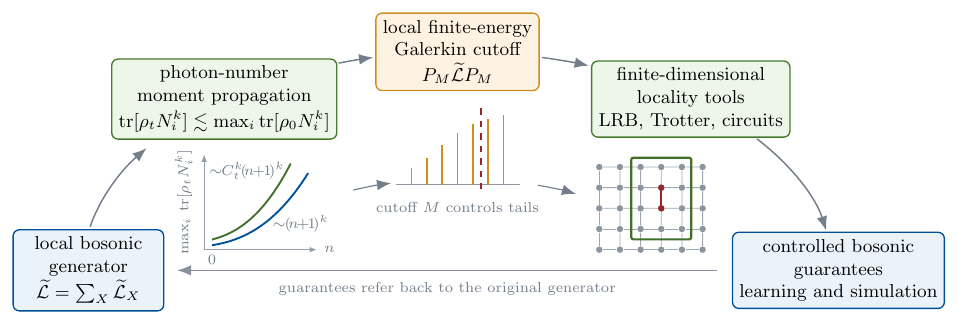}
		\caption{Finite-energy locality principle: Moment propagation controls local photon-number growth, allowing unbounded bosonic generators to be compared with finite-energy Galerkin approximations. Finite-dimensional locality, product-formula and circuit tools then transfer learning and simulation guarantees back to the original dynamics.}
		\label{fig:finite-energy-principle}
	\end{figure}

	The main payoff is experimentally natural Hamiltonian learning. With coherent-state preparation, evolution under the unknown local Hamiltonian for times that depend only polylogarithmically on the target precision, known dissipative channels when needed, and local heterodyne detection, we estimate the coefficients of a known graph-local bounded-degree polynomial Hamiltonian ansatz and achieve $\varepsilon^{-2}$ sample complexity and evolution time scaling up to polylogarithmic factors and logarithmic system-size dependence, as in the best locality-based protocols for finite-dimensional systems under these weak assumptions \cite{Stilckfranca.2024}. Bose--Hubbard systems need no engineered dissipation, while cat-code oscillator arrays supply the required stabilization as part of the device: the protocol then learns the coefficients of detuning, Kerr, beam-splitter and cross-Kerr terms, with the stabilizing dissipation treated as known. The same finite-energy reduction also yields simulation guarantees beyond previously tractable model classes \cite{Kuwahara.2021,Kuwahara.2024digital,Faupin.2022}: digitization into qubit circuits, decomposition into local bosonic evolutions, and system-size-independent approximation of local observables at constant time. The shared principle is summarized in \Cref{fig:finite-energy-principle}.

\section*{From moment control to algorithms}
	We consider an $m$-mode bosonic system, with modes indexed by a finite set $V$ of size $m=|V|$ and equipped with a metric $\dist$. In the graph-based examples and in the learning theorem below, $\dist$ is the graph distance of a known interaction graph $G=(V,E)$; the general locality statements only require the metric and a family of local supports. We assume that the evolution of the system is described by a Liouvillian of GKSL form \cite{Gorini.1976, Lindblad.1976} that is local and finitely connected, i.e.
	\begin{equation}\label{eq:GKSL-main}
		\widetilde{\cL} = \sum_{X\in\cV^{(r)}}\widetilde{\cL}_X\qquad\text{with}\qquad\widetilde{\cL}_X=\cH[H_X] + \sum_{j=1}^{\nu(X)}\cL[L_{X,j}] \, .
	\end{equation}
	Here, $\cH[H](\cdot)\coloneqq -i[H,\cdot]$ and $\cL[L](\cdot)\coloneqq L(\cdot)L^\dagger-\frac12\{L^\dagger L,\cdot\}$ are the Hamiltonian and dissipative parts of the generator. The family $\cV^{(r)}$ contains local supports, or hyperedges, of diameter at most $r$ and cardinality at most $r_{\mathrm{exp}}$; $\nu(X)\leq \nu\in\N$ for all $X\in\cV^{(r)}$ is a bound on the number of Lindblad operators per support; $\gamma_i=|\{X\in \cV^{(r)}:i\in X\}|\leq \gamma$ bounds the support degree; and the local Hamiltonian operators $H_X$ and Lindblad operators $L_{X,j}$ are given by polynomials in annihilation and creation operators $a_i,\ad_i$ for $i\in X$ with maximal degree $d\in\N$. We write $N_i\coloneqq \ad_i a_i$ for the local number operator of mode $i$.
	Our main technical contributions apply to GKSL evolutions satisfying the photon-number moment assumptions: stability and propagation, defined precisely in the Methods under \textbf{Moment propagation and locality}. Informally, stability means that finite local photon-number moments remain controlled in time, while propagation means that their growth at one mode is governed only by nearby modes up to controlled source terms. These assumptions capture natural finite-energy scaling conditions for infinite-dimensional systems, are verified for Bose--Hubbard and quadratic dynamics without engineered dissipation, and provide a class-level entry point for a broad family of bosonic open-system dynamics.

	\subsection*{Weak locality beyond Bose--Hubbard}
		For finite-dimensional many-body systems, Lieb--Robinson bounds \cite{LiebRobinson.1972,Nachtergaele.2011lrb} and Trotter error bounds \cite{Haah.2021,Childs.2022,Childs.2019} are standard algorithmic tools.
		For suitably well-behaved local generators (Hamiltonian or dissipative) Lieb--Robinson bounds confine information propagation to an effective light cone. Local observable expectation values can then be approximated by truncating the evolution to a neighbourhood of the observable's support \cite{Barthel.2012,Kliesch.2014}, enabling localization for simulation \cite{Kuwahara.2024digital} and learning \cite{Stilckfranca.2024}.
		Trotter bounds give the complementary discretization step: they replace continuous-time evolution generated by a sum of local terms with a circuit of local operations \cite{Lloyd.1996,Haah.2021,Childs.2022,Low.2019}.

		While well-established for many-qubit systems, extending Lieb--Robinson and Trotter bounds to infinite-dimensional bosonic systems with multiple modes is challenging. In fact, such extensions are not possible in full generality \cite{Eisert.2009,Eisert.2020,Becker.2025}. 
		Existing positive results therefore rely on additional structure, most notably for Bose--Hubbard-type evolutions \cite{Kuwahara.2021,Kuwahara.2024digital,Kuwahara.2024Lemm,Faupin.2022}. This leaves many relevant bosonic Hamiltonians and open-system generators outside the available theory, including non-Gaussian polynomial interactions used in bosonic encodings \cite{Michael.2016new,Albert.2020} and Kerr-cat systems, where quartic Hamiltonian terms are combined with two-photon driving and dissipation \cite{Mirrahimi.2014dynamically,Guillaud.2023}. Since universal bosonic quantum computation requires non-quadratic resources \cite{Lloyd.1999,Gottesman.2001encoding}, rigorous locality and discretization guarantees for unbounded nonlinear dynamics are a basic missing ingredient.

		Our first contribution is a weak bosonic Lieb--Robinson bound that is model-independent within the moment-propagation class and covers the mentioned examples as concrete instances.

\begin{figure}[H]
			\centering
			\includegraphics[width=0.92\textwidth]{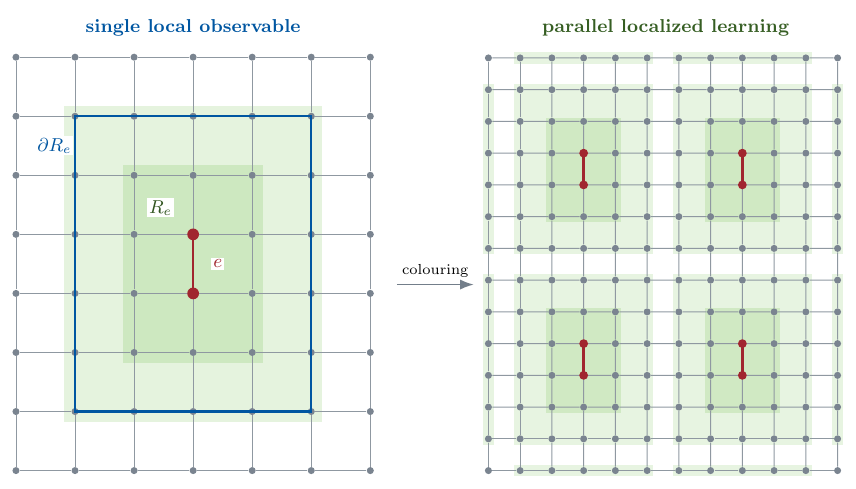}
			\caption{Weak Lieb--Robinson localization. A local observable supported on a set $T$ (shown as an edge $e$ for learning) is approximated by evolving only inside a neighbourhood $R_T$, with an error controlled by the distance from $T$ to the boundary and by the finite-energy cutoff. For learning, well-separated regions can be assigned to the same colour class and probed in parallel.}
			\label{fig:lrb-regions-main}
		\end{figure}

		\begin{result}[Weak bosonic Lieb--Robinson bound]\label{thm:LRB-informal}
			Assume that $\widetilde{\cL}$ is a local, finitely connected GKSL generator that satisfies these photon-number moment assumptions. For any bounded local observable $O_T$ and any finite-moment input state, the expectation value of $O_T$ under the global bosonic evolution can be compared with the expectation value obtained by evolving only on a neighbourhood $R$ of $T$ with a local Fock cutoff $M$, with an explicit error controlled by the input moments, $M$, and the distance from $T$ to the boundary of $R$.

			In particular, let $\ket{\alpha}$ be a coherent product state with bounded amplitudes, and let $O_T$ be supported on $T\subset V$. For fixed time $t=\mathcal O(1)$ and every $\epsilon\in(0,1)$, one can choose a neighbourhood $R$ of $T$ and a local Fock cutoff $M$, both of polylogarithmic size in $\epsilon^{-1}$, such that, with $P_R^{(M)}$ the cutoff projector, $\ket{\psi_\alpha}=P_R^{(M)}\ket{\alpha}$, and $\widetilde{\cL}^{(M)}_R$ the restricted cutoff generator,
		    \begin{align*}
				\Big|\tr\big[O_T e^{t\widetilde{\cL}}(\ketbra{\alpha}{\alpha})\big]-\tr\big[O_T e^{t\widetilde{\cL}^{(M)}_R}(\ketbra{\psi_\alpha}{\psi_\alpha})\big]\Big|\le \epsilon\|O_T\|_\infty\,.
    		\end{align*}
		\end{result}
		Result \ref{thm:LRB-informal} is the bosonic, weak analogue of the Lieb--Robinson bounds used in learning and simulation on finite-dimensional spaces, but its proof does not depend on the algebraic special structure of the Bose--Hubbard model. Instead, photon-number moment propagation justifies a finite-energy Galerkin approximation of the local generator; the projected dynamics is finite dimensional, admits standard locality estimates, and can then be compared back to the original evolution with explicit moment-dependent errors. This produces a reusable toolbox for localization, cutoff approximation, local-observable simulation, product-formula analysis and perturbative estimates. It recovers known locality guarantees for the Bose--Hubbard model \cite{Kuwahara.2021,Kuwahara.2024digital}, covers quadratic bosonic Hamiltonians, and extends to the regularized GKSL dynamics of bounded-degree polynomial Hamiltonians with suitable known multi-photon dissipation. The localization geometry used later for learning is shown in \Cref{fig:lrb-regions-main}.

		Having established an infinite-dimensional bosonic counterpart of the weak Lieb--Robinson bounds, we turn our attention to Trotter product formulas. While of independent interest as a tool in quantum simulation, Trotter formulas become particularly important for dissipatively regularized Hamiltonians, because they give a realizable route to implementing the regularization. We prove:

		\begin{result}[Bosonic Trotter approximation]\label{thm:trotter-product-informal}
			Consider a polynomial two-body Hamiltonian together with known $p$-photon dissipation, with $p\geq4d+2$ for Hamiltonian degree $d$ and bounded dissipation shifts $|\alpha_i|=\mathcal O(1)$. Assume that the Hamiltonian and dissipative parts generate quantum Markov semigroups on the required finite-moment domain. Then, for $n\in\N$ product-formula steps and every sufficiently regular input state $\rho$,
    		\begin{align*}
				\bigl\|\left(e^{\frac{t}{n}\cH^{(d)}}e^{\frac{t}{n}\cL^{(\alpha,p)}}\right)^n(\rho)-e^{t(\cH^{(d)}+\cL^{(\alpha,p)})}(\rho)\bigr\|_1
				\leq \frac{C_1(t,\rho)m}{\sqrt n}+\frac{C_2(t,\rho)m^2}{n}\,,
    		\end{align*} 
		where $m=|V|$ and the constants are explicit in the Methods in terms of the coefficient bounds, evolution time and photon-number moments of $\rho$.
		\end{result}

\begin{figure}[H]
			\centering
			\includegraphics[width=0.95\textwidth]{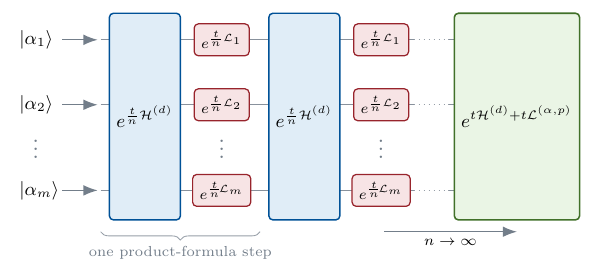}
			\caption{Dissipative Trotterization. A product-formula step alternates Hamiltonian evolution generated by $\cH^{(d)}$ with on-site $p$-photon dissipation $\cL_j^{(\alpha,p)}$. Repeating the step approximates the regularized evolution $e^{t(\cH^{(d)}+\cL^{(\alpha,p)})}$, which is the dynamics entering the locality and learning guarantees for general polynomial Hamiltonians.}
			\label{fig:trotter-scheme-main}
		\end{figure}

		This product formula has two roles. It is a discretization primitive for the regularized bosonic evolution. It also gives a physically explicit way of interleaving Hamiltonian evolution with engineered dissipation to realize the regularized dynamics used in the learning and locality results. This alternating implementation is illustrated in \Cref{fig:trotter-scheme-main}.

	\subsection*{Simulation from finite-energy control}
\begin{figure}[H]
			\centering
			\includegraphics[width=\textwidth]{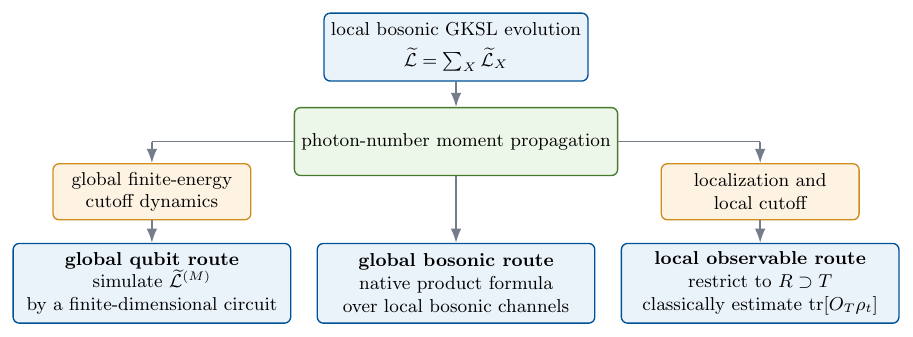}
			\caption{Simulation routes from moment-controlled finite-energy approximation. Photon-number moment propagation gives finite-energy control of a local bosonic GKSL evolution. This yields three complementary outputs: a global qubit route via cutoff dynamics and a finite-dimensional circuit, a global native bosonic route via a product formula over local bosonic channels, and a local observable route in which the dynamics is localized around the measured observable before the expectation value is approximated on a finite-dimensional local region.}
			\label{fig:simulation-routes-main}
		\end{figure}

		We prove a finite-energy dissipative Church--Turing-type simulation guarantee, expressed through three complementary routes out of the same moment-controlled finite-energy toolbox. The first is a global qubit route: after a finite-energy cutoff, the output state of a dissipative bosonic evolution is approximated by a finite-dimensional circuit. The second is a global native bosonic route: the same global evolution is approximated by a concatenation of local bosonic channels. The third is a local observable route: expectation values are localized around the observable and then approximated using a finite-dimensional local evolution.
		We summarize our simulation guarantees as follows:

		\begin{result}[Simulating bosonic evolutions] \label{thm:informal-simulation}
			Assume that $\widetilde{\cL}$ is a local, finitely connected GKSL generator that satisfies the photon-number moment stability and photon-number propagation assumptions. Let $\rho$ satisfy the moment assumptions required by the respective statement used in each item below. Then, the following approximation guarantees hold:
			\begin{enumerate}
        		\item \textbf{Global qubit route.} Up to times $t=\mathcal{O}(\log |V|)$, or more generally whenever the exponential-in-time factor in the cutoff estimate remains polynomial, the state $e^{t\widetilde{\cL}}(\rho)$ can be approximated by a multi-qubit circuit with a number of gates polynomial in the system size $|V|$ and the inverse target precision.
				\item \textbf{Global bosonic route.} If the local generators separately define quantum Markov semigroups, then for fixed simulation time $t$, the state $e^{t\widetilde{\cL}}(\rho)$ can be approximated by a concatenation of local bosonic channels generated by the terms $\{\widetilde{\cL}_X\}_{X\in\cV^{(r)}}$.
				\item \textbf{Local observable route.} On fixed-dimensional lattices, for fixed time $t=\mathcal{O}(1)$, expectation values of local observables measured on the state $e^{t\widetilde{\cL}}(\rho)$ can be reduced to a finite-dimensional simulation on a neighbourhood of the observable whose size is independent of the total number of modes.
    		\end{enumerate}
		\end{result}
		These guarantees address complementary simulation targets, but they are all consequences of the same finite-energy approximation toolbox. The first gives a route from oscillator dynamics to a finite-dimensional qubit processor after an energy cutoff. The second is native to bosonic hardware and quantifies the overhead for replacing a many-mode evolution by a sequence of local evolutions. The third is the local-observable regime most directly tied to Lieb--Robinson bounds and is independent of the total number of modes at fixed time. For cat-code oscillator arrays, the result applies to the full known GKSL model, including stabilizing multi-photon dissipation and Hamiltonian nonlinearities; if the physical goal is to simulate the undamped Hamiltonian alone, adding dissipation instead defines a different, regularized dynamics. The three simulation routes are collected in \Cref{fig:simulation-routes-main}.

	\subsection*{Learning with qubit-like scaling}
		Hamiltonian learning is the main application of our locality framework. The aim is to recover the coefficients of a known graph-local bounded-degree polynomial Hamiltonian from coherent-state probes, evolution under the unknown dynamics together with known dissipative channels when present, and local heterodyne readout. A short-time finite-difference scheme already shows why this experimental primitive is informative, but its Taylor remainder forces a vanishing time step. The locality-assisted protocol below keeps the same simple data acquisition while using weak Lieb--Robinson bounds to recover qubit-like scaling.

		\begin{figure}[H]
			\centering
			\includegraphics[width=\textwidth]{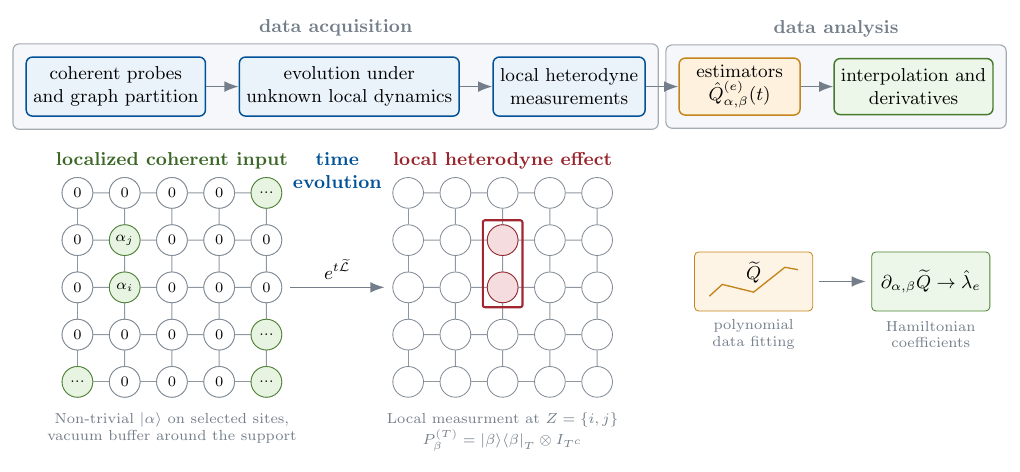}
			\caption{Hamiltonian learning pipeline and local readout structure. Top: coherent states are prepared on well-separated local supports, evolved for chosen short or constant times, and measured by local heterodyne detection. The empirical estimators are assembled into polynomial data; interpolation and differentiation then recover the local Hamiltonian coefficients. Bottom: the local coherent preparation and heterodyne effect used inside the data-acquisition step. The input is a coherent product state with non-vacuum amplitudes on selected modes and vacuum buffers around the active support, while the measured effect is local, $P_\beta^{(T)}=\ketbra{\beta}{\beta}_T\otimes I_{T^c}$. The lower-right boxes indicate the polynomial data and recovered Hamiltonian coefficients.}
			\label{fig:learning-pipeline-main}
		\end{figure}

		Our finite-difference baseline serves as a proof-of-principle demonstration. For polynomial two-body Hamiltonians, coherent probes and local heterodyne measurements turn infinitesimal changes of heterodyne probabilities into polynomial data whose derivatives determine the Hamiltonian coefficients. In finite-dimensional systems the Taylor error is controlled by a local norm bound; in continuous variables the Taylor remainder involves unbounded double commutators and photon-number tails, controlled here by moment propagation. The scheme is rigorous, and for the Bose--Hubbard model needs no added dissipation, but it requires a short evolution time $t=\mathcal O(\varepsilon)$ to keep the Taylor bias below the target accuracy while the empirical estimator is divided by $t$. Consequently, estimating a polynomial two-body Hamiltonian to accuracy $\varepsilon$ gives the sample complexity $N_{\mathrm{samp}}=\mathcal{O}(\varepsilon^{-4}\ln(|V|+|E|))$ and total evolution time $T_{\mathrm{evo}}=\mathcal{O}(\varepsilon^{-3}\ln(|V|+|E|))$. Thus the method verifies that coherent probes and heterodyne readout contain the desired polynomial information, while also exposing the short-time penalty that motivates the locality-assisted scheme.

		The main learning result removes this finite-difference penalty. Ref.~\cite{Stilckfranca.2024} showed that weak Lieb--Robinson bounds improve both the sample complexity and the time resolution for finite-dimensional Hamiltonian learning. In our bosonic setting, we give an improved scheme that first localizes the dynamics to a finite-energy neighbourhood, then fits the measured probabilities as a multivariate polynomial in coherent amplitudes, heterodyne outcomes and time; derivatives of this interpolant recover the local coefficients, compare \Cref{fig:learning-pipeline-main}. Our bosonic weak Lieb--Robinson bounds lift these improvements to continuous variables:

		\begin{result}[Locality-assisted Hamiltonian learning]\label{thm:main-improved-learning}
			Consider a known bounded-degree polynomial Hamiltonian ansatz local on a bounded-growth interaction graph, including fixed-dimensional lattices. This includes the Bose--Hubbard model and regularized polynomial two-body Hamiltonians, and it also covers other known dissipative backgrounds whenever the combined GKSL generator satisfies the stated moment, locality and projected-generator boundedness assumptions. Using coherent-state preparation, the known dissipative channel when present, and heterodyne measurements, we can learn the Hamiltonian coefficients to accuracy $\varepsilon$ and failure probability $\delta$ with
			\begin{equation*}
				N_{\mathrm{samp}}=\widetilde{\mathcal{O}}\!\left(\varepsilon^{-2}\ln\!\left(\frac{|V|+|E|}{\delta}\right)\right),
				\qquad
				T_{\mathrm{evo}} = \widetilde{\mathcal{O}}\!\left(\varepsilon^{-2}\ln\!\left(\frac{|V|+|E|}{\delta}\right)\right).
			\end{equation*}
			Here each experiment uses evolution time $t=\widetilde{\mathcal{O}}(1)$.
		\end{result}

		In this result, our weak Lieb--Robinson bounds play a crucial role: they extend the improved guarantees of Ref.~\cite{Stilckfranca.2024} to infinite-dimensional local Hilbert spaces. The main point is that in the weak-access model of coherent probes---unknown bosonic evolution, known dissipative channels when present, and heterodyne readout---continuous-variable systems recover the locality-assisted finite-dimensional benchmark: standard-quantum-limit total evolution time, logarithmic dependence on the number of local terms, and no system-size-dependent data-acquisition time grid. The time-resolution statement concerns the learning protocol itself; if a non-native regularized generator is implemented by globally trotterizing Hamiltonian and dissipative pieces, the separate implementation overhead is the product-formula cost quantified above. For bounded-degree Bose--Hubbard systems, no added dissipation is required and the logarithmic dependence on $|V|+|E|$ matches the best finite-dimensional locality-based rates. For bosonic cat-code hardware, the same protocol can characterize the Hamiltonian sector of a stabilized oscillator array, including unwanted detunings, Kerr shifts and inter-mode couplings, while the dissipative stabilization itself is treated as calibrated and known.

		This is the sense in which the theorem reaches the qubit benchmark: it matches the locality-assisted finite-dimensional scaling available under the same weak-access, no-control assumptions. Heisenberg-limited protocols address a different access model. Finite-dimensional qubit protocols can reach total evolution time $\mathcal O(\varepsilon^{-1})$ by using additional coherent control during the evolution \cite{Huang.2023learning,Dutkiewicz.2023,Brahmachari.2026heisenberg}, while Ref.~\cite{Dutkiewicz.2023} shows that, without such control, broad classes of many-body Hamiltonian-learning problems are standard-quantum-limited. Recent bosonic Heisenberg-limited protocols similarly use stronger resources: Ref.~\cite{Li.2024heisenberg} interleaves the unknown Bose--Hubbard evolution with frequent random Gaussian controls; its global effective-dynamics error estimate scales as $\mathcal O(N^2t^2/r)$ for $r=t/\tau$ inserted controls, so at Heisenberg times $t=\Theta(\varepsilon^{-1})$ the required interleaving resolution scales as $\tau=\widetilde{\mathcal O}(\varepsilon/N^2)$. Ref.~\cite{Moebus.2025heisenberg} obtains Heisenberg scaling for more general bosonic Hamiltonians through engineered dissipation whose required strength grows polynomially with the number of modes. Thus these works are important comparators, but they do not yet recover the qubit locality-assisted resource profile on the bosonic side: Heisenberg precision is obtained through control resolution or dissipation strength with system-size dependence, rather than through a weak Lieb--Robinson reduction. Our focus is complementary: a weak-access bosonic theorem attaining the locality-assisted qubit scaling that is optimal for broad no-control finite-dimensional settings, up to polylogarithmic factors.

\section*{Physical scope and outlook}
	The finite-energy viewpoint turns moment propagation into a general-purpose approximation principle: unbounded local generators can be replaced, on finite-energy states and for local observables, by controlled finite-energy Galerkin approximations. Once this control is available, either intrinsically as in the Bose--Hubbard model or through adding engineered dissipation, bosonic systems recover enough locality to support efficient learning and simulation guarantees.

	For simulation, the results go beyond existing Bose--Hubbard-centred locality and simulation guarantees \cite{Kuwahara.2021,Kuwahara.2024digital,Kuwahara.2024Lemm,Faupin.2022} by treating a moment-controlled class that includes quadratic dynamics, regularized polynomial bosonic GKSL generators and other local open-system generators satisfying the same assumptions. Quadratic, or Gaussian, dynamics provide a baseline class where the assumptions hold without added dissipation; higher-degree polynomial interactions require either intrinsic stability or known dissipative control strong enough to regularize photon-number growth. For learning, the results show that the sample-complexity and system-size advantages of finite-dimensional locality-based protocols survive in the continuous-variable setting, up to polylogarithmic factors. This is particularly clean for the Bose--Hubbard model, where no added dissipation is needed and the protocol learns all local interaction parameters with total evolution time $\widetilde{\mathcal O}(\varepsilon^{-2}\log(|V|+|E|))$.

	The most immediate experimental target is therefore Hamiltonian characterization in oscillator arrays. In cold-atom or circuit-QED realizations of Bose--Hubbard physics, the protocol gives a rigorous route to learning hopping, detuning and interaction coefficients from coherent probes and heterodyne-type measurements. In bosonic cat-code devices \cite{Mirrahimi.2014dynamically,Ofek.2016extending,Guillaud.2019repetition,Guillaud.2023}, the stabilizing multi-photon dissipation is part of the calibrated device model, while the Hamiltonian sector contains coherent errors and engineered couplings that must be characterized. The same proof structure also accommodates fixed known loss channels, such as local $p$-photon loss, whenever the resulting GKSL dynamics satisfies the stated moment and locality assumptions. Our results apply directly to these known-dissipation settings; learning unknown dissipative rates is a different task and is not claimed here.

	Several natural questions remain. Our learning result prioritizes locality and system-size scaling in a weak-access model rather than Heisenberg scaling in $\varepsilon$, consistent with finite-dimensional lower bounds showing that broad no-control settings are standard-quantum-limited \cite{Dutkiewicz.2023}. Existing bosonic Heisenberg-limited protocols achieve the stronger precision dependence by adding control resources, such as frequent Gaussian controls or engineered dissipation, which in the cited bosonic settings introduce system-size-dependent overheads. Likewise, the dissipative regularization used for general polynomial Hamiltonians changes the physical dynamics being simulated or learned; it is a tool for stabilized or deliberately regularized devices, not an approximation to arbitrary closed-system dynamics. These limitations point to a broader programme: developing a full algorithmic theory of continuous-variable many-body systems in which energy, locality and control are treated on equal footing. The present work supplies one part of that programme: a locality and finite-energy approximation framework, model-independent within the moment-controlled class, that can be reused beyond the specific learning and simulation protocols developed here.
\section*{Methods}
	\subsection*{Moment propagation and locality}
		The proof of the weak Lieb--Robinson bound uses two assumptions. The first is structural. Given a finite metric space $(V,\dist)$, the Hamiltonian and dissipative terms are supported on sets $X\in\cV^{(r)}$ of diameter at most $r$, with at most $\gamma$ such sets meeting any given vertex. Thus the support geometry has bounded degree. Each Hamiltonian term $H_X$ is a polynomial of degree at most $d$ in creation and annihilation operators, and each jump operator $L_{X,j}$ has degree at most $d/2$. The latter convention is only a bookkeeping choice: equivalently, one may introduce separate Hamiltonian and jump degrees and choose the common parameter $d$ large enough. Together, these terms define a local GKSL generator
		\begin{equation*}
    		\widetilde{\cL} = \sum_{X \in \cV^{(r)}} \cH[H_X] + \sum_{j=1}^{\nu(X)} \cL[L_{X,j}] \,.
    	\end{equation*}
		Here,
		\begin{equation*}
    		\cL[L] \coloneqq L(\cdot) L^\dagger - \frac{1}{2} \, \{L^\dagger L, \, \cdot\} \qquad \text{and} \qquad \cH[H] \coloneqq -i[H, \, \cdot \,]\,.
		\end{equation*}
		The second assumption is a quantitative photon-number propagation estimate. For a fixed integer $p\geq1$ and a moment order $k$, let $f$ be defined by $f(0)=f(1)=\dots=f(p-1)=0$ and $f(m+p)=f(m)+k m^{k-1}$. In regularized polynomial models, the same integer $p$ is the order of the photon-loss operator introduced below. We assume that there are constants $\Gamma'$, $\Gamma''$, $\Gamma'''$ and $\xi$ such that, for all sufficiently large $k$,
		\begin{equation*}
    		\frac{d}{dt} \tr[\rho_t f(N_i)] \leq k \Gamma' \tr[\rho_t f(N_i)] + k \Gamma'' \sum_{1 \leq \dist(i, j) \leq r} \tr[\rho_t f(N_j)] + k^{\xi k} \Gamma''',
    	\end{equation*}
		where $N_i$ is the number operator for the $i$-th mode and $\rho_t=e^{t \widetilde{\cL}}(\rho)$. This photon-number propagation bound, together with locality, is the class-level input used to prove the bosonic weak Lieb--Robinson bound summarized in Result \ref{thm:LRB-informal}.

		The first step of the proof is to prove a moment propagation bound (see Lemma \ref{lem:moment-prop-bound} in the Supplementary Material) of the following form: for all $t \geq 0$ and all sufficiently large $k$
		\begin{align*}
			\tr[\rho_t (N_j+I)^{k}]\le p4^{p+1} e^{k \Gamma t} \max_{i \in V} \tr[\rho (N_i+I)^{k}] + \Upsilon_k(t)\,,
		\end{align*}
		where $\Gamma = \Gamma' + \gamma \Gamma''$ and $\Upsilon_k(t) = k^{\mathcal{O}(k)}$. This shows that finite local moments remain finite under the evolution, with constants that are explicit enough for the subsequent cutoff estimates.

		This moment propagation bound is crucial, because it allows us to introduce an energy cutoff and approximate the semigroup on finite-moment states by a Galerkin-type semigroup with bounded generator. Related energy-controlled perturbation and admissibility techniques for bosonic quantum Markov semigroups appear in Ref.~\cite{Gondolf.2024}. For such projected generators we can apply known Lieb--Robinson bounds \cite{Nachtergaele.2011lrb}; the moment estimate then controls the error made by removing the cutoff. The structural assumption bounds the Hamiltonian terms and jump operators composing the projected generator. This cutoff-and-compare argument is the main technical device behind the weak Lieb--Robinson bound, the simulation estimates and the perturbative/product-formula bounds.

		Applying the weak Lieb--Robinson bound therefore reduces to verifying the structural and moment assumptions. For the Bose--Hubbard Hamiltonian and for quadratic Hamiltonians, they hold without added dissipation. For general bounded-degree Hamiltonians on a graph $(V,E)$, we consider
		\begin{align} \label{eq:methods-bounded-deg-H}
			H^{(d)} = \sum_{e\in E}H_e\qquad\text{with}\qquad H_e \coloneqq \sum_{k,\ell,k',\ell'=0}^d\,\lambda^{(e)}_{k\ell k'\ell'}\,(a_i^\dagger)^k\,a_i^\ell\,(a_j^\dagger)^{k'}\,a_j^{\ell'}\,,
		\end{align}
		for some complex coefficients $\lambda^{(e)}_{k\ell k'\ell'}=\overline{\lambda}^{(e)}_{\ell k \ell' k'}$ with $|\lambda^{(e)}_{k\ell k'\ell'}|\le \Lambda$. Here, $a_i^\dagger$ and $a_i$ are the creation and annihilation operators acting on the $i$-th mode, respectively. Note that we can fix $\lambda_{0000}=0$ due to gauge freedom, and we denote $\cH^{(d)}\coloneqq -i[H^{(d)},\cdot]$. These Hamiltonians in general need to be regularized in order to fulfill the second assumption. Given $\alpha \in \mathbb{C}^m$ and $p \in \mathbb{N}$, a dissipative regularizer is defined via the $p$-photon dissipation by 
		\begin{align}\label{eq:methods-p-photon-dissipation}
			\cL^{(\alpha,p)} \coloneqq \sum_{j \in V} \cL[L_j^{(\alpha_j,p)}], \qquad \text{where} \quad L_j^{(\alpha_j,p)} \coloneqq a_j^p - \alpha_j^p I \,.
		\end{align}
		The regularized generator is
		\begin{equation*}
			\widetilde{\cL}^{(\alpha)} = \cH^{(d)} + \cL^{(\alpha,p)}\,.
		\end{equation*}
		The shifted dissipator is the version used in the polynomial learning proof because it is matched to the coherent amplitudes used as inputs. The same locality-assisted proof extends to other known dissipative backgrounds, for instance $\cH^{(d)}+\sum_{j\in V}\cL[a_j^p]$, provided the combined GKSL generator satisfies the same moment propagation and projected-generator boundedness conditions.

	\subsection*{Trotter-Kato product formula}
		Hamiltonians of the form in Equation \eqref{eq:methods-bounded-deg-H} require regularization before the weak Lieb--Robinson bound applies. If the available primitive is Hamiltonian time evolution, we must justify the insertion of this regularizing dissipation. Trotter-Kato product formulas provide this step: alternating Hamiltonian evolution and dissipation approximates the regularized evolution. Result \ref{thm:trotter-product-informal} is supported by Proposition \ref{thm:trotter} in the Supplementary Material. There, we show that for sufficiently regular initial states $\rho$, namely states with finite local moments of order $4p$ with $p \geq 4d+2$, one can control the error after $n\in\N$ Trotter steps:
	   \begin{equation*}
            \bigl\|\left(e^{\frac{t}{n}\cH^{(d)}}e^{\frac{t}{n}\cL^{(\alpha,p)}}\right)^n(\rho)-e^{t(\cH^{(d)}+\cL^{(\alpha,p)})}(\rho)\bigr\|_1\\
            \leq\frac{C_1 m}{\sqrt{n}}+\frac{C_2 m^2}{n}\,,
        \end{equation*} 
		where $m = |V|$ and $C_1$, $C_2$ are constants when $t$, $\max_i|\alpha_i|$, $d$ and $p$ are treated as order-one parameters. To prove this statement, we make use of the Chernoff $\sqrt{n}$-Lemma \cite{Chernoff.1968}, the moment propagation bound, and bounds on concatenations of $\cH^{(d)}$ and $\cL^{(\alpha,p)}$ in different combinations; related strong-topology Trotter and Zeno product-formula bounds with bosonic applications are developed in Ref.~\cite{Moebus.2024}. The concatenations can be controlled by the number operator, because both $\cH^{(d)}$ and $\cL^{(\alpha,p)}$ are relatively bounded with respect to $\sum_{j \in V} (N_j+I)^q$ for sufficiently large $q$. The point needed for the present work is a state-dependent trace-norm estimate with explicit dependence on the relevant photon-number moments and the number of modes.

	\subsection*{Simulation algorithms}
		Result \ref{thm:informal-simulation} contains three simulation guarantees, each using a different consequence of the finite-energy approximation framework. The first concerns simulating a bosonic evolution on a multi-qubit system and appears as Theorem \ref{thm:pauli-approx} in the Supplementary Material. As in the proof of the weak Lieb--Robinson bound, we introduce an energy cutoff. Lemma \ref{lem:approximation} shows that a sufficiently regular bosonic evolution satisfying the structural and moment assumptions can be approximated by a finite-dimensional evolution, while Corollary \ref{cor:state-approximation} approximates sufficiently regular states by cutoff states. The finite-dimensional simulation algorithm of Ref.~\cite{Cleve.2017} can then be applied.

		The second simulation result concerns simulating bosonic time evolution by local bosonic time evolutions; that is, we want to control
		\begin{equation*}
    		\biggl\|e^{t\widetilde{\cL}}(\rho)-\Bigl(\prod_{X\in\cV^{(r)}}e^{\frac{t}{n}\widetilde{\cL}_X}\Bigr)^n(\rho)\biggr\|_{1},
		\end{equation*}
		where $\widetilde{\cL}_X$ are the local generators composing $\widetilde{\cL}$. It can be found as Proposition \ref{prop:church-turing-infinite} in the Supplementary Material. To prove this statement, we make use of the fact that generators satisfying the first and second assumptions can be controlled by powers of the local number operators as in the proof of the Trotter-Kato product formula and also of the moment propagation bound.

		Unlike the first two simulation results, which are global, the last one is local. It shows that the expectation value of a local observable can be approximated using a local time evolution if the original time evolution satisfies the first and second assumptions, which can subsequently be approximated using a Lie-Trotter product formula. The important aspect is that the approximation does not depend on the number of modes. The proof makes use of our new weak Lieb--Robinson bounds for the first approximation. Since the approximating local time evolution is bounded, we can subsequently make use of the standard Lie-Trotter product formula. The statement of the result can be found in Proposition \ref{prop:simulation-local-observables} of the Supplementary Material.

	\subsection*{Locality-assisted learning}
		The main learning theorem is Result \ref{thm:main-improved-learning}. Its purpose is to avoid the short-time penalty of finite differences. In finite-dimensional systems, this is achieved by replacing the global Hamiltonian with its restriction to a Lieb--Robinson neighbourhood and fitting constant-time Taylor data \cite{Stilckfranca.2024}. Our weak bosonic Lieb--Robinson bound supplies the same reduction after a finite-energy projection.

		For each edge or vertex $e\in E\cup V$, let $R_e$ be the polylogarithmic-radius neighbourhood supplied by Result \ref{thm:LRB-informal}, and let $P=P_{R_e}^{(M)}$ be the corresponding Fock cutoff. The experiment uses coherent states supported on well-separated regions, evolves for a set of constant times, and performs local heterodyne measurements. Independently of the amplitude and measurement grids, the supports are coloured so that all probes in the same colour class have disjoint neighbourhoods $R_e$, allowing parallel data acquisition. The main routine is:

		\begin{algorithm}[H]
			\caption{Locality-assisted bosonic Hamiltonian learning}\label{alg:main-locality-assisted-learning}
			\KwIn{Accuracy parameters, time grid $\cN_t$, coherent-state grid $\cN_{\cS}$, measurement grid $\cN_{\cM}$, and sample number $N_{\mathrm{samp}}$.}
			\KwOut{Estimates of the coefficients in the known graph-local Hamiltonian ansatz on $V\cup E$.}
			Choose a polylogarithmic-radius neighbourhood $R_e$ and cutoff $M$ for each $e\in V\cup E$ using the weak Lieb--Robinson bound.\;
			Partition the supports $e\in V\cup E$ into colour classes whose neighbourhoods $R_e$ are disjoint.\;
			For each colour class, each coherent amplitude $\alpha\in\cN_{\cS}$, and each time $t\in\cN_t$, prepare coherent states on the active supports and vacuum elsewhere.\;
			Evolve under the unknown Hamiltonian together with the known dissipative channel when required, and perform local heterodyne measurements.\;
			Form the empirical functions $\hat L^{(e)}_{\alpha,\beta}(t)$ from the measurement outcomes.\;
			Use outlier-robust multivariate polynomial interpolation in $(\alpha,\beta,t)$ and differentiate at $t=0$ to recover local Fock-basis matrix elements, then convert these matrix elements into the coefficients of $H_e$.\;
		\end{algorithm}

		The full version of this routine is Algorithm~\ref{protocollearningLRprojected} in the Supplementary Material. The estimator has the form
		\begin{equation*}
			\hat{L}^{(e,N_{\mathrm{samp}})}_{(\alpha,\beta)}(t)
			=\frac{1}{N_{\mathrm{samp}}}\sum_{i=1}^{N_{\mathrm{samp}}}
			e^{|\beta^{(i)}_e|^2+|\alpha|^2}1_{R_\beta}(\beta^{(i)}_e),
		\end{equation*}
		where $\beta_e^{(i)}$ is the local heterodyne outcome on $e$ and $R_\beta\subset\mathbb R^{2|e|}$ is the measurement cell associated with the grid point $\beta$. Thus $\hat L$ estimates a weighted cell integral of the heterodyne probability density, with separate nets for vertex and edge terms. Hoeffding concentration controls the statistical error with $N_{\mathrm{samp}}=\widetilde{\mathcal O}(\varepsilon^{-2}\log((|V|+|E|)/\delta))$ samples, because the estimator is not divided by a vanishing time step. The weak Lieb--Robinson bound controls the replacement of the global evolution by the projected local evolution on $R_e$, and a Taylor approximation of the projected local dynamics turns the constant-time data into a multivariate polynomial in $(\alpha,\beta,t)$. We sample this polynomial on shifted Chebyshev grids and use the outlier-robust multivariate interpolation primitive from Ref.~\cite{Arora.2024}, as stated in the proof in the Supplementary Material. After projection, coherent-state amplitudes and heterodyne outcomes no longer act as exact eigenvalues of the creation and annihilation operators; the proof therefore relates higher-order derivatives of the interpolated functions to Fock-basis matrix elements of $H_e$ and recovers the polynomial coefficients iteratively. This is the mechanism behind the $\widetilde{\mathcal O}(\varepsilon^{-2}\log(|V|+|E|))$ sample and total-evolution-time scaling in Result \ref{thm:main-improved-learning}.

	\subsection*{Finite-difference baseline}
		The finite-difference protocol is a useful baseline and a diagnostic stepping stone, but it is not the rate-optimal scheme. It uses the coherent-state identity
		\begin{align}\label{aonalpha}
			a_i|\alpha\rangle=\alpha_i|\alpha\rangle\,,\qquad i\in\{1,...,m\},\,\alpha_i\in\mathbb{C}\,,
		\end{align}
		to convert local heterodyne probabilities into polynomial data. For bounded-degree Hamiltonians as in Equation \eqref{eq:methods-bounded-deg-H} and two-mode heterodyne effects $P_\beta^{(e)}=\pi^{-2}\ketbra{\beta}{\beta}_e\otimes I_{e^c}$, define
		\begin{equation*}
			\frac{\pi^2}{\abs{\braket{\alpha,\beta}}^2}\tr[P_\beta^{(e)}\widetilde{\cL}^{(\alpha)}(\ketbra{\alpha}{\alpha})]\eqqcolon g_e(\alpha, \beta),
		\end{equation*}
			where the coefficients of $g_e$ determine the Hamiltonian coefficients; the derivative identities are proved in \Cref{Lemmacoeffpoly}.

		\begin{algorithm}[H]
			\caption{Short-time finite-difference baseline}\label{alg:main-finite-difference-learning}
			\KwIn{Short time $t=\mathcal O(\varepsilon)$, amplitude grid $\cN_{\cS}$, measurement grid $\cN_{\cM}$, and sample number $N_{\mathrm{samp}}$.}
			\KwOut{Polynomial data used to estimate the local coefficients of each two-mode interaction.}
			Partition edges into constantly many well-separated colour classes.\;
			For each class and each $\alpha\in\cN_{\cS}$, prepare coherent states on the active edges with vacuum buffers around them.\;
			Evolve for time $t$ under the target dynamics, including the known dissipative channel when needed.\;
			Perform local heterodyne measurements and form the finite-difference estimator $\hat Q^{(e)}_{\alpha,\beta}(t)$.\;
			Interpolate the polynomial $Q^{(e)}_\alpha(\beta)$ from the grid values, using the same robust interpolation primitive when outliers are allowed, and differentiate the interpolant to estimate the coefficients of $H_e$.\;
		\end{algorithm}

		The full finite-difference routine is Algorithm~\ref{alg:poly-interaction-finite-diff} in the Supplementary Material. Its error analysis starts from
		\begin{equation} \label{eqboundmomentfinitediff}
			\big| \frac{1}{\pi^2}\,g_e(\alpha,\beta)|\langle{\alpha}|{\beta}\rangle|^2-t^{-1}\tr\big[(\rho_t-\rho)P_\beta^{(e)}\big]\big|\le \frac{t}{2}\max_{s\le t} \big|\tr\big[\rho_s(\widetilde{\cL}^{(\alpha)})^2(P_{\beta}^{(e)})\big]\big| \,,
		\end{equation}
		for $\rho_s=e^{s\widetilde{\cL}^{(\alpha)}}(\ketbra{\alpha}{\alpha})$. Moment propagation bounds the right-hand side by $\mathcal O(t)$. The estimator contains a factor $1/t$, so Hoeffding concentration gives
		\begin{equation*}
			N_{\operatorname{samp}}=\mathcal{O}\left(\frac{1}{(t\epsilon_{\operatorname{stat}})^2}\log\left(\frac{|E|\,|\cN_{\cS}|\,|\cN_{\cM}|}{\delta}\right)\right).
		\end{equation*}
		Choosing $t=\mathcal O(\varepsilon)$ and $\epsilon_{\operatorname{stat}}=\mathcal O(\varepsilon)$ leads to the baseline $\mathcal O(\varepsilon^{-4}\log(|V|+|E|))$ sample scaling discussed in the main text. Thus the finite-difference argument verifies that coherent probes and heterodyne measurements contain the desired polynomial information, while the locality-assisted protocol above is the scheme that removes the short-time penalty.

\clearpage
\thispagestyle{firstpage}

\begin{center}
	{\bfseries
	SUPPLEMENTARY INFORMATION FOR\\
	LEARNING AND SIMULATING BOSONIC SYSTEMS VIA\\
	FINITE-ENERGY LOCALITY\par}

	\vspace{1.5em}

	{\normalfont\normalsize
	\begin{tabular}{c}
		Tim M\"obus\textsuperscript{1,2}, Andreas Bluhm\textsuperscript{3}, Matthias C. Caro\textsuperscript{4,5},\\
		Albert H. Werner\textsuperscript{6}, Cambyse Rouz\'{e}\textsuperscript{1,2,7}
	\end{tabular}\par}

	\vspace{0.75em}

	\begin{minipage}{0.96\textwidth}
		\centering\footnotesize
		\textsuperscript{1}Department of Mathematics, Technical University of Munich, M\"unchen, Germany\\
		\textsuperscript{2}Munich Center for Quantum Science and Technology (MCQST), M\"unchen, Germany\\
		\textsuperscript{3}Univ.\ Grenoble Alpes, CNRS, Grenoble INP, LIG, 38000 Grenoble, France\\
		\textsuperscript{4}Department of Computer Science, University of Warwick, Coventry, UK\\
		\textsuperscript{5}Dahlem Center for Complex Quantum Systems, Freie Universit\"at Berlin, Berlin, Germany\\
		\textsuperscript{6}Department of Mathematical Sciences, University of Copenhagen, Universitetsparken 5, 2100 Copenhagen, Denmark\\
		\textsuperscript{7}Inria, T\'{e}l\'{e}com Paris -- LTCI, Institut Polytechnique de Paris, 91120 Palaiseau, France\\[0.5em]
		\texttt{moebustim@gmail.com, andreas.bluhm@univ-grenoble-alpes.fr, matthias.caro@warwick.ac.uk}\\
		\texttt{werner@math.ku.dk, rouzecambyse@gmail.com}
	\end{minipage}
\end{center}

\setcounter{tocdepth}{2}
\addtocontents{toc}{\protect\setcounter{tocdepth}{2}}
\setcounter{section}{0}
\setcounter{subsection}{0}
\setcounter{subsubsection}{0}
\setcounter{equation}{0}
\setcounter{figure}{0}
\setcounter{table}{0}
\setcounter{algocf}{0}
\setcounter{thm}{0}
\setcounter{assum}{0}
\setcounter{assumst}{0}
\setcounter{ex}{0}
\renewcommand{\thesection}{S\arabic{section}}
\renewcommand{\thesubsection}{\thesection.\arabic{subsection}}
\renewcommand{\thesubsubsection}{\thesubsection.\arabic{subsubsection}}
\renewcommand{\theequation}{S\arabic{equation}}
\renewcommand{\thefigure}{S\arabic{figure}}
\renewcommand{\thetable}{S\arabic{table}}
\renewcommand{\thealgocf}{S\arabic{algocf}}
\renewcommand{\thethm}{S\arabic{thm}}
\renewcommand{\theconj}{S\arabic{thm}}
\renewcommand{\thecor}{S\arabic{thm}}
\renewcommand{\thelem}{S\arabic{thm}}
\renewcommand{\theprop}{S\arabic{thm}}
\renewcommand{\thedefi}{S\arabic{thm}}
\renewcommand{\theassum}{S\arabic{assum}}
\renewcommand{\theassumst}{S\arabic{assumst}*}
\renewcommand{\theex}{S\arabic{ex}}
\renewcommand{\thermk}{S\arabic{ex}}

\def\theHsection{supplementary.\arabic{section}}
\def\theHsubsection{\theHsection.\arabic{subsection}}
\def\theHsubsubsection{\theHsubsection.\arabic{subsubsection}}
\def\theHequation{supplementary.\arabic{equation}}
\def\theHfigure{supplementary.\arabic{figure}}
\def\theHtable{supplementary.\arabic{table}}
\def\theHalgocf{supplementary.\arabic{algocf}}

\tableofcontents

\section{Preliminaries}\label{sec:prelim}
	\subsection{Operators and norms}\label{sec:prelim-op}
		Given a separable Hilbert space $\mathcal{H}$, we denote by $\mathcal{B}(\mathcal{H})$ the space of bounded linear operators on $\mathcal{H}$, and by $\mathcal{T}_p(\mathcal{H})$ the \emph{Schatten $p$-class}, which is the Banach subspace of $\mathcal{B}(\mathcal{H})$ formed by all bounded linear operators whose Schatten $p$-norm, defined as $\|X\|_{p} = \left(\mathrm{tr} |X|^p\right)^{1/p}$, is finite. The identities on both $\mathcal{B}(\mathcal{H})$ and $\mathcal{B}(\mathcal{T}_1)$ are denoted by $I$. Henceforth, we refer to $\mathcal{T}_1(\mathcal{H})\equiv\mathcal{T}_1$ as the set of \emph{trace class} operators. The set of quantum states (or density operators) --- that is, positive semi-definite operators $\rho \in \mathcal{T}_1(\mathcal{H})$ of unit trace --- is denoted by $\mathcal{S}(\mathcal{H})$. The Schatten $1$-norm $\|\cdot\|_1$ is called the \emph{trace norm} and the corresponding induced distance (e.g., between quantum states) is the \emph{trace distance}. The Schatten $2$-norm, $\|\cdot\|_2$, coincides with the \emph{Hilbert--Schmidt norm}. An unbounded operator on $\cH$ or $\cT_1$ is a linear map $A:\cD(A)\subset\cH\rightarrow\cH$ defined on a domain $\cD(A)$. For a pair of positive semi-definite operators $A$ and $B$ with domains $\operatorname{dom}(A), \operatorname{dom}(B) \subseteq \mathcal{H}$, we write $A \geq B$ if and only if $\operatorname{dom}\bigl(A^{1/2}\bigr) \subseteq \operatorname{dom}\bigl(B^{1/2}\bigr)$ and $\bigl\|A^{1/2} \ket{\psi}\bigr\|^2 \geq \bigl\|B^{1/2} \ket{\psi}\bigr\|^2$ for all $\ket{\psi} \in \operatorname{dom}\bigl(A^{1/2}\bigr)$. If $\rho$ is a quantum state with spectral decomposition $\rho = \sum_i p_i \ket{\phi_i} \bra{\phi_i}$ and $A$ is a positive semi-definite operator, the \emph{expected value} of $A$ on $\rho$ is defined as
		\begin{equation}\label{expected positive}
			\operatorname{tr}[\rho A] \coloneqq \sum_{i:\, p_i > 0} p_i \left\|A^{1/2} \ket{\phi_i} \right\|^2 \in \mathbb{R}_+ \cup \{+\infty\} \,;
		\end{equation}
		here, we use the convention that $\operatorname{tr}[\rho A] = +\infty$ if the above series diverges or if there exists an index $i$ for which $p_i > 0$ and $\ket{\phi_i} \notin \operatorname{dom}\left(A^{1/2}\right)$. 
		
		This definition can be extended to a generic densely defined self-adjoint operator $A$ on $\mathcal{H}$, by considering its decomposition $A = A_+ - A_-$ into positive and negative parts, where $A_\pm$ are positive semi-definite operators with mutually orthogonal supports. The operator $A$ is said to have a \emph{finite expected value on $\rho$} if $(i)$ $\ket{\phi_i} \in \operatorname{dom}\big(A_+^{1/2}\big) \cap \operatorname{dom}\big(A_-^{1/2}\big)$ for all $i$ for which $p_i > 0$, and $(ii)$ the two series $\sum_i p_i \big\|A_\pm^{1/2} \ket{\phi_i}\big\|^2$ both converge. In this case, the following quantity is called the \emph{expected value} of $A$ on $\rho$:
		\begin{equation}\label{expected}
			\operatorname{tr}[\rho A] \coloneqq \sum_{i:\, p_i > 0} p_i \left\|A_+^{1/2} \ket{\phi_i}\right\|^2 - \sum_{i:\, p_i > 0} p_i \left\|A_-^{1/2} \ket{\phi_i}\right\|^2 \,.
		\end{equation}
		Obviously, for a pair of operators $A, B$ satisfying $A \geq B$, we have that $\operatorname{tr}[\rho A] \geq \operatorname{tr}[\rho B]$.
		
		Here, we adopt standard notation from quantum information theory: given a multipartite quantum system with associated Hilbert space $\mathcal{H} = \mathcal{H}_A \otimes \mathcal{H}_B$ and a state $\rho \in \mathcal{S}(\mathcal{H})$, we denote by $\rho_A \coloneqq \operatorname{tr}_{\mathcal{H}_B}(\rho)$ the marginal state on system $A$. Similarly, an observable on $A$, i.e.~a self-adjoint bounded operator $X$ on $\mathcal{H}_A$, is identified with the observable $X_A \otimes I_B$ on the system $AB$, i.e., over the joint Hilbert space $\mathcal{H}_A \otimes \mathcal{H}_B$, where $I_B$ denotes the identity operator on $\mathcal{H}_B$. 
		
		A quantum channel with input system $A$ and output system $B$ is any completely positive, trace-preserving (CPTP) linear map $\mathcal{N} : \mathcal{T}_1(\mathcal{H}_A) \to \mathcal{T}_1(\mathcal{H}_B)$, where $\mathcal{H}_A, \mathcal{H}_B$ are the Hilbert spaces corresponding to $A$ and $B$, respectively. We denote the identity superoperator over a system $A$ by $I_A$.

	\subsection{Phase-space formalism}\label{secphasespaceformalism}
		In this paper, given $m \in \mathbb{N}$, we are concerned with the Hilbert space $\mathcal{H}_m \coloneqq L^2(\mathbb{R}^m)$ of a so-called $m$-mode oscillator, which is the space of square-integrable functions on $\mathbb{R}^m$. For more details, we refer to \cite[Section 12]{Holevo.2012}. We denote by $x_j$ and $p_j$ the canonical position and momentum operators on the $j^{\text{th}}$ mode. The $j^{\text{th}}$ creation and annihilation operators $a_j = (x_j + i p_j) / \sqrt{2}$ and $a_j^\dagger = (x_j - i p_j) / \sqrt{2}$ satisfy the well-known \emph{canonical commutation relations} (CCR):
		\begin{align}\label{eq:ccr}
			[a_j, a_k] = 0 \,, \qquad [a_j, a_k^\dagger] = \delta_{jk} I \,.
		\end{align}
		In terms of the vector of canonical operators $R \coloneqq (x_1, p_1, \dots, x_m, p_m)$, the above relations take the compact form $[R_j, R_{k}] = i (\Omega_{m})_{jk}$, where $\Omega_m$ denotes the $2m \times 2m$ standard symplectic form defined as
		\begin{equation} \label{comm}
			\Omega_{m} \coloneqq \begin{pmatrix} 0 & 1 \\ -1 & 0 \end{pmatrix}^{\oplus m} \,.
		\end{equation}
		We will often omit the subscript $m$ if the number of modes is fixed. The \emph{total photon number} operator is defined by
		\begin{equation}\label{eq:total-photon-number}
			N \coloneqq \sum_{j=1}^m a_j^\dag a_j = \sum_{j=1}^m N_j = \sum_{j=1}^m \frac{x_j^2 + p_j^2}{2} - \frac{m}{2} \,,
		\end{equation}
		where $N_j \coloneqq a_j^\dagger a_j$. The operator $N$ is diagonal in the multi-mode Fock basis $\{|\mathbf{k}\rangle\}_{\mathbf{k} \in \mathbb{N}^m}$, with
		\begin{equation*}
			N \,|\mathbf{k}\rangle = \left(\sum_{i=1}^m k_i \right) \,|\mathbf{k}\rangle \,, \qquad \mathbf{k} \equiv (k_1, \dots, k_m) \,.
		\end{equation*}
		The following simple identities will prove useful in our derivations \cite{Gondolf.2024}: given a real-valued measurable function $f:\mathbb{Z} \to \mathbb{R}$, any $j \in \{1, \dots, m\}$, and any integer $\ell$,
		\begin{equation}\label{eq:symmetry-function}
			\begin{aligned}
				a_j f(N_j + \ell I) = f(N_j + (\ell+1)I) a_j\qquad\text{and}\qquad f(N_j - \ell I) a_j = a_j f(N_j - (\ell+1)I)\,,
			\end{aligned}
		\end{equation}
		where the identities are satisfied, e.g., on the subspace $\mathcal{H}_f$ of finite linear combinations of Fock states. We will often deal with functions defined on $\mathbb{N}$ and extend them by $0$ to the entire set $\mathbb{Z}$. In what follows, we denote by $\mathcal{T}_f$ the set of states that can be expressed as finite linear combinations of rank-one operators of support and range in $\mathcal{H}_f$. These two spaces are common invariant cores for polynomials in annihilation and creation operators. We also use the canonical commutation relation to write $(\ad)^l a^l$ as a function of $N$:
		\begin{equation}\label{higher-order-product-aadag}
			\begin{aligned}
				(\ad_j)^l a_j^l &= (N_j - (l-1)I)(N_j - (l-2)I) \cdots (N_j - I) N_j \equiv N_j[-l+1:0] \,, \\
				a_j^l (\ad_j)^l &= (N_j + I)(N_j + 2I) \cdots (N_j + (l-1)I)(N_j + lI) \equiv N_j[1:l] \,,
			\end{aligned}
		\end{equation}
		where, given two integers $r \le j$, we denote by convention
		\begin{equation}\label{eq:notation-increasing-sequence}
			\begin{aligned}
				N_j[r:j] &\coloneqq (N_j + rI) \cdots (N_j + jI) \quad &\text{with} \quad N_j[r:j] &= I \quad \text{for} \quad r > j \,, \\
				n_j[r:j] &\coloneqq (n_j + r) \cdots (n_j + j) \quad &\text{with} \quad n_j[r:j] &= 1 \quad \text{for} \quad r > j \,.
			\end{aligned}
		\end{equation}
		In what follows, we use coherent states defined for a given $\alpha = (\alpha_1, \dots, \alpha_m) \in \mathbb{C}^m$ by
		\begin{align}\label{eq:coherent-state}
			|\alpha\rangle \coloneqq \bigotimes_{i} |\alpha_i\rangle \,, \qquad \text{where} \qquad |\alpha_i\rangle \coloneqq e^{-\frac{1}{2}|\alpha_i|^2} \sum_{n=0}^\infty \frac{\alpha_i^n }{\sqrt{n!}} \,|n\rangle \,.
		\end{align}
		The inner product between two single-mode coherent states satisfies
		\begin{equation}\label{eq:coherent-state-overlap}
			\langle \beta | \alpha \rangle = e^{-\frac{1}{2}(|\alpha|^2 + |\beta|^2) + \alpha^\intercal \overline{\beta}} \quad \text{and} \quad |\langle \beta | \alpha \rangle|^2 = e^{-|\alpha - \beta|^2} \,.
		\end{equation}
		Finally, we recall that a heterodyne measurement is a continuous-valued POVM whose effects correspond to rank-one projections onto coherent states, i.e., for any Borel set $A \subset \mathbb{R}^2$,
		\begin{equation}\label{eq:heterodyne-measurement}
			M(A) = \frac{1}{\pi} \int_{A} \ketbra{\beta}{\beta} \, d^2\beta
		\end{equation}
		determines a positive operator-valued measurement (POVM) because $M(\mathbb{R}^2) = I$.

	\subsection{Sobolev preserving quantum Markov semigroups}
		In the following, we are concerned with Markov semigroups, i.e., semigroups of completely positive, trace-preserving maps over bosonic $m$-mode systems. A semigroup $(\cP_t)_{t\geq0}$ satisfies the following three defining properties:
		\begin{equation*}
			\cP_t\cP_s = \cP_{s+t}, \qquad \cP_0 = I, \qquad \text{and} \qquad \lim_{t \rightarrow 0} \cP_t x = x
		\end{equation*}
		for all $t, s \geq 0$ and $x \in \cT_1(\cH_m)$. Assuming that the Markov semigroup $\cP_t$ in the Schr\"{o}dinger picture satisfies the above strongly continuous semigroup properties, there exists a densely defined and closed operator $(\widetilde{\cL},\cD)$ with domain $\cD\subset\cT_1$ such that $\cP_t$ is the unique solution of the following initial value problem:
		\begin{equation}\label{eq:master-equation}
			\frac{\partial}{\partial t}\rho(t) = \widetilde{\cL} \rho(t), \qquad \rho(0) = \rho_0 \in \cD,
		\end{equation}
		where $\widetilde{\cL}$ is an unbounded superoperator --- the generator --- with domain $\cD$ \cite{Engel.2000}, and the initial quantum state $\rho_0 \in \cD \subset \cT_1$. This allows the notation $e^{t\widetilde{\cL}}$. Analogous to the GKSL form in finite dimensions \cite{Gorini.1976, Lindblad.1976}, we define the following superoperators on $\cT_f$ for an unbounded operator $(L, \cH_f)$ and a symmetric operator $(H, \cH_f)$, both of which can be expressed as polynomials of the creation and annihilation operators on a suitable domain,
		\begin{equation}\label{eq:notation-super-operators}
			\cL[L] \coloneqq L(\cdot) L^\dagger - \frac{1}{2} \, \{L^\dagger L, \, \cdot\} \qquad \text{and} \qquad \cH[H] \coloneqq -i[H, \, \cdot \,]\,.
		\end{equation}
		Note that a common strategy for handling unbounded operators is to carry out the analysis on finite approximation spaces, such as $\cH_f$ and $\cT_f$, and then extend the results via the closure. Next, we assume that the generator $(\widetilde{\cL}, \cD)$ is in \emph{GKSL form}, i.e.~the generator $\widetilde{\cL}$ is given by the closure of
		\begin{equation}\label{eq:GKSL}
			\widetilde{\cL} = \cH[H] + \sum_{i=1}^{\nu} \cL[L_i] \qquad \text{on} \qquad \cT_f(\cH_m),
		\end{equation}
		for a set of unbounded operators $L_i$ and a symmetric operator $H$, each defined by polynomials of the annihilation and creation operators on $\cH_m$. In addition to the GKSL structure, in \Cref{assum:local-GKSL}, we introduce further locality assumptions.
		
		Interestingly, in infinite dimensions the GKSL form does not by itself imply the existence of a quantum Markov semigroup. For example, \cite[Ex.~3]{Davies.1977} demonstrates that there exist generators in GKSL form which do not generate a quantum Markov semigroup. Therefore, we impose an additional assumption from \cite{Gondolf.2024}, which is sufficient for the existence and uniqueness of a Markov semigroup:
		
		\begin{assum*}[Photon-number moment stability]
			There exists $K \in \mathbb{N}$ and a sequence defined by $\omega_{k} \ge 0$ for all $k \in \mathbb{N}$ such that for all $k \geq K$ and positive semi-definite $x \in \cT_f$,
			\begin{equation}\label{eq:moment-stability}
				\tr[\cW^k(\cL(x))] \le \omega_{k} \tr[\cW^k(x)],
			\end{equation}
			where for $R \subset \{1, \ldots, m\}$,
			\begin{equation}\label{eq:local-bosonic-sobolev-operator}
				\cW^{k}_R(X) \coloneqq \bigotimes_{j \in R} (N_j + I)^{k/4} X \bigotimes_{j \in R} (N_j + I)^{k/4} \equiv \bigotimes_{j \in R} \cW_j^{k}(X),
			\end{equation}
			and for $R = \{1, \ldots, m\}$, we simply write $\cW^{k}$.
		\end{assum*}
		In analogy to classical function spaces, the vector space $W^{k,1}_R=\cW^{-k}_R(\cT_{1})$ equipped with the norm 
		\begin{equation}\label{eq:local-bosonic-sobolev-norm}
			\|\cdot\|_{W^{k,1}_R}\coloneqq\|\cW^{k}_R(\cdot)\|_{1}
		\end{equation}
		is called Sobolev space. It is proven in \cite{Gondolf.2024} that $W^{k,1}_R$ (or $W^{k,1}$ if $R=\{1,\ldots,m\}$) is a Banach space and more importantly \Cref{eq:moment-stability} induces that the closure of the operator $\widetilde{\cL}$ generates a strongly continuous semigroup of quantum channels, i.e., a quantum Markov semigroup, $e^{t\widetilde{\cL}}$. 
		\begin{thm}[Generation theorem \texorpdfstring{\cite[Thm.~4.4]{Gondolf.2024}}{Gondolf 2024}]\label{thm:generation-theory}
			Let $(\widetilde{\cL}, \cT_f)$ be an operator of GKSL form \eqref{eq:GKSL} satisfying the photon-moment stability condition \eqref{eq:moment-stability}. Then, the closure $\widetilde{\cL}$ generates a strongly continuous, positivity-preserving semigroup $(\cP_t)_{t \ge 0}$ on $W^{k,1}$ for all $k \in \N$ with the additional \emph{Sobolev preservation property}:
			\begin{equation}\label{eq:sobolev-bound}
				\|e^{t\widetilde{\cL}}(x)\|_{W^{k,1}} \leq e^{t\omega_k}\|x\|_{W^{k,1}}\,,
			\end{equation}
			for all $x \in W^{k,1}$ and where $\omega_k = \frac{k}{K}\omega_{K}$ for $k < K$. In the special case $k = 0$, the semigroup is a quantum Markov semigroup.
		\end{thm}
		\begin{proof}
			In \cite[Thm.~4.4]{Gondolf.2024}, the above statement is proven for the restriction of self-adjoint operators in $W^{k,1}$. In \cite[Thm.~4.1.1]{Moebus.2025thesis}, this result is extended to $W^{k,1}$ by following the ideas of \cite[Lem.~3.18]{Wolf.2023LectureInference}, i.e.,
			\begin{equation*}
				\begin{aligned}
					\|T_t(x)\|_{W^{k,1}} &\leq \|(\cW^{k} \circ T_t) \otimes I_2(\hat{x}^+)\|_1 + \|(\cW^{k} \circ T_t) \otimes I_2(\hat{x}^-)\|_1 \\
					&= \|\cW^{k} \circ T_t(\tr_2[\hat{x}^+])\|_1 + \|\cW^{k} \circ T_t(\tr_2[\hat{x}^-])\|_1 \\
					&\leq e^{\omega_{k} t} \Bigl(\|\cW^{k}(\tr_2[\hat{x}^+])\|_1 + \|\cW^{k}(\tr_2[\hat{x}^-])\|_1\Bigr) \\
					&= e^{\omega_{k} t}\|x\|_{W^{k,1}}\,,
				\end{aligned}
			\end{equation*}
			where $x \in W^{k,1}$, $\hat{x} = \frac{1}{2}(x \otimes \ketbra{0}{1} + x^\dagger \otimes \ketbra{1}{0})$, and $\hat{x} = \hat{x}^+ - \hat{x}^-$.
		\end{proof}
		For certain examples (see, for instance, \Cref{ex:reg-hamiltonian}), an even stronger assumption than \Cref{eq:moment-stability} can be shown to hold. Specifically, for any $k > 0$, if there exist constants $c_k, \mu_k > 0$ such that for any state $\rho \in \cT_f$:
		\begin{equation}\label{eq:moment-stability-strengthen}
			\tr\Big[\widetilde{\cL}(\rho) \bigotimes_{j=1}^m (N_j + I)^{k/2}\Big] \leq - c_k \tr\Big[\rho \bigotimes_{j=1}^m (N_j + I)^{k/2}\Big] + \mu_k\,,
		\end{equation}
		then \cite[Prop.~5.1]{Gondolf.2024} shows that for all $\rho \in W^{k,1}$:
		\begin{equation}\label{eq:sobolev-bound-strengthen}
			\|e^{t\widetilde{\cL}}(\rho)\|_{W^{k,1}} \leq \max\Big\{\frac{\mu_k}{c_k}, \|\rho\|_{W^{k,1}}\Big\}\,.
		\end{equation}
		The Sobolev norm defined above in a product structure can be related, via Young's inequality, to a sum of locally defined Sobolev norms:
		\begin{equation}\label{eq:sobolev-young-inequality}
			\big\|\cW^{k/m}(\rho)\big\|_1 = \tr\Bigl[\sqrt{\rho} \prod_{j=1}^{m}(N_j + I)^{k/(2m)} \sqrt{\rho}\Bigr] \leq \frac{1}{m} \tr\Bigl[\rho \sum_{j=1}^{m}(N_j + I)^{k/2}\Bigr] \coloneqq \frac{1}{m} \big\|\cW_{\Sigma}^{k}(\rho)\big\|_1
		\end{equation}
		for any quantum state $\rho \in \cW^{k,1}$. Next, we consider four examples defined on an $m$-mode bosonic system, where each mode is associated with a vertex $v \in V$ of a graph $G = (V, E)$ with $|V| = m$. In \Cref{subsec:model-specific-lrb}, the following examples are analyzed to determine whether they are Sobolev-preserving semigroups. We define the degree of a vertex by $\gamma_i = |\{j \in V \mid (i,j)\text{ or } (j,i) \in E\}|$ and the maximum degree by $\gamma = \max_{i \in \{1, \dots, m\}} \gamma_i$. We start with the celebrated Bose--Hubbard model \cite{Gersch.1963}:
		
		\begin{ex}[Bose--Hubbard model]\label{ex:Bose-Hubbard}
			Given a graph $(V,E)$, the Bose--Hubbard model is defined by the Hamiltonian
			\begin{equation*}
				H\coloneqq \sum_{(i,j)\in E}\lambda^{(i,j)}\,a_i^\dagger a_j+\sum_{i\in V}u^{(i)}N_i(N_i-I)+\sum_{i\in V}\mu^{(i)}N_i\,,
			\end{equation*}
			for coefficients $\lambda^{(i,j)}=\overline{\lambda}^{(j,i)}\in\C$ and $u^{(i)}$, $\mu^{(i)}\in\R$, which are bounded in absolute value by $\Lambda \geq 0$ for all $i\in V$ and $(i,j)\in E$.
		\end{ex}
		
		The next example introduces quadratic Hamiltonians, characterized by a self-adjoint coefficient matrix $\lambda \in \mathbb{C}^{n \times n}$ and two vectors $b, c \in \mathbb{C}^n$. For simplicity, we allow for loops in the graph below, i.e., edges of the form $(i, i)$ for some $i \in V$.
		\begin{ex}[Quadratic Hamiltonian]\label{ex:Gaussian}
			Given a graph $(V,E)$, a self-adjoint coefficient matrix $\lambda\in\C^{n\times n}$ respecting the graph structure, i.e., $\lambda_{i,j}=0$ if $(i,j)\notin E$, and two vectors $b,c\in\C^n$, a quadratic Hamiltonian is defined by
			\begin{equation*}
				H\coloneqq \sum_{(i,j)\in E}\lambda^{(i,j)}\,a_i^\dagger a_j+\sum_{i\in V}b^{(i)}a_i^2+\overline{b}^{(i)}(\ad_i)^2+c^{(i)}a_i+\overline{c}^{(i)}\ad_i\,,
			\end{equation*}
			with the absolute values of the coefficients upper-bounded by $\Lambda \geq 0$.
		\end{ex}
		
		In the next two examples, we focus on the two-body interaction and extend it to higher-order polynomials:
		\begin{ex}[Higher-order two-body interactions]\label{ex:general-hamiltonian}
			Given a graph $(V,E)$, we consider the Hamiltonian defined by
			\begin{align}\label{eq:bosonic-hamiltonian}
				H^{(d)} = \sum_{e\in E}H_e\qquad\text{with}\qquad H_e \coloneqq \sum_{k,\ell,k',\ell'=0}^d\,\lambda^{(e)}_{k\ell k'\ell'}\,(a_i^\dagger)^k\,a_i^\ell\,(a_j^\dagger)^{k'}\,a_j^{\ell'}\,,
			\end{align}
			for some complex coefficients $\lambda^{(e)}_{k\ell k'\ell'}=\overline{\lambda}^{(e)}_{\ell k \ell' k'}$ with $|\lambda^{(e)}_{k\ell k'\ell'}|\le \Lambda$. Note that we can fix $\lambda_{0000}=0$, due to gauge freedom, and we denote $\cH^{(d)}\coloneqq -i[H^{(d)},\cdot]$.
		\end{ex}
		
		Importantly, the operator $H^{(d)}$ does not necessarily satisfy \Cref{eq:moment-stability}, so it is unclear whether it generates a quantum Markov semigroup. Even if this is assumed, several ambiguities remain: for example, it is not clear whether its Taylor expansion converges or whether finite propagation bounds, such as cluster expansions and Lieb--Robinson estimates, apply to such Hamiltonians. As these tools play a crucial role in the analysis of sample-optimal protocols like those in \cite{Stilckfranca.2024, Huang.2023learning}, their proofs do not directly extend to the present continuous-variable setting. To address this issue, we introduce a class of local, Sobolev-preserving dissipative generators, which we call \emph{dissipative regularizers}. These generators, which arise in the theory of continuous-variable quantum error-correcting codes \cite{Guillaud.2019repetition, Guillaud.2023}, take the following form:
		\begin{ex}[Regularized Hamiltonian]\label{ex:reg-hamiltonian}
			Given $\alpha \in \mathbb{C}^m$ and $p \in \mathbb{N}$, the dissipative regularizer is defined via the $p$-photon dissipation by 
			\begin{align}\label{eq:p-photon-dissipation}
				\cL^{(\alpha,p)} \coloneqq \sum_{j \in V} \cL[L_j^{(\alpha_j,p)}], \qquad \text{where} \quad L_j^{(\alpha_j,p)} \coloneqq a_j^p - \alpha_j^p I \,.
			\end{align}
			Then, the higher-order two-body interaction Hamiltonian in combination with the dissipative regularizer is denoted by
			\begin{equation}\label{eq:regularized-hamiltonian}
				\widetilde{\cL}^{(\alpha)} = \cH^{(d)} + \cL^{(\alpha,p)}\,.
			\end{equation}
		\end{ex}
		
		As we will see in \Cref{sec:lrb}, the added dissipation can enforce a weak Lieb--Robinson-type information propagation bound, allowing us to analyze separated regions independently by choosing a suitable parameter vector $\alpha$ for the input coherent state as well as for the jump operators $L_j^{(\alpha_j,p)}$ in our learning scheme (see \Cref{sec:learning}). Note that all examples are well-defined on the domain $\cH_f$.

\section{Weak Bosonic Lieb--Robinson bounds}\label{sec:lrb}
    In this section, we demonstrate that the combined use of locality and a moment propagation bound enables us to prove a weak Lieb--Robinson-type information propagation bound, allowing for approximation by a local evolution. On a technical level, we generalize the methods used in the recently derived bounds within the context of the Bose--Hubbard model in \cite{Kuwahara.2021} to models with locality assumptions similar to those in \cite{Nachtergaele.2011lrb}.
    
    \begin{assum}[Local GKSL generators]\label{assum:local-GKSL}
    	Let $(V, \dist)$ be a finite metric space and $\cV^{(r)} \subset 2^V$ a subset of the power set satisfying 
    	\begin{itemize}
    		\item (r-locality) for a given $r \in \mathbb{N}$ and all $X \in \cV^{(r)}$, we have $\diam(X) \leq r$, and 
    		\item (finite connectivity) for all $i \in V$, there is a $\gamma \in \mathbb{N}$ so that $\gamma_i\coloneqq|\{ X \in \cV^{(r)} \mid i \in X \}| \leq \gamma$.
    	\end{itemize}
    	By increasing $\gamma$ if necessary, we also assume that $|\{j\in V\,|\,1\leq\dist(i,j)\leq r\}|\leq\gamma$ and $|\{j\in V\,|\,\dist(i,j)\leq s\}|\leq \gamma^s$ for $1\leq s\leq r$. Then, there is a $r_{\exp}\in\N$ such that $|X|\leq r_{\exp}$ for all $X\in\cV^{(r)}$. For $X \in \cV^{(r)}$, let $H_X$ be a symmetric operator defined by a polynomial $p_{H_X} \in \mathbb{C}[x_1, \dots, x_{2|X|}]$ in $a_i, \ad_i$ with maximal degree $d \in \mathbb{N}$ acting non-trivially on $i \in X$. Similarly, $\{ L_{X,j} \}_{j=1}^{\nu(X)}$ are defined by $p_{L_{X,j}} \in \mathbb{C}[x_1, \dots, x_{2|X|}]$ with maximal degree $d/2$. For definiteness, every monomial is written as $\lambda_{\mathbf{k}\Bell}^{(X,j)} \bigotimes_{i \in X} (\ad_i)^{k_i} a_i^{\ell_i}$, where $\mathbf{k}, \Bell \in \{0, \dots, d\}^{|X|}$ and the coefficients $\lambda_{\mathbf{k}\Bell}^{(X,j)} \in \mathbb{C}$ are bounded in absolute value by $\Lambda$ for $H$ and $\sqrt{\Lambda}$ for $\{ L_{X,j} \}_{j=1}^{\nu(X)}$, with $\nu(X) \leq \nu \in \mathbb{N}$ for all $X \in \cV^{(r)}$. Then, we call $(\widetilde{\cL}, \cT_f)$ a generator in local GKSL form if the closure generates a quantum Markov semigroup and
    	\begin{equation}\label{eq:lindblad}
    		\widetilde{\cL} = \sum_{X \in \cV^{(r)}}\left(\cH[H_X] + \sum_{j=1}^{\nu(X)} \cL[L_{X,j}]\right) \,.
    	\end{equation}
    	Moreover, we denote the reduction of the operator to $R \subset V$ by
    	\begin{equation}\label{eq:lindblad-reduced}
    		\widetilde{\cL}_R = \sum_{X \in \cV^{(r)}_R}\left(\cH[H_X] + \sum_{j=1}^{\nu(X)} \cL[L_{X,j}]\right) \,,
    	\end{equation}
    	where $\cV^{(r)}_R=\{X\in\cV^{(r)}\,|\,X\subset R\}$.
    \end{assum}
    
    Note that, in addition to the structure, the fundamental assumption that the closure defines a generator of a strongly continuous semigroup is crucial. For our examples, we provide a proof in \Cref{subsec:model-specific-lrb}. To achieve a weak Lieb--Robinson-type bound for operators of the above form, the following assumption allows us to control the propagation of the moments (see \Cref{lem:moment-prop-bound}).
    
    \begin{assum}[Photon-number propagation]\label{assum-lrb}
	Let $(\widetilde{\cL}, \cT_f)$ satisfy the local GKSL \Cref{assum:local-GKSL} on $(V, \dist)$ with $m = |V|$. We assume that there are constants $\Gamma'>0$, $\Gamma''$, $\Gamma'''\geq0$, $\xi \geq 2$ and $p, K \in \mathbb{N}$ such that for all $k \in \mathbb{N}$ with $k \geq K$, states $\rho \in \cT_f$, and all $i \in V$,
    	\begin{equation*}
    		\frac{d}{dt} F^{(k)}_i(t) \coloneqq \frac{d}{dt} \tr[\rho_t f(N_i)] \leq k \Gamma' F^{(k)}_i(t) + k \Gamma'' \sum_{1 \leq \dist(i, j) \leq r} F^{(k)}_j(t) + k^{\xi k} \Gamma''',
    	\end{equation*}
    	where $\rho_t \coloneqq e^{t\widetilde{\cL}}(\rho)$ and $F^{(k)}_i(t) \coloneqq \tr[\rho_t f(N_i)]$. The function $f : \mathbb{N} \to \mathbb{R}$ is defined recursively by $f(0) = f(1) = \dots = f(p-1) = 0$ and $f(m + p) = f(m) + k m^{k-1}$ for all $m\geq0$.
    \end{assum}
    
    Note that the functions $F_j^{(k)}(t)$ are not the moments of $\rho(t)$, which is one of the crucial differences between the bounds found in \Cref{prop:LRB-bose-hubbard,prop:LRB-regularized-hamiltonian} and those in \cite{Gondolf.2024}. The choice of $f$ allows us to better track the dependence of the bounds on the parameter $k$.
    
	\subsection{Weak Bosonic Lieb--Robinson information propagation bound}\label{subsec:general-lrb}
		Under the two assumptions given above, we prove a weak Lieb--Robinson-type information propagation bound in \Cref{thm:bosonicLRB} that is satisfied by the Bose--Hubbard model (\ref{ex:Bose-Hubbard}), any quadratic Hamiltonian (\ref{ex:Gaussian}), and a Hamiltonian regularized by a power of the photon number operator (\ref{ex:reg-hamiltonian}). Before doing so, we introduce a few standard objects that will be useful in stating our bound. Given a finite metric space $(V,\operatorname{dist})$ and $F^{\downarrow}:[0,\infty)\to (0,\infty)$ a non-increasing function, we denote
		\begin{align*}
			&\|F^{\downarrow}\|\coloneqq \max_{x\in V}\sum_{y\in V}F^{\downarrow}(\operatorname{dist}(x,y))\\
			&C\coloneqq \max_{x,y\in V}\,\sum_{z\in V}\,\frac{F^{\downarrow}(\operatorname{dist}(x,z))\,F^{\downarrow}(\operatorname{dist}(y,z))}{F^{\downarrow}(\operatorname{dist}(x,y))}\,.
		\end{align*}
		and define $F^{\downarrow}_\mu(x)=e^{-\mu x}F^{\downarrow}(x)$ with corresponding constants $\|F^{\downarrow}_\mu\|\le \|F^{\downarrow}\|$ and $C_\mu\le C$ for any $\mu>0$. If $(V,E)$ is a graph, $\operatorname{dist}(x,y)$ denotes the shortest path between two vertices $x,y\in V$. In particular, for the finite-size $D$-dimensional lattice $V=\{-n,...,n\}^D$, a reasonable choice for $F^{\downarrow}$ is $F^{\downarrow}(x)\coloneqq (1+x)^{-D-1}$ \cite{Nachtergaele.2006propagation}. In that case, the constant $C$ defined above can be controlled as $C\le 2^{D+2}\sum_{x\in\mathbb{Z}^{D}}\frac{1}{(1+|x|)^{D+1}}<\infty$, where $|x|$ denotes the $\ell_1$-norm of $x\in \mathbb{Z}^D$. Moreover, one can give a simple upper bound on the cardinality of the sets $X\in\cV^{(r)}$ given by $r_{\exp}=r^{D}$
        
		\begin{thm}[Weak Bosonic Lieb--Robinson-type bound]\label{thm:bosonicLRB}
			Let $(\widetilde{\cL},\cT_f)$ be an operator on $\cT_1$ satisfying \Cref{assum:local-GKSL,assum-lrb} with respect to a $K\in\N$ and a power $p\in\N$. Given $M\in\N$, we denote the finite-rank projection on a region $R\subset V$ as $P\coloneqq \prod_{j\in R}P^{(M)}_{j}$, where $P^{(M)}_{i}=\sum_{n\le M}|n\rangle\langle n|_i$. Next, we denote 
			\begin{equation*}
				\widetilde{\cL}^{(M)}_{R} = \sum_{X\in\cV^{(r)}_R}\left(\cH[PH_XP] + \sum_{j=1}^{\nu(X)}\cL[L_{X,j}P]\right).
			\end{equation*}
			Then, for $t\geq0$, $\hat{k}\coloneqq \max\{K,4dr_{\exp}\}$, $M\in\N$ and any bounded observable $O_T$ supported on a region $T\subset R[-r]\coloneqq \{v\in V\,|\,\dist(v,R^c)> r\}$, there are constants $c_1,c_2,\xi_2\ge 0$ independent of $|T|$, $|R|$ and $|V|$ (see \Cref{eq:proof-lrb-constant-c1c2}) such that for all states $\rho\in\cT_1$ with finite local moments $\max_{j\in V}\tr[\rho (N_j+I)^{\hat{k}}]<\infty$
			\begin{equation*}
				\begin{aligned}
					\frac{\left|\tr\left[ O_T\Big(e^{t\widetilde{\cL}}-e^{t\widetilde{\cL}^{(M)}_{{R}}}\Big)(\rho) \right]\right|}{\|O_T\|_\infty} &\leq c_1 |R|^{r+1}e^{\hat{k} \Gamma t}\frac{\sqrt{\max_{j\in V}\tr[\rho (N_j+I)^{\hat{k}}] + \hat{k}^{\xi  \hat{k}}}}{(M-d+1)^{\hat{k}/4-r_{\exp}d/2}}\\
                    &\qquad+c_2|R|^r\frac{F^\downarrow_\mu(r)}{C_\mu^2}\,e^{t\xi_2(M+1)^{r_{\exp}d}\frac{C_\mu}{F^\downarrow_\mu(r)}}\sum_{x\in \partial_rR}\sum_{y\in T}F^\downarrow_{\mu}(\operatorname{dist}(x,y))
				\end{aligned}
			\end{equation*}
			Moreover, for $\rho=\ketbra{\alpha}{\alpha}$ with $\alpha\in\C^m$, set $\eta=\max_{i\in V}|\alpha_i|$ and $\dist(\partial E_R,T)\coloneqq\dist(\bigcup_{X\in\partial E_R}X,T)$. If $M-d+1=  \left(\frac{\mu F_\mu^\downarrow(r) \dist(\partial E_R,T)}{2^{r_{\exp}d/2+1}t \xi_2C_\mu}\right)^{1/{r_{\exp}d}}$ and $\widetilde{k}=\biggl(\frac{(M-d+1)^{1/4-\varphi}}{e^{\Gamma t+1}}\biggr)^{\frac{2}{\xi }}\geq \max\{\hat{k},2,2(e-1)\eta^2\}$ with $\varphi$ satisfying
			\begin{equation*}
				0<\varphi\coloneqq \frac{r_{\exp}d}{\ln(\frac{\mu F_\mu^\downarrow(r) \dist(\partial E_R,T)}{2^{r_{\exp}d/2+1}t \xi_2C_\mu})}<\frac{1}{8}
			\end{equation*}
			(see Equation (\ref{eq:proof-lrb-assumptions})), there are constants $c_3,\xi_3,\kappa>0$ (see Equation (\ref{eq:proof-lrb-constants})) so that
			\begin{equation}\label{LRBsimple}
                \begin{aligned}
                    &\frac{\Bigl|\tr\left[ O_T\Big(e^{t\widetilde{\cL}}-e^{t\widetilde{\cL}^{(M)}_{{R}}}\Big)(\ketbra{\alpha}{\alpha}) \right]\Bigr|}{\|O_T\|_\infty}\\
                    &\qquad\leq c_3 |R|^{r}(|R|+\min\{|T|,|\partial_rR|\})\sqrt{\frac{\dist(\partial E_R,T)}{t}}e^{-\xi_3e^{-\kappa t}\Bigl(\frac{\dist(\partial E_R,T)}{t}\Bigr)^{1/{(4\xi r_{\exp}d)}}}\,.
                \end{aligned}
			\end{equation}
            with the annulus $\partial_rR\coloneqq\{v\in V\,|\,\dist(v,R)\leq r\wedge\dist(v,R^c)\leq r\}$ and the boundary interaction supports $\partial E_{R}\coloneqq\{X\in\cV^{(r)}\,|\,X\cap R\neq\emptyset\neq X\cap R^c\}$; these are hyperedges in the general support family, not necessarily graph edges. 
		\end{thm}
		Before proving the result, we state a corollary that allows us to approximate the time evolution on a coherent state by a time evolution on a finite Fock space.
		\begin{cor}\label{cor:lrb-projective-input}
			Let $(\widetilde{\cL},\cT_f)$ satisfy the assumptions of \Cref{thm:bosonicLRB} with constant time. If additionally
			\begin{equation*}
				\ln(M+1)\geq4(\ln(\eta\sqrt[4]{2})+1)\,,
			\end{equation*}
			where $\eta=\max_{i\in V}|\alpha_i|$, then for $|\psi\rangle=P\ket{\alpha}$ there are constants $\widetilde{c}_3$, $\xi_3$, $\kappa\geq0$ so that
			\begin{equation*}
                \begin{aligned}
                    &\frac{\biggl|\tr\biggl[ O_T\Big(e^{t\widetilde{\cL}}(\ketbra{\alpha}{\alpha})-e^{t\widetilde{\cL}^{(M)}_{{R}}}(\ketbra{\psi}{\psi})\Big) \biggr]\biggr|}{\|O_T\|_\infty}\\
                    &\qquad\leq \widetilde{c}_3 |R|^{r}(|R|+\min\{|T|,|\partial_rR|\})\sqrt{\frac{\dist(\partial E_R,T)}{t}}e^{-\xi_3e^{-\kappa t}\Bigl(\frac{\dist(\partial E_R,T)}{t}\Bigr)^{1/{(4\xi r_{\exp}d)}}}\,.
                \end{aligned}
            \end{equation*}
            The constants are defined in \Cref{thm:bosonicLRB} and in Equation (\ref{eq:proof-lrb-assumptions}).
		\end{cor}
		
		\begin{proof}
			First, we apply the triangle inequality to show
			\begin{equation*}
				\begin{aligned}
					\biggl|\tr\biggl[ O_T&\Big(e^{t\widetilde{\cL}}(\ketbra{\alpha}{\alpha})-e^{t\widetilde{\cL}^{(M)}_{{R}}}(\ketbra{\psi}{\psi})\Big) \biggr]\biggr|\\
					&\le\left|\tr\left[ O_T\Big(e^{t\widetilde{\cL}}-e^{t\widetilde{\cL}^{(M)}_{{R}}}\Big)(\ketbra{\alpha}{\alpha}) \right]\right|+\left|\tr\left[ O_Te^{t\widetilde{\cL}^{(M)}_{{R}}}(\ketbra{\psi}{\psi}-\ketbra{\alpha}{\alpha})\right]\right|\\
					&\le\left|\tr\left[ O_T\Big(e^{t\widetilde{\cL}}-e^{t\widetilde{\cL}^{(M)}_{{R}}}\Big)(\ketbra{\alpha}{\alpha}) \right]\right|+\|O_T\|_{\infty}\,\|\ketbra{\psi}{\psi}-\ketbra{\alpha}{\alpha}\|_1\,.
				\end{aligned} 
			\end{equation*}
			As a shorthand, we introduce $\overline{P} = I - P$. Then, using the relations 
            \begin{equation*}
                \begin{aligned}
                    \|\ketbra{\phi_1}{\phi_2}\|_1&=\tr[\sqrt{\ketbra{\phi_1}{\phi_2}\ketbra{\phi_2}{\phi_1}}]=\|{\phi_1}\|\,\|{\phi_2}\|\\
                    \|P\ket{\alpha}\|,\|\overline{P}\ket{\alpha}\|&\leq1,
                \end{aligned}
            \end{equation*}
            we find
			\begin{equation*}
				\begin{aligned}
					\|\ketbra{\psi}{\psi}-\ketbra{\alpha}{\alpha}\|_1&\leq \|\overline{P}\ket{\alpha}\|^2+\|\overline{P}\ketbra{\alpha}{\alpha}P\|_1+\|P\ketbra{\alpha}{\alpha}\overline{P}\|_1\leq3\|\overline{P}\ket{\alpha}\|\,.
				\end{aligned}
			\end{equation*}
			Next, we upper bound $\|\overline{P}\ket{\alpha}\|$ using $(M+1)!\geq \bigl(\frac{M+1}{2}\bigr)^{\frac{M+1}{2}}$:
			\begin{equation*}
				\begin{aligned}
					\|\overline{P}\ket{\alpha}\|&\leq|R|\max_{i\in {R}}\|(I-P_i^{(M)})\ket{\alpha_i}\|\\
					&=|R|\max_{i\in {R}}\sqrt{e^{-|\alpha_i|^2}\sum_{n_i=M+1}^{\infty}\frac{|\alpha_i|^{2n_i}}{n_i!}}\\
					&\leq|R|\frac{\eta^{M+1}}{\sqrt{(M+1)!}}\\
					&\leq|R|e^{(M+1)\left(\ln(\eta)-\frac{1}{4}\ln\left(\frac{M+1}{2}\right)\right)}\,,
				\end{aligned}
			\end{equation*}
			where $\eta=\max_{i\in V}|\alpha_i|$. Then, the assumption of the statement, i.e.
            \begin{equation*}
				\ln(M+1)\geq4(\ln(\eta\sqrt[4]{2})+1)\,,
            \end{equation*}
            implies that
            \begin{equation*}
				\begin{aligned}
					\|\ketbra{\psi}{\psi}-\ketbra{\alpha}{\alpha}\|_1&\leq3|R|e^{(M+1)\left(\ln(\eta)-\frac{1}{4}\ln\left(\frac{M+1}{2}\right)\right)}\\
                    &\leq3|R|e^{-(M+1)}\\
                    &\leq3|R|e^{-(M-d+1)}
				\end{aligned}
			\end{equation*}
            By the choice of $M-d+1$ in \Cref{thm:bosonicLRB}, i.e.
			\begin{equation*}
				M-d+1=  \left(\frac{\mu F_\mu^\downarrow(r) \dist(\partial E_R,T)}{2^{r_{\exp}d/2+1}t \xi_2C_\mu}\right)^{1/{r_{\exp}d}}\,,
			\end{equation*}
			we achieve the bound
			\begin{equation*}
                \begin{aligned}
                    &\frac{\biggl|\tr\biggl[ O_T\Big(e^{t\widetilde{\cL}}(\ketbra{\alpha}{\alpha})-e^{t\widetilde{\cL}^{(M)}_{{R}}}(\ketbra{\psi}{\psi})\Big) \biggr]\biggr|}{\|O_T\|_\infty}\\
                    &\qquad\leq \widetilde{c}_3 |R|^{r}(|R|+\min\{|T|,|\partial_rR|\})\sqrt{\frac{\dist(\partial E_R,T)}{t}}e^{-\xi_3e^{-\kappa t}\Bigl(\frac{\dist(\partial E_R,T)}{t}\Bigr)^{1/{(4\xi r_{\exp}d)}}}\,.
                \end{aligned}
			\end{equation*}
			by inserting the above third term into \Cref{eq:proof-lrb-combine1}, using \Cref{eq:proof-lrb-combine2} and adapting the constant $\widetilde{c}_3=5c_3$ (see \Cref{eq:proof-lrb-constants}).
		\end{proof}
		
		Next, we prove \Cref{thm:bosonicLRB}. The proof is divided into three steps. First, we lift \Cref{assum-lrb} to establish a moment propagation bound, which serves as a key tool in the second step, in which we approximate the quantum Markov evolution by a locally bounded generator. Finally, we conclude the proof using the finite-dimensional result from \cite{Nachtergaele.2011lrb}. To begin, we define the $k^{\text{th}}$-moment at $j \in V$ of an evolved state $\rho_t = e^{t\widetilde{\cL}}(\rho)$ by
		\begin{equation}\label{eq:i-local-k-moment}
			M_j^{(k)}(t) \coloneqq \tr[\rho_t (N_j+I)^{k}] = \tr[\cW_j^{2k}(\rho_t)]\,.
		\end{equation}
		\begin{lem}[Moment propagation bound]\label{lem:moment-prop-bound}
			Let $(\widetilde{\cL},\cT_f)$ be an operator satisfying \Cref{assum:local-GKSL,assum-lrb} with respect to $K \in \mathbb{N}$ and period $p \in \mathbb{N}$. Then, for all $t \ge 0$, $j \in V$, and $k \geq K$,
			\begin{align}\label{eq:moment-prop-bound}
				M_j^{(k)}(t) \le p4^{p+1} e^{k \Gamma t} \max_{i \in V} M_i^{(k)}(0) + \Upsilon_k(t)\,,
			\end{align}
			where $\Gamma = \Gamma' + \gamma \Gamma''$ and $\Upsilon_k(t) = k^{\mathcal{O}(k)}$, as defined explicitly in \Cref{eq:dgl-linear-upper-bound}. Moreover, for any region $R \subset V$, the local Sobolev norm satisfies
			\begin{align}\label{eq:sobolev-prop-bound}
				\|\rho(t)\|_{W^{2k,1}_R} \le \max_{j \in R} M_j^{(k|R|)}(t) \le p4^{p+1} e^{k |R| \Gamma t} \max_{i \in V} \|\rho\|_{W^{2k|R|,1}_i} + \Upsilon_{k|R|}(t)\,.
			\end{align}
		\end{lem}
        Before diving into the proof, we briefly demonstrate how these moments can be bounded in the case of a coherent state: 
        \begin{lem}\label{lem:coherent-state-sobolev-norm}
			For any $\alpha\in\mathbb{C}$, the coherent state $|\alpha\rangle$ in one mode satisfies
			\begin{align*}
				\||\alpha\rangle\langle\alpha|\|_{W^{2k,1}}\le 2^k \max\left\{1,\left(\frac{k}{\ln(k/|\alpha|^2+1)}\right)^k\right\}\,,
			\end{align*}
		\end{lem}
		\begin{proof}
			It is well-known that $|\alpha\rangle$ generates a Poisson distribution over the spectrum of $N$, showing
			\begin{align*}
				\||\alpha\rangle\langle\alpha|\|_{W^{2k,1}} &= \langle \alpha|(N+I)^k |\alpha\rangle\\
				&= e^{-|\alpha|^2} \sum_{n=0}^\infty \frac{|\alpha|^{2n}}{n!}\, (n+1)^k\\
				&= \sum_{l=0}^k\binom{k}{l} \sum_{n=0}^\infty n^l\,\frac{|\alpha|^{2n}}{n!}e^{-|\alpha|^2}\\
				&\le 2^k \left(\frac{k}{\ln(k/|\alpha|^2+1)}\right)^k\,,
			\end{align*}
			using a simple bound on the $k^{\text{th}}$ moment of the Poisson distribution; see, e.g., \cite{Ahle.2022}. For $\alpha=0$, the right-hand side is understood through the displayed maximum and gives the trivial bound by $2^k$.
		\end{proof}
    	\begin{proof}[Proof of \Cref{lem:moment-prop-bound}]
    		Denote the vector $F^{(k)}(t) \coloneqq (F_1^{(k)}(t), \dots, F_m^{(k)}(t))^T$ with entries defined in \Cref{assum-lrb}, i.e.~$F_i^{(k)}(t) \coloneqq \tr[e^{t\widetilde{\cL}}(\rho)f(N_i)]$. This expression translates to
    		\begin{align*}
    			\frac{d}{dt}F^{(k)}(t) \le A F^{(k)}(t) + Y_k\,,
    		\end{align*}
    		where the inequality is meant coordinate-wise, and $A_{ij} \coloneqq k\Gamma'\delta_{ij} + k\Gamma''E_{ij}$ with the $m \times m$ matrix $[E]_{ij} = 1$ if $1\leq\dist(i,j) \leq r$ and $[E]_{ij}=0$ otherwise. Additionally, we define the constant vector $Y_k \coloneqq k^{\xi k}\Gamma'''(1, \dots, 1)^T$ of dimension $m$. Recall that $\gamma$ bounds the number of sites within distance $r$ of any fixed site, and let $Z_i^{(k)}(t) := F_i^{(k)}(t) + \frac{k^{\xi k}\Gamma'''}{k \Gamma' }$. Then, for all $k \geq K$,
    		\begin{align*}
    			\frac{d}{dt}Z_i^{(k)}(t) &\le k\Gamma' F^{(k)}_i(t) + k\Gamma'' \sum_{1 \leq \dist(i,j) \leq r}F^{(k)}_j(t) + k^{\xi k}\Gamma''' \\
    			&= k\Gamma' Z_i^{(k)}(t) + k\Gamma'' \sum_{1 \leq \dist(i,j) \leq r}F^{(k)}_j(t) \\
    			&\le k\Gamma' Z_i^{(k)}(t) + k\Gamma'' \sum_{1 \leq \dist(i,j) \leq r}Z^{(k)}_j(t) \\
    			&\le (A Z^{(k)}(t))_{i}\, .
    		\end{align*}
    		Standard results on differential inequalities and the fact that $A$ is entry-wise positive \cite{Sremr.2006} show
    		\begin{align*}
    			F^{(k)}(t) &\le Z^{(k)}(t) \\
    			&\le e^{At}Z^{(k)}(0) \\
    			&\le e^{\|A\|_{\ell^\infty \to \ell^\infty}t} \, \Big\|F^{(k)}(0) + (k\Gamma')^{-1}Y_k\Big\|_{\ell^\infty}(1, \dots, 1)^T \\
    			&\le e^{k\Gamma t} \, \Big\|F^{(k)}(0) + (k\Gamma')^{-1}Y_k\Big\|_{\ell^\infty}(1, \dots, 1)^T\,,
    		\end{align*}
    		where $\gamma = \max_{i \in V} \{\gamma_i\}$ and $\Gamma=\Gamma' + \gamma \Gamma''$. One application of \Cref{lem:connection-moments} gives
    		\begin{equation}\label{eq:dgl-linear-upper-bound}
    			\begin{aligned}
    				M^{(k)}(t) &\le p4^{p+1}\, F^{(k)}(t) + 4^{p+1} \frac{k^{ k}}{3}(1, \dots, 1)^T \\
    				&\le \biggl( p4^{p+1} e^{k \Gamma t} \, \Big\|F^{(k)}(0) + (k\Gamma')^{-1} Y_k\Big\|_{\ell^\infty} + 4^{p+1}\frac{k^{k}}{3} \biggr)(1, \dots, 1)^T \\
    				&\le p4^{p+1} e^{k \Gamma t} \, \|M^{(k)}(0)\|_{\ell^\infty}(1, \dots, 1)^T + \underbrace{4^{p+1} \Big( e^{k \Gamma t} \frac{k^{\xi k}\Gamma'''}{k\Gamma'} + \frac{k^{ k}}{3}\Big)}_{=:\Upsilon_k(t)}(1, \dots, 1)^T\,,
    			\end{aligned}
    		\end{equation}
    		which proves the first inequality of the statement. Next, we use Young's inequality similar to \Cref{eq:sobolev-young-inequality}, which shows 
    		\begin{equation*}
    			\|\rho(t)\|_{W_R^{2k,1}}=\tr\Bigl[\sqrt{\rho(t)}\prod_{i\in R}(N_i+I)^k\sqrt{\rho(t)}\Bigr]\leq\frac{1}{|R|}\sum_{i\in R}\tr[\rho(t)(N_i+I)^{k|R|}]
    		\end{equation*}
    		so that the inequality in \Cref{eq:dgl-linear-upper-bound} shows 
    		\begin{equation}
    			\|\rho(t)\|_{W_R^{2k,1}}\leq p4^{p+1} e^{k|R| \Gamma t}\max_{j\in V}\|\rho\|_{W_j^{2k|R|,1}}+\Upsilon_{k|R|}(t)
    		\end{equation}
    		and finishes the proof.
    	\end{proof}
    	
    	In what follows, we denote the enlargement and reduction of $R$ by a distance $s \in \mathbb{N}$ as
    	\begin{equation}\label{eq:r-closure-interior-R}
    		R[s] \coloneqq \{ v \in V \, | \, \operatorname{dist}(v, R) \leq s \} \qquad \text{and} \qquad R[-s] \coloneqq \{ v \in V \, | \, \operatorname{dist}(v, R^c) > s \} \,,
    	\end{equation}
    	where $R^c=V\backslash R$ denotes the complement of $R$. The projections on the first $M \in \mathbb{N}$ Fock states on the sets above are defined by
    	\begin{equation}\label{eq:projection-extension}
    		P_{\pm s} \equiv P_{\pm s}^{(M)} \coloneqq \prod_{j \in R[\pm s]} P_j^{(M)} \,, \qquad \overline{P}_{\pm s} = I - P_{\pm s} \qquad \text{and} \qquad \mathcal{P}_{\pm s} \equiv \mathcal{P}_{\pm s}^{(M)} \coloneqq P_{\pm s} \cdot P_{\pm s} 
    	\end{equation}
    	with $ P_j^{(M)} \coloneqq \sum_{n \leq M} \ketbra{n}{n}_j $. The special case $ P_0 = P $ and $ \mathcal{P}_0 = \mathcal{P}$ is obviously included. Moreover, note that the generator splits into
    	\begin{equation}\label{eq:split-generator}
    		\widetilde{\mathcal{L}} = \widetilde{\mathcal{L}}_{R^c} + \widetilde{\mathcal{L}}_{\partial E_{R}} + \widetilde{\mathcal{L}}_{R} \,,
    	\end{equation}
	where $ \partial E_{R} \coloneqq \{ X \in \mathcal{V}^{(r)} \, | \, X \cap R \neq \emptyset \neq X \cap R^c \} $ denotes the boundary interaction supports.
    	\begin{figure}[ht]
    		\begin{center}
				\includegraphics[width=0.72\textwidth]{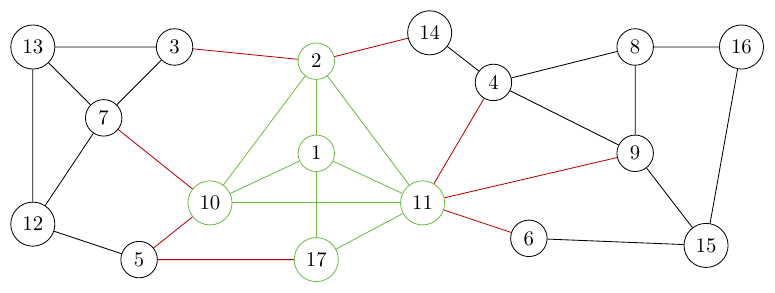}
    		\end{center}    
    		\caption{The different shadings illustrate the decomposition $(\ref{eq:split-generator})$ in the case of nearest-neighbor interactions. Here, the \textcolor{mygreen}{terms supported inside $ R $} correspond to the reduced operator $ \widetilde{\mathcal{L}}_R $, the \textcolor{black!30!red}{boundary interactions} connecting $ R $ and $ R^c $ to $ \widetilde{\mathcal{L}}_{\partial E_{R}} $, and the terms supported inside $ R^c $ to $ \widetilde{\mathcal{L}}_{R^c} $.}
    		\label{supp:fig1}
    	\end{figure}
    	With the help of the above splitting, the moment bound of \Cref{lem:moment-prop-bound} allows us to prove \Cref{thm:bosonicLRB}.
    	
		\begin{proof}[Proof of \Cref{thm:bosonicLRB}]
        	For a region $R$ large enough so that $T\subset R[-r]= \{v\in V\,|\,\dist(v,R^c)> r\}$ with $T\neq\emptyset$, we consider the difference 
        	\begin{equation*}
        		|\tr\Bigl[O_T\bigl(e^{t\widetilde{\cL}_R^{(M)}}-e^{t\widetilde{\cL}}\bigr)(\rho)\Bigr]|
        	\end{equation*}
        	for a local observable $O_T\in\cB(\cH)$ and the state $\rho\in\cT_f$. Note that by definition $\widetilde{\cL}_R^{(M)}$ is given by a Hamiltonian and Lindblad operators which are projected to a finite-dimensional subspace (see \Cref{eq:swap-projections}) so that \Cref{thm:generation-theory} is trivially satisfied and $e^{\widetilde{\cL}_R^{(M)}}$ defines a quantum Markov semigroup. In a first step, we apply Duhamel's formula, which is well-defined because of the boundedness of $\cL_R^{(M)}$ and the property $\widetilde{\cL}e^{st\widetilde{\cL}}=e^{st\widetilde{\cL}}\widetilde{\cL}$ imply the strong continuity of the integrand \cite[Lem.~B.15]{Engel.2000}. Then
        	\begin{equation}\label{eq:thm-main-difference}
        		\begin{aligned}
        			|\tr\Bigl[O_T\bigl(e^{t\widetilde{\cL}_R^{(M)}}-e^{t\widetilde{\cL}}\bigr)(\rho)\Bigr]|&=|t\int_{0}^{1}\tr\Bigl[O_Te^{(1-s)t\widetilde{\cL}_R^{(M)}}\bigl(\widetilde{\cL}_R^{(M)}-\widetilde{\cL}\bigr)e^{st\widetilde{\cL}}(\rho)\Bigr]ds|\\
        			&\overset{(1)}{=}|t\tr\Bigl[\int_{0}^{1}e^{(1-s)t(\widetilde{\cL}_R^{(M)})^\dagger}(O_T)\bigl(\widetilde{\cL}_R^{(M)}-\widetilde{\cL}\bigr)e^{st\widetilde{\cL}}(\rho)ds\Bigr]|\\
        			&\overset{(2)}{=}|t\int_{0}^{1}\tr\Bigl[e^{(1-s)t(\widetilde{\cL}_R^{(M)})^\dagger}(O_T)\bigl(\widetilde{\cL}_R^{(M)}-\widetilde{\cL}_{\partial E_R}-\widetilde{\cL}_{R}\bigr)e^{st\widetilde{\cL}}(\rho)\Bigr]ds|\\
        			&\leq t\int_{0}^{1}|\tr\Bigl[e^{(1-s)t(\widetilde{\cL}_R^{(M)})^\dagger}(O_T)\bigl(\widetilde{\cL}_R^{(M)}-\widetilde{\cL}_{\partial E_R}-\widetilde{\cL}_{R}\bigr)e^{st\widetilde{\cL}}(\rho)\Bigr]|ds\,,\\
        		\end{aligned}
        	\end{equation}
        	where we used in $(1)$ that $\widetilde{\cL}_R^{(M)}$ is bounded and admits a unique adjoint with respect to the Hilbert-Schmidt norm, and in $(2)$ that $e^{(1-s)t(\widetilde{\cL}_R^{(M)})^\dagger}(O_T)$ is local on $R$ as well as the decomposition presented in \Cref{eq:split-generator}. Next, we insert projections on the region $R[r]$:
        	\begin{equation}\label{eq:thm-main-steps}
        		\begin{aligned}
        			&|\tr\Bigl[O_Te^{(1-s)t\widetilde{\cL}_R^{(M)}}\bigl(\widetilde{\cL}_R^{(M)}-\widetilde{\cL} _{\partial E_R}-\widetilde{\cL}_{R}\bigr)e^{st\widetilde{\cL}}(\rho)\Bigr]|\\
        			&\qquad\leq|\tr\Bigl[O_Te^{(1-s)t\widetilde{\cL}_R^{(M)}}(I-\cP_r)\bigl(\widetilde{\cL}_R^{(M)}-\widetilde{\cL}_{\partial E_R}-\widetilde{\cL}_{R}\bigr)\cP_re^{st\widetilde{\cL}}(\rho)\Bigr]|&\qquad& (I)\\
        			&\qquad\qquad+|\tr\Bigl[O_Te^{(1-s)t\widetilde{\cL}_R^{(M)}}\bigl(\widetilde{\cL}_R^{(M)}-\widetilde{\cL}_{\partial E_R}-\widetilde{\cL}_{R}\bigr)(I-\cP_r)e^{st\widetilde{\cL}}(\rho)\Bigr]|&\qquad& (II)\\
        			&\qquad\qquad+|\tr\Bigl[O_Te^{(1-s)t\widetilde{\cL}_R^{(M)}}\cP_r\bigl(\widetilde{\cL}_R^{(M)}-\widetilde{\cL}_{\partial E_R}-\widetilde{\cL}_{R}\bigr)\cP_re^{st\widetilde{\cL}}(\rho)\Bigr]|\,.&\qquad& (III)
        		\end{aligned}
        	\end{equation}
        	
        	Then, we start bounding Term $(I)$: Here, note first that the operator $\bigl(\widetilde{\cL}_R^{(M)}-\widetilde{\cL}_{\partial E_R}-\widetilde{\cL}_{R}\bigr)\cP_r$ is bounded. This can be seen by the following identities for all $i\in V$ and $k,\ell\in\N$ with $k,\ell\leq d$
        	\begin{equation}\label{eq:swap-projections}
        		\begin{aligned}
        			P_i^{(M)}(\ad_i)^ka_i^\ell&=P_i^{(M)}(\ad_i)^ka_i^\ell P_i^{(M+\ell-k)} \qquad&\text{and}\qquad (\ad_i)^ka_i^\ell P_i^{(M)}&=P_i^{(M+k-\ell)}(\ad_i)^ka_i^\ell P_i^{(M)}\,,\\
        			\overline{P}_i^{(M)}(\ad_i)^ka_i^\ell&=\overline{P}_i^{(M)}(\ad_i)^ka_i^\ell \overline{P}_i^{(M-\ell+k)} \qquad&\text{and}\qquad (\ad_i)^ka_i^\ell \overline{P}_i^{(M)}&=\overline{P}_i^{(M-k+\ell)}(\ad_i)^ka_i^\ell \overline{P}_i^{(M)}\,,
        		\end{aligned}
        	\end{equation}
        	which implies that every Hamiltonian and Lindblad operator is projected to a finite-dimensional subspace and, in particular, $\bigl(\widetilde{\cL}_R^{(M)}-\widetilde{\cL}_{\partial E_R}-\widetilde{\cL}_{R}\bigr)\cP_r$ is bounded. Next, we use the notation $\rho_t=e^{t\widetilde{\cL}}(\rho)$ as well as $\cK=\widetilde{\cL}_R^{(M)}-\widetilde{\cL}_{\partial E_R}-\widetilde{\cL}_{R}$ and apply Hölder's inequality to show
        	\begin{equation*}
        		\begin{aligned}
        			|&\tr[O_Te^{(1-s)t\widetilde{\cL}_R^{(M)}}(I-\cP_r)\bigl(\widetilde{\cL}_R^{(M)}-\widetilde{\cL}_{\partial E_R}-\widetilde{\cL}_{R}\bigr)\cP_re^{st\widetilde{\cL}}(\rho)]|\\
        			&\qquad\leq\|O_T\|_{\infty}\|(I-\cP_r)\cK\cP_r(\rho_{st})\|_1\\
        			&\qquad\leq\|O_T\|_{\infty}\left(\|\overline{P_r}\cK(P_r\rho_{st}P_r)\overline{P_r}\|_1+\|\overline{P_r}\cK(P_r\rho_{st}P_r)P_r\|_1+\|P_r\cK(P_r\rho_{st}P_r)\overline{P_r}\|_1\right)\,.
        		\end{aligned}
        	\end{equation*}
        	Then, we label all element in $R[r]$ by $i\in\{1,...,|R[r]|\}$ and use
        	\begin{equation}\label{eq:comp-projection-union-bound}
        		\overline{P}_r^{(M)}=I-\prod_{i=1}^{|R[r]|}P_i^{(M)}=I-P_1+P_1\Bigl(I-\prod_{i=2}^{|R[r]|}P_i^{(M)}\Bigr)=\sum_{i=1}^{|R[r]|}\prod_{j=1}^{i-1} P_j^{(M)}\overline{P}_i^{(M)}\,,
        	\end{equation}
        	where the empty product is the identity, to apply the identities in \Cref{eq:swap-projections}:
        	\begin{equation}\label{eq:proof-lrb-I-block1}
        		\begin{aligned}
        			|&\tr[O_Te^{(1-s)t\widetilde{\cL}_R^{(M)}}(I-\cP_r)\bigl(\widetilde{\cL}_R^{(M)}-\widetilde{\cL}_{\partial E_R}-\widetilde{\cL}_{R}\bigr)\cP_re^{st\widetilde{\cL}}(\rho)]|\\
        			&\qquad\leq\|O_T\|_{\infty}\sum_{i=1}^{|R[r]|}\biggl(\|\prod_{j=1}^{i-1}P_j\overline{P}_i\cK(P_r\overline{P}_i^{(M-d)}\rho_{st}P_r)\overline{P}_r\|_1\\
        			&\qquad\qquad\qquad\qquad\qquad+\|\prod_{j=1}^{i-1}P_j\overline{P}_i\cK(P_r\overline{P}_i^{(M-d)}\rho_{st}P_r)P_r\|_1\\   &\qquad\qquad\qquad\qquad\qquad+\|P_r\cK(P_r\rho_{st}\overline{P}_i^{(M-d)}P_r)\prod_{j=1}^{i-1}P_j\overline{P}_i\|_1\biggr)\\
        			&\qquad\leq\|O_T\|_{\infty}\|\cK\cP_r\|_{1\rightarrow1}\sum_{i=1}^{|R[r]|}\biggl(2\|\overline{P}_i^{(M-d)}\rho_{st}\|_1+\|\rho_{st}\overline{P}_i^{(M-d)}\|_1\biggr)\,,
        		\end{aligned}
        	\end{equation}
        	where we applied the identities in \Cref{eq:swap-projections} using the degree assumption on $\cK$ given in \Cref{assum:local-GKSL}. Next, we apply the Cauchy–Schwarz inequality to show
        	\begin{equation}\label{eq:1-norm-to-probability}
        		\begin{aligned}
        			\|\rho_{st}\overline{P}_i^{(M-d)}\|_1\leq\|\sqrt{\rho_{st}}\|_2\,\|\sqrt{\rho_{st}}\overline{P}_i^{(M-d)}\|_2\leq\|\sqrt{\rho_{st}}\overline{P}_i^{(M-d)}\|_2=\sqrt{\tr\big[\overline{P}^{(M-d)}_i \rho_{st}\big]}\,,\\
        			\|\overline{P}_i^{(M-d)}\rho_{st}\|_1\leq\|\overline{P}_i^{(M-d)}\sqrt{\rho_{st}}\|_2\,\|\sqrt{\rho_{st}}\|_2\leq\|\overline{P}_i^{(M-d)}\sqrt{\rho_{st}}\|_2=\sqrt{\tr\big[\overline{P}_i^{(M-d)} \rho_{st}\big]}\,.
        		\end{aligned}
        	\end{equation}
        	Next, we denote by $N_i(t)$ the random variable corresponding to the measurement of the local number operator $N_i$, i.e.~$\mathbb{P}(N_i(t)=n)=\tr[\rho_t\ket{n}\bra{n}_i]$. Then, Markov's inequality proves
        	\begin{equation}\label{eq:markov-union-bound}
        		\mathbb{P}_{i,M}^{(t)}\coloneqq \mathbb{P}\Big(N_i(t)> M\Big)=\mathbb{P}\Big((N_i(t)+1)^k> (M+1)^k\Big)\le \frac{M_i^{(k)}(t)}{(M+1)^k}\,.
        	\end{equation}
        	Inserting the above bounds into \Cref{eq:proof-lrb-I-block1} shows
        	\begin{equation}\label{eq:proof-lrb-I-block2}
        		\begin{aligned}
        			|&\tr[O_Te^{(1-s)t\widetilde{\cL}_R^{(M)}}(I-\cP_r)\bigl(\widetilde{\cL}_R^{(M)}-\widetilde{\cL}_{\partial E_R}-\widetilde{\cL}_{R}\bigr)\cP_re^{st\widetilde{\cL}}(\rho)]|\\
        			&\qquad\leq3\|O_T\|_{\infty}\|\cK\cP_r\|_{1\rightarrow1}|R[r]|\max_{i\in R[r]}\sqrt{\tr[\overline{P_i}^{(M-d)}\rho_{st}]}\\
        			&\qquad\leq3\|O_T\|_{\infty}\|\cK\cP_r\|_{1\rightarrow1}|R[r]|\frac{\sqrt{\max_{i\in V}M_i^{(k)}(st)}}{(M-d+1)^{k/2}}\\
        			&\qquad\leq\|O_T\|_{\infty}|R[r]|6(1+\nu)|\cV_R^{(r)}|\Lambda d!^{r_{\exp}}(d/2+1)^{2r_{\exp}(d+2)}\frac{\sqrt{\max_{i\in V}M_i^{(k)}(st)}}{(M-d+1)^{k/2-r_{\exp}d}}
        		\end{aligned}
        	\end{equation}
        	where we used \Cref{lem:boundedness-projective-generator} as well as 
        	\begin{equation}\label{eq:pull-out-d}
        		\begin{aligned}
        			(M+1)^{r_{\exp}d}&=(d+1)^{r_{\exp}d}\Bigl(\frac{M-d}{d+1}+1\Bigr)^{r_{\exp}d}\\
        			&\leq(d+1)^{r_{\exp}d}(M-d+1)^{r_{\exp}d}\leq(d/2+1)^{2r_{\exp}d}(M-d+1)^{r_{\exp}d}
        		\end{aligned}
        	\end{equation}
        	in the last line. Next, we define the constant
        	\begin{equation}\label{eq:proof-lrb-constant-I}
        		c_{I}\coloneqq 6(1+\nu)\Lambda d!^{r_{\exp}}(d/2+1)^{2r_{\exp}(d+2)}
        	\end{equation}
        	so that the bounds in \Cref{eq:proof-lrb-I-block2} and \eqref{eq:proof-lrb-I-block1} show
        	\begin{equation}\label{eq:proof-lrb-bound-I}
        		\begin{aligned}
        			|&\tr[O_Te^{(1-s)t\widetilde{\cL}_R^{(M)}}(I-\cP_r)\bigl(\widetilde{\cL}_R^{(M)}-\widetilde{\cL}_{\partial E_R}-\widetilde{\cL}_{R}\bigr)\cP_re^{st\widetilde{\cL}}(\rho)]|\\
        			&\qquad\leq \|O_T\|_{\infty}|R[r]|\,|\cV_R^{(r)}|c_{I}\frac{\sqrt{\max_{i\in V}M_i^{(k)}(st)}}{(M-d+1)^{k/2-r_{\exp}d}}\,,
        		\end{aligned}
        	\end{equation}
        	which finishes the bound on Term I. 
        	
        	In the next step, we prove an upper bound on Term II: Again, we start applying Hölder's inequality and use the notation $\rho_t=e^{t\widetilde{\cL}}(\rho)$ and $\overline{\rho}_t=(I-\cP_r)(\rho_t)$ to show
        	\begin{equation}\label{eq:proof-lrb-II-block0}
        		\begin{aligned}
        			|&\tr[O_Te^{(1-s)t\widetilde{\cL}_R^{(M)}}\bigl(\widetilde{\cL}_R^{(M)}-\widetilde{\cL}_{\partial E_R}-\widetilde{\cL}_{R}\bigr)(I-\cP_r)e^{st\widetilde{\cL}}(\rho)]|\\
        			&\qquad\leq\|O_T\|_{\infty}\|\bigl(\widetilde{\cL}_R^{(M)}-\widetilde{\cL}_{\partial E_R}-\widetilde{\cL}_{R}\bigr)(\overline{\rho}_{st})\|_1\\
        			&\qquad\leq\|O_T\|_{\infty}\Bigl(\|\widetilde{\cL}_R^{(M)}(\overline{\rho}_{st})\|_1+\|(\widetilde{\cL}_{\partial E_R}+\widetilde{\cL}_{R})(\overline{\rho}_{st})\|_1\Bigr)\,.
        		\end{aligned}
        	\end{equation}
        	We start with bounding the first term. Here, we first apply Hölder's inequality and then \Cref{lem:boundedness-projective-generator} to show
        	\begin{equation*}
        		\begin{aligned}
        			\frac{\|\widetilde{\cL}_R^{(M)}(\overline{\rho}_{st})\|_1}{\|\overline{\rho}_{st}\|_1}&\leq \sum_{X\in\cV^{(r)}_R}\|\cH[PH_XP](\overline{\rho}_{st})\|_1 + \sum_{j=1}^{\nu(X)}\|\cL[L_{X,j}P](\overline{\rho}_{st})\|_1\\
        			&\leq \sum_{X\in\cV^{(r)}_R}2\|PH_XP\|_{\infty} + \sum_{j=1}^{\nu(X)}\|PL_{X,j}^\dagger L_{X,j}P\|_\infty+\|L_{X,j}^\dagger P\|_\infty\|PL_{X,j}\|_\infty\\
        			&\leq \sum_{X\in\cV^{(r)}_R}2\|PH_XP\|_{\infty} + \sum_{j=1}^{\nu(X)}\|PL_{X,j}^\dagger L_{X,j}P\|_\infty+\|L_{X,j}^\dagger P\|_\infty\|PL_{X,j}\|_\infty\\
        			&\leq2(1+\nu)|\cV_R^{(r)}|\Lambda d!^{r_{\exp}}(M+1)^{r_{\exp}d}(d/2+1)^{4r_{\exp}}
        		\end{aligned}
        	\end{equation*}
        	Then, we briefly recall that the procedure used in \Cref{eq:1-norm-to-probability} to \Cref{eq:markov-union-bound} shows
        	\begin{equation}\label{eq:comp-moment-prop}
        		\|\rho_{st}\overline{P}_i^{(M)}\|_1,\|\overline{P}_i^{(M)}\rho_{st}\|_1\leq\frac{\sqrt{\max_{i\in V}M_i^{(k)}(st)}}{(M+1)^{k/2}}
        	\end{equation}
        	so that an application of \Cref{eq:comp-projection-union-bound} and \eqref{eq:proof-lrb-constant-I} in combination with the above bounds prove
        	\begin{equation}\label{eq:proof-lrb-II-block1}
        		\|\widetilde{\cL}_R^{(M)}(\overline{\rho}_{st})\|_1\leq|R[r]|\,|\cV_R^{(r)}|2(1+\nu)\Lambda d!^{r_{\exp}}(d/2+1)^{4r_{\exp}}\frac{\sqrt{\max_{i\in V}M_i^{(k)}(st)}}{(M+1)^{k/2-r_{\exp}d}}\,.
        	\end{equation}
        	Next, we consider the second term in \Cref{eq:proof-lrb-II-block0}:
        	\begin{equation}\label{eq:proof-lrb-II-block2.1}
        		\begin{aligned}
        			\|&(\widetilde{\cL}_{\partial E_R}+\widetilde{\cL}_{R[r]})(\overline{\rho}_{st})\|_1\\
        			&\leq \Lambda(d/2+1)^{4r_{\exp}}d!^{r_{\exp}}|\cV_{R[r]}^{(r)}|\max_{X\in\cV^{(r)}_R}\Bigl((1+\frac{\nu}{2})\bigl(\|\overline{\rho}_{st} (N+I)_X^{\dd}\|_{1}+\|(N+I)_X^{\dd}\overline{\rho}_{st} \|_{1}\bigr)\\
        			&\qquad\qquad\qquad\qquad\qquad\qquad\qquad\qquad\qquad\qquad\qquad+\nu\|(N+I)_X^{\dd/2}\overline{\rho}_{st} (N+I)_X^{\dd/2}\|_{1}\Bigr)\,.
        		\end{aligned}
        	\end{equation}
        	where, we use the relative boundedness given in \Cref{lem:relative-bounds-number-op}. Next, we apply \Cref{eq:comp-projection-union-bound} to reduce the terms $\|\overline{\rho}_{st} (N+I)_X^{\dd}\|_{1}$, $\|(N+I)_X^{\dd}\overline{\rho}_{st} \|_{1}$ and $\|(N+I)_X^{\dd/2}\overline{\rho}_{st} (N+I)_X^{\dd/2}\|$ to a single mode:
        	\begin{equation}
        		\begin{aligned}
        			\|&\overline{\rho}_{st} (N+I)_X^{\dd}\|_{1}\\
        			&=\|P_r\rho_{st}\overline{P}_r (N+I)_X^{\dd}+\overline{P}_r\rho_{st}P_r(N+I)_X^{\dd}+\overline{P}_r\rho_{st}\overline{P}_r(N+I)_X^{\dd}\|_{1}\\
        			&\leq|R[r]|\max_{i\in R[r]}\Bigl(\|\rho_{st}\overline{P}_i (N+I)_X^{\dd}\|_{1}+\|\overline{P}_i\rho_{st}P_r(N+I)_X^{\dd}\|_{1}+\|\overline{P}_i\rho_{st}(N+I)_X^{\dd}\|_{1}\Bigr)\,.
        		\end{aligned}
        	\end{equation}
        	Then, we follow the procedure of \Cref{eq:1-norm-to-probability,eq:proof-lrb-I-block2}:
        	\begin{equation}\label{eq:bound-58}
        		\begin{aligned}
        			\|&\rho_{st}\overline{P}_i (N+I)_X^{\dd}\|_{1}+\|\overline{P}_i\rho_{st}P_r(N+I)_X^{\dd}\|_{1}+\|\overline{P}_i\rho_{st}(N+I)_X^{\dd}\|_{1}\\
        			&\overset{(1)}{\leq}\|\sqrt{\rho_{st}}\overline{P}_i (N+I)_X^{\dd}\|_{2}+(M+1)^{d|X|}\|\overline{P}_i\sqrt{\rho_{st}}\|_{2}+\|\overline{P}_i\sqrt{\rho_{st}}\|_{2}\|\sqrt{\rho_{st}}(N+I)_X^{\dd}\|_{2}\\
        			&=\sqrt{\tr[\overline{P}_i\rho_{st}(N+I)_X^{2\dd}]}+\sqrt{\tr[\rho_{st}]}\Bigl((M+1)^{d|X|}+\sqrt{\tr[\overline{P}_i\rho_{st}(N+I)_X^{2\dd}]}\Bigr)\\
        			&\overset{(2)}{\leq}\sqrt{\|\overline{P}_i\sqrt{\rho_{st}}\|_2\|\sqrt{\rho_{st}}(N+I)_X^{2\dd}\|_2}+\sqrt{\tr[\overline{P}_i\rho_{st}]}\Bigl((M+1)^{d|X|}+\sqrt{\tr[\rho_{st}(N+I)_X^{2\dd}]}\Bigr)\\
        			&=\sqrt[4]{\tr[\overline{P}_i\rho_{st}]\tr[\rho_{st}(N+I)_X^{4\dd}]}+\sqrt{\tr[\overline{P}_i\rho_{st}]}\Bigl((M+1)^{d|X|}+\sqrt{\tr[\rho_{st}(N+I)_X^{2\dd}]}\Bigr)\\
        			&\overset{(3)}{\leq}\sqrt[4]{\tr[\overline{P}_i\rho_{st}]}\Bigl(2\sqrt[4]{\tr[\rho_{st}(N+I)_X^{4\dd}]}+(M+1)^{r_{\exp}\,d}\Bigr)\,,
        		\end{aligned}
        	\end{equation}
        	where we used Hölder's inequality in $(1,3)$ and Cauchy's inequality in $(2)$. Similarly, 
        	\begin{equation*}
        		\begin{aligned}
        			\|(N+I)_X^{\dd}\overline{\rho}_{st}\|_{1}\leq|R[r]|\max_{i\in R[r]}\sqrt[4]{\tr[\overline{P}_i\rho_{st}]}\Bigl(2\sqrt[4]{\tr[\rho_{st}(N+I)_X^{4\dd}]}+(M+1)^{r_{\exp}\,d}\Bigr)
        		\end{aligned}
        	\end{equation*}
        	and
        	\begin{equation*}
        		\begin{aligned}
        			\|&(N+I)_X^{\dd/2}\overline{\rho}_{st} (N+I)_X^{\dd/2}\|_{1}\\
        			&\leq|R[r]|\max_{i\in R[r]}\biggl((M+1)^{d|X|/2}\Bigl(\|(N+I)_X^{\dd/2}\overline{P}_i\rho_{st}\|_{1}+\|\rho_{st}\overline{P}_i (N+I)_X^{\dd/2}\|_{1}\Bigr)\\
        			&\qquad\qquad\qquad\qquad\qquad\qquad\qquad\qquad\qquad\qquad\qquad\qquad\qquad+\tr[(N+I)_X^{\dd}\rho_{st}\overline{P}_i]\biggr)\\
        			&\leq|R[r]|\max_{i\in R[r]}\sqrt[4]{\tr[\overline{P}_i\rho_{st}]}\biggl(2(M+1)^{d|X|/2}\sqrt[4]{\tr[\rho_{st}(N+I)_X^{2\dd}]}+\sqrt[4]{\tr[\rho_{st}(N+I)_X^{4\dd}]}\biggr)\\
        			&\leq|R[r]|\max_{i\in R[r]}\sqrt[4]{\tr[\overline{P}_i\rho_{st}]\tr[\rho_{st}(N+I)_X^{4\dd}]}\biggl(2(M+1)^{dr_{\exp}/2}+1\biggr)\,.
        		\end{aligned}
        	\end{equation*}
        	To sum up, we achieve
        	\begin{equation}\label{eq:proof-lrb-II-block2.2}
        		\begin{aligned}
        			(1&+\frac{\nu}{2})\bigl(\|\overline{\rho}_{st} (N+I)_X^{\dd}\|_{1}+\|(N+I)_X^{\dd}\overline{\rho}_{st} \|_{1}\bigr)+\nu\|(N+I)_X^{\dd/2}\overline{\rho}_{st} (N+I)_X^{\dd/2}\|_{1}\\
        			&\leq (1+\frac{\nu}{2})2|R[r]|\max_{i\in R[r]}\sqrt[4]{\tr[\overline{P}_i\rho_{st}]}\Bigl(2\sqrt[4]{\tr[\rho_{st}(N+I)_X^{4\dd}]}+(M+1)^{r_{\exp}\,d}\Bigr)\\
        			&\qquad+\nu|R[r]|\max_{i\in R[r]}\sqrt[4]{\tr[\overline{P}_i\rho_{st}]\tr[\rho_{st}(N+I)_X^{4\dd}]}\biggl(2(M+1)^{dr_{\exp}/2}+1\biggr)\\
        			&\leq 4(1+\nu)((M+1)^{dr_{\exp}/2}+1)|R[r]|\max_{i\in R[r]}\sqrt[4]{\tr[\overline{P}_i\rho_{st}]\tr[\rho_{st}(N+I)_X^{4\dd}]}\\
        			&\overset{(1)}{\leq} 4(1+\nu)((M+1)^{dr_{\exp}/2}+1)|R[r]|\max_{i\in R[r]}\sqrt[4]{\tr[\overline{P}_i\rho_{st}]\max_{j\in X}M_j^{(4dr_{\exp})}(st)}\\
        			&\overset{(2)}{\leq} 4(1+\nu)((M+1)^{dr_{\exp}/2}+1)|R[r]|\max_{i\in R[r]}\sqrt[4]{\frac{\max_{i\in V}M_i^{(k)}(st)}{(M+1)^{k}}\max_{j\in X}M_j^{(4dr_{\exp})}(st)}\,,
        		\end{aligned}
        	\end{equation}
        	where, we used Young's inequality in step $(1)$, i.e.
        	\begin{equation*}
        		\begin{aligned}
        			\tr[\rho_{st}(N+I)_X^{4\dd}]&\leq\frac{1}{|X|}\sum_{j\in X}\tr[\rho_{st}(N_j+I)^{|X|4\dd}]\\
        			&\leq\max_{j\in X}\tr[\rho_{st}(N_j+I)^{4r_{\exp}d}]\\
        			&=\max_{j\in X}M_j^{(4r_{\exp}d)}(st)\,.
        		\end{aligned}
        	\end{equation*}
        	and the procedure in \Cref{eq:1-norm-to-probability} to \Cref{eq:markov-union-bound} (similar to \Cref{eq:comp-moment-prop}) in step $(2)$. Finally, we combine \Cref{eq:proof-lrb-II-block0}, \eqref{eq:proof-lrb-II-block1}, \eqref{eq:proof-lrb-II-block2.1} and \eqref{eq:proof-lrb-II-block2.2} to show
        	\begin{equation}\label{eq:proof-lrb-II-block3}
        		\begin{aligned}
        			|&\tr[O_Te^{(1-s)t\widetilde{\cL}_R^{(M)}}\bigl(\widetilde{\cL}_R^{(M)}-\widetilde{\cL}_{\partial E_R}-\widetilde{\cL}_{R}\bigr)(I-\cP_r)e^{st\widetilde{\cL}}(\rho)]|/\|O_T\|_{\infty}\\
        			&\leq|R[r]|\,|\cV_R^{(r)}|2(1+\nu)\Lambda d!^{r_{\exp}}(d/2+1)^{4r_{\exp}}\frac{\sqrt{\max_{i\in V}M_i^{(k)}(st)}}{(M+1)^{k/2-r_{\exp}d}}\\
        			&\qquad+\Lambda(d/2+1)^{4r_{\exp}}d!^{r_{\exp}}|\cV_{R[r]}^{(r)}|4(1+\nu)((M+1)^{dr_{\exp}/2}+1)|R[r]|\\
        			&\qquad\qquad\cdot \max_{i\in R[r]}\sqrt[4]{\frac{\max_{i\in V}M_i^{(k)}(st)}{(M+1)^{k}}\max_{j\in X}M_j^{(4dr_{\exp})}(st)}\\
        			&\leq4\nu\Lambda d!^{r_{\exp}}(d/2+1)^{4r_{\exp}}|\cV_R^{(r)}|\,|R[r]|\frac{\sqrt{\max_{i\in V}M_i^{(k)}(st)}}{(M+1)^{k/2-r_{\exp}d}}\\
        			&\qquad+16\nu\Lambda(d/2+1)^{4r_{\exp}}d!^{r_{\exp}}|\cV_{R[r]}^{(r)}|\,|R[r]| \frac{\sqrt[4]{\max_{i\in V}M_i^{(k)}(st)\max_{j\in X}M_j^{(4dr_{\exp})}(st)}}{(M+1)^{k/4-dr_{\exp}/2}}    
        		\end{aligned}
        	\end{equation}
        	With the following constant
        	\begin{equation}\label{eq:proof-lrb-constant-II}
        		\begin{aligned}
        			c_{II}&\coloneqq4\nu\Lambda d!^{r_{\exp}}(d/2+1)^{4r_{\exp}}
        		\end{aligned}
        	\end{equation}
        	we achieve the bound
        	\begin{equation}\label{eq:proof-lrb-bound-II}
        		\begin{aligned}
        			|&\tr[O_Te^{(1-s)t\widetilde{\cL}_R^{(M)}}\bigl(\widetilde{\cL}_R^{(M)}-\widetilde{\cL}_{\partial E_R}-\widetilde{\cL}_{R}\bigr)(I-\cP_r)e^{st\widetilde{\cL}}(\rho)]|/\|O_T\|_{\infty}\\
        			&\leq c_{II}|\cV_{R[r]}^{(r)}|\,|R[r]|\biggl(\frac{\sqrt{\max_{j\in V}M_j^{(k)}(st)}}{(M+1)^{k/2-r_{\exp}d}}+4\frac{\sqrt[4]{\max_{i\in V}M_i^{(k)}(st)\max_{j\in X}M_j^{(4dr_{\exp})}(st)}}{(M+1)^{k/4-dr_{\exp}/2}} \biggr)           
        		\end{aligned}
        	\end{equation}	
			Next, we upper bound the final Term III.
			\begin{equation*}
				|\tr[O_Te^{(1-s)t\widetilde{\cL}_R^{(M)}}\cP_r\bigl(\widetilde{\cL}_R^{(M)}-\widetilde{\cL}_{\partial E_R}-\widetilde{\cL}_{R}\bigr)\cP_re^{st\widetilde{\cL}}(\rho)]|\,.
			\end{equation*}
			First we observe that 
			\begin{equation*}
				\cP_r\bigl(\widetilde{\cL}_R^{(M)}-\widetilde{\cL}_{\partial E_R}-\widetilde{\cL}_{R}\bigr)\cP_r=-\cP_r\widetilde{\cL}_{\partial E_R}\cP_r
			\end{equation*}
			by definition of $\widetilde{\cL}_R^{(M)}$. Moreover (see \Cref{eq:projection-extension}), 
			\begin{equation*}
				\cP_r=\cP_{\partial r}\circ\cP_{-r}\qquad\text{defined via}\qquad\cP_{\partial r}=P_{\partial}\cdot P_{\partial}\quad\text{with}\quad P_{\partial}\coloneqq\prod_{j\in \partial_rR}P_j^{(M)}
			\end{equation*}
			where $\partial_rR\coloneqq R[r]\backslash R[-r]$ shows that
			\begin{equation*}
				\begin{aligned}
					\cP_r\widetilde{\cL}_{\partial E_R}\cP_r=\cP_\partial\widetilde{\cL}_{\partial E_R}\cP_\partial\cP_r\,.
				\end{aligned}
			\end{equation*}
			Then, we compare the above term $\cP_r\widetilde{\cL}_{\partial E_R}\cP_r$ to the following bounded generator defined by the interactions connecting $R[r]$ and $R[-r]$:
			\begin{equation}\label{eq:projected-generator-annulus}
				\widetilde{\cL}_{\partial E_R}^{(M,\mathrm{sym})}=\sum_{\substack{X\in\cV^{(r)}\\ X\cap R\neq\emptyset\neq X\cap R^c}}\left(\cH[P_{\partial}H_XP_{\partial}] + \sum_{j=1}^{\nu(X)}\cL[P_{\partial}L_{X,j}P_{\partial}]\right)\,,
			\end{equation}
			which results for $\rho^{(M)}_r\coloneqq \cP_r(\rho_{st})$ in 
			\begin{equation*}
				\begin{aligned}
					\bigl(\widetilde{\cL}_{\partial E_R}^{(M,\mathrm{sym})}&-\cP_\partial{\widetilde{\cL}}_{\partial E_R}\cP_\partial\bigr)\cP_r(\rho_{st})\\
					&=\bigl(\widetilde{\cL}_{\partial E_R}^{(M,\mathrm{sym})}\cP_\partial-\cP_\partial\widetilde{\cL}_{\partial E_R}\cP_\partial\bigr)(\rho^{(M)}_r)\\
					&=-\frac{1}{2}\sum_{\substack{X\in\cV^{(r)}\\ X\cap R\neq\emptyset\neq X\cap R^c}}\sum_{j=1}^{\nu(X)}\biggl(\{P_{\partial}L_{X,j}^\dagger P_{\partial} L_{X,j}P_{\partial},\,\rho^{(M)}_r\}-\{P_{\partial}L_{X,j}^\dagger L_{X,j}P_{\partial},\,\rho^{(M)}_r\}\biggr)\\
					&=\frac{1}{2}\sum_{\substack{X\in\cV^{(r)}\\ X\cap R\neq\emptyset\neq X\cap R^c}}\sum_{j=1}^{\nu(X)}\{P_{\partial}L_{X,j}^\dagger\overline{P}_{\partial} L_{X,j}P_{\partial},\,\rho^{(M)}_r\}\\
					&=\frac{1}{2}\sum_{\substack{X\in\cV^{(r)}\\ X\cap R\neq\emptyset\neq X\cap R^c}}\sum_{j=1}^{\nu(X)}\sum_{i_1,i_2=1}^{|\partial_rR|}\Bigl(P_{\partial}L_{X,j}^\dagger\prod_{\ell=1}^{i_1-1} P_\ell^{(M)}\overline{P}_{i_1}^{(M)} L_{X,j}P_{\partial}\overline{P}_{i_1}^{(M-d/2)}\rho^{(M)}_r\\
					&\qquad\qquad\qquad\qquad\qquad\qquad\qquad+\rho^{(M)}_r\overline{P}_{i_2}^{(M-d/2)}P_{\partial}L_{X,j}^\dagger\prod_{\ell=1}^{i_2-1} P_\ell^{(M)}\overline{P}_{i_2}^{(M)} L_{X,j}P_{\partial}\Bigr)
				\end{aligned}
			\end{equation*}
			where, we used in the last line \Cref{eq:swap-projections} in combination with \Cref{eq:comp-projection-union-bound}. Then, we upper bound the above difference by the same procedure used to upper bound Term I (see \Cref{eq:proof-lrb-I-block1} to \eqref{eq:markov-union-bound}):
			\begin{equation*}
				\begin{aligned}
					\|\bigl(&\widetilde{\cL}_{\partial E_R}^{(M,\mathrm{sym})}-\cP_\partial\widetilde{\cL}_{\partial E_R}\cP_\partial\bigr)\cP_r(\rho_{st})\|_1\\
					&\leq\frac{1}{2}\sum_{\substack{X\in\cV^{(r)}\\ X\cap R\neq\emptyset\neq X\cap R^c}}\sum_{j=1}^{\nu(X)}\sum_{i_1,i_2=1}^{|\partial_rR|}\|P_{\partial}L_{X,j}^\dagger\|_\infty\|L_{X,j}P_{\partial}\|_\infty\Bigl(\|\overline{P}_{i_1}^{(M-d/2)}\,\rho^{(M)}_r\|_1+\|\rho^{(M)}_r\overline{P}_{i_2}^{(M-d/2)}\|_{\infty}\Bigr)\\
					&\overset{(1)}{\leq}\frac{1}{2}\nu|\partial E_R|\,|\partial_rR|\Lambda(d/2+1)^{4r_{\exp}}(d/2)!^{2r_{\exp}}(M+1)^{r_{\exp}d}\\
					&\qquad\qquad\qquad\qquad\qquad\qquad\cdot\max_{i_1,i_2\in \partial_rR}\Bigl(\|\overline{P}_{i_1}^{(M-d/2)}\,\rho^{(M)}_r\|_1+\|\rho^{(M)}_r\overline{P}_{i_2}^{(M-d/2)}\|_{1}\Bigr)\\
					&\overset{(2)}{\leq}|\partial E_R|\,|\partial_rR|\nu\Lambda(d/2+1)^{r_{\exp}(4+d)}(d/2)!^{2r_{\exp}}\frac{\sqrt{\max_{i\in V}M_i^{(k)}(st)}}{(M-d/2+1)^{k/2-r_{\exp}d}}\\
					&{\leq}|\partial E_R|\,|\partial_rR|c_{III}\frac{\sqrt{\max_{i\in V}M_i^{(k)}(st)}}{(M-d/2+1)^{k/2-r_{\exp}d}}\,.
				\end{aligned}
			\end{equation*}
			with
			\begin{equation}\label{eq:proof-lrb-constant-III.1}
				c_{III}\coloneqq \nu\Lambda(d/2+1)^{r_{\exp}(4+d)}(d/2)!^{2r_{\exp}}\,.
			\end{equation}
			In detail, we applied \Cref{lem:boundedness-projective-generator} in $(1)$ and  \Cref{eq:comp-moment-prop} combined with \Cref{eq:pull-out-d} in $(2)$. With the help of the above inequality, we show
			\begin{equation}\label{eq:proof-lrb-III-block1}
				\begin{aligned}
					|&\tr[O_Te^{(1-s)t\widetilde{\cL}_R^{(M)}}\cP_r\bigl(\widetilde{\cL}_R^{(M)}-\widetilde{\cL}_{\partial E_R}-\widetilde{\cL}_{R}\bigr)\cP_re^{st\widetilde{\cL}}(\rho)-O_Te^{(1-s)t\widetilde{\cL}_R^{(M)}}\widetilde{\cL}_{\partial E_R}^{(M,\mathrm{sym})}\cP_re^{st\widetilde{\cL}}(\rho)]|\\
					&=|\tr[O_Te^{(1-s)t\widetilde{\cL}_R^{(M)}}\cP_\partial\widetilde{\cL}_{\partial E_R}\cP_\partial\cP_re^{st\widetilde{\cL}}(\rho)]-\tr[O_Te^{(1-s)t\widetilde{\cL}_R^{(M)}}\widetilde{\cL}_{\partial E_R}^{(M,\mathrm{sym})}\cP_re^{st\widetilde{\cL}}(\rho)]|\\
					&\leq\|O_T\|_\infty\|\bigl(\widetilde{\cL}_{\partial E_R}^{(M,\mathrm{sym})}-\cP_\partial\widetilde{\cL}_{\partial E_R}\cP_\partial\bigr)\cP_re^{st\widetilde{\cL}}(\rho)\|_1\\
					&\leq\|O_T\|_\infty|\partial E_R|\,|\partial_rR|\,c_{III}\frac{\sqrt{\max_{i\in V}M_i^{(k)}(st)}}{(M-d/2+1)^{k/2-r_{\exp}d}}\,.
				\end{aligned}
			\end{equation}
			This brings us to the final tool used in the proof of the present theorem. First, we note that the Hamiltonian and Lindblad operators of $(\widetilde{\cL}_{\partial E_R}^{(M,\mathrm{sym})})^\dagger$ are projected to a finite-dimensional subspace, which directly implies the boundedness of the operator. Moreover, it satisfies
			\begin{equation*}
				\widetilde{\cL}_{\partial E_R}^{(M,\mathrm{sym})\dagger}(O_T)=0
			\end{equation*}
			for all $T\subset R[-r]$.
			Since we have now reduced our problem to a bounded operator in GKSL form, we can use the dissipative Lieb--Robinson bound in \cite[Theorem 2]{Nachtergaele.2011lrb}. We have that
			\begin{equation}\label{eq:first-lrb}
				\Big\| \widetilde{\cL}_{\partial E_R}^{(M,\mathrm{sym})\dagger} e^{s\widetilde{\cL}^{(M)\dagger}_{R}}(O_T)\Big\|_\infty\leq \frac{\|\widetilde{\cL}_{\partial E_R}^{(M,\mathrm{sym})\dagger}\|_{\operatorname{cb}}\|O_T\|_\infty}{C_\mu}\,e^{s\,\|\widetilde{\cL}_{R}^{(M)\dagger}\|_{\mu}C_\mu}\sum_{x\in \cup\partial E_R}\sum_{y\in T}F^\downarrow_{\mu}(\operatorname{dist}(x,y))\,,
			\end{equation}
			where $\cup\partial E_R=\bigcup_{y\in \partial E_R}y$,
			\begin{equation*}
				\|\widetilde{\cL}_{\partial E_R}^{(M,\mathrm{sym})\dagger}\|_{\operatorname{cb}}=\max_{n\in\N}\|\widetilde{\cL}_{\partial E_R}^{(M,\mathrm{sym})\dagger}\otimes I_n\|_{\infty\rightarrow\infty}
			\end{equation*}
			is the completely bounded norm and 
			\begin{equation*}
				\|\widetilde{\cL}_{R}^{(M)\dagger}\|_{\mu}=\max_{x,y\in V}\sum_{X\in\cV_R^{(r)}\,|\,x,y\in X}\frac{\|\Psi_{X}\|_{\operatorname{cb}}}{F^\downarrow_\mu(d(x,y))}
			\end{equation*}
			defined by 
			\begin{equation*}
				\widetilde{\cL}_{R}^{(M)\dagger}=\sum_{X\in\cV^{(r)}_R}\Bigl(\cH[PH_XP] + \sum_{j=1}^{\nu(X)}\cL[L_{X,j}P]\Bigr)\eqqcolon\sum_{X\in\cV_R^{(r)}}\Psi_X \,.
			\end{equation*}
			Moreover, recall that $F^{\downarrow}:[0,\infty)\to (0,\infty)$ is a non-increasing function satisfying
			\begin{align*}
				&\|F^{\downarrow}\|\coloneqq \max_{x\in V}\sum_{y\in V}F^{\downarrow}(\operatorname{dist}(x,y))<\infty\qquad C\coloneqq \max_{x,y\in V}\,\sum_{z\in V}\,\frac{F^{\downarrow}(\operatorname{dist}(x,z))\,F^{\downarrow}(\operatorname{dist}(y,z))}{F^{\downarrow}(\operatorname{dist}(x,y))}<\infty\,.
			\end{align*}
			For any $\mu > 0$, we define the function $F^{\downarrow}_\mu(x) = e^{-\mu x} F^{\downarrow}(x)$, with $\|F^{\downarrow}_\mu\| \leq \|F^{\downarrow}\|$ and $C_\mu \leq C$ (see Assumption 1 in \cite{Nachtergaele.2011lrb} for details). By definition of $F^\downarrow_\mu$ and $\cup\partial E_R\subset\partial_rR$, the bound in \Cref{eq:first-lrb} can be rewritten to
			\begin{equation}\label{eq:second-lrb}
				\begin{aligned}
					\Big\|\widetilde{\cL}_{\partial E_R}^{(M,\mathrm{sym})\dagger} &e^{s\widetilde{\cL}^{(M)\dagger}_{R}}(O_T)\Big\|_\infty\\
					&\leq \frac{\|\widetilde{\cL}_{\partial E_R}^{(M,\mathrm{sym})\dagger}\|_{\operatorname{cb}}\|O_T\|_\infty}{C_\mu}\,\|F^\downarrow\|\min\{|T|,|\partial_rR|\}e^{-\mu \bigl(d(T,\partial_rR)-\frac{\|\widetilde{\cL}_{R}^{(M)\dagger}\|_{\mu} C_\mu}{\mu}s\bigr)}\,.
				\end{aligned}
			\end{equation}
			In the next steps, we upper bound the above quantities. First, the cb-norm can be upper bounded by using the triangle inequality as well as the fact that $\|\cdot\|_{\infty}$ is an operator norm with respect to the induced scalar product norm on $\cH$. Therefore, $\|\cdot\|_\infty$ is sub-multiplicative, so that for all $X \in \cB(\cH \otimes \cH_n)$ for $n \in \N$ and any $n$-dimensional Hilbert space, we have
			\begin{equation*}
				\begin{aligned}
					\| \widetilde{\cL}_{\partial E_R}^{(M,\mathrm{sym})\dagger} & \otimes I_n(Y) \|_{\infty} \\
					& = \Bigl\| \sum_{X \in \partial E_R} \cH[P_{\partial} H_X P_{\partial} \otimes I_n](Y) 
					+ \sum_{j=1}^{\nu(X)} \cL[P_{\partial} L_{X,j} P_{\partial} \otimes I_n](Y) \Bigr\|_\infty \\
					& \leq 2|\partial E_R| \max_{X \in \partial E_R} \Bigl( 
					\|P_{\partial} H_X P_{\partial} \otimes I_n\|_\infty \\
					& \qquad\qquad\qquad\qquad + \nu\max_{j \in \{1, \dots, \nu(X)\}} 
					\|P_{\partial} L_{X,j} P_{\partial} \otimes I_n\|_\infty 
					\|P_{\partial} L_{X,j}^\dagger P_{\partial} \otimes I_n\|_\infty 
					\Bigr) \|Y\|_\infty \\
					& \leq 2|\partial E_R| \max_{X \in \partial E_R} \Bigl( 
					\|P_{\partial} H_X P_{\partial}\|_\infty 
					+ \nu\max_{j \in \{1, \dots, \nu(X)\}} 
					\|P_{\partial} L_{X,j} P_{\partial}\|_\infty 
					\|P_{\partial} L_{X,j}^\dagger P_{\partial}\|_\infty 
					\Bigr) \|Y\|_\infty,
				\end{aligned}
			\end{equation*}
			where we used the definition in \Cref{eq:projected-generator-annulus}. Then, we can apply \Cref{lem:boundedness-projective-generator} to find
			\begin{equation}\label{eq:proof-lrb-constant-III.21}
				\begin{aligned}
					\|\widetilde{\cL}_{\partial E_R}^{(M,\mathrm{sym})\dagger}\|_{\mathrm{cb}}&\leq2(1+\nu)|\partial E_R|\Lambda(d/2+1)^{4r_{\exp}}d!^{r_{\exp}}(M+1)^{r_{\exp}d}\\
					&\eqqcolon|\partial E_R|\kappa(M+1)^{r_{\exp}d}
					\,.
				\end{aligned}
			\end{equation}
            for a constant $\kappa\geq0$. Next, we continue with the norm $\|\widetilde{\cL}_{R}^{(M)\dagger}\|_{\mu}$ for which we repeat the same calculation above to apply  \Cref{lem:boundedness-projective-generator} and show
			\begin{equation}\label{eq:proof-lrb-constant-III.22}
				\begin{aligned}
					\|\widetilde{\cL}_{R}^{(M)\dagger}\|_{\mu}&=\max_{x,y\in V}\sum_{X\in\cV_R^{(r)}\,|\,x,y\in X}\frac{\|\Psi_{X}\|_{\operatorname{cb}}}{F^\downarrow_\mu(d(x,y))}\\
					&\leq\max_{x,y\in V}\sum_{X\in\cV_R^{(r)}\,|\,x,y\in X}\frac{2(1+\nu)\Lambda(d/2+1)^{4r_{\exp}}d!^{r_{\exp}}(M+1)^{r_{\exp}d}}{F_\mu^\downarrow(d(x,y))}\\
					&\leq\gamma\frac{2(1+\nu)\Lambda(d/2+1)^{4r_{\exp}}d!^{r_{\exp}}(M+1)^{r_{\exp}d}}{F_\mu^\downarrow(r)}\\
					&\eqqcolon \frac{\gamma\,\kappa}{F_\mu^\downarrow(r)}(M+1)^{r_{\exp}d}
					\,.
				\end{aligned}
			\end{equation}
            Here, $\gamma$ is the constant from \Cref{assum:local-GKSL}; it bounds both the number of local supports meeting any site and, after enlargement if necessary, the number of sites within distance $r$ of any fixed site. Combining the bounds in \Cref{eq:first-lrb}, \eqref{eq:second-lrb}, \eqref{eq:proof-lrb-constant-III.21} and \eqref{eq:proof-lrb-constant-III.22}, we obtain the general weak Lieb--Robinson information propagation bound of the statement:
			\begin{equation}\label{eq:proof-lrb-III-block2}
				\begin{aligned}
					\Big\| &\widetilde{\cL}_{\partial E_R}^{(M,\mathrm{sym})\dagger}e^{st\widetilde{\cL}^{(M)\dagger}_{R}}(O_T)\Big\|_\infty\\
					&\leq \frac{|\partial E_R|\kappa\,(M+1)^{r_{\exp}d}\|O_T\|_\infty}{C_\mu}\,e^{st\,\frac{\gamma\kappa(M+1)^{r_{\exp}d}}{F^\downarrow_\mu(r)}C_\mu}\sum_{x\in \partial_rR}\sum_{y\in T}F^\downarrow_{\mu}(\operatorname{dist}(x,y))\\
					&\leq \frac{|\partial E_R|\kappa\,(M+1)^{r_{\exp}d}\|O_T\|_\infty}{C_\mu}\,\|F^\downarrow\|\,\min\{|T|,|\partial_rR|\}\,e^{-\mu \bigl(d(T,\partial_rR)-\frac{\gamma\kappa C_\mu}{\mu F_\mu^\downarrow(r)}(M+1)^{r_{\exp}d}s\bigr)}\,.
				\end{aligned}
			\end{equation}
			with constant
			\begin{equation}\label{eq:proof-lrb-constant-III.2}
				\begin{aligned}
					\kappa=2(1+\nu)\Lambda(d/2+1)^{4r_{\exp}}d!^{r_{\exp}}\,.
				\end{aligned}
			\end{equation}
			Together with \Cref{eq:proof-lrb-III-block1}, this proves the following upper bound on Term III in \Cref{eq:thm-main-steps}:
			\begin{equation}\label{eq:proof-lrb-bound-III}
				\begin{aligned}
					|&\tr[O_Te^{(1-s)t\widetilde{\cL}_R^{(M)}}\cP_r\bigl(\widetilde{\cL}_R^{(M)}-\widetilde{\cL}_{\partial E_R}-\widetilde{\cL}_{R}\bigr)\cP_re^{st\widetilde{\cL}}(\rho)]|\\
					&\qquad\leq\|O_T\|_\infty|\partial E_R|\,|\partial_rR|\,c_{III}\frac{\sqrt{\max_{i\in V}M_i^{(k)}(st)}}{(M-d/2+1)^{k/2-r_{\exp}d}}\\
					&\qquad\qquad+\frac{|\partial E_R|\kappa\,(M+1)^{r_{\exp}d}\|O_T\|_\infty}{C_\mu}\,e^{st\,\frac{\gamma\kappa(M+1)^{r_{\exp}d}}{F^\downarrow_\mu(r)}}\sum_{x\in \partial_rR}\sum_{y\in T}F^\downarrow_{\mu}(\operatorname{dist}(x,y))\,.
				\end{aligned}
			\end{equation}
			Finally, we combine the bounds of \Cref{eq:proof-lrb-bound-I}, \eqref{eq:proof-lrb-bound-II} and \eqref{eq:proof-lrb-bound-III} via \Cref{eq:thm-main-steps}, which show
			\begin{equation*}
				\begin{aligned}
					|&\tr[O_Te^{(1-s)t\widetilde{\cL}_R^{(M)}}\bigl(\widetilde{\cL}_R^{(M)}-\widetilde{\cL} _{\partial E_R}-\widetilde{\cL}_{R}\bigr)e^{st\widetilde{\cL}}(\rho)]|/\|O_T\|_{\infty}\\
					&\leq|R[r]|\,|\cV_R^{(r)}|c_{I}\frac{\sqrt{\max_{i\in V}M_i^{(k)}(st)}}{(M-d+1)^{k/2-r_{\exp}d}}\\
					&\qquad+c_{II}|\cV_{R[r]}^{(r)}|\,|R[r]|\biggl(\frac{\sqrt{\max_{j\in V}M_j^{(k)}(st)}}{(M+1)^{k/2-r_{\exp}d}}+4\frac{\sqrt[4]{\max_{i\in V}M_i^{(k)}(st)\max_{j\in X}M_j^{(4dr_{\exp})}(st)}}{(M+1)^{k/4-dr_{\exp}/2}} \biggr)\\
					&\qquad+|\partial E_R|\,|\partial_rR|\,c_{III}\frac{\sqrt{\max_{i\in V}M_i^{(k)}(st)}}{(M-d/2+1)^{k/2-r_{\exp}d}}\\
					&\qquad\qquad+\frac{|\partial E_R|\kappa\,(M+1)^{r_{\exp}d}}{C_\mu}\,e^{st\,\frac{\gamma\kappa(M+1)^{r_{\exp}d}}{F^\downarrow_\mu(r)}}\sum_{x\in \partial_rR}\sum_{y\in T}F^\downarrow_{\mu}(\operatorname{dist}(x,y))\\
					&\leq\max\Bigl\{|R[r]|\,|\cV_R^{(r)}|c_{I}, 4c_{II}|\cV_{R[r]}^{(r)}|\,|R[r]|, |\partial E_R|\,|\partial_rR|\,c_{III}\Bigr\}\\
					&\qquad\cdot\biggl(\frac{\sqrt{\max_{j\in V}M_j^{(k)}(st)}}{(M-d+1)^{k/2-r_{\exp}d}}+\frac{\sqrt[4]{\max_{i\in V}M_i^{(k)}(st)\max_{j\in X}M_j^{(4dr_{\exp})}(st)}}{(M+1)^{k/4-dr_{\exp}/2}} \biggr)\\
					&\qquad\qquad+\frac{|\partial E_R|\kappa\,(M+1)^{r_{\exp}d}}{C_\mu}\,e^{st\,\frac{\gamma\kappa(M+1)^{r_{\exp}d}}{F^\downarrow_\mu(r)}}\sum_{x\in \partial_rR}\sum_{y\in T}F^\downarrow_{\mu}(\operatorname{dist}(x,y))\\
					&\leq c\max\{|\cV_{R[r]}^{(r)}|\,|R[r]|, |\partial E_R|\,|\partial_rR|\}\frac{\sqrt{\max\bigl\{\max_{j\in V}M_j^{(k)}(st),\,\max_{j\in X}M_j^{(4dr_{\exp})}(st)\bigr\}}}{(M-d+1)^{k/4-r_{\exp}d/2}}\\
					&\qquad\qquad+\frac{|\partial E_R|\kappa\,(M+1)^{r_{\exp}d}}{C_\mu}\,e^{st\,\frac{\gamma\kappa(M+1)^{r_{\exp}d}}{F^\downarrow_\mu(r)}C_\mu}\sum_{x\in \partial_rR}\sum_{y\in T}F^\downarrow_{\mu}(\operatorname{dist}(x,y))
				\end{aligned}
			\end{equation*}
			with constants given in \Cref{eq:proof-lrb-constant-I}, \eqref{eq:proof-lrb-constant-II}, \eqref{eq:proof-lrb-constant-III.1} and \eqref{eq:proof-lrb-constant-III.2}
			\begin{equation}
				\begin{aligned}
					c_{I}&= 6(1+\nu)\Lambda d!^{r_{\exp}}(d/2+1)^{2r_{\exp}(d+2)}\\
					c_{II}&=4\nu\Lambda d!^{r_{\exp}}(d/2+1)^{4r_{\exp}}\\
					c_{III}&= \nu\Lambda(d/2+1)^{r_{\exp}(4+d)}(d/2)!^{2r_{\exp}}\\
					\kappa&=2(1+\nu)\Lambda(d/2+1)^{4r_{\exp}}d!^{r_{\exp}}\,.
				\end{aligned}
			\end{equation}
			satisfying
			\begin{equation}\label{eq:proof-lrb-constant-summary-c}
				c_{I},4c_{II},c_{III}\leq16(1+\nu)\Lambda d!^{r_{\exp}}(d/2+1)^{2r_{\exp}(d+2)}\eqqcolon c\,.
			\end{equation}
			Next, we apply the moment propagation bound proven in \Cref{lem:moment-prop-bound}, i.e,
			\begin{align*}
				M_j^{(k)}(st) \le p4^{p+1} e^{k \Gamma st} \max_{i \in V} M_i^{(k)}(0) + \Upsilon_k(st)\,,
			\end{align*}
			with the constant 
			\begin{equation}\label{eq:proof-lrb-constant-IV}
				\begin{aligned}
					\Upsilon_k(st)=4^{p+1} \Big( e^{k \Gamma st} \frac{k^{\xi k}\Gamma'''}{k\Gamma'} + \frac{k^{ k}}{3}\Big)\leq \frac{c_{IV}}{p}e^{k \Gamma st}k^{\xi k}
				\end{aligned}
			\end{equation}
			for constants $c_{IV}$ so that
			\begin{equation}\label{eq:moment-prop-bound-adapted}
				M_j^{(k)}(st) \le c_{IV} e^{k \Gamma st} \bigl(\max_{i \in V} M_i^{(k)}(0) + k^{\xi k}\bigr)\,.
			\end{equation}
			Taking $k=\hat{k}\coloneqq \max\{K,4dr_{\exp}\}$ and using $(N_i+I)^{\hat{k}}\ge (N_i+I)^{4dr_{\exp}}$, this proves
			\begin{equation*}
				\begin{aligned}
					|&\tr[O_Te^{(1-s)t\widetilde{\cL}_R^{(M)}}\bigl(\widetilde{\cL}_R^{(M)}-\widetilde{\cL} _{\partial E_R}-\widetilde{\cL}_{R}\bigr)e^{st\widetilde{\cL}}(\rho)]|/\|O_T\|_{\infty}\\
					&\leq c\max\{|\cV_{R[r]}^{(r)}|\,|R[r]|, |\partial E_R|\,|\partial_rR|\}\frac{\sqrt{c_{IV} e^{\hat{k} \Gamma st} \bigl(\max_{i \in V} M_i^{(\hat{k})}(0) + \hat{k}^{\xi \hat{k}}\bigr)}}{(M-d+1)^{\hat{k}/4-r_{\exp}d/2}}\\
					&\qquad\qquad+\frac{|\partial E_R|\kappa\,(M+1)^{r_{\exp}d}}{C_\mu}\,e^{st\,\frac{\gamma\kappa(M+1)^{r_{\exp}d}}{F^\downarrow_\mu(r)}C_\mu}\sum_{x\in \partial_rR}\sum_{y\in T}F^\downarrow_{\mu}(\operatorname{dist}(x,y))\,.
				\end{aligned}
			\end{equation*}
            In the next step, we find upper bounds on the sets $|\cV_{R[r]}^{(r)}|$, $|R[r]|$, $|\partial E_R|$ and $|\partial_rR|$:
            \begin{equation}\label{eq:hypergraph-bounds}
				\begin{aligned}
					|R[r]|&=|\{v\in V\,|\,\dist(v,R)\leq r\}|\leq|R|\gamma^r\\
					|\cV_{R[r]}^{(r)}|&=|\{W\in\cV^{(r)} \subset 2^V\,|\,W\subset R[r]\}|\leq \gamma |R[r]|\leq\gamma^{r+1}|R|\leq\gamma^{r+1}|R|^r\\
					|\partial E_R|&=\bigl|\{X\in\cV^{(r)}\,|\,X\cap R\neq\emptyset\neq X\cap R^c\}\bigr|\leq|\cV_{R[r]}^{(r)}|\leq\gamma^{r+1}|R|\leq\gamma^{r+1}|R|^r\\
					|\partial_r R|&=|R[r]\setminus R[-r]|\leq|R[r]|\leq|R|\gamma^r
				\end{aligned}
			\end{equation}
            With the help of the above bounds, we prove the first bound of \Cref{thm:bosonicLRB}
			\begin{equation*}
				\begin{aligned}
                    |\tr\Bigl[&O_T\bigl(e^{t\widetilde{\cL}_R^{(M)}}-e^{t\widetilde{\cL}}\bigr)(\rho)\Bigr]|/\|O_T\|_{\infty}\\
                    &\leq t\int_0^1 c\,\gamma^{2r+1}|R|^{r+1}\sqrt{c_{IV}}e^{\hat{k} \Gamma st/2} \frac{\sqrt{\max_{i \in V} M_i^{(\hat{k})}(0) + \hat{k}^{\xi \hat{k}}}}{(M-d+1)^{\hat{k}/4-r_{\exp}d/2}}ds\\
					&\qquad\qquad+t\int_0^1\frac{\gamma^{r+1}|R|^r\kappa\,(M+1)^{r_{\exp}d}}{C_\mu}\,e^{st\,\frac{\gamma\kappa(M+1)^{r_{\exp}d}}{F^\downarrow_\mu(r)}C_\mu}\sum_{x\in \partial_rR}\sum_{y\in T}F^\downarrow_{\mu}(\operatorname{dist}(x,y))ds\\
					&\leq c_1 |R|^{r+1}(e^{\hat{k} \Gamma t}-1)\frac{\sqrt{\max_{i \in V} M_i^{(\hat{k})}(0) + \hat{k}^{\xi \hat{k}}}}{(M-d+1)^{\hat{k}/4-r_{\exp}d/2}}\\
                    &\qquad\qquad+c_2|R|^r\frac{F^\downarrow_\mu(r)}{C_\mu^2}\,e^{t\xi_2(M+1)^{r_{\exp}d}\frac{C_\mu}{F^\downarrow_\mu(r)}}\sum_{x\in \partial_rR}\sum_{y\in T}F^\downarrow_{\mu}(\operatorname{dist}(x,y))\,.
				\end{aligned}
			\end{equation*}
            with the constants
            \begin{equation}\label{eq:proof-lrb-constant-c1c2}
                \begin{aligned}
                    c_1&=c\,\gamma^{2r+1}\sqrt{c_{IV}}=16(1+\nu)\Lambda d!^{r_{\exp}}(d/2+1)^{2r_{\exp}(d+2)}\gamma^{2r+1}\sqrt{p4^{p+1}\max\{\Gamma'''/\Gamma',1/3\}}\\
                    c_2&=\frac{\gamma^{r+1}\kappa}{\xi_2}=\gamma^{r}\\
                    \xi_2&=\gamma\kappa=\gamma 2(1+\nu)\Lambda(d/2+1)^{4r_{\exp}}d!^{r_{\exp}}
                \end{aligned}
            \end{equation}
            calculated from \Cref{eq:proof-lrb-constant-IV}, \eqref{eq:proof-lrb-constant-summary-c} and \eqref{eq:proof-lrb-constant-III.2}. Assuming that $V\subset \Z^{D}$ is equipped with the lattice graph distance and choosing $F^\downarrow(x)=(1+x)^{-(D+1)}$ (see \cite{Nachtergaele.2006propagation}), we have
			\begin{equation}\label{eq:lattice-constants}
				\begin{aligned}
					C_{\mu}&\leq 2^{D+2}\sum_{n\in\Z^D}\frac{1}{(1+|n|)^{D+1}}<\infty\,.
				\end{aligned}
			\end{equation}
            Next, we use the definition of $F^\downarrow_\mu$ from the underlying metric and fix a coherent input state $\rho=\ketbra{\alpha}{\alpha}$. Write $\eta=\max_{i\in V}|\alpha_i|$. The associated moments are bounded by \Cref{lem:coherent-state-sobolev-norm}:
            \begin{equation*}
                \begin{aligned}
                    \max_{i\in V}\big\{M_i^{(k)}\big\}&\leq\max_{i\in V}2^{k} \max\left\{1,\left(\frac{k}{\ln(k/|\alpha_i|^2+1)}\right)^k\right\}\\
                    &\leq 2^k\max\left\{1,\left(\frac{k}{\ln(k/\max_{i\in V}|\alpha_i|^2+1)}\right)^k\right\}\\
                    &\leq k^{\xi k}
                \end{aligned}
            \end{equation*}
            where we used the assumptions that $\xi\geq2$, $k\geq2$ and $k\geq2(e-1)\max_{i\in V}|\alpha_i|^2$. This shows
            \begin{equation}\label{eq:thm-exponential-proof-step}
				\begin{aligned}
                    |\tr\Bigl[&O_T\bigl(e^{t\widetilde{\cL}_R^{(M)}}-e^{t\widetilde{\cL}}\bigr)(\rho)\Bigr]|/\|O_T\|_{\infty}\\
                    &\leq c_1 |R|^{r+1}(e^{\hat{k} \Gamma t}-1)\frac{\sqrt{2}\hat{k}^{\xi/2 \hat{k}}}{(M-d+1)^{\hat{k}/4-r_{\exp}d/2}}\\
                    &\qquad+c_2|R|^r\frac{F^\downarrow_\mu(r)}{C_\mu^2}\,\|F^\downarrow\|\,\min\{|T|,|\partial_rR|\}\,e^{-\mu \bigl(d(T,\partial_rR)-t\frac{\xi_2 C_\mu}{\mu F_\mu^\downarrow(r)}2^{r_{\exp}d/2}(M-d+1)^{r_{\exp}d}\bigr)}\,,
				\end{aligned}
			\end{equation}
            where we additionally assumed that $M+1\geq 2d$ to use the following bound 
            \begin{equation*}
                \begin{aligned}
                    \Bigl((1/2)(M+1)\Bigr)^{r_{\exp}d/2}\leq\Bigl(\frac{M+1-d}{M+1}(M+1)\Bigr)^{r_{\exp}d/2}=(M+1-d)^{r_{\exp}d/2}\,.
                \end{aligned}
            \end{equation*}
            Next, we choose
            \begin{equation}\label{eq:proof-lrb-assumption-M}
                M-d+1=  \left(\frac{\mu F_\mu^\downarrow(r) \dist(\partial E_R,T)}{2^{r_{\exp}d/2+1}t \xi_2C_\mu}\right)^{1/{r_{\exp}d}} \,, 
		      \end{equation}
            we achieve
            \begin{equation*}
				\begin{aligned}
                    |\tr\Bigl[&O_T\bigl(e^{t\widetilde{\cL}_R^{(M)}}-e^{t\widetilde{\cL}}\bigr)(\rho)\Bigr]|/\|O_T\|_{\infty}\\
                    &\leq c_1 \sqrt{2}|R|^{r+1}e^{\hat{k} \Gamma t}\frac{\hat{k}^{\hat{k}\xi/2}}{\Bigl(\frac{\mu F_\mu^\downarrow(r) \dist(\partial E_R,T)}{2^{r_{\exp}d/2+1}t \xi_2C_\mu}\Bigr)^{\hat{k}/(4r_{\exp}d)-1/2}}\\
                    &\qquad\qquad+c_2|R|^r\frac{F^\downarrow_\mu(r)}{C_\mu^2}\,\|F^\downarrow\|\,\min\{|T|,|\partial_rR|\}\,e^{-\frac{\mu}{2}d(T,\partial_rR)}\\
                    &\leq c_1 \sqrt{2}|R|^{r+1}\frac{\Bigl(e^{\Gamma t+1}\hat{k}^{\xi/2}\Bigr)^{\hat{k}}}{\Bigl(\frac{\mu F_\mu^\downarrow(r) \dist(\partial E_R,T)}{2^{r_{\exp}d/2+1}t \xi_2C_\mu}\Bigr)^{\hat{k}/(4r_{\exp}d)-1/2}}\\
                    &\qquad\qquad+c_2|R|^r\frac{F^\downarrow_\mu(r)}{C_\mu^2}\,\|F^\downarrow\|\,\min\{|T|,|\partial_rR|\}\,e^{-\frac{\mu}{2}d(T,\partial_rR)}\,.
				\end{aligned}
			\end{equation*}
            In the next step, we choose for any $0<\varphi<1/8$
		      \begin{equation}\label{eq:proof-lrb-assumption-k}
                \widetilde{k} = \biggl(\frac{(M-d+1)^{1/4-\varphi}}{e^{\Gamma t+1}}\biggr)^{\frac{2}{\xi}} > \biggl(\frac{(M-d+1)^{1/8}}{e^{\Gamma t+1}}\biggr)^{\frac{2}{\xi}}
            \end{equation}
            and under the assumption $\widetilde{k}\geq\hat{k}$, it implies
            \begin{equation*}
                \Bigl(e^{\Gamma t+1}\widetilde{k}^{\xi/2}\Bigr)^{\widetilde{k}}=(M-d+1)^{\widetilde{k}(1/4-\varphi)}=\left(\frac{\mu F_\mu^\downarrow(r) \dist(\partial E_R,T)}{2^{r_{\exp}d/2+1}t \xi_2C_\mu}\right)^{\widetilde{k}(1/4-\varphi)/{r_{\exp}d}}
            \end{equation*}
            so that \Cref{eq:proof-lrb-assumption-M} implies
            \begin{equation*}
				\begin{aligned}
                    |\tr\Bigl[&O_T\bigl(e^{t\widetilde{\cL}_R^{(M)}}-e^{t\widetilde{\cL}}\bigr)(\rho)\Bigr]|/\|O_T\|_{\infty}\\
                    &\leq c_1 \sqrt{2}|R|^{r+1}\left(\frac{\mu F_\mu^\downarrow(r) \dist(\partial E_R,T)}{2^{r_{\exp}d/2+1}t \xi_2C_\mu}\right)^{-\widetilde{k}\varphi/(r_{\exp}d)+1/2}\\
                    &\qquad\qquad+c_2|R|^r\frac{F^\downarrow_\mu(r)}{C_\mu^2}\,\|F^\downarrow\|\,\min\{|T|,|\partial_rR|\}\,e^{-\frac{\mu}{2}d(T,\partial_rR)}\\
                    &\leq c_1 \sqrt{2}|R|^{r+1}e^{-(\widetilde{k}\varphi/(r_{\exp}d)-1/2)\ln(\frac{\mu F_\mu^\downarrow(r) \dist(\partial E_R,T)}{2^{r_{\exp}d/2+1}t \xi_2C_\mu})}\\
                    &\qquad\qquad+c_2|R|^r\frac{F^\downarrow_\mu(r)}{C_\mu^2}\,\|F^\downarrow\|\,\min\{|T|,|\partial_rR|\}\,e^{-\frac{\mu}{2}d(T,\partial_rR)}\\
				\end{aligned}
			\end{equation*}
            Finally, for 
            \begin{equation}\label{eq:proof-lrb-assumption-dist}
                0<\varphi\coloneqq \frac{r_{\exp}d}{\ln(\frac{\mu F_\mu^\downarrow(r) \dist(\partial E_R,T)}{2^{r_{\exp}d/2+1}t \xi_2C_\mu})}<\frac{1}{8}
            \end{equation}
            we can show by using \Cref{eq:proof-lrb-assumption-k} again that
            \begin{equation}\label{eq:proof-lrb-combine1}
				\begin{aligned}
                    |\tr\Bigl[&O_T\bigl(e^{t\widetilde{\cL}_R^{(M)}}-e^{t\widetilde{\cL}}\bigr)(\rho)\Bigr]|/\|O_T\|_{\infty}\\
                    &\leq c_1 \sqrt{2}|R|^{r+1}e^{-\Bigl(e^{-\Gamma t}(M-d+1)^{1/8}\Bigr)^{\frac{2}{\xi}}}\sqrt{\frac{\mu F_\mu^\downarrow(r) \dist(\partial E_R,T)}{2^{r_{\exp}d/2+1}t \xi_2C_\mu}}\\
                    &\qquad\qquad+c_2|R|^r\frac{F^\downarrow_\mu(r)}{C_\mu^2}\,\|F^\downarrow\|\,\min\{|T|,|\partial_rR|\}\,e^{-\frac{\mu}{2}d(T,\partial_rR)}\\
                    &\leq c_1 \sqrt{2}|R|^{r+1}e^{-e^{-2\Gamma t/\xi}\Bigl(\frac{\mu F_\mu^\downarrow(r) \dist(\partial E_R,T)}{2^{r_{\exp}d/2+1}t \xi_2C_\mu}\Bigr)^{1/{(4\xi r_{\exp}d)}}}\sqrt{\frac{\mu F_\mu^\downarrow(r) \dist(\partial E_R,T)}{2^{r_{\exp}d/2+1}t \xi_2C_\mu}}\\
                    &\qquad\qquad+c_2|R|^r\frac{F^\downarrow_\mu(r)}{C_\mu^2}\,\|F^\downarrow\|\,\min\{|T|,|\partial_rR|\}\,e^{-\frac{\mu}{2}d(T,\partial_rR)}\\
                    &\leq c_3 |R|^{r}(|R|+\min\{|T|,|\partial_rR|\})\sqrt{\frac{\dist(\partial E_R,T)}{t}}e^{-\xi_3e^{-\kappa t}\Bigl(\frac{\dist(\partial E_R,T)}{t}\Bigr)^{1/{(4\xi r_{\exp}d)}}}\,,
				\end{aligned}
			\end{equation}
            where we used that \Cref{eq:proof-lrb-assumption-dist} implies
            \begin{equation}\label{eq:proof-lrb-combine2}
                \begin{aligned}
                    e^{8r_{\exp}d}&<\frac{\mu F_\mu^\downarrow(r) \dist(\partial E_R,T)}{2^{r_{\exp}d/2+1}t \xi_2C_\mu}\\
                \end{aligned}
            \end{equation}
            for the following constants (see also \Cref{eq:proof-lrb-constant-c1c2}):
            \begin{equation}\label{eq:proof-lrb-constants}
                \begin{aligned}
                    c_1&=16(1+\nu)\Lambda d!^{r_{\exp}}(d/2+1)^{2r_{\exp}(d+2)}\gamma^{2r+1}\sqrt{p4^{p+1}\max\{\Gamma'''/\Gamma',1/3\}}\\
                    c_2&=\gamma^{r}\\
                    \xi_2&=\gamma 2(1+\nu)\Lambda(d/2+1)^{4r_{\exp}}d!^{r_{\exp}}\\
                    c_3&=\sqrt{\frac{\mu F_\mu^\downarrow(r)}{2^{r_{\exp}d/2} \xi_2C_\mu}}\max\biggl\{c_1, c_2\frac{F^\downarrow_\mu(r)}{C_\mu^2}\,\|F^\downarrow\|\biggr\}\\
                    \xi_3&=\min\Bigl\{1/2,\Bigl(\frac{\mu F_\mu^\downarrow(r)}{2^{r_{\exp}d/2+1}\xi_2C_\mu}\Bigr)^{1/{(4\xi r_{\exp}d)}}\Bigr\}\\
                    \kappa&=2\Gamma/\xi\,.
                \end{aligned}
            \end{equation}
            Moreover, we used the following assumptions (see \Cref{eq:proof-lrb-assumption-M} and \eqref{eq:proof-lrb-assumption-k})
            \begin{equation}\label{eq:proof-lrb-assumptions}
                \begin{gathered}
                    M-d+1\geq d\\
                    M-d+1=  \left(\frac{\mu F_\mu^\downarrow(r) \dist(\partial E_R,T)}{2^{r_{\exp}d/2+1}t \xi_2C_\mu}\right)^{1/{r_{\exp}d}}\\
                    \widetilde{k}=\biggl(\frac{(M-d+1)^{1/4-\varphi}}{e^{\Gamma t+1}}\biggr)^{\frac{2}{\xi }}\geq \max\{\hat{k},2,2(e-1)\eta^2\},\qquad \hat{k}=\max\{K,4dr_{\exp}\}\\
                    0<\varphi\coloneqq \frac{r_{\exp}d}{\ln(\frac{\mu F_\mu^\downarrow(r) \dist(\partial E_R,T)}{2^{r_{\exp}d/2+1}t \xi_2C_\mu})}<\frac{1}{8}
                \end{gathered}
            \end{equation}
            which finishes the proof.
        \end{proof}
        
	\subsection{Model-specific information propagation bounds}\label{subsec:model-specific-lrb}
		In this section, we prove the required assumptions for the examples given in \Cref{ex:Bose-Hubbard}-\ref{ex:reg-hamiltonian},
		so that \Cref{thm:bosonicLRB} then yields weak Lieb--Robinson-type bounds. To this end, we first introduce the following function:  
		\begin{equation}\label{eq:f-g-l-function}
			g_\ell(m) = 
			\begin{cases} 
				f(m) - f(m - \ell), & m \geq \ell; \\
				f(m), & 0 \leq m < \ell; \\
				0, & m < 0.
			\end{cases}
		\end{equation}
		
		We then proceed to prove the assumptions for the seminal Bose--Hubbard model (\Cref{ex:Bose-Hubbard}), which is also considered in \cite{Kuwahara.2021} and later improved in \cite{Kuwahara.2024digital} and \cite{Kuwahara.2024Lemm}.
		\begin{prop}[LRB--Bose--Hubbard model]\label{prop:LRB-bose-hubbard}
			Let $(H,\cH_f)$ be the Bose--Hubbard Hamiltonian on a graph $(V,E)$ defined in \Cref{ex:Bose-Hubbard} by 
			\begin{equation*}
				H\coloneqq \sum_{(i,j)\in E}\lambda^{(i,j)}\,a_i^\dagger\,a_j+\sum_{i\in V}u^{(i)}N_i(N_i-1)+\sum_{i\in V}\mu^{(i)}N_i\,.
			\end{equation*}
			Then, it satisfies \Cref{assum:local-GKSL,assum-lrb} for period $p=1$ and all $k\in\N_{\geq1}$, 
			\begin{equation*}
				\frac{\partial}{\partial t}F^{(k)}_i(t) \le 32\,k\,\Lambda\gamma F^{(k)}_i(t)+32\,k\,\Lambda\sum_{(i,j)\in E}F^{(k)}_j(t)+20\,\Lambda\,\gamma\,k\,k^{k}
			\end{equation*}
			where $\Lambda$ bounds the absolute values of the coefficients and $\max_{j\in V}\gamma_j\leq\gamma$. Then, the Bose--Hubbard model satisfies the weak Lieb--Robinson-type information propagation bound proven in \Cref{thm:bosonicLRB} with $\Gamma'=32\Lambda\gamma$, $\Gamma''=32\Lambda$, $\Gamma'''=20\Lambda\gamma$ and $\xi=2$.
		\end{prop}
        Note that the above result can be further improved following the approach in \cite{Kuwahara.2024digital} and \cite{Kuwahara.2024Lemm}. Here the main idea is that the moment propagation bound which is key for the proof of the weak Lieb--Robinson bound can be improved for the Bose--Hubbard model.
		\begin{proof}
			First, note that the closure of $(H,\cH_f)$ defines a generator of a quantum Markov semigroup \cite[Appx.~A]{Faupin.2022BoseHubbard} and satisfies the structural assumption in \Cref{assum:local-GKSL} by definition. Then, we continue with verifying \Cref{assum-lrb}. Therefore, assume that $\rho\in\cT_f$ and $\rho$ is a state. Then,
			\begin{equation*}
				\frac{\partial}{\partial t}\tr[\rho_tf(N_s)]=\tr[\rho_{t}\,\cL^\dagger f(N_s)]
			\end{equation*}
			for $s\in V$ motivates to consider the following operator with $f$ given in \Cref{assum-lrb} with $p=1$:
			\begin{align*}
				\cL^\dagger\left(f(N_s)\right)&=i[H,\,f(N_s)]\\
				&=i\sum_{(i,j)\in E}\lambda^{(i,j)}\left[a_i^\dagger\,a_j,\,f(N_s)\right]\\
				&=-i\sum_{(i,j)\in E}\delta_{s,i} (\lambda^{(i,j)}a_jg_1(N_s)\ad_i-\lambda^{(j,i)}a_ig_1(N_s)\ad_j)
			\end{align*}
			where we used the commutation relation stated in \Cref{eq:symmetry-function}
			\begin{equation*}
				\begin{aligned}
					a_if(N_i)&=f(N_i + I)a_i\quad\text{ as well as }\quad a^\dagger_i f(N_i)=f(N_i - I)a_i^\dagger\,.
				\end{aligned}
			\end{equation*}
			and $g_1(x)=f(x)-f(x-1)$ (see \Cref{eq:f-g-l-function}) in the last equality. Next, we express the above discrete operator in the Fock basis and represent it using matrices of rank two:
			\begin{align*}
				\cL^\dagger\left(f(N_s)\right)&=-i\sum_{(i,j)\in E} \delta_{s,i}\sum_{n_i,n_j=0}^{\infty}\lambda^{(i,j)}g(n_i)a_j\ket{n_i}\bra{n_i}\otimes\ket{n_j}\bra{n_j}\ad_i+ h.c.\\
				&=-i\sum_{(i,j)\in E} \delta_{i,s}\sum_{n_i,n_j=1}^{\infty}\lambda^{(i,j)}g(n_i)\sqrt{n_i n_j}\,\ket{n_i,n_j-1}\bra{n_i-1,n_j}+ h.c.
			\end{align*}
			Interpreting for all $i,j\in\N_{\geq1}$ the summand as a $2\times2$ matrix defined by 
			\begin{equation*}
				\begin{aligned}
					i\ket{e_1}\bra{e_2}-i\ket{e_2}\bra{e_1}&=-i\ket{n_i,n_j-1}\bra{n_i-1,n_j}+i\ket{n_i-1,n_j}\bra{n_i,n_j-1}\\
					&\leq \ket{e_1}\bra{e_1}+\ket{e_2}\bra{e_2}.
				\end{aligned}
			\end{equation*}
			The last inequality directly follows by its eigenvalues decomposition. Then, the matrix trick in combination with the monotonicity of the coefficients show
			\begin{equation*}
				\begin{aligned}
					\cL^\dagger\left(f(N_s)\right)&\leq \sum_{(i,j)\in E} \delta_{i,s}\sum_{n_i,n_j=1}^{\infty}|\lambda^{(i,j)}|g(n_i)\sqrt{n_i n_j}\,\Bigl(\ket{n_i-1,n_j}\bra{n_i-1,n_j}\\
					&\hspace*{40ex}+\ket{n_i,n_j-1}\bra{n_i,n_j-1}\Bigr)\\
					&\leq \sum_{(i,j)\in E}\delta_{i,s}\sum_{n_i,n_j=0}^{\infty}2|\lambda^{(i,j)}|g(n_i+1)\sqrt{(n_i+1)(n_j+1)}\ket{n_i,n_j}\bra{n_i,n_j}\\
					&=\sum_{(i,j)\in E}\delta_{i,s}\sum_{n_i,n_j=0}^{\infty}2|\lambda^{(i,j)}|kn_i^{k-1}\sqrt{(n_i+1)(n_j+1)}\ket{n_i,n_j}\bra{n_i,n_j},
				\end{aligned}
			\end{equation*}
			where the last identity directly follows by the definition of $f$. Then, Young's inequality proves:
			\begin{align*}
				\begin{aligned}
					\cL^\dagger\left(f(N_s)\right)&\leq \sum_{(i,j)\in E}\delta_{i,s}\sum_{n_i,n_j=0}^{\infty}2|\lambda^{(i,j)}|k(n_i+1)^{k-1/2}(n_j+1)^{1/2}\ket{n_i,n_j}\bra{n_i,n_j}\\
					&\leq \sum_{(i,j)\in E}\delta_{i,s}\sum_{n_i,n_j=0}^{\infty}2k|\lambda^{(i,j)}|\left((n_i+1)^{k}+(n_j+1)^{k}\right)\ket{n_i,n_j}\bra{n_i,n_j}\\
					&=2 k\sum_{(i,j)\in E}\delta_{i,s}|\lambda^{(i,j)}|\left((N_i+I)^k+(N_j+I)^k\right)\,.
				\end{aligned}
			\end{align*}
			Finally, we apply \Cref{lem:connection-moments} to upper bound the right-hand side by $f$. We repeat that $f(m)$ is increasing in $m$ and satisfies for $m\geq p=1$
			\begin{equation}\label{eq:connect-f-power}
				(m+\ell)^k\leq4^{p+\ell}pf(m)+k^k\frac{4^{p+\ell}-1}{3}
			\end{equation}
			for any integer $\ell$ so that
			\begin{equation*}
				(m+1)^k\leq16f(m)+5k^k\,.
			\end{equation*}
			for all $m,k\geq 0$. Then, we can show
			\begin{align*}
				\begin{aligned}
					\cL^\dagger\left(f(N_i)\right)&\leq k\Lambda\left(2\gamma_s(N_s+I)^k+2\sum_{(s,j)\in E}(N_j+I)^k\right)\\
					&\leq k\Lambda\left(32\gamma_if(N_i)+32\sum_{(s,j)\in E}f(N_j)+20\gamma_sk^k\right)\\
					&\leq 32\,k\,\Lambda\gamma f(N_i)+32\,k\,\Lambda\sum_{(s,j)\in E}f(N_j)+20\,\Lambda\,\gamma\,k\,k^{k}\,.
				\end{aligned}
			\end{align*}
			Using the notation  $F^{(k)}_i(t)\coloneqq \tr[\rho_tf(N_i)]$, we achieve \Cref{assum-lrb} with
			\begin{equation*}
				\frac{\partial}{\partial t}F^{(k)}_i(t) \le 32\,k\,\Lambda\gamma F^{(k)}_i(t)+32\,k\,\Lambda\sum_{(i,j)\in E}F^{(k)}_j(t)+20\,\Lambda\,\gamma\,k\,k^{k}
			\end{equation*}
			which allows us to apply the \Cref{thm:bosonicLRB} in combination with \Cref{cor:boundedness-projective-generator-bose-hubbard} to finish the proof.
		\end{proof}
		
		Next, we consider any quadratic Hamiltonian as defined in \Cref{ex:Gaussian}. Note that, in contrast to the example above, we allow for loops in the graph in the following example.
		\begin{prop}[Quadratic Hamiltonian]\label{prop:propagation-quadratic-Hamiltonian} 
			Let $H\in\cH_f$ be the quadratic Hamiltonian defined in \Cref{ex:Gaussian} by
			\begin{equation*}
				H\coloneqq \sum_{(i,j)\in E}\lambda^{(i,j)}\,a_i^\dagger\,a_j+\sum_{i\in V}b^{(i)}a_i^2+\overline{b}^{(i)}(\ad_i)^2+c^{(i)}a_i+\overline{c}^{(i)}\ad_i\,.
			\end{equation*}
			Then, it satisfies \Cref{assum:local-GKSL,assum-lrb} with $p=2$ by 
			\begin{equation*}
				\frac{\partial}{\partial t}F^{(k)}_i(t) \le 256\,k\,\Lambda(\gamma+3)F^{(k)}_i(t)+256\,k\,\Lambda\sum_{(i,j)\in E}F^{(k)}_j(t)+42\,\Lambda\,(\gamma+3)\,k\,k^{k}
			\end{equation*}
			where $\Lambda$ upper bounds all absolute values of the coefficients. Therefore, \Cref{thm:bosonicLRB} holds true with $\Gamma'=256\,\Lambda(\gamma+3)$, $\Gamma''=256\,\Lambda$, $\Gamma'''=42\,\Lambda\,(\gamma+3)$ and $\xi=2$.
		\end{prop} 
		
		\begin{proof}
			In a first step, we start with the generation theory of the Hamiltonian $H$. By \cite[Thm.~1]{Fagnola.2023} and \cite[Thm.~I.5.8]{Engel.2000}, it directly follows that $-i[H,\cdot]$ generates a quantum dynamical semigroup. Since \Cref{assum:local-GKSL} is satisfied by definition, we continue proving \Cref{assum-lrb}. Therefore, let $\rho\in\cT_f$ be a state. Then, for $\rho_t=e^{t\cL}(\rho)$ with $\cL=-i[H,\cdot]$
			\begin{equation*}
				\frac{\partial}{\partial t}\tr[\rho_tf(N_i)]=\tr[\rho_{t}\,\cL^\dagger f(N_i)]
			\end{equation*} 
			motivates to only consider the following operator with $f$ given in \Cref{assum-lrb} for $p=2$: 
			\begin{align*}
				\cL^\dagger\left(f(N_s)\right)&= i[H,\,f(N_s)]\\
				&=i\sum_{(i,j)\in E}\lambda^{(i,j)}\left[a_i^\dagger\,a_j,\,f(N_s)\right]+i b^{(s)}\left[a_s^2,f(N_s)\right]+i\overline{b}^{(s)}\left[(\ad_s)^2,f(N_s)\right]\\
				&\qquad\qquad\qquad\qquad\qquad\qquad\qquad\qquad+ic^{(s)}\left[a_s,f(N_s)\right]+i\overline{c}^{(s)}\left[\ad_s,\,f(N_s)\right]\\
				&={i}\sum_{(i,j)\in E}\lambda^{(j,i)}a_ig_1(N_s)\ad_j \delta_{i,s}-\lambda^{(i,j)}a_jg_1(N_s)\ad_i \delta_{i,s}\\
				&\qquad+i b^{(s)}a_s^2g_2(N_s)-i \overline{b}^{(s)}g_2(N_s)(\ad_s)^2+i c^{(s)}a_sg_1(N_s)-i \overline{c}^{(s)}g_1(N_s)\ad_s
			\end{align*}
			where we used the commutation relation stated in \Cref{eq:symmetry-function}
			\begin{equation*}
				\begin{aligned}
					a_if(N_i)&=f(N_i + I)a_i\quad\text{ as well as }\quad a^\dagger_i f(N_i)=f(N_i - I)a_i^\dagger\,.
				\end{aligned}
			\end{equation*}
			and $g_j(x)=f(x)-\delta_{x\geq j}f(x-j)$ for $j\in\{1,2\}$ and $x\geq0$ (see \Cref{eq:f-g-l-function}). Next, we express the above discrete operator in the Fock basis and represent it using matrices of rank two:
			\begin{align*}
				\cL^\dagger\left(f(N_s)\right)&=\sum_{(i,j)\in E}\delta_{i,s}\sum_{n_i,n_j=0}^{\infty}{i}\lambda^{(j,i)}g_1(n_i)a_i\ket{n_i}\bra{n_i}\otimes\ket{n_j}\bra{n_j}\ad_j+ h.c.\\
				&\qquad+\sum_{n_s=0}^{\infty}ib^{(s)}g_2(n_s)a_s^2\ketbra{n_s}{n_s}+ic^{(s)}g_1(n_s)a_s\ketbra{n_s}{n_s}+h.c.\\
				&=\sum_{(i,j)\in E} \delta_{i,s}\sum_{n_i,n_j=1}^{\infty}{i}\lambda^{(j,i)}g_1(n_i)\sqrt{n_i n_j}\,\ket{n_i-1,n_j}\bra{n_i,n_j-1}+ h.c.\\
				&\qquad+\sum_{n_s=0}^{\infty}ib^{(s)}g_2(n_s)\sqrt{n_s(n_s-1)}\ketbra{n_s-2}{n_s}+h.c.\\
				&\qquad\qquad\qquad\qquad+ic^{(s)}\sqrt{n_s}g_1(n_s)\ketbra{n_s-1}{n_s}+h.c.\\
			\end{align*}
			Interpreting for all $i,j\in\N_{\geq1}$ the summand as a $2\times2$ matrix defined by 
			\begin{equation*}
				\begin{aligned}
					i\ket{e_1}\bra{e_2}-i\ket{e_2}\bra{e_1}&=i\ket{n_i,n_j-1}\bra{n_i-1,n_j}-i\ket{n_i-1,n_j}\bra{n_i,n_j-1}\\
					&\leq \ket{e_1}\bra{e_1}+\ket{e_2}\bra{e_2}.
				\end{aligned}
			\end{equation*}
			The last inequality directly follows from its eigenvalue decomposition. Then, the matrix trick, combined with the monotonicity of $g_{\ell}(m)$ in $\ell$ and $g_2(m)$ in $m$ (see \Cref{lem:connection-moments}), shows
			\begin{equation*}
				\begin{aligned}
					\cL^\dagger&\left(f(N_s)\right)\\
					&\leq \sum_{(i,j)\in E} \delta_{i,s}\sum_{n_i,n_j=1}^{\infty}|\lambda^{(i,j)}|g_1(n_i)\sqrt{n_i n_j}\,\Bigl(\ket{n_i-1,n_j}\bra{n_i-1,n_j}+\ket{n_i,n_j-1}\bra{n_i,n_j-1}\Bigr)\\
					&\qquad+\sum_{n_s=0}^{\infty}|b^{(s)}|g_2(n_s)\sqrt{n_s(n_s-1)}\bigl(\ketbra{n_s-2}{n_s-2}+\ketbra{n_s}{n_s}\bigr)\\
					&\qquad\qquad\qquad\qquad+|c^{(s)}|\sqrt{n_s}g_1(n_s)\bigl(\ketbra{n_s-1}{n_s-1}+\ketbra{n_s}{n_s}\bigr)\\
					&\leq \sum_{(i,j)\in E}\delta_{i,s}\sum_{n_i,n_j=0}^{\infty}2|\lambda^{(i,j)}|g_2(n_i+2)\sqrt{(n_i+1)(n_j+1)}\ket{n_i,n_j}\bra{n_i,n_j}\\
					&\qquad+2\sum_{n_i=0}^{\infty}\Bigl(|b^{(s)}|g_2(n_s+2)(n_s+1)+|c^{(s)}|\sqrt{n_s}g_2(n_s+2)\Bigr)\ketbra{n_s}{n_s}\,.
				\end{aligned}
			\end{equation*}
			Then, we use for $p=2$ the identity $f(n+p)-f(n)=kn^{k-1}$ given by definition of $f$ (see \Cref{lem:connection-moments}):
			\begin{equation*}
				\begin{aligned}
					\cL^\dagger\left(f(N_s)\right)&\leq\sum_{(i,j)\in E} \delta_{i,s}\sum_{n_i,n_j=0}^{\infty}2|\lambda^{(i,j)}|kn_i^{k-1}\sqrt{(n_i+1)(n_j+1)}\ket{n_i,n_j}\bra{n_i,n_j}\\
					&\qquad+2\sum_{n_i=0}^{\infty}kn_s^{k-1}\Bigl(2|b^{(s)}|(n_s+1)+|c^{(s)}|\sqrt{n_s}\Bigr)\ketbra{n_s}{n_s}\,.
				\end{aligned}
			\end{equation*}
			In the next step, Young's inequality helps to prove:
			\begin{align*}
				\begin{aligned}
					\cL^\dagger\left(f(N_s)\right)&\leq \sum_{(i,j)\in E} \delta_{i,s}\sum_{n_i,n_j=0}^{\infty}2k|\lambda^{(i,j)}|\left((n_i+1)^{k}+(n_j+1)^{k}\right)\ket{n_i,n_j}\bra{n_i,n_j}\\
					&\qquad+2\sum_{n_i=0}^{\infty}kn_s^{k-1}\Bigl(2|b^{(s)}|(n_s+1)+|c^{(s)}|\sqrt{n_s}\Bigr)\ketbra{n_s}{n_s}\\
					&\leq 2k\sum_{(i,j)\in E} \delta_{i,s}|\lambda^{(i,j)}|\left((N_i+I)^k+(N_j+I)^k\right)+2k\Bigl(2|b^{(s)}|+|c^{(s)}|\Bigr)(N_s+I)^k\,.
				\end{aligned}
			\end{align*}
			Finally, we apply \Cref{lem:connection-moments}, in particular \Cref{eq:connect-f-power}, to upper bound the above right hand side by $f$, i.e.
			\begin{equation*}
				(m+1)^k\leq128f(m)+21k^k
			\end{equation*}
			for all $m,k\geq 0$. Then, we can show
			\begin{align*}
				\begin{aligned}
					\cL^\dagger\left(f(N_s)\right)&\leq 2k\Lambda\left((\gamma_s+3)(N_s+I)^k+\sum_{(s,j)\in E}(N_j+I)^k\right)\\
					&\leq 2k\Lambda\left(128(\gamma_s+3)f(N_s)+128\sum_{(s,j)\in E}f(N_j)+21(2\gamma_s+3)k^k\right)\\
					&\leq 256\,k\,\Lambda(\gamma+3)f(N_s)+256\,k\,\Lambda\sum_{(s,j)\in E}f(N_j)+42\,\Lambda\,(\gamma+3)\,k\,k^{k}\,.
				\end{aligned}
			\end{align*}
			Using the notation  $F^{(k)}_i(t)\coloneqq \tr[\rho_tf(N_i)]$, we achieve \Cref{assum-lrb} with
			\begin{equation*}
				\frac{\partial}{\partial t}F^{(k)}_i(t) \le 256\,k\,\Lambda(\gamma+3)F^{(k)}_i(t)+256\,k\,\Lambda\sum_{(i,j)\in E}F^{(k)}_j(t)+42\,\Lambda\,(\gamma+3)\,k\,k^{k}
			\end{equation*}
			which allows us to apply the \Cref{thm:bosonicLRB} in combination with \Cref{cor:boundedness-projective-generator-bose-hubbard} to finish the proof.
		\end{proof}
		
		In the next step, we prove the assumptions for the regularized two-body interaction Hamiltonian introduced in \Cref{eq:regularized-hamiltonian} defined by 
		\begin{equation*}
			\widetilde{\cL}^{(\alpha)} = \cH^{(d)} + \cL^{(\alpha,p)}
		\end{equation*}
		with the Hamiltonian (\ref{eq:bosonic-hamiltonian})
		\begin{align*}
			H^{(d)} = \sum_{e\in E}H_e\qquad\text{with}\qquad H_e \coloneqq \sum_{k,\ell,k',\ell'=0}^d\,\lambda^{(e)}_{k\ell k'\ell'}\,(a_i^\dagger)^k\,a_i^\ell\,(a_j^\dagger)^{k'}\,a_j^{\ell'}\,,
		\end{align*}
		for some complex coefficients $|\lambda^{(e)}_{k\ell k'\ell'}|\le \Lambda$, and the photon dissipation \eqref{eq:p-photon-dissipation}
		\begin{align*}
			\cL^{(\alpha,p)} \coloneqq \sum_{j \in V} \cL[L_j^{(\alpha_j,p)}], \qquad \text{where} \quad L_j^{(\alpha_j,p)} \coloneqq a_j^p - \alpha_j^p I \,.
		\end{align*}
		for $\alpha\in\mathbb{C}^{m}$ with $|\alpha_i|\leq\eta$ and $p\in\mathbb{N}$. Note that we can fix $\lambda_{0000}=0$, due to gauge freedom.
		
		\begin{prop}[LRB-Regularized Hamiltonian]\label{prop:LRB-regularized-hamiltonian}
			Let $\widetilde{\cL}^{(\alpha)}$ be the regularized Hamiltonian, which is defined in \Cref{eq:regularized-hamiltonian}. Then, it satisfies \Cref{assum:local-GKSL,assum-lrb} for $p\geq 2(d+1)$ and $k-2+p\geq\frac{k(k+d)}{k-d}$, i.e.
			\begin{equation*}
				\begin{aligned}
					\widetilde{\cL}^{(\alpha)\dagger}(f(N_i))\leq k\,\Gamma'f(N_i) + k\Gamma''\sum_{(i,j)\in E}f(N_j)+\Gamma'''k^{\xi k}
				\end{aligned}
			\end{equation*}
			where the constants are defined in \Cref{eq:assum-constant-reg-hamiltonian}. Therefore, it also satisfies the weak Lieb--Robinson-type information propagation bound in \Cref{thm:bosonicLRB}.
		\end{prop}
		\begin{proof}
			The proof relies on the two bounds proven in \Cref{lem:assumption-hamiltonian,lem:assumption-dissipation}:
			\begin{equation*}
				\begin{aligned}
					i[H^{(d)},f(N_i)]&\leq \Lambda\,(d+1)^4\,d!\,\left((k-d)\,(N_i+I)^{\frac{(k+d)k}{(k-d)}}+d\,(N_j+I)^{k}\right)\\
					\cL[a^p_i-\alpha^p_i]^\dagger(f(N_i))&\le -k1_{N_i\geq p}\,\frac{(N_i+I)^{k-1+p}}{2\times 4^{p+1}}+k (N_i+I)^k\,\widetilde{\Delta}+\Delta_k
				\end{aligned}
			\end{equation*}
			for $e\in E$, $i\in V$ and $\widetilde{\Delta},\Delta_k\geq0$ defined in \Cref{lem:assumption-dissipation}. To apply both bound at the same time, we choose $f:\mathbb{N}\to\mathbb{R}$ as follows: $f(0)=f(1)=\dots =f(p-1)=0$ and for any $m\ge 0$, $f(m+p)=f(m)+km^{k-1}$ for any $k\geq1$. Then, 
			\begin{equation*}
				\begin{aligned}
					\widetilde{\cL}^{(\alpha)\dagger}(f(N_i))&= \sum_{e\ni i}  (\cH^{(d)})^\dagger(f(N_i))+  \cL[a_i^p-\alpha_i^p]^\dagger(f(N_i))\\
					&\leq \,\Lambda\,(d+1)^4\,d!\,\left(\gamma(k-d)\,(N_i+I)^{\frac{(k+d)k}{k-d}}+\sum_{(i,j)\in E}d\,(N_j+I)^{k}\right)\\
					&\qquad-k1_{N_i\geq p}\,\frac{(N_i+I)^{k-1+p}}{2\times 4^{p+1}}+k (N_i+I)^k\,\widetilde{\Delta}+\Delta_k\\
					&\leq 1_{N_i\geq p}\left(-\frac{k}{2\times 4^{p+1}}(N_i+I)^{k+p-1}+\gamma\,\Lambda\,(d+1)^4\,d!\,k\,(N_i+I)^{\frac{(k+d)k}{k-d}}\right)\\
					&\qquad+\sum_{(i,j)\in E} \Lambda\,(d+1)^4\,d!\,d\,(N_j+I)^{k}+k (N_i+I)^k\,\widetilde{\Delta}\\
					&\qquad+\Delta_k+\gamma\Lambda\,(d+1)^4\,d!\,(k-d)\,(p+1)^{\frac{(k+d)k}{(k-d)}}\,.
				\end{aligned}
			\end{equation*}
			Using $k-2+p\geq\frac{k(k+d)}{k-d}$, a simple polynomial optimization gives (see \Cref{polymax})
			\begin{align*}
				-\frac{k}{2\times 4^{p+1}}(N_i+I)^{k+p-1}&+\gamma\,\Lambda\,(d+1)^4\,d!\,k\,(N_i+I)^{\frac{(k+d)k}{k-d}}\\
				&\le \left({2\times 4^{p+1}\gamma\,\Lambda\,(d+1)^4\,d!}\right)^{\frac{(k+d)k}{k-d}}\gamma\,\Lambda\,(d+1)^4\,d!\,k\\
				&\le \left({32\times4^{p+1}\gamma\,\Lambda\,d^4\,d!}\right)^{\frac{(k+d)k}{k-d}}\gamma\,\Lambda\,16\,d^4\,d!\,k\,,
			\end{align*}
			which shows in combination with \Cref{lem:connection-moments}
			\begin{equation}\label{eq:assum-constant-reg-hamiltonian}
				\begin{aligned}
					&\widetilde{\cL}^{(\alpha)\dagger}(f(N_i))\\
					&\quad\leq\,\Lambda\,(d+1)^4\,d!\,d\sum_{(i,j)\in E}(N_j+I)^{k}+k\,\widetilde{\Delta}(N_i+I)^k\\
					&\quad\quad+\Delta_k+\Lambda\,(d+1)^4\,d!\,(k-d)\,(p+1)^{\frac{(k+d)k}{(k-d)}+1}+\left({32\times 4^{p+1}\gamma\,\Lambda\,d^4\,d!}\right)^{\frac{(k+d)k}{k-d}}\gamma\,\Lambda\,16\,d^4\,d!\,k\\
					&\quad\leq \underbrace{4^{p+1}\,p\,\Lambda\,(d+1)^4\,d!\,d}_{\Gamma''}\sum_{(i,j)\in E}f(N_j)+k\,\underbrace{4^{p+1}\,p\,\widetilde{\Delta}}_{\Gamma'}f(N_i)\\
					&\quad\quad+\left(\gamma\,\Lambda\,(d+1)^4\,d!\,d+k\,\widetilde{\Delta}\right)k^k\frac{4^{p+1}-1}{3}\\
					&\quad\quad+\underbrace{\Delta_k+\Lambda\,(d+1)^4\,d!\,(k-d)\,(p+1)^{\frac{(k+d)k}{(k-d)}+1}+\left({32\gamma\,\Lambda\,d^4\,d!}\right)^{\frac{(k+d)k}{k-d}}\gamma\,\Lambda\,16\,d^4\,d!\,k}_{\eqqcolon R_k\leq\Gamma'''k^{\xi k}}\\
					&\quad\leq\colon k\,\Gamma'f(N_i) + k\Gamma''\sum_{(i,j)\in E}f(N_j)+\Gamma'''k^{\xi k}
				\end{aligned}
			\end{equation}
			Repeating the calculation for $f(x)=(x+1)^k$ proves that $\widetilde{\cL}^{(\alpha)}$ satisfies \Cref{eq:moment-stability}. Hence \Cref{thm:generation-theory} applies and shows that the closure of $(\widetilde{\cL}^{(\alpha)},\cT_f)$ generates a quantum Markov semigroup for $p\geq2(d+1)$ (see \cite[Lem.~5.3-5.7]{Gondolf.2024}).
		\end{proof}
	
		In the next step, we prove the two main ingredients of the above proof. First, we start with the operator bound on the Hamiltonian:
		\begin{lem}\label{lem:assumption-hamiltonian}
			Let $k>d$ and $H^{(d)}_e$ be a two-body interaction Hamiltonian on an edge $e=(i,j)\in E$ with degree $d$ defined in \Cref{eq:regularized-hamiltonian}. For the function $f:\mathbb{N}\to\mathbb{R}$ defined by $f(0)=f(1)=\dots =f(p-1)=0$ with $p\geq d$ and $f(m+p)=f(m)+km^{k-1}$ for any $m\ge 0$, we have
			\begin{align*}
				i[H^{(d)}_e,f(N_i)]&\leq \Lambda\,(d+1)^4\,d!\,\left((k-d)\,(N_i+I)^{\frac{(k+d)k}{(k-d)}}+d\,(N_j+I)^{k}\right)\,,
			\end{align*}
			where $\Lambda\geq0$ denotes the upper bound on the coefficients of $H^{(d)}$ (see \Cref{assum:local-GKSL}). 
		\end{lem}
		\begin{proof}
			Extending the ideas of \Cref{prop:LRB-bose-hubbard}, we prove a similar bound for more general two-body interactions. For $d=0$ the bound is trivial, so assume $d\geq1$ and without loss of generality $e=(i,j)\in E$. By \Cref{eq:symmetry-function}, \Cref{higher-order-product-aadag} and the symmetry $\lambda^{(e)}_{k\ell k'\ell'}=\overline{\lambda}^{(e)}_{\ell k \ell'k'}$ for all $k,\ell,k',\ell'$:
			\begin{equation*}
				\begin{aligned}
					i[H^{(d)}_e,f(N_i)]&=i\sum_{\substack{0\leq k<\ell\leq d\\0\leq k',\,\ell'\leq d}}\lambda^{(e)}_{k\ell k'\ell'}\,\Bigl(N_i[-k+1:0]a_i^{\ell-k}\,f(N_i)\\
					&\qquad\qquad\qquad\qquad\qquad-f(N_i)N_i[-k+1:0]a_i^{\ell-k}\Bigr)\,(\ad_j)^{k'}\,a_j^{\ell'}+h.c.\\
					&=i\sum_{\substack{0\leq k<\ell\leq d\\0\leq k',\,\ell'\leq d}}\lambda^{(e)}_{k\ell k'\ell'}\,(a_i)^{\ell-k}\,\widetilde{g}_{\ell-k}(N_i)\,(\ad_j)^{k'}\,a_j^{\ell'}+h.c.\,,
				\end{aligned}
			\end{equation*}
			where $\widetilde{g}_{\ell-k}(N_i)\coloneqq N_i[-\ell+1:k-\ell]\,g_{\ell-k}(N_i)$. Then, we proceed as before and represent the above operator in the Fock basis.
			\begin{equation*}
				\begin{aligned}
					i[H^{(d)}_e,f(N_i)]&=i\sum_{n_i,n_j=0}^{\infty}\sum_{\substack{0\leq k<\ell\leq d\\0\leq k',\,\ell'\leq d}}\lambda^{(e)}_{k\ell k'\ell'}\,\widetilde{g}_{\ell-k}(n_i)\,(\ad_j)^{k'}\,(a_i)^{\ell-k}\ketbra{n_i,n_j}{n_i,n_j}\,a_j^{\ell'}+h.c.\\
					&=i\sum_{n_i,n_j=0}^{\infty}\sum_{\substack{0\leq k<\ell\leq d\\0\leq k',\,\ell'\leq d\\n_i-\ell+k\geq0}}\lambda^{(e)}_{k\ell k'\ell'}\,\hat{g}_{k\ell k'\ell'}(n_i,n_j)\,\ketbra{n_i-\ell+k,n_j+k'}{n_i,n_j+\ell'}+h.c.\,,
				\end{aligned}
			\end{equation*}
			where $\hat{g}_{k\ell k'\ell'}(n_i,n_j)\coloneqq\sqrt{n_i[-\ell+k+1:0]}\sqrt{n_j[1:k']\,n_j[1:\ell']}\widetilde{g}_{\ell-k}(n_i)$ and $h.c.$ the Hermitian conjugate of the preceding term. Then, the matrix trick shows with $\ket{e_1}=\ket{n_i-\ell+k,n_j+k'}$ and $\ket{e_2}=\ket{n_i,n_j+\ell'}$
			\begin{equation*}
				\begin{aligned}
					i[H^{(d)}_e&,f(N_i)]\\
					&\leq\sum_{n_i,n_j=0}^{\infty}\sum_{\substack{0\leq k<\ell\leq d\\0\leq k',\,\ell'\leq d\\n_i-\ell+k\geq0}} |\lambda^{(e)}_{k\ell k'\ell'}|\,\hat{g}_{k\ell k'\ell'}(n_i,n_j)\Bigl(\ketbra{e_1}{e_1}+\ketbra{e_2}{e_2}\Bigr)\\
					&=\sum_{n_i,n_j=0}^{\infty}\sum_{\substack{0\leq k<\ell\leq d\\0\leq k',\,\ell'\leq d}} |\lambda^{(e)}_{k\ell k'\ell'}|\,\Bigl(\hat{g}_{k\ell k'\ell'}(n_i+\ell-k,n_j-k')+\hat{g}_{k\ell k'\ell'}(n_i,n_j-\ell')\Bigr)\ketbra{n_i,n_j}{n_i,n_j}\,.
				\end{aligned}
			\end{equation*}
			Then, we first use $g_{\ell-k}(n)\leq g_{p}(n)$ and redefine $\hat{g}_{k\ell k'\ell'}(n_i,n_j)$ with $\widetilde{g}_{\ell-k}(n_i)\coloneqq N_i[-\ell+1:k-\ell]\,g_{d}(n_i)$. Then, the monotonicity of $g_{d}(n)$ in $n$ (see \Cref{lem:connection-moments}) shows
			\begin{equation*}
				\begin{aligned}
					\hat{g}_{k\ell k'\ell'}(n_i+\ell-k,n_j-k')&\leq\hat{g}_{k\ell k'\ell'}(n_i+\ell-k,n_j-\min\{\ell',k'\})\\
					\hat{g}_{k\ell k'\ell'}(n_i,n_j-\ell')&\leq\hat{g}_{k\ell k'\ell'}(n_i,n_j-\min\{\ell',k'\})\\
					&\leq\hat{g}_{k\ell k'\ell'}(n_i+\ell-k,n_j-\min\{\ell',k'\})\,.
				\end{aligned}
			\end{equation*}
			This proves together with $|\lambda^{(e)}_{k\ell k'\ell'}|\leq \Lambda$ for all $0\leq k,\ell, k',\ell'\leq d$ the following inequality
			\begin{equation*}
				\begin{aligned}
					i[H^{(d)}_e,f(N_i)]&\leq\sum_{n_i,n_j=0}^{\infty}\sum_{\substack{0\leq k<\ell\leq d\\0\leq k',\,\ell'\leq d}}2\,\Lambda\,\,\hat{g}_{k\ell k'\ell'}(n_i+\ell-k,n_j-\min\{\ell',k'\})\,\ketbra{n_i,n_j}{n_i,n_j}\\
					&\leq\sum_{\substack{0\leq k<\ell\leq d\\0\leq k',\,\ell'\leq d}}2\,\Lambda\,\,\,\hat{g}_{k\ell k'\ell'}(N_i+(\ell-k)I,N_j-\min\{\ell',k'\}I)\,.
				\end{aligned}
			\end{equation*}
			Collecting all terms coming up in $\hat{g}$, using the monotonicity of $f$, the assumption $p\geq d$ and the bounds shown in \Cref{lem:bounds-ccr-l-product}, then
			\begin{equation*}
				\begin{aligned}
					&\hat{g}_{k\ell k'\ell'}(N_i+(\ell-k)I,N_j-\min\{\ell',k'\}I)\\
					&\quad=N_j[1-\min\{\ell',k'\}:0]\sqrt{N_j[1:|k'-\ell'|]}\sqrt{N_i[1:\ell-k]}N_i[-k+1:0]\,g_{p}(N_i+(\ell-k))\\
					&\quad\leq (N_j+I)^{\min\{\ell',k'\}}\sqrt{|k'-\ell'|!}(N_j+I)^{\frac{|k'-\ell'|}{2}}\sqrt{(\ell-k)!}(N_i+I)^{\frac{\ell-k}{2}}(N_i+I)^{k}\,g_{p}(N_i+(\ell-k))\\
					&\quad\leq d!\,(N_i+I)^d\,(N_j+I)^d\,g_{p}(N_i+p)\,,
				\end{aligned}
			\end{equation*}
			which implies
			\begin{equation}\label{eq-proof:hamiltonian-bound}
				\begin{aligned}
					i[H^{(d)}_e,f(N_i)]&\leq \Lambda\,(d+1)^4\,d!\,(N_i+I)^d\,(N_j+I)^d\,g_{p}(N_i+p)\\
					&\leq \Lambda k\,(d+1)^4\,d!\,(N_i+I)^d\,(N_j+I)^d\,N_i^{k-1}\\
					&\leq \Lambda k\,(d+1)^4\,d!\,(N_i+I)^{k+d-1}\,(N_j+I)^d\,.
				\end{aligned}
			\end{equation}
			by \Cref{lem:connection-moments}. Since $N_i+I\ge I$, we may replace the exponent $k+d-1$ by $k+d$. Finally, we use Young's inequality, which is
			\begin{equation*}
				ab\leq\frac{a^p}{p}+\frac{b^q}{q}
			\end{equation*}			
			for $a,b\geq0$ and $p,q>1$ satisfying $\frac{1}{p}+\frac{1}{q}=1$. Here, we choose $q=\frac{k}{d}$ and $p=\frac{k}{k-d}$
			to prove
			\begin{align*}
				(N_i+I)^{k+d}(N_j+I)^{d}\le \frac{k-d}{k}\,(N_i+I)^{\frac{(k+d)k}{(k-d)}}+\frac{d}{k}\,(N_j+I)^{k}\,,
			\end{align*}
			which finishes the proof of the lemma.
		\end{proof}
		
		In the next step, we prove a similar bound for the photon-dissipation. Since the dissipation is defined on-site, it is enough to consider the one-mode case. 
		\begin{lem}\label{lem:assumption-dissipation}
			For any two integers $k\ge 1$, $p\ge 2$, let the function $f:\mathbb{N}\to\mathbb{R}$ be defined as follows: $f(0)=f(1)=\dots =f(p-1)=0$ and for any $m\ge 0$, $f(m+p)=f(m)+km^{k-1}$. Then,
			\begin{equation*}
				\cL[a^p-\alpha^p]^\dagger(f(N))\le -k1_{N\geq p}\,\frac{(N+I)^{k-1+p}}{2\times 4^{p+1}}+k (N+I)^k\,\widetilde{\Delta}+\Delta_k
			\end{equation*}
			for any $\alpha\in\mathbb{C}$, $\widetilde{\Delta}$ only depending on $p$ and $|\alpha|$ (see Equation (\ref{eq:def-tilde-delta})) and $\Delta_k=k^{\mathcal{O}(k)}$ (see Equation (\ref{eq:def-delta-k})).
		\end{lem}
		\begin{proof}
			As in \Cref{prop:LRB-bose-hubbard}, we start with an operator bound on the $p$-photon dissipation similar to the proof of Lemma 5.3 in \cite{Gondolf.2024}. However, we define $f$ as in \Cref{lem:connection-moments} and consider
			\begin{equation*}
				\begin{aligned}
					(\cL^{(\alpha,p)})^\dagger(f(N))=-N[-p+1:0]g_p(N)+\frac{\overline{\alpha}^p}{2}a^pg_{p}(N)+\frac{\alpha^p}{2}g_{p}(N)(\ad)^p\,.
				\end{aligned}
			\end{equation*}
			In the next step, we denote $\widetilde{g}_p(n)\coloneqq 1_{n\geq p}\sqrt{n[-p+1:0]}g_p(n)$, where $g_p(n)\coloneqq f(n)-f(n-p)$ (see \Cref{eq:f-g-l-function}), and represent the operator in the Fock basis
			\begin{equation*}
				\begin{aligned}
					(&\cL^{(\alpha,p)})^\dagger(f(N))\\
					&=\sum_{n=p}^\infty \frac{\widetilde{g}_p(n)}{2}\left(\overline{\alpha}^p\ketbra{n-p}{n}+\alpha^p\ketbra{n}{n-p}\right)-\sqrt{n[-p+1:0]}\widetilde{g}_p(n)\ketbra{n}{n}\\
					&=\sum_{n=0}^\infty \frac{\widetilde{g}_p(n+p)}{2}\left(\overline{\alpha}^p\ketbra{n}{n+p}+\alpha^p\ketbra{n+p}{n}-2\sqrt{n[1:p]}\ketbra{n+p}{n+p}\right)\,.
				\end{aligned}
			\end{equation*}
			Similar to the proof of \Cref{prop:LRB-bose-hubbard}, we consider the following bound on a $2\times2$-matrix
			\begin{equation}\label{eq:2x2}
				-x_1\ketbra{e_1}{e_1}-x_2\ketbra{e_2}{e_2}+y\ketbra{e_2}{e_1}+\overline{y}\ketbra{e_1}{e_2}\leq\left(-\min\{x_1,x_2\}+|y|\right)\left(\ketbra{e_1}{e_1}+\ketbra{e_2}{e_2}\right),
			\end{equation}
			which holds true because of the following bound on the eigenvalues:
			\begin{equation*}
				\frac{-x_1-x_2\pm\sqrt{(x_1-x_2)^2+4|y|^2}}{2}\leq -\min\{x_1,x_2\}+|y|\,.
			\end{equation*}
			Then, we apply the bound to each block  defined by $\ket{e_1}\coloneqq \ket{n}$ and $\ket{e_2}\coloneqq \ket{n+p}$ with coefficients $x_1=\frac{1}{2}\sqrt{n[-p+1:0]}\widetilde{g}_p(n)$, $x_2=\frac{1}{2}\sqrt{n[1:p]}\widetilde{g}_p(n+p)$ and $y=\frac{1}{2}\widetilde{g}_p(n+p)\alpha^p$:
			\begin{equation*}
				\begin{aligned}
					(\cL^{(\alpha,p)})^\dagger(f(N))&\leq\sum_{n=0}^\infty\frac{1}{2}\left(\widetilde{g}_p(n+p)|\alpha|^p-\sqrt{n[-p+1:0]}\widetilde{g}_p(n)\right)\left(\ketbra{n}{n}+\ketbra{n+p}{n+p}\right)\\
					&\leq\widetilde{g}_p(N+pI)|\alpha|^p-\sqrt{N[-p+1:0]}\widetilde{g}_p(N)\\
					&=k|\alpha|^p\sqrt{N[1:p]}N^{k-1}-kN[-p+1:0](N-pI)^{k-1}\\
					&\leq |\alpha|^p k\sqrt{p!}\,(N+I)^{p/2+k-1}-kN[-p+1:0](N-pI)^{k-1}
				\end{aligned}
			\end{equation*}
			where we used the definition of $f$ in \Cref{lem:connection-moments} and the bounds of \Cref{lem:bounds-ccr-l-product} for the positive term. Next, we upper bound the negative term above: 
			\begin{equation*}
				\begin{aligned}
					N[-p+1:0]\,g_p(N)&=kN[-p+1:0](N-pI)^{k-1}\\
					&\overset{(3)}{\geq}k\,1_{N\geq p}\,(N-pI)^{k-1}\big((N+I)^p-\frac{p(p+1)}{2}(N+I)^{p-1}\big)\\
					&\ge k\,\,1_{N\geq p}\,(N-pI)^{k-1+p}-k\frac{p(p+1)}{2} 1_{N\geq p}(N+I)^{k+p-2}\, .
				\end{aligned}
			\end{equation*}
			Here, we used in the first line that the kernel of $N[-p+1:0]$ simplifies the expression for $g_p(N)$ and in $(3)$ \Cref{lem:bounds-ccr-l-product}. Then, \Cref{lem:connection-moments} shows
			\begin{align*}
				N[-p+1:0]\,g_p(N)&\geq k 1_{N\geq p}\frac{(N+I)^{k+p-1}}{4^{p+1}}-1_{N\geq p} \frac{k (k-1+p)^{k-1+p}}{3}\\
				&\qquad-k\frac{p(p+1)}{2}1_{N\geq p}(N+1)^{k+p-2}\,.
			\end{align*}
			To sum up, we showed that
			\begin{align}
				\cL[a^p-\alpha^p]^\dagger&(f(N))\leq |\alpha|^p k\sqrt{p!}\,(N+I)^{p/2+k-1}-kN[-p+1:0](N-pI)^{k-1}\nonumber\\
				&\le k1_{N\geq p}\,\Big(-\frac{(N+I)^{k-1+p}}{4^{p+1}}+\frac{p(p+1)}{2}(N+I)^{k+p-2}+|\alpha|^p\sqrt{p!}(N+I)^{k-1+\frac{p}{2}}\Big)\nonumber\\
				&\qquad+|\alpha|^pk\sqrt{p!}\,1_{N< p}(N+I)^{k-1+\frac{p}{2}}+ k1_{N\geq p}\frac{ (k-1+p)^{k-1+p}}{3}\nonumber\\
				&\le k1_{N\geq p}(N+I)^k\,\Big(-\frac{(N+I)^{p-1}}{4^{p+1}}+\frac{p(p+1)}{2}(N+I)^{p-2}+|\alpha|^p\sqrt{p!}(N+1)^{\frac{p}{2}-1}\Big)\nonumber\\
				&\qquad+\underbrace{|\alpha|^pk\sqrt{p!}\,(p+1)^{k-1+\frac{p}{2}}+k\frac{ (k-1+p)^{k-1+p}}{3}}_{=:\Delta_{k}}\label{eq:def-delta-k}\\
				&\le k1_{N\geq p}(N+I)^k\,\Big(-\frac{(N+I)^{p-1}}{4^{p+1}}+\underbrace{\Big(\frac{p(p+1)}{2}+|\alpha|^p\sqrt{p!}\Big)}_{=:\Delta}(N+I)^{p-2}\Big)+\Delta_k\label{eq:def-delta}
			\end{align}
			Next, by \Cref{polymax} applied to the polynomial $-\frac{1}{2\times 4^{p+1}}X^{p-1}+\Delta X^{p-2}$, we find 
			\begin{equation*}
				\begin{aligned}
					\cL[a^p-\alpha^p]^\dagger(f(N))&\le -k1_{N\geq p}\,\frac{(N+I)^{k-1+p}}{2\times 4^{p+1}}+k (N+I)^k\,\Delta\left(\frac{2\times 4^{p+1}\Delta(p-2)}{p-1}\right)^{p-2}+\Delta_k\\
					&= -k1_{N\geq p}\,\frac{(N+I)^{k-1+p}}{2\times 4^{p+1}}+k (N+I)^k\,\widetilde{\Delta}+\Delta_k
				\end{aligned}
			\end{equation*}
			with 
			\begin{equation}\label{eq:def-tilde-delta}
				\widetilde{\Delta}=
				\begin{cases}
					\Delta, & p=2,\\
					\Delta\left(\frac{2\times 4^{p+1}\Delta(p-2)}{p-1}\right)^{p-2}, & p>2,
				\end{cases}
			\end{equation}
			which finishes the proof.
		\end{proof}

	\subsection{Auxiliary results}
	
		\subsubsection{Auxiliary results: Bounds on projected generators}
			In this section, we calculate bounds for generators concatenated with a projection onto a finite-dimensional subspace given via the Fock basis. This auxiliary result is used in the proof of \Cref{thm:bosonicLRB}. For that, we briefly recall the definition of the projections involved
			\begin{equation*}
				P_{\pm s}=P^{(M)}_{\pm s}\coloneqq\prod_{j\in R[\pm s]}P_j^{(M)}\,,\qquad\overline{P}_{\pm s}=I-P_{\pm s} \qquad\text{and}\qquad\cP_{\pm s}=\cP_{\pm s}^{(M)}\coloneqq P_{\pm s}\cdot P_{\pm s} 
			\end{equation*}
			with $R[s]\coloneqq\{v\in V\,|\,\dist(v,R)\leq s\}$, $R[-s]\coloneqq\{v\in V\,|\,\dist(v,R^c)> s\}$ and $P_j^{(M)}\coloneqq\sum_{n\leq M}\ketbra{n}{n}_j$. The special case $s=0$ is denoted by $P_0=P$ and $\cP_0=\cP$. In the following, we prove an explicit bound depending on $M$ as well as the size of the region $R$.
		
			\begin{lem}\label{lem:boundedness-projective-generator}
				Let $\widetilde{\cL}_R$ be a local generator satisfying \Cref{assum:local-GKSL} and defined on the subset $R\subset V$. Then,
				\begin{equation*}
					\begin{aligned}
						\|H_XP\|_{\infty}, \|PH_X\|_{\infty}&\leq \Lambda (d+1)^{2r_{\exp}}d!^{r_{\exp}}(M+1)^{r_{\exp}d}\\
						\|L_{X,j}P\|_{\infty}\|P(L_{X,j})^\dagger\|_{\infty}&\leq\Lambda(d/2+1)^{4r_{\exp}}(d/2)!^{2r_{\exp}}(M+1)^{r_{\exp}d}\\
						\|(L_{X,j})^\dagger L_{X,j}P\|_{\infty},\|P(L_{X,j})^\dagger L_{X,j}\|_{\infty}&\leq \Lambda(d/2+1)^{4r_{\exp}}d!^{r_{\exp}}(M+1)^{r_{\exp}d}\,.
					\end{aligned}
				\end{equation*}
				for all $X\in\cV^{(r)}_R$ satisfying $j\in\{1,...,\nu(X)\}$. Moreover, 
				\begin{equation*}
					\begin{aligned}
						\|\widetilde{\cL}_R\circ\cP\|_{1\rightarrow 1}\leq 2(1+\nu)|\cV_R^{(r)}|\Lambda d!^{r_{\exp}}(M+1)^{r_{\exp}d}(d/2+1)^{4r_{\exp}}\,.
					\end{aligned}
				\end{equation*}
			\end{lem}
			\begin{proof}
				Let $Y \in \mathcal{T}_f$. Then, Hölder's inequality shows
				\begin{equation*}
					\begin{aligned}
						\|\widetilde{\cL}_R&\circ\cP(Y)\|_{1}\\
						&\leq 2(1+\nu)|\cV_R^{(r)}|\max_{X\in\cV^{(r)}_R}\max_{j\in\{1,...,\nu(X)\}}\Bigl\{\|H_XP\|_{\infty},\|PH_X\|_{\infty},\|L_{X,j}P\|_{\infty}\|P(L_{X,j})^\dagger\|_{\infty},\\
						&\qquad\qquad\qquad\qquad\qquad\qquad\qquad\qquad\qquad\|(L_{X,j})^\dagger L_{X,j}P\|_{\infty},\|P(L_{X,j})^\dagger L_{X,j}\|_{\infty}\Bigr\}\|Y\|_1\,.
					\end{aligned}
				\end{equation*}
				Next, we consider the polynomials $H_X,L_{X,j}$ defined on $r_{\exp}$-local subsets consisting of at most $(d+1)^{2|X|}\leq (d+1)^{2r_{\exp}}$ Hamiltonian monomials and $(d/2+1)^{2|X|}\leq (d/2+1)^{2r_{\exp}}$ Lindbladian monomials. Moreover, all coefficients can be upper bounded by $\Lambda$ and all monomials are by definitions represented by a $|X|$-tensor product of terms of the form $(a_j^\dagger)^{k'}a_j^{\ell'}$ for $\ell',k'\leq d$ (or $\leq d/2$). To calculate the norm $\|PH_X\|_{\infty}$, we first consider an eigenvector of $P$ and calculate
				\begin{equation*}
					\begin{aligned}
						(a_j^\dagger)^{\ell'}a_j^{k'}\ket{n_j}&=\sqrt{c_{\ell'k'}(n_j)}\ket{n_j-k'+\ell'}
					\end{aligned}
				\end{equation*}
				and $c_{\ell' k'}(n_j)\coloneqq \delta_{n_j\geq \ell'-k'}n_j[-k'+1:0]n_j[-k'+1:-k'+\ell'+1]$ (see \Cref{eq:notation-increasing-sequence}). Then, 
				\begin{equation*}
					\begin{aligned}
						\|Pa_j^{k'}(a_j^\dagger)^{\ell'}\ket{\psi}\|_{2}^2&=\sum_{n_j=0}^{M}{c_{\ell'k'}(n_j)}\,|\braket{\psi\,|\,n_j-k'+\ell'}|^2\\
						&\leq \sum_{n_j=0}^{M}\ell'!^2(n_j+1)^{k'+\ell'}|\braket{\psi\,|\,n_j-k'+\ell'}|^2\\
						&\leq d!^2(M+1)^{2d}\sum_{n_j=0}^{\infty}|\braket{\psi\,|\,n_j}|^2\\
						&\leq d!^2(M+1)^{2d}\|\ket{\psi}\|_2^2\,.
					\end{aligned}
				\end{equation*}
				By tensorization, i.e.
				\begin{equation*}
					\bigotimes_{j\in X} (a_j^\dagger)^{k'_j}a_j^{\ell'_j}\ket{n_j}=\bigotimes_{j\in X}\sqrt{c_{\ell_j'k'_j}(n_j)}\ket{n_j-k'_j+\ell'_j}\,
				\end{equation*}
				and the CCR, the above upper bound implies
				\begin{equation*}
					\begin{aligned}
						\|P\bigotimes_{j\in X} (a_j^\dagger)^{k'_j}a_j^{\ell'_j}\ket{\psi}\|_{2}^2\leq (d!)^{2|X|}(M+1)^{2|X|d}\|\ket{\psi}\|_2^2\leq (d!)^{2r_{\exp}}(M+1)^{2r_{\exp}d}\|\ket{\psi}\|_2^2\,.
					\end{aligned}
				\end{equation*}
				which proves $\|PH_X\|_{\infty}\leq \Lambda(d+1)^{2r_{\exp}}(d!)^{r_{\exp}}(M+1)^{r_{\exp}d}$. This calculation can be extended to monomials coming up in the product $\|P(L_{X,j})^\dagger L_{X,j}\|_{\infty}$ by
				\begin{equation*}
					\begin{aligned}
						(a_j^\dagger)^{\ell'_2}&a_j^{k'_2}(a_j^\dagger)^{k'_1}a_j^{\ell'_1}\ket{n_j}\\
						&=\sqrt{c_{k'_1\ell'_1}(n_j)}a_j^{\ell'_2}(a_j^\dagger)^{k'_2}\ket{n_j-\ell'_1+k'_1}\\
						&=\sqrt{c_{\ell'_2k'_2}(n_j-\ell'_1+k'_1)c_{k'_1\ell'_1}(n_j)}\ket{n_j+\ell_2'-\ell'_1+k'_1-k_2'}\\
						&=\sqrt{c_{\ell'_2k'_2}(n_j+\ell'_1-k'_1)c_{k'_1\ell'_1}(n_j)}\ket{n_j-\overline{k}'+\overline{\ell}'}\\
						&\coloneqq \sqrt{c}\ket{n_j-\overline{k}'+\overline{\ell}'}
					\end{aligned}
				\end{equation*}
				where $\overline{k}'\coloneqq k_2'-k_1'$ and $\overline{\ell}'$ in the same way. Then, 
				\begin{equation*}
					\begin{aligned}
						\|P(a_j^\dagger)^{\ell'_1}&a_j^{k'_1}(a_j^\dagger)^{k'_2}a_j^{\ell'_2}\ket{\psi}\|_{2}^2\\
						&=\sum_{n_j=0}^{M}c\,|\braket{\psi\,|\,n_j-\overline{k}'+\overline{\ell}'}|^2\\
						&\leq \sum_{n_j=0}^{M}(\ell_2'+k_1')!^2(n_j+1)^{k_1'+\ell_1'+k_2'+\ell_2'}\,|\braket{\psi\,|\,n_j-\overline{k}'+\overline{\ell}'}|^2\\
						&\leq d!^2(M+1)^{2d}\sum_{n_j=0}^{\infty}|\braket{\psi\,|\,n_j}|^2\\
						&\leq d!^2(M+1)^{2d}\|\ket{\psi}\|_2^2\,.
					\end{aligned}
				\end{equation*}
				Again by tensorizing the monomial and using the CCR to achieve a $|X|$-local monomial, the above calculation shows
				\begin{equation*}
					\|P\bigotimes_{j\in X} (a_j^\dagger)^{\ell'_{j,1}}a_j^{k'_{j,1}}(a_j^\dagger)^{k'_{j,2}}a_j^{\ell'_{j,2}}\ket{\psi}\|_{2}^2\leq d!^{2|X|}(M+1)^{2d |X|}\|\ket{\psi}\|_2^2\leq d!^{2r_{\exp}}(M+1)^{2r_{\exp}d}\|\ket{\psi}\|_2^2\,.
				\end{equation*}
				In a similar way, the other bounds follow, so that for $X\in\cV^{(r)}_R$ satisfying $j\in[\nu(X)]$
				\begin{equation*}
					\begin{aligned}
						\|H_XP\|_{\infty}, \|PH_X\|_{\infty}&\leq \Lambda (d+1)^{2r_{\exp}}d!^{r_{\exp}}(M+1)^{r_{\exp}d}\\
						\|L_{X,j}P\|_{\infty}\|P(L_{X,j})^\dagger\|_{\infty}&\leq\Lambda(d/2+1)^{4r_{\exp}}(d/2)!^{2r_{\exp}}(M+1)^{r_{\exp}d}\\
						\|(L_{X,j})^\dagger L_{X,j}P\|_{\infty},\|P(L_{X,j})^\dagger L_{X,j}\|_{\infty}&\leq \Lambda(d/2+1)^{4r_{\exp}}d!^{r_{\exp}}(M+1)^{r_{\exp}d}\,.
					\end{aligned}
				\end{equation*}
				Therefore,
				\begin{equation*}
					\begin{aligned}
						\|\widetilde{\cL}_R\circ\cP(Y)\|_{1}&\leq 2(1+\nu)|\cV_R^{(r)}|\Lambda d!^{r_{\exp}}(M+1)^{r_{\exp}d}\max\Bigl\{(d+1)^{2r_{\exp}},(d/2+1)^{4r_{\exp}}\Bigr\}\|Y\|_1\\
						&\leq 2(1+\nu)|\cV_R^{(r)}|\Lambda d!^{r_{\exp}}(M+1)^{r_{\exp}d}(d/2+1)^{4r_{\exp}}\|Y\|_1\,,
					\end{aligned}
				\end{equation*}
				where we used $d+1\leq (d/2+1)^2$ in the last line. By continuity of the trace norm, the inequality extends to all $Y\in\cT_1$ finishing the proof.
			\end{proof}
		
			Since the explicit examples of the Bose--Hubbard model \Cref{ex:Bose-Hubbard}, any quadratic Hamiltonian \Cref{ex:Gaussian}, and the regularized Hamiltonian \Cref{ex:reg-hamiltonian} are important for our application, we present tighter error bounds adapted to these examples.
			\begin{cor}\label{cor:boundedness-projective-generator-bose-hubbard}
				Let $(V,E)$ be a graph and $H$ the Bose--Hubbard Hamiltonian (\ref{ex:Bose-Hubbard})
				\begin{equation*}
					H=\sum_{(i,j)\in E}\lambda^{(i,j)}\,a_i^\dagger\,a_j+\sum_{i\in V}u^{(i)}N_i(N_i-I)+\sum_{i\in V}\mu^{(i)}N_i\,.
				\end{equation*}
				For the projection $P$ on a region $R\subset V$ as defined in \Cref{thm:bosonicLRB}, with induced edge set $E_R$, 
				\begin{equation*}
					\|[H,\cdot]\circ\cP\|_{1\rightarrow1}\leq2\Lambda(|E_R|+|R|)(M+1)^2\,,
				\end{equation*}
				where $\cP=P\cdot P$ and $\Lambda$ the upper bound on the absolute values of the coefficients.
			\end{cor}
			\begin{proof}
				By following the proof of \Cref{lem:boundedness-projective-generator}, for $X\in\cT_f$ 
				\begin{equation*}
					\begin{aligned}
						\|\cP\circ[H_R,\cdot](X)\|_1&\leq (\|PH_R\|_{\infty}+\|H_RP\|_{\infty})\|X\|_1\\
						&\leq\Lambda\Bigl(|E_R|\max_{(i,j)\in E_R}\{\|Pa_i^\dagger\,a_j\|_{\infty}+\|a_i^\dagger\,a_jP\|_{\infty}\}\\
						&\qquad\quad+|R|\max_{v\in R}\{\|PN_v(N_v-I)\|_{\infty}+\|PN_v\|_{\infty}\\
						&\qquad\qquad\qquad\qquad\quad+\|N_v(N_v-I)P\|_{\infty}+\|N_vP\|_{\infty}\}\Bigr)\|X\|_1\\
						&\leq2\Lambda\Bigl(|E_R|(M+1)+|R|\bigl(M(M-1)+M\bigr)\Bigr)\|X\|_1\\
						&\leq2\Lambda(|E_R|+|R|)(M+1)^2\|X\|_1\,.
					\end{aligned}
				\end{equation*}
				Then, continuity of the norm extends the bound from $\cT_f$ to $\cT_1$.
			\end{proof}
		
			In the next step, we generalize the above bound to any quadratic Hamiltonian:
			\begin{cor}\label{cor:boundedness-projective-generator-quadratic-generalized}
				Let $(V,E)$ be a graph and $H$ a quadratic Hamiltonian \Cref{ex:Gaussian}
				\begin{equation*}
					H\coloneqq \sum_{(i,j)\in E}\lambda^{(i,j)}\,a_i^\dagger\,a_j+\sum_{i\in V}b^{(i)}a_i^2+\overline{b}^{(i)}(\ad_i)^2+c^{(i)}a_i+\overline{c}^{(i)}\ad_i\,.
				\end{equation*}
				For the projection $P$ on a region $R\subset V$ as defined in \Cref{thm:bosonicLRB}, with induced edge set $E_R$, 
				\begin{equation*}
					\|[H,\cdot]\circ\cP\|_{1\rightarrow1}\leq2\Lambda\Bigl(|E_R|(M+1)+5|R|(M+1)\Bigr)\,,
				\end{equation*}
				where $\cP=P\cdot P$ and $\Lambda$ the upper bound on the absolute values of the coefficients.
			\end{cor}
			\begin{proof}
				By following the proof of \Cref{lem:boundedness-projective-generator}, for $X\in\cT_f$
				\begin{equation*}
					\begin{aligned}
						\|\cP\circ[H_R,\cdot](X)\|_1&\leq(\|PH_R\|_{\infty}+\|H_RP\|_{\infty})\|X\|_1\\
						&\leq\Lambda\Bigl(|E_R| \max_{(i,j)\in E_R}\{\|Pa_i^\dagger\,a_j\|_{\infty}+\|a_i^\dagger\,a_jP\|_{\infty}\}\\
						&\qquad\quad+|R| \max_{v\in R}\{\|Pa_v^2\|_{\infty}+\|Pa_v\|_{\infty}+\|a_v^2P\|_{\infty}+\|a_vP\|_{\infty}\\
						&\qquad\qquad\qquad\qquad+\|P(\ad_v)^2\|_{\infty}+\|P\ad_v\|_{\infty}+\|(\ad_v)^2P\|_{\infty}+\|\ad_vP\|_{\infty}\Bigr)\|X\|_1\\
						&\leq2\Lambda\Bigl(|E_R|(M+1)+|R|\bigl(3(M+1)+2\sqrt{M+1}\bigr)\Bigr)\|X\|_1\\
						&\leq2\Lambda\Bigl(|E_R|(M+1)+|R|5(M+1)\Bigr)\|X\|_1\,.
					\end{aligned}
				\end{equation*}
				Then, continuity of the norm extends the bound from $\cT_f$ to $\cT_1$.
			\end{proof}
		
			Finally, we present an explicit bound on the regularized Hamiltonian given in \Cref{ex:reg-hamiltonian}.
			\begin{cor}\label{cor:boundedness-projective-generator-reg-hamiltonian}
				For $\alpha\in\C^{|R|}$, $p\in\N$, the regularized Hamiltonian (\ref{ex:reg-hamiltonian}) is given by
				\begin{equation*}
					\widetilde{\cL}^{(\alpha)}_R=\cH^{(d)}+\cL^{(\alpha,p)}\,,
				\end{equation*}
				on a region $R\subset V$ with induced edge set $E_R$ and the projection $P$ as defined in \Cref{thm:bosonicLRB}. Then,
				\begin{equation*}
					\begin{aligned}
						\|Pa_j^p\|_\infty,\|a_j^pP\|_\infty,\|P(\ad_j)^p\|_\infty,\|(\ad_j)^pP\|_\infty&\leq\sqrt{p!}(M+1)^{p/2}\\
						\|P(\ad_j)^pa_j^p\|_\infty,\|(\ad_j)^pa_j^pP\|_\infty&\leq (M+1)^{p}
					\end{aligned}
				\end{equation*}
				and
				\begin{equation*}
					\begin{aligned}
						\|\widetilde{\cL}_R^{(\alpha)}\circ \cP\|_{1\rightarrow1}&\leq|R|(1+2\eta)(p!+1)(M+1)^{p}+2|E_R|\Lambda d^{4}(d!)^{2}(M+1)^{2d}\,.
					\end{aligned}
				\end{equation*}
				for $\eta=\max_{j\in R}\{|\alpha_j|^p\}$, $\cP=P\cdot P$ and $\Lambda$ the upper bound on the absolute values of the coefficients of the two-body interaction Hamiltonian $H$.
			\end{cor}
			\begin{proof}
				Again, we follow the proof of \Cref{lem:boundedness-projective-generator}, for $X\in\cT_f$ 
				\begin{equation*}
					\begin{aligned}
						\|[H^{(d)},\cdot]\circ\cP\|_{1\rightarrow 1}\leq2|E_R|\Lambda d^{4}(d!)^{2}(M+1)^{2d}\,.
					\end{aligned}
				\end{equation*}
				Next, we consider the dissipation $\cL_R^{(\alpha,p)}=\sum_{j\in R}\cL[L_j^{(\alpha_j,p)}]$ with $L_j^{(\alpha_j,p)}=a_j^p-\alpha_j^p$. 
				\begin{equation*}
					\begin{aligned}
						\|\cL_R^{(\alpha,p)}\circ\cP(X)\|_{1}&\leq |R|\max_{j\in R}\|\cL[L_j^{(\alpha_j,p)}]\circ\cP(X)\|_1\\
						&\leq |R|\max_{j\in R}\|\Bigl(\cL[a_j^p]+\frac{1}{2}\overline{\alpha}_j^pi\cH[a^p_j]+\frac{1}{2}{\alpha}_j^pi\cH[(\ad_j)^p]\Bigr)\circ\cP(X)\|_1\\
						&\leq |R|\max_{j\in R}\Bigl(\|Pa_j^p\|_\infty\|(\ad_j)^pP\|_\infty+\frac{
							1}{2}\|P(\ad_j)^pa_j^p\|_\infty+\frac{1}{2}\|(\ad_j)^pa_j^pP\|_\infty\\
						&\qquad\qquad\quad+\frac{1}{2}|\alpha_j|^p\,\bigl(\|Pa_j^p\|_\infty+\|a_j^pP\|_\infty+\|P(\ad_j)^p\|_\infty+\|(\ad_j)^pP\|_\infty\bigr)\Bigr)\|X\|_1\,.
					\end{aligned}
				\end{equation*}
				By the same arguments as in the proof of \Cref{lem:boundedness-projective-generator}
				\begin{equation*}
					\begin{aligned}
						\|Pa_j^p\|_\infty,\|a_j^pP\|_\infty,\|P(\ad_j)^p\|_\infty,\|(\ad_j)^pP\|_\infty&\leq\sqrt{p!}(M+1)^{p/2}\\
						\|P(\ad_j)^pa_j^p\|_\infty,\|(\ad_j)^pa_j^pP\|_\infty&\leq M^{p}
					\end{aligned}
				\end{equation*}
				which show
				\begin{equation*}
					\begin{aligned}
						\|\cP\cL_R^{(\alpha,p)}(X)\|_{1}&\leq |R|\Bigl((p!+1)(M+1)^{p}+2\max_{j\in R}\{|\alpha_j|^p\}\sqrt{p!}(M+1)^{p/2}\Bigr)\|X\|_1\,.
					\end{aligned}
				\end{equation*}
				and finishes the proof by again extending $\cT_f$ to $\cT_1$ via continuity. 
			\end{proof}

		\subsubsection{Auxiliary results: relative bounds by the photon-number operator}
			In this section, we calculate relative bounds for polynomials in annihilation and creation operators on a region $R$ by the photon-number operator. This auxiliary result is used in the proof of \Cref{thm:bosonicLRB}.
		
			\begin{lem}\label{lem:relative-bounds-number-op}
				Let $\widetilde{\cL}_R$ be an $r$-local generator satisfying \Cref{assum:local-GKSL} and defined on the subset $R\subset V$. Then,
				\begin{equation*}
					\begin{aligned}
						\|H_X\rho\|_{1}&\leq \Lambda (d+1)^{2r_{\exp}}d!^{r_{\exp}}\|(N+I)^{\dd}_X\rho\|_{1}\,,\\
						\|\rho H_X\|_{1}&\leq \Lambda (d+1)^{2r_{\exp}}d!^{r_{\exp}}\|\rho(N+I)^{\dd}_X\|_{1}\,,\\
						\|L_{X,j}\rho L_{X,j}^\dagger\|_{1}&\leq \Lambda(d/2+1)^{4r_{\exp}}(d/2)!^{2r_{\exp}}\|(N+I)^{\dd/2}_X\rho(N+I)^{\dd/2}_X\|_{1}\,,\\
						\|L_{X,j}^\dagger L_{X,j}\rho\|_{1}&\leq \Lambda(d/2+1)^{4r_{\exp}}d!^{r_{\exp}}\|(N+I)^{\dd}_X\rho\|_{1}\,,\\
						\|\rho L_{X,j}^\dagger L_{X,j}\|_{1}&\leq \Lambda(d/2+1)^{4r_{\exp}}d!^{r_{\exp}}\|\rho(N+I)^{\dd}_X\|_{1}\,.
					\end{aligned}
				\end{equation*}
				for all $\rho\in W^{d,1}$, $X\in\cV^{(r)}_R$ satisfying $j\in\{1,...,\nu(X)\}$ and $(N+I)^{\dd}_X\coloneqq\bigotimes_{j\in X}(N_j+I)^d$ for $\dd=(d,...,d)\in\N^{|X|}$. Moreover,
				\begin{equation*}
					\begin{aligned}
						\|\widetilde{\cL}_R(\rho)\|_{1}\leq \Lambda(d/2+1)^{4r_{\exp}}d!^{r_{\exp}}|\cV_R^{(r)}|\max_{X\in\cV^{(r)}_R}\Bigl((1+\frac{\nu}{2})\bigl(\|\rho (N&+I)_X^{\dd}\|_{1}+\|(N+I)_X^{\dd}\rho \|_{1}\bigr)\\
						&+\nu\|(N+I)_X^{\dd/2}\rho (N+I)_X^{\dd/2}\|_{1}\Bigr)\,.
					\end{aligned}
				\end{equation*}
			\end{lem}
			\begin{proof}
				To prove the above result, we follow the proof of \Cref{lem:boundedness-projective-generator} for a state $\rho\in\cT_f$. First, we bound the Hamiltonian terms as well as Lindblad terms by
				\begin{equation*}
					\begin{aligned}
						\|\cH[H_X](\rho)\|_1 &\leq\|H_X\rho\|_1+\|\rho H_X\|_1\,,\\
						\|\cL[L_{X,j}](\rho)\|_1 &\leq\|L_{X,j}\rho L_{X,j}^\dagger\|_1+\frac{1}{2}\|L_{X,j}^\dagger L_{X,j}\rho\|_1+\frac{1}{2}\|\rho L_{X,j}^\dagger L_{X,j}\|_1
					\end{aligned}
				\end{equation*}
				for $X\in\cV^{(r)}_R$ and $j\in\{1,...,\nu(X)\}$. Next, we consider the polynomials $H_X,L_{X,j}$ defined on local subsets $X\in\cV_R^{(r)}$ consisting of at most $(d+1)^{2|X|}\leq (d+1)^{2r_{\exp}}$ Hamiltonian monomials and $(d/2+1)^{2|X|}\leq(d/2+1)^{2r_{\exp}}$  Lindbladian monomials. Moreover, all coefficients can be upper bounded by $\Lambda$ and all monomials are by definition represented by a $|X|$-tensor product of terms of the form $a_j^{k'}(a_j^\dagger)^{\ell'}$ for $\ell',k'\leq d$ (or $\leq d/2$). To calculate the relative boundedness, we use the invertibility of $(N_j+I)^{d}$ as well as Hölder's inequality to show for any $\rho\in\cT_f$
				\begin{equation*}
					\begin{aligned}
						\|\rho (a_j^\dagger)^{k'}a_j^{\ell'}\|_1&\leq\|\rho(N_j+I)^{d}\|_1\,\|(N_j+I)^{-d}(a_j^\dagger)^{k'}a_j^{\ell'}\|_{\infty}\,.
					\end{aligned}
				\end{equation*}
				Next, we bound $\|(N_j+I)^{-d}(a_j^\dagger)^{k'}a_j^{\ell'}\|_{\infty}$ by first considering an arbitrary eigenvector of $(N_j+I)^{-d}$
				\begin{equation*}
					\begin{aligned}
						(a_j^\dagger)^{\ell'}a_j^{k'}\frac{1}{(n_j+1)^d}\ket{n_j}&=\sqrt{c_{\ell'k'}(n_j)}\frac{1}{(n_j+1)^d}\ket{n_j-k'+\ell'}
					\end{aligned}
				\end{equation*}
				with $c_{\ell' k'}(n_j)\coloneqq \delta_{n_j\geq \ell'-k'}n_j[-k'+1:0]n_j[-k'+1:-k'+\ell'+1]$ (see \Cref{eq:notation-increasing-sequence}). Then,
				\begin{equation*}
					\begin{aligned}
						\|(N_j+I)^{-d}(a_j^\dagger)^{k'}a_j^{\ell'}\ket{\psi}\|_{2}^2&=\sum_{n_j=0}^{\infty}\frac{c_{\ell'k'}(n_j)}{(n_j+1)^{2d}}\,|\braket{\psi\,|\,n_j-k'+\ell'}|^2\\
						&\leq \sum_{n_j=0}^{\infty}k'!^2\frac{(n_j+1)^{k'+\ell'}}{(n_j+1)^{2d}}|\braket{\psi\,|\,n_j-k'+\ell'}|^2\\
						&\leq d!^2 \sum_{n_j=0}^{\infty}|\braket{\psi\,|\,n_j}|^2\\
						&= d!^2\|\ket{\psi}\|_2^2
					\end{aligned}
				\end{equation*}
				for $\psi\in\cT_f$. By repeating the above calculations the following bound holds true
				\begin{equation}\label{eq:rel-boundedness-single-mode-term}
					\begin{aligned}
						\|(a_j^\dagger)^{k'}a_j^{\ell'}(N_j+I)^{-d}\ket{\psi}\|_{2}^2&\leq d!^2\|\ket{\psi}\|_2^2\,.
					\end{aligned}
				\end{equation}
				On a region $X$, we consider polynomials of the form $p_X(a,\ad)\coloneqq\bigotimes_{j\in X} (a_j^\dagger)^{k'_j}a_j^{\ell'_j}$ and the number operator $(N+I)^{\dd}_X\coloneqq\bigotimes_{j\in X}(N_j+I)^d$ for $\dd=(d,...,d)\in\N^{|X|}$. Then, the above single-mode bounds directly imply
				\begin{equation*}
					\begin{aligned}
						\|(N+I)^{-\dd}_Xp_{X}(a,\ad)\|_{\infty}&\leq d!^{|X|}\leq d!^{r_{\exp}}\,,\\
						\|p_{X}(a,\ad)(N+I)^{-\dd}_X\|_{\infty}&\leq d!^{|X|}\leq d!^{r_{\exp}}\,,
					\end{aligned}
				\end{equation*}
				where we also used continuity of the norms to take to closure of $\cH_f$ to $\cH$. Then, by triangle inequality, we achieve the bounds
				\begin{equation}\label{eq:rel-bounded-bound-part1}
					\begin{aligned}
						\|H_X\rho\|_{1}&\leq \Lambda (d+1)^{2r_{\exp}}d!^{r_{\exp}}\|(N+I)^{\dd}_X\rho\|_{1}\\
						\|\rho H_X\|_{1}&\leq \Lambda (d+1)^{2r_{\exp}}d!^{r_{\exp}}\|\rho(N+I)^{\dd}_X\|_{1}\\
						\|L_{X,j}\rho L_{X,j}^\dagger\|_{1}&\leq \Lambda(d/2+1)^{4r_{\exp}}(d/2)!^{2r_{\exp}}\|(N+I)^{\dd/2}_X\rho(N+I)^{\dd/2}_X\|_{1}\,.
					\end{aligned}
				\end{equation}
				For the terms $L_{X,j}^\dagger L_{X,j}\rho$ and $\rho L_{X,j}^\dagger L_{X,j}$, we consider the monomials
				\begin{equation*}
					\begin{aligned}
						(a_j^\dagger)^{\ell'_2}&a_j^{k'_2}(a_j^\dagger)^{k'_1}a_j^{\ell'_1}\frac{1}{(n_j+1)^d}\ket{n_j}\\
						&=\sqrt{c_{k'_1\ell'_1}(n_j)}\frac{1}{(n_j+1)^d}(a_j^\dagger)^{\ell'_2}a_j^{k'_2}\ket{n_j-\ell'_1+k'_1}\\
						&=\sqrt{c_{\ell'_2k'_2}(n_j-\ell'_1+k'_1)c_{k'_1\ell'_1}(n_j)}\frac{1}{(n_j+1)^d}\ket{n_j-\overline{k}'+\overline{\ell}'}\\
						&\coloneqq \frac{\sqrt{c}}{(n_j+1)^d}\ket{n_j-\overline{k}'+\overline{\ell}'}
					\end{aligned}
				\end{equation*}
				for $k_1'$, $k_2'$, $\ell_1'$, $\ell_2'\leq d/2$ and $\overline{k}'\coloneqq k_2'-k_1'$ and $\overline{\ell}'$ in the same way. Then, 
				\begin{equation*}
					\begin{aligned}
						\|(N_j+I)^{-d} (a_j^\dagger)^{\ell'_1}&a_j^{k'_1}(a_j^\dagger)^{k'_2}a_j^{\ell'_2}\ket{\psi}\|_{2}^2\\
						&=\sum_{n_j=0}^{\infty}\frac{c}{(n_j+1)^{2d}}\,|\braket{\psi\,|\,n_j-\overline{k}'+\overline{\ell}'}|^2\\
						&\leq \sum_{n_j=0}^{\infty}(\ell_1'+k_2')!^2\frac{(n_j+1)^{k_1'+\ell_1'+k_2'+\ell_2'}}{(n_j+1)^{2d}}\,|\braket{\psi\,|\,n_j-\overline{k}'+\overline{\ell}'}|^2\\
						&\leq d!^2\sum_{n_j=0}^{\infty}|\braket{\psi\,|\,n_j}|^2\\
						&\leq d!^2\|\ket{\psi}\|_2^2\,.
					\end{aligned}
				\end{equation*}
				Again by Hölder's inequality, triangle inequality and the continuity of the norms, we achieve
				\begin{equation}\label{eq:rel-bounded-bound-part2}
					\begin{aligned}
						\|L_{X,j}^\dagger L_{X,j}\rho\|_{1}&\leq \Lambda(d/2+1)^{4r_{\exp}}d!^{r_{\exp}}\|(N+I)^{\dd}_X\rho\|_{1}\,,\\
						\|\rho L_{X,j}^\dagger L_{X,j}\|_{1}&\leq \Lambda(d/2+1)^{4r_{\exp}}d!^{r_{\exp}}\|\rho(N+I)^{\dd}_X\|_{1}\,.
					\end{aligned}
				\end{equation}
				Combining \Cref{eq:rel-bounded-bound-part1} and \eqref{eq:rel-bounded-bound-part2} and using the inequality $d+1\leq (d/2+1)^2$ show
				\begin{equation*}
					\begin{aligned}
						\|&\widetilde{\cL}_R(\rho)\|_{1}\\
						&\leq \Lambda(d/2+1)^{4r_{\exp}}d!^{r_{\exp}}|\cV_R^{(r)}|\max_{X\in\cV^{(r)}_R}\Bigl((1+\frac{\nu}{2})\|\rho (N+I)_X^{\dd}\|_{1}+(1+\frac{\nu}{2})\|(N+I)_X^{\dd}\rho \|_{1}\\
						&\qquad\qquad\qquad\qquad\qquad\qquad\qquad\qquad\qquad\qquad\qquad\qquad+\nu\|(N+I)_X^{\dd/2}\rho (N+I)_X^{\dd/2}\|_{1}\Bigr)\,,
					\end{aligned}
				\end{equation*}
				which finishes the proof for $\rho\in\cT_f$. The estimates extend from states in $\cT_f$ to finite-rank, not necessarily self-adjoint operators by linearity, and then to $W^{d,1}$ by continuity of the norms.
			\end{proof}

		\subsubsection{Auxiliary results: Tools to prove moment propagation bounds}
			The main difference between the bounds in \Cref{prop:LRB-bose-hubbard,prop:LRB-regularized-hamiltonian} and those found in \cite{Gondolf.2024} lies in the choice of the function $f$, which allows us to better track the dependence on the parameter $k$. However, as we show in \Cref{lem:connection-moments} below, such bounds can be translated into bounds of the form required by \Cref{eq:moment-stability}. First, we recall the function $g_\ell$ (see \Cref{eq:f-g-l-function}):
			\begin{equation*}
				g_\ell(m) = \begin{cases}
					f(m) - f(m - \ell) & m \ge \ell;\\
					f(m) & \ell > m \ge 0;\\
					0 & m < 0.
				\end{cases}
			\end{equation*}
			
			\begin{lem}\label{lem:connection-moments}
				Define $f:\mathbb{N} \to \mathbb{R}$ such that it satisfies $f(0) = f(1) = \dots = f(p-1) = 0$ and
				\begin{equation*}
					g_p(m+p) = f(m+p) - f(m) = k m^{k-1}
				\end{equation*}
				for any $m \ge 0$. Then, $f(m)$ is increasing in $m$, and for any integer $\ell$ and $m\geq p$,
				\begin{equation*}
					\frac{1}{p} \left( \frac{(m+\ell)^k}{4^{p+\ell}} - k^k \frac{4^{p+\ell}-1}{3 \times 4^{p+\ell}}\right)
					\le \frac{1}{p}(m-p)^k \le f(m) \le m^k
				\end{equation*}
				and
				\begin{equation*}
					(m+\ell)^k\leq 4^{p+\ell}pf(m)+k^{k}\frac{4^{p+\ell}-1}{3}
				\end{equation*}
				Furthermore, $g_\ell(m)$ is increasing in $\ell$ and $g_p(m)$ is increasing in $m$.
			\end{lem}
			
			\begin{proof}
				In a first step, we prove the explicit formula for the function $f$ by induction, which is 
				\begin{equation}\label{eq:explicit-f}
					f(m) = \begin{cases}
						0 & \text{if } 0 \leq m < p, \\
						k \sum_{j=0}^{q-1} (r+jp)^{k-1} & \text{if } m \geq p
					\end{cases}
				\end{equation}
				for $m = qp + r$ with $r \in \{0, \dots, p-1\}$. The identity is clear for all $0 \leq m \leq p$ because $f(m) = 0$, and for $p < m \leq 2p$, we have $f(m) = k (m-p)^{k-1}$ by definition, which settles the induction start. 
				
				Next, we assume that the formula holds for all $m$ satisfying $0 \leq m \leq \widetilde{m}$ with $\widetilde{m} = pq + r$ and $\tilde m \geq 2p$. Here, we consider two cases: First, $r \in \{0, \dots, p-2\}$, for which 
				\begin{equation*}
					\begin{aligned}
						f(\widetilde{m}+1) &= k (\widetilde{m} + 1 - p)^{k-1} + f(\widetilde{m} + 1 - p) \\
						&= k (p(q-1) + r + 1)^{k-1} + k \sum_{j=0}^{q-2} (r+1 + jp)^{k-1} \\
						&= k \sum_{j=0}^{q-1} (r+1 + jp)^{k-1}
					\end{aligned}
				\end{equation*}
				holds, and second, $r = p-1$:
				\begin{equation*}
					\begin{aligned}
						f(\widetilde{m}+1) = f(p(q+1)) = f(pq) + k(pq)^{k-1} = k \sum_{j=0}^{q} (jp)^{k-1}.
					\end{aligned}
				\end{equation*}
				This proves the explicit formula of $f$ given in \Cref{eq:explicit-f}. By the explicit formula in \Cref{eq:explicit-f}, the monotonicity is directly given in the case $0 \leq m \leq p$ and $m = pq + r$ with $r \in \{0, \dots, p-2\}$. In the case $r = p-1$, the inequalities
				\begin{equation*}
					f(m) = k \sum_{j=0}^{q-1} (r+jp)^{k-1} \leq k \sum_{j=0}^{q-1} (p + jp)^{k-1} \leq k \sum_{j=0}^{q} (jp)^{k-1} = f(m+1)
				\end{equation*}
				show that $f$ is increasing in $m$. Next, we prove the main inequality of the statement. We start with a lower bound on $(m-p)^k$, which is iteratively proven by 
				\begin{equation}\label{eq:from-m-to-m-1}
					(m-1)^k + \frac{k^k}{4} \ge \frac{m^k}{4}
				\end{equation}
				for $m \geq 1$, as shown in \cite[Equation (S.135)]{Kuwahara.2021}. Iterating this inequality, we find:
				\begin{equation}\label{eq:proof-connection-moments-bound1}
					(m-p)^k \ge \frac{(m+\ell)^k}{4^{p+\ell}} - k^k \frac{4^{p+\ell} - 1}{3 \times 4^{p+\ell}},
				\end{equation}
				for $m \geq p$. Assuming $m \geq p$ and $m = qp + r$ as above, \Cref{eq:explicit-f} shows
				\begin{equation}\label{eq:proof-connection-moments-bound3}
					\begin{aligned}
						f(m) &= k \sum_{j=0}^{q-1} (r + jp)^{k-1}\\
						&\le k \int_0^q (r + xp)^{k-1} \, dx \\
						&\le (r + qp)^k - r^k \\
						&\le m^k
					\end{aligned}
				\end{equation}
				and the same bound trivially holds for $0 \leq m \leq p$. Moreover,
				\begin{equation}\label{eq:proof-connection-moments-bound2}
					\begin{aligned}
						f(m) &\ge k \int_0^{q-1} (r + xp)^{k-1} \, dx + k r^{k-1} \\
						&= \frac{1}{p} \big((r + (q-1)p)^k - r^k\big) + k r^{k-1} \\
						&\ge \frac{1}{p} (m-p)^k.
					\end{aligned}
				\end{equation}
				Combining the bounds in \Cref{eq:proof-connection-moments-bound1}-\eqref{eq:proof-connection-moments-bound3}  proves the inequality in the statement. For the second inequality, we still need to consider the inequality for $m\in\{1,...,p\}$ for which $f(m)=0$ by definition. Then, \Cref{eq:proof-connection-moments-bound1} shows
				\begin{equation*}
					(m+\ell)^k\leq (p+\ell)^k\leq k^{k}\frac{4^{p+\ell}-1}{3}\,.
				\end{equation*}
				Finally, we show that the function $g_\ell(m)$ is increasing in $\ell$ and $g_p(m)$ in $m$. The monotonicity in $\ell$ is directly clear by definition and monotonicity of $f(m)$ in $m$. Similarly, the monotonicity of $g_p(m)$ in $m$ for $m \leq 2p$ also follows immediately from the monotonicity of $f$ and the definition of $g_p(m)$. Finally, we consider the monotonicity of $g_p(m)$ in $m$ for $m > 2p$. By definition,
				\begin{equation*}
					g_p(m) = k (m - p)^{k-1},
				\end{equation*}
				which finishes the proof.
			\end{proof}
			
			Next, we repeat the following upper and lower bounds on increasing product formulas.
			\begin{lem}[\texorpdfstring{\cite[Lem.~C.3]{Gondolf.2024}}{Gondolf 2024}]\label{lem:bounds-ccr-l-product}
				Let $\ell \in \mathbb{N}$ and $x \geq \ell$, then
				\begin{equation*}
					\begin{aligned}
						(x+1)^\ell-\frac{(\ell+1)\ell}{2}(x+1)^{\ell-1}&\leq ((x+1)-\ell)\cdots ((x+1)-1) \leq (x+1)^\ell\\
						(x+1)^\ell &\leq (x+1)\cdots (x+1+\ell-1) \leq \ell!(x+1)^\ell
					\end{aligned}
				\end{equation*}
			\end{lem}
			
			Finally, we state a simple optimization tool for upper bounding a sum of two polynomials with negative leading order:
			\begin{lem}\label{polymax}
				For $\alpha, \beta > 0$ and $a > b > 0$, the polynomial $p(X) = -\alpha X^a + \beta X^{b}$ satisfies
				\begin{align*}
					\max_{X \geq 0}\,p(X) \le \left(\frac{\beta b}{\alpha a}\right)^{\frac{b}{a-b}}\,\beta.
				\end{align*}
			\end{lem}
			\begin{proof}
				Searching for zeros of the derivative of $p$, we see that it attains its maximum on $[0,\infty)$ at $X_0 = \bigl(\frac{\beta b}{\alpha a}\bigr)^{1/(a-b)}$. Plugging this into the polynomial, we get
				\begin{equation*}
					p(X_0) = \left(\frac{\beta b}{\alpha a}\right)^{\frac{b}{a-b}}\,\beta\, \left(1 - \frac{b}{a}\right) \le \left(\frac{\beta b}{\alpha a}\right)^{\frac{b}{a-b}}\,\beta
				\end{equation*}
				which finishes the proof.
			\end{proof}
            
\section{Trotter-Kato product formula}\label{sec:trotter}
	In the previous section, we obtained information propagation bounds for general Hamiltonians and Lindbladians. Next, we discuss a Trotter-Kato scheme to construct the generator presented in \Cref{ex:reg-hamiltonian}, which is a fundamental step in one of our protocols. For this, we assume access not only to the evolution under the unknown Hamiltonian (see \Cref{ex:general-hamiltonian}), but also to an evolution with added engineered photon dissipation (see \Cref{ex:reg-hamiltonian}). The Trotter-Kato scheme justifies access to this overall dissipative evolution by verifying that it can be well approximated in terms of a Trotter-product formula. First, we recall \Cref{eq:sobolev-young-inequality}
    \begin{equation*}
        \cW_{\Sigma}^{2k}=\sum_{j\in V}(N_j+I)^{k/2}\cdot(N_j+I)^{k/2}\,.
    \end{equation*}
    \begin{prop}\label{thm:trotter}
    	Let $(\cL^{(\alpha,p)},\cT_f)$ be the $p$-photon dissipation defined in \Cref{eq:p-photon-dissipation}, $(\cH^{(d)},\cT_f)$ be the Hamiltonian considered in \Cref{eq:bosonic-hamiltonian} and assume that $(\cH^{(d)},\cT_f)$ generates a quantum Markov semigroup. If $p\geq 4d+2$ and $|\alpha_i|=\mathcal{O}(1)$ for all $i \in V$, then for all $\rho\in W^{8p,1}$, $t\ge 0$ and $n\in\N$
	    \begin{equation*}
            \begin{aligned}
                \bigl\|\left(e^{\frac{t}{n}\cH^{(d)}}e^{\frac{t}{n}\cL^{(\alpha,p)}}\right)^n(\rho)-&e^{t(\cH^{(d)}+\cL^{(\alpha,p)})}(\rho)\bigr\|_1\\
                &\leq\frac{2ct}{\sqrt{n}}\|\cW_\Sigma^{4p}(\rho)\|_1+\frac{3cm^2t^2}{n}\left(4^{p+1}e^{4p \Gamma''t}\,\max_{\substack{i\in V}}\big\{\|\rho\|_{W^{8p,1}_i}\big\}+\Upsilon_{4p}(t)\right)\,,
            \end{aligned}
        \end{equation*}
        where we recall that $m=|V|$, $c$ is the relative bound given in \Cref{lem:rel-bounds}, and $\Gamma'',\Upsilon_{4p}(t)$ were introduced in \Cref{lem:moment-prop-bound}.
    \end{prop}

    The proof follows Chernoff's proof for the convergence of the Trotter product formula in the strong topology (see \cite{Chernoff.1968} or Section 4 in \cite{Moebus.2023}) and uses the following three ingredients. First, we restate the Chernoff $\sqrt{n}$-Lemma \cite{Chernoff.1968}:
    \begin{lem}[Chernoff $\sqrt{n}$-Lemma]\label{lem:chnernoff}
        Let $C$ be a contraction on $\cT_1(\cH_m)$. Then for all $n\in\N$, the operator $e^{n(C-I)}$ is a contraction and for all $\rho\in \cT_1$
        \begin{equation*}
            \|C^n\rho-e^{n(C-I)}\rho\|_1\leq\sqrt{n}\|(C-I)\rho\|_1\,.
        \end{equation*}
    \end{lem}
    Second, we use relative bounds on concatenations of $\cH^{(d)}$ and $\cL^{(\alpha,p)}$ with respect to $\mathcal{W}_\Sigma^{2k}$:
    
    \begin{lem}[Relative bounds]\label{lem:rel-bounds}
		Let $(\cH^{(d)},\cT_f)$ be the Hamiltonian considered in \Cref{ex:general-hamiltonian} and $(\cL^{(\alpha,p)},\cT_f)$ the $p$-photon dissipation introduced in \Cref{eq:p-photon-dissipation} with  $|\alpha_i|\leq\eta$ for all $i\in V$. Then, for all states $\rho\in\dom(\cW_{\Sigma}^{2k})$ for $k$ appropriately, i.e.~such that the r.h.s.~of the following bounds are well-defined,
		\begin{align}
			\|\cH^{(d)}\rho\|_1&\leq\underbrace{2\gamma \Lambda(d+1)^4d!}_{=:c_1}\,\|\cW_{\Sigma}^{8d}(\rho)\|_1\label{eq:rel-bound-cH}\\
			\|(\cH^{(d)})^2\rho\|_1&\leq\underbrace{4\gamma^2\Lambda^2(1+d)^8(d!)^2(1+d)^{2d}}_{=:c_2}\|(\cW_\Sigma^{8d})^2(\rho)\|_1\,,\label{eq:rel-bound-cH2}\\
			\|\cL^{(\alpha,p)}(\rho)\|_1&\leq\underbrace{8\max\{\sqrt{p!},\eta^p\sqrt{p!},\eta^{2p}\}}_{=:c_3}\,\|\cW_{\Sigma}^{4p}(\rho)\|_1\,,\label{eq:rel-bound-cV}\\
			\|(\cL^{(\alpha,p)})^2(\rho)\|_1&\leq\underbrace{64\max\{1,\eta^{4p}\}p!(1+p)^{p}}_{=:c_4}\|(\cW_{\Sigma}^{4p})^2(\rho)\|_1\,,\label{eq:rel-bound-cV2}\\
			\|\cH^{(d)}\cL^{(\alpha,p)}(\rho)\|_1&\leq\underbrace{16\gamma \Lambda(1+d)^4d!\sqrt{p!}(1+p)^d\max\{1,\eta^{2p}\}}_{=:c_5}\|\cW_{\Sigma}^{8d}\cW_{\Sigma}^{4p}(\rho)\|_1\,,\label{eq:rel-bound-cHcV}
		\end{align}
		and we define $c=\max\{c_1,c_2,c_3,c_4,c_5\}$.
	\end{lem}
	
	For our third ingredient, we assume $p\geq4d+2$ and $k\geq2d$, which implies $d<k$ and $p\ge \lceil\frac{2kd}{k-d}+2\rceil$, so that we can use the bound given in \Cref{lem:moment-prop-bound} due to the bound found in \Cref{prop:LRB-regularized-hamiltonian}:
	\begin{equation*}
		\tr[e^{s\widetilde{\cL}}(\rho) (N_i+I)^k]\le 4^{p+1}e^{k \Gamma''s}\,\max_{i\in V}\big\{\|\rho\|_{W^{k,1}_i}\big\}+\Upsilon_k(s)
	\end{equation*}
	for some constant $\Gamma''$ and a function $\Upsilon_k$ such that, for $s,p,d,|\alpha_i|=\mathcal{O}(1)$, $\Upsilon_k(s)=k^{\mathcal{O}(k)}$ and $\Gamma''=\mathcal{O}(1)$.
	Combining these three tools, we can now prove the Trotter product formula.

	Note that the statements given here are strongly adapted to the setup of the paper. However, they can be generalized to contractive semigroups on Banach spaces satisfying a relative boundedness condition with respect to a reference operator (here $\cW_{\Sigma}^k$) satisfying certain conditions.

	\begin{proof}[Proof of \Cref{thm:trotter}]
		First, define $F(n)\coloneqq e^{\frac{t}{n}\cH^{(d)}}e^{\frac{t}{n}\cL^{(\alpha,p)}}$. Then, Chernoff's $\sqrt{n}$-Lemma in combination with Lemma II.1.3 in \cite{Engel.2000} imply
		\begin{equation*}
			\begin{aligned}
				\big\|F(n)^n(\rho)-e^{n(F(n)-I)}(\rho)\|_1&\leq\sqrt{n}\,\big\|(F(n)-I)(\rho)\big\|_1\\
				&\leq\sqrt{n}\,\left\|\left(e^{\frac{t}{n}\cH^{(d)}}\left(e^{\frac{t}{n}\cL^{(\alpha,p)}}-I\right)+e^{\frac{t}{n}\cH^{(d)}}-I\right)(\rho)\right\|_1\\
				&\leq\frac{t}{\sqrt{n}}\left(\|\cL^{(\alpha,p)}(\rho)\|_1+\|\cH^{(d)}(\rho)\|_1\right)\,.
			\end{aligned}
		\end{equation*}
		Next, by \Cref{lem:rel-bounds}, there is a constant $c\geq0$ such that 
		\begin{equation}\label{eq-trotter:chernoff-bound}
			\big\|F(n)^n(\rho)-e^{n(F(n)-I)}(\rho)\big\|_1\leq\frac{ct}{\sqrt{n}}\left(\|\cW_{\Sigma}^{4p}(\rho)\|_1+\|\cW_{\Sigma}^{8d}(\rho)\|_1\right)\,.
		\end{equation}
		In the next, we apply the fundamental theorem of calculus for Bochner integrals: 
		\begin{equation*}
			\begin{aligned}
				\big\|e^{n(F(n)-I)}(\rho)&-e^{t(\cH^{(d)}+\cL^{(\alpha,p)})}(\rho)\big\|_1\\
                &\leq\int_0^1\big\|e^{(1-s)n(F(n)-I)}\left(n(F(n)-I)-t\left(\cH^{(d)}+\cL^{(\alpha,p)}\right)\right)e^{st(\cH^{(d)}+\cL^{(\alpha,p)})}(\rho)\big\|_1\,ds\\
				&\leq\int_0^1\big\|\left(n(F(n)-I)-t\left(\cH^{(d)}+\cL^{(\alpha,p)}\right)\right)e^{st(\cH^{(d)}+\cL^{(\alpha,p)})}(\rho)\big\|_1\,ds\,.
			\end{aligned}
		\end{equation*}
		where we used in the last line that $s\mapsto e^{(1-s)n(F(n)-I)}$ is a contraction semigroup by \Cref{lem:chnernoff} and the definition of $F(n)$. Next, we define $\rho_s\coloneqq e^{st(\cH^{(d)}+\cL^{(\alpha,p)})}(\rho)$ and use Lemma II.1.3 in \cite{Engel.2000} again to show
		\begin{equation*}
			\begin{aligned}
				\Bigl(n&(F(n)-I)-t\cH^{(d)}-t\cL^{(\alpha,p)}\Bigr)(\rho_s)\\
                &=\left(n\left(e^{\frac{t}{n}\cH^{(d)}}e^{\frac{t}{n}\cL^{(\alpha,p)}}-e^{\frac{t}{n}\cH^{(d)}}+e^{\frac{t}{n}\cH^{(d)}}-I\right)-t\cH^{(d)}-t\cL^{(\alpha,p)}\right)(\rho_s)\\
				&=t\int_0^1\left(\left(e^{\frac{t}{n}\cH^{(d)}}e^{\frac{tv_1}{n}\cL^{(\alpha,p)}}-I\right)\cL^{(\alpha,p)}+\left(e^{\frac{tv_1}{n}\cH^{(d)}}-I\right)\cH^{(d)}\right)(\rho_s)dv_1\\
				&=t\int_0^1\left(\left(e^{\frac{t}{n}\cH^{(d)}}e^{\frac{tv_1}{n}\cL^{(\alpha,p)}}-e^{\frac{t}{n}\cH^{(d)}}+e^{\frac{t}{n}\cH^{(d)}}-I\right)\cL^{(\alpha,p)}+\left(e^{\frac{tv_1}{n}\cH^{(d)}}-I\right)\cH^{(d)}\right)(\rho_s)dv_1\\
				&=\frac{t^2}{n}\int_{[0,1]^2}\left(v_1e^{\frac{t}{n}\cH^{(d)}}e^{\frac{tv_1v_2}{n}\cL^{(\alpha,p)}}(\cL^{(\alpha,p)})^2+e^{\frac{tv_2}{n}\cH^{(d)}}\cH^{(d)}\cL^{(\alpha,p)}+v_1e^{\frac{tv_1v_2}{n}\cH^{(d)}}(\cH^{(d)})^2\right)(\rho_s)d^2v\,,
			\end{aligned}
		\end{equation*}
		so that 
		\begin{equation*}
			\begin{aligned}
				\big\|e^{n(F(n)-I)}(\rho)&-e^{t(\cH^{(d)}+\cL^{(\alpha,p)})}(\rho)\big\|_1\\
                &\leq\frac{t^2}{n}\int_0^1\Bigl(\big\|(\cL^{(\alpha,p)})^2(\rho_s)\big\|_1+\big\|\cH^{(d)}\cL^{(\alpha,p)}(\rho_s)\big\|_1+\big\|(\cH^{(d)})^2(\rho_s)\big\|_1\Bigr)\,ds\\
				&\leq\frac{ct^2}{n}\int_0^1\Bigl(\big\|(\cW_{\Sigma}^{4p})^2(\rho_s)\big\|_1+\big\|\cW_{\Sigma}^{8d}\cW_{\Sigma}^{4p}(\rho_s)\big\|_1+\big\|(\cW_\Sigma^{8d})^2(\rho_s)\big\|_1\Bigr)\,ds\\
				&\leq\frac{3ct^2}{n}\int_0^1\big\|(\cW_{\Sigma}^{4p})^2(\rho_s)\big\|_1ds\,,
			\end{aligned}
		\end{equation*}
		by the relative bounds in \Cref{lem:rel-bounds}. Moreover, note that $\rho_s$ is a state and $\cW_\Sigma$ is positivity preserving, so that all norms can be written as a trace and the assumption $2d+2\leq p$ explains the last inequality. Next, by Young's inequality:
		\begin{equation*}
			\begin{aligned}
				\big\|(\cW_{\Sigma}^{4p})^2(\rho_s)\big\|_1&=\sum_{i,j\in V}\tr\big[\sqrt{\rho_s}(N_i+I)^{2p}(N_j+I)^{2p}\sqrt{\rho_s}\,\big]\\
				&\le \frac{1}{2}\sum_{i,j\in V}\tr\big[\sqrt{\rho_s}\left((N_i+I)^{4p}+(N_j+I)^{4p}\right)\sqrt{\rho_s}\,\big]\\
				&=m\,\|\cW_{\Sigma}^{8p}(\rho_s)\|_1\,.
			\end{aligned}
		\end{equation*}
		Finally, we use the assumption $p\geq 4d+2$ and \Cref{lem:moment-prop-bound} to show 
		\begin{equation}\label{eq-trotter:diff}
			\begin{aligned}
				\big\|e^{n(F(n)-I)}(\rho)&-e^{t(\cH^{(d)}+\cL^{(\alpha,p)})}(\rho)\big\|_1\\
                &\leq\frac{3\,c\,m\,t^2}{n}\int_0^1\big\|\cW_{\Sigma}^{8p}(\rho_s)\big\|_1ds\\
				&=\frac{3\,c\,m\,t^2}{n}\int_0^1\left(\sum_{j\in V}\tr\big[\rho_s(N_j+I)^{4p}\big]\right)ds\\
				&\leq\frac{3\,c\,m\,t^2}{n}\int_0^1m\left(4^{p+1}e^{4p \Gamma''st}\,\max_{i\in V}\big\{\|\rho\|_{W^{8p,1}_i}\big\}+\Upsilon_{4p}(st)\right)ds\\
				&\leq\frac{3\,c\,m^2t^2}{n}\left(4^{p+1}e^{4p \Gamma''t}\,\max_{\substack{i\in V}}\big\{\|\rho\|_{W^{8p,1}_i}\big\}+\Upsilon_{4p}(t)\right)\,.
			\end{aligned}
		\end{equation}
		\Cref{lem:moment-prop-bound} is applicable because $p\geq 4d+2$ and $k=4p$ shows $d<k$ and $p\ge \big\lceil\frac{2kd}{k-d}+2\big\rceil$. Combining the bound in \eqref{eq-trotter:chernoff-bound} and in \eqref{eq-trotter:diff} finishes the proof.
	\end{proof}
	In the next step, we prove the relative bounds of \Cref{lem:rel-bounds} used in the proof above: 
	\begin{proof}[Proof of \Cref{lem:rel-bounds}]
		To prove the relative boundedness, we follow the proof of \Cref{lem:relative-bounds-number-op}, but specify the bound due to the explicit structure of the Hamiltonian and the dissipation (see \Cref{ex:general-hamiltonian} and \ref{ex:reg-hamiltonian}). First, we show for $q\in\R_{\geq 0}$ and $i\in V$ as well as $(i,j),(j,h)\in E$      
		\begin{align}
			\|(N_i+I)^{q}L_i^{(\alpha_i,p)}&(N_i+I)^{-q-p/2}\|_\infty\nonumber\\
            &\leq2\max\{1,\eta^{p}\}\,,\label{eq:rel-NL}\\
			\|(N_i+I)^{q}(L_i^{(\alpha_i,p)})^\dagger&(N_i+I)^{-q-p/2}\|_\infty\nonumber\\
            &\leq2\max\{\sqrt{p!}(1+p)^q,\eta^{p}\}\,,\label{eq:rel-NLdag}\\
			\|(N_i+I)^q(N_j+I)^qH_{(j,h)}&(N_i+I)^{-q}(N_j+I)^{-d-q}(N_h+I)^{-d}\|_\infty\nonumber\\
            &\leq \Lambda(1+d)^4d!(1+d)^{q+1_{i=h}q}\,.\label{eq:rel-NH}
		\end{align}
		The first bound \eqref{eq:rel-NL} reduces for $\ket{\phi}\in\cH^{(d)}$ to 
		\begin{equation*}
			\begin{aligned}
				\|(N_i+I)^q&a_i^p(N_i+I)^{-q-p/2}\ket{\varphi}\|^2\\
                &= \bra{\varphi}(N_i+(1-p)I)^q(N_i+I)^{-q-p/2}(\ad_i)^p a_i^p(N_i+I)^{-q-p/2}(N_i+(1-p)I)^q\ket{\varphi}\\
				&= \bra{\varphi}N_i[1-p:0](N_i+I)^{-2q-p}(N_i+(1-p)I)^{2q}\ket{\varphi}\\
				&\leq\|\varphi\|^2
			\end{aligned}
		\end{equation*}
		where we used the commutation relation \eqref{eq:symmetry-function} for $f(n)=(n+1)^q1_{n\geq0}$ so that $(N+I)^qa^p=a^p(N+(1-p)I)^q$, $N_i[1-p:0]\leq(N_i+I)^p$ and $(N_i+(1-p)I)^q\leq (N_i+I)^q$. Note that $(N_i+(1-p)I)^q$ is defined by $f(n)=(n+1)^q1_{n\geq0}$ and thereby Fock basis elements less than $p-1$ are set to be zero. Similarly, the bound in \Cref{eq:rel-NLdag} follows from
		\begin{equation*}
			\begin{aligned}
				\|(N_i+I)^q&(\ad_i)^p(N_i+I)^{-q-p/2}\ket{\varphi}\|^2\\
                &= \bra{\varphi}(N_i+I)^{-q-p/2} (N_i+(1+p)I)^qa_i^p(\ad_i)^p(N_i+(1+p)I)^q(N_i+I)^{-q-p/2}\ket{\varphi}\\
				&= \bra{\varphi}N_i[1:p](N_i+I)^{-2q-p}(N_i+(1+p)I)^{2q}\ket{\varphi}\\
				&\leq p!(1+p)^{2q}\|\varphi\|^2
			\end{aligned}
		\end{equation*}
		where we have additionally used $N_i[1:p]\leq p!(N_i+I)^p$ and $(N_i+(1+p)I)^q\leq(1+p)^q(N_i+I)^q$. With help of these two bounds, we can directly prove \Cref{eq:rel-NL} and \eqref{eq:rel-NLdag}. i.e.~
		\begin{equation*}
			\begin{aligned}
				\|(N_i+I)^{q}&L_i^{(\alpha_i,p)}(N_i+I)^{-q-p/2}\|_\infty\\
                &\leq\|(N_i+I)^{q}a_i^p(N_i+I)^{-q-p/2}\|_\infty+\eta^p\leq2\max\{1,\eta^{p}\}\,,\\
				\|(N_i+I)^{q}&(L_i^{(\alpha_i,p)})^\dagger(N_i+I)^{-q-p/2}\|_\infty\\
                &\leq\|(N_i+I)^{q}(\ad)_i^p(N_i+I)^{-q-p/2}\|_\infty+\eta^p\leq2\max\{\sqrt{p!}(1+p)^q,\eta^{p}\}\,,
			\end{aligned}
		\end{equation*}
		because $\|(N+I)^{-q}\|_\infty\leq1$, which we will use from time to time. Next, we consider the Hamiltonian inequality \eqref{eq:rel-NH}. For this, we follow the derivation in the proof of \Cref{lem:boundedness-projective-generator}: By definition 
		\begin{equation*}
			H_{\{j,h\}}=\sum_{k,\ell, k', \ell'=0}^d\lambda^{(j,h)}_{k\ell k' \ell'}\,(a_j^\dagger)^{k}a_j^{\ell}(a_h^\dagger)^{k'}a_h^{\ell'}\,.
		\end{equation*}
		Then, the same tricks as before (\ref{eq:symmetry-function}, \ref{higher-order-product-aadag}) show
		\begin{equation*}
			\begin{aligned}
				\|&(N_i+I)^q(N_j+I)^q(a_j^\dagger)^{k}a_j^{\ell}(a_h^\dagger)^{k'}a_h^{\ell'}(N_i+I)^{-q}(N_j+I)^{-d-q}(N_h+I)^{-d}\ket{\phi}\|^2\\
				&=\bra{\phi}N_j[1-\ell:0]N_j[1-\ell:k-\ell]N_h[1-\ell':0]N_h[1-\ell':k'-\ell']\bigl(N_i+(1+1_{i=h}(k'-\ell'))I\bigr)^{2q}\\
				&\qquad\quad\cdot(N_j+(1+k-\ell)I)^{2q}(N_i+I)^{-2q}(N_j+I)^{-2d-2q}(N_h+I)^{-2d}\ket{\phi}\\
				&\leq(d!)^2(1+d)^{2q+1_{i=h}2q}
			\end{aligned}
		\end{equation*}
		so that
		\begin{equation*}
			\begin{aligned}
				\|&(N_i+I)^q(N_j+I)^q\,H_{(j,h)}\,(N_i+I)^{-q}(N_j+I)^{-d-q}(N_h+I)^{-d}\|_\infty\leq \Lambda(1+d)^4d!(1+d)^{q+1_{i=h}q}
			\end{aligned}
		\end{equation*}
		proves \Cref{eq:rel-NH} and finishes the first part of the proof. In the second part, we use the bounds in \Cref{eq:rel-NL}, \eqref{eq:rel-NLdag} and \eqref{eq:rel-NH} to prove relative bounds on $\cL^{(\alpha,p)}$, $\cH^{(d)}$, $(\cL^{(\alpha,p)})^2$, $\cH^{(d)}\cL^{(\alpha,p)}$ and $(\cH^{(d)})^2$, i.e.~\Cref{eq:rel-bound-cH}-\eqref{eq:rel-bound-cHcV}. We start with the bound \eqref{eq:rel-bound-cH} for which Hölder's inequality and \Cref{eq:rel-NH} with $q=0$ show for $\rho\in\dom(\cW_{\Sigma}^{8d})$
		\begin{equation*}
			\begin{aligned}
				\|\cH^{(d)}\rho\|_1&\leq \sum_{e=(i,j)\in E}\|H_{e}(N_i+I)^{-d}(N_{{j}}+I)^{-d}(N_{{i}}+I)^{d}(N_j+I)^{d}\rho\|_1+\|\rho H_{e}\|_1\\
				&\leq \Lambda(d+1)^4d!\sum_{(i,j)\in E}\|(N_i+I)^{d}(N_j+I)^{d}\rho(N_i+I)^{d}(N_j+I)^{d}(N_i+I)^{-d}(N_j+I)^{-d}\|_1\\
                &\qquad+\|\rho H_{e}\|_1\\
				&\leq 2\Lambda(d+1)^4d!\sum_{(i,j)\in E}\tr[(N_i+I)^{d}(N_j+I)^{d}\rho(N_i+I)^{d}(N_j+I)^{d}]\\
				&= 2\Lambda(d+1)^4d!\sum_{(i,j)\in E}\tr[\sqrt{\rho}(N_i+I)^{2d}(N_j+I)^{2d}\sqrt{\rho}]\\
				&\leq \Lambda(d+1)^4d!\sum_{(i,j)\in E}\left(\|(N_i+I)^{2d}\rho(N_i+I)^{2d}\|_1+\|(N_j+I)^{2d}\rho(N_j+I)^{2d}\|_1\right)\\
				&\leq 2\gamma \Lambda(d+1)^4d! \,\|\cW_{\Sigma}^{8d}(\rho)\|_1,
			\end{aligned}
		\end{equation*}
		where we used Young's inequality in the fifth step. In the last inequality, we used that, as the graph has degree $\gamma$, every vertex in $V$ has at most $\gamma$ neighbors and the positivity of $(N_i+I)^{2d}\rho(N_i+I)^{2d}$ to rewrite the 1-norm as a trace so that
		\begin{equation*}
			\sum_{i\in V}\,\|(N_i+I)^{2d}\rho(N_i+I)^{2d}\|_1=\sum_{i\in V}\,\tr[(N_i+I)^{2d}\rho(N_i+I)^{2d}]=\|\cW_{\Sigma}^{8d}(\rho)\|_1\,.
		\end{equation*}
		
		Next, we continue proving the bound on the dissipation in \Cref{eq:rel-bound-cV} using again Hölder's inequality and \Cref{eq:rel-NL} for $q=0$. In the following auxiliary bound, we will see that one advantage of the proven bounds \eqref{eq:rel-NL}-\eqref{eq:rel-NH} is that these are iteratively applicable. For $i\in V$,
		\begin{equation}\label{eq:rel-LdagL}
			\begin{aligned}
				\|(L_i^{(\alpha_i,p)})^\dagger L_i^{(\alpha_i,p)}(N_i+I)^{-p}\|_\infty&=\|(L_i^{(\alpha_i,p)})^\dagger (N_i+I)^{-p/2}(N_i+I)^{p/2}L_i^{(\alpha_i,p)}(N_i+I)^{-p}\|_\infty\\
				&\leq4\max\{\sqrt{p!},\eta^p\sqrt{p!},\eta^{2p}\}
			\end{aligned}
		\end{equation}
		by applying \Cref{eq:rel-NLdag} with $q=0$ and \eqref{eq:rel-NL} with $q=p/2$. 
		Let $\rho\in\dom(\cW_{\Sigma}^{4p})$ 
		\begin{equation}\label{eq:rel-boung-cL}
			\begin{aligned}
				\|&\cL^{(\alpha,p)}\rho\|_1\\
                &\leq\sum_{i\in V}\|L_i^{(\alpha_i,p)}(N_i+I)^{-p+p}\rho(N_i+I)^{p-p}(L_i^{(\alpha_i,p)})^\dagger\|_1+\|(L_i^{(\alpha_i,p)})^\dagger L_i^{(\alpha_i,p)}(N_i+I)^{-p+p}\rho\|_1\\
				&\leq\sum_{i\in V}\|L_i^{(\alpha_i,p)}(N_i+I)^{-p/2}\|_\infty\,\|(N_i+I)^{p}\rho(N_i+I)^{p}\|_1\,\|(N_i+I)^{-p/2}(L_i^{(\alpha_i,p)})^\dagger\|_\infty\\
				&\qquad+\|(L_i^{(\alpha_i,p)})^\dagger L_i^{(\alpha_i,p)}(N_i+I)^{-p}\|_\infty\,\|(N_i+I)^{p}\rho(N_i+I)^{p}\|_1\,\|(N_i+I)^{-p}\|_\infty\,.
			\end{aligned}
		\end{equation}
		Then, we use $\|A\|_\infty=\|A^\dagger\|_\infty$ for $A\in\cT_1$, apply the bounds \eqref{eq:rel-NL} with $q=0$ and \eqref{eq:rel-LdagL} and use $\|(N_i+I)^{-p}\|_\infty\leq1$ to show
		\begin{equation*}
			\begin{aligned}
				\|\cL^{(\alpha,p)}\rho\|_1
				&\leq\sum_{i\in V}4\max\{1,\eta^{2p}\}\,\|(N_i+I)^{p}\rho(N_i+I)^{p}\|_1\\
                &\qquad+4\max\{\sqrt{p!},\eta^p\sqrt{p!},\eta^{2p}\}\,\|(N_i+I)^{p}\rho(N_i+I)^{p}\|_1\\
				&\leq8\max\{\sqrt{p!},\eta^p\sqrt{p!},\eta^{2p}\}\sum_{i\in V}\,\|(N_i+I)^{p}\rho(N_i+I)^{p}\|_1\\
				&\leq8\max\{\sqrt{p!},\eta^p\sqrt{p!},\eta^{2p}\}\,\|\cW_{\Sigma}^{4p}(\rho)\|_1\,,
			\end{aligned}
		\end{equation*}
		where we have used again the positivity of $(N_i+I)^p\rho(N_i+I)^p$.
		
		Next, we prove the relative bound on $(\cH^{(d)})^2$ (see \Cref{eq:rel-bound-cH2}) using the bound \eqref{eq:rel-NH}. For $\rho\in\dom((\cW_{\Sigma}^{2d})^2)$, \Cref{eq:rel-NH} with $q=0$ shows
		\begin{equation*}
			\begin{aligned}
				\|(\cH^{(d)})^2(\rho)\|_1&\leq\sum_{e=(i,j)\in E}\|H_e\cH^{(d)}(\rho)\|_1+\|\cH^{(d)}(\rho)H_e\|_1\\
				&\leq 2\Lambda(1+d)^4d!\sum_{e=(i,j)\in E}\|(N_i+I)^d(N_j+I)^d\cH^{(d)}(\rho)(N_i+I)^d(N_j+I)^d\|_1\,.
			\end{aligned}
		\end{equation*}
		Then, \Cref{eq:rel-NH} with $q=d$ and $A\coloneqq(N_i+I)^d(N_j+I)^d(N_{i'}+I)^d(N_{j'}+I)^d$ proves for $e'=\{i',j'\}$
		\begin{equation*}
			\begin{aligned}
				\|(N_i&+I)^d(N_j+I)^dH_{e'}\rho(N_i+I)^d(N_j+I)^d\|_1\\
				&\leq \Lambda(1+d)^4d!(1+d)^{2d}\|A\rho A\|_1\\
				&=\Lambda(1+d)^4d!(1+d)^{2d}\tr[\sqrt{\rho}(N_i+I)^{2d}(N_{i'}+I)^{2d}(N_j+I)^{2d}(N_{j'}+I)^{2d}\sqrt{\rho}]\,.
			\end{aligned}
		\end{equation*}
		Then, Young's inequality shows 
		\begin{equation*}
			\begin{aligned}
				(N_i+I)^{2d}(N_{i'}+I)^{2d}(N_j+I)^{2d}(N_{j'}+I)^{2d}\leq\frac{1}{2}\left((N_i+I)^{4d}(N_{i'}+I)^{4d}+(N_j+I)^{4d}(N_{j'}+I)^{4d}\right)
			\end{aligned}
		\end{equation*}
		so that
		\begin{equation*}
			\begin{aligned}
				\|&(\cH^{(d)})^2(\rho)\|_1\\
                &\leq4\Lambda^2(1+d)^8(d!)^2(1+d)^{2d}\sum_{e,e'\in E}\tr[\sqrt{\rho}(N_i+I)^{2d}(N_{i'}+I)^{2d}(N_j+I)^{2d}(N_{j'}+I)^{2d}\sqrt{\rho}]\\
				&\leq2\Lambda^2(1+d)^8(d!)^2(1+d)^{2d}\sum_{e,e'\in E}\tr[\sqrt{\rho}\left((N_i+I)^{4d}(N_{i'}+I)^{4d}+(N_j+I)^{4d}(N_{j'}+I)^{4d}\right)\sqrt{\rho}]\\
				&\leq4\gamma^2\Lambda^2(1+d)^8(d!)^2(1+d)^{2d}\|(\cW_\Sigma^{8d})^2(\rho)\|_1
			\end{aligned}
		\end{equation*}
		where we used again $|E|\leq\gamma|V|$. Next, we continue with the case $(\cL^{(\alpha,p)})^2$ (see \Cref{eq:rel-bound-cV2}). Here, we use again \Cref{eq:rel-NL} and \eqref{eq:rel-NLdag} in the same iterative way as before:
		\begin{equation*}
			\begin{aligned}
				\|(\cL^{(\alpha,p)})^2(\rho)\|_1&\leq64\max\{1,\eta^{4p}\}p!(1+p)^{p}\sum_{i,j\in V}\|(N_i+I)^{p}(N_j+I)^{p}(\rho)(N_i+I)^{p}(N_j+I)^{p}\|_1\\
				&\leq64\max\{1,\eta^{4p}\}p!(1+p)^{p}\|(\cW_{\Sigma}^{4p})^2(\rho)\|_1\,.
			\end{aligned}
		\end{equation*}
		Finally, we consider $\cH^{(d)}\cL^{(\alpha,p)}$ which is a combination from both proofs above, i.e.~we use \Cref{eq:rel-NH} with $q=0$, \eqref{eq:rel-NLdag} with $q=d$ and \eqref{eq:rel-NL}:
		\begin{equation*}
			\begin{aligned}
				\|&\cH^{(d)}\cL^{(\alpha,p)}(\rho)\|_1\\
                &\leq 16\Lambda(1+d)^4d!\max\{\sqrt{p!}(1+p)^d,\sqrt{p!}(1+p)^d\eta^p,\eta^{2p}\}\\
				&\qquad\qquad\cdot\sum_{e=(i,j)\in E}\sum_{h\in V}\|(N_i+I)^d(N_j+I)^d(N_h+I)^p\rho(N_i+I)^d(N_j+I)^d(N_h+I)^p\|_1\\
				&\leq 16\gamma \Lambda(1+d)^4d!\max\{\sqrt{p!}(1+p)^d,\sqrt{p!}(1+p)^d\eta^p,\eta^{2p}\}\|\cW_{\Sigma}^{8d}\cW_{\Sigma}^{4p}(\rho)\|_1\\
				&\leq 16\gamma \Lambda(1+d)^4d!\sqrt{p!}(1+p)^d\max\{1,\eta^{2p}\}\|\cW_{\Sigma}^{8d}\cW_{\Sigma}^{4p}(\rho)\|_1
			\end{aligned}
		\end{equation*}
		where we have used Young's inequality on 
		\begin{equation*}
			(N_i+I)^{2d}(N_j+I)^{2d}(N_h+I)^{2p}\leq\frac{1}{2}\left((N_i+I)^{4d}+(N_j+I)^{4d}\right)(N_h+I)^{2p}\,.
		\end{equation*}
		This finishes the proof. 
	\end{proof}

\section{Simulation of bosonic Lindbladians}\label{sec:simulation}
	One of the fundamental applications of quantum computers is the simulation of quantum mechanical systems \cite{Feynman.1982}. Since the pioneering work \cite{Lloyd.1996}, the field of quantum simulations of locally defined quantum Markov semigroups has seen remarkable progress \cite{Aharonov.2007,Berry.2006,Childs.2011,Wiebe.2011,Poulin.2011,Pocrnic.2023,Low.2019}.

    \subsection{Global simulation of bosonic Lindbladian}
    	In our first statements, we focus on a bosonic extension of the works \cite{Cleve.2017,Barthel.2012, Kliesch.2011} in the sense that the global bosonic system is simulated via a finite-dimensional circuit of local quantum dynamical semigroups in trace norm. Second, we turn our attention to scenarios in which the global bosonic evolution is approximated by locally acting bosonic quantum Markov semigroups using a Trotter product formula as explored in \cite{Moebus.2024}. In finite dimensions, this Trotter decomposition scheme \cite[Thm.~1]{Kliesch.2011} is claimed to imply the so-called dissipative Church-Turing theorem \cite[Impl.~1]{Kliesch.2011}. Here, we consider bosonic quantum Markov semigroups in GKSL form satisfying \Cref{assum:local-GKSL,assum-lrb}.
    	
    	First, we present approximation bounds used in our simulation schemes to approximate a quantum Markov semigroup acting on a state using a low-energy Fock space. Next, we provide a result demonstrating how to approximate a quantum Markov semigroup, given an appropriate input state, using the Pauli-circuit scheme discussed in \cite{Cleve.2017}. Finally, we prove the approximation of quantum Markov semigroups by locally defined bosonic quantum Markov semigroups.
    	
    	\subsubsection{Finite approximation of quantum Markov semigroups and states}\label{subsec:apprximation-qmss-states}
    		We start with a fundamental result on how to approximate a bosonic quantum Markov semigroup satisfying \Cref{assum-lrb}:
    		\begin{lem}\label{lem:approximation}
    			Let $(\widetilde{\cL},\cT_f)$ be an operator satisfying \Cref{assum:local-GKSL,assum-lrb} (with constants $K$, $\xi$, $\Gamma$ and degree $d$) and define the finite cutoff in terms of the Fock basis via $P_{X}=P^{(M)}_{X}\coloneqq\prod_{j\in X}P_j^{(M)}$ with $P_j^{(M)}\coloneqq\sum_{n\leq M}\ketbra{n}{n}_j$ by
    			\begin{equation*}
    				\widetilde{\cL}^{(M)} = \sum_{\substack{X\in\cV^{(r)}}}\left(\cH[P_XH_XP_X] + \sum_{j=1}^{\nu(X)}\cL[L_{X,j}P_X]\right)
    			\end{equation*}
    			following the structure of \Cref{assum:local-GKSL}. For $\hat{k}=\max\{k,4dr_{\exp},K\}$, there is a constant $c\geq0$ (see \Cref{eq:lem-approx-constant}) such that for every state $\rho\in\cT_1$ with $\max_{i\in V}\|\rho\|_{W^{2\hat{k},1}_i}<\infty$
    			\begin{equation*}
    				\begin{aligned}
    					\|e^{t\widetilde{\cL}}(\rho)-e^{t\widetilde{\cL}^{(M)}}(\rho)\|_1&\leq \mathcal{K}c \hat{k}^{\xi\hat{k}}te^{\hat{k} \Gamma t/2}\frac{\sqrt{\max_{i \in V}\|\rho\|_{W^{2\hat{k},1}_i}}}{(M-d+1)^{\hat{k}/4}}\,,
    				\end{aligned}
    			\end{equation*}
    			holds true with $\mathcal{K}=\sum_{X\in\cV^{(r)}}(1+\nu(X))$.
    		\end{lem}
    		\begin{rmk*}
    			Note that the constant $\mathcal{K}$ depends on the system size if the considered quantum Markov semigroup acts, for instance, everywhere non-trivially. However, this is natural for global bounds (compare to \cite{Kliesch.2011}). Moreover, one choice of $r_{\exp}$ could be $|X|\leq r_{\exp}=\gamma^{\lceil r/2 \rceil}$ due to the finite connectivity assumed in \cref{assum:local-GKSL}, but better bounds can be obtained if more is known about the support geometry, such as in the case of a $2$-dimensional lattice, i.e.~$|X|\leq r^2$.
    		\end{rmk*}
    		\begin{proof}
    			We fix the following notation for the projections onto a low-energy Fock space: $\overline{P}_X=I-P_X$, $\cP_X=P_X\cdot P_X$ and $\overline{\cP}_X=I-{\cP}_X$. To reduce the bosonic, unbounded generator of the quantum Markov semigroup to finite dimensions, we use Duhamel's formula:
    			\begin{equation}\label{eq:approximation-lem-step1}
    				\begin{aligned}
    					\|e^{t\widetilde{\cL}}(\rho)&-e^{t\widetilde{\cL}^{(M)}}(\rho)\|_{1}\\
    					&=t\|\int_{0}^1e^{(1-s)t\widetilde{\cL}^{(M)}}(\widetilde{\cL}-\widetilde{\cL}^{(M)})e^{st\widetilde{\cL}}(\rho)ds\|_1\\
    					&=t\|\int_{0}^1e^{(1-s)t\widetilde{\cL}^{(M)}}\sum_{X\in\cV^{(r)}}\biggl(\cH[\overline{\cP}_X(H_X)] \\
    					&\quad\qquad\qquad\qquad\qquad\qquad+ \sum_{j=1}^{\nu(X)}\Bigl(L_{X,j}\overline{\cP}_X(\cdot)L_{X,j}^\dagger-\frac{1}{2}\{\overline{\cP}_X(L_{X,j}^\dagger L_{X,j}),\cdot\}\Bigr)\biggr)e^{st\widetilde{\cL}}(\rho)ds\|_1\\
    					&\leq t\int_{0}^1\sum_{X\in\cV^{(r)}}\Bigl\|\biggl(\cH[\overline{\cP}_X(H_X)] \\
    					&\qquad\qquad\qquad\qquad+ \sum_{j=1}^{\nu(X)}\Bigl(L_{X,j}\overline{\cP}_X(\cdot)L_{X,j}^\dagger-\frac{1}{2}\{\overline{\cP}_X(L_{X,j}^\dagger L_{X,j}),\cdot\}\Bigr)\biggr)e^{st\widetilde{\cL}}(\rho)\bigr\|_1ds
    				\end{aligned}
    			\end{equation}
    			Next, we consider one term in the above sum for fixed $X \in \cV^{(r)}$ and $j\in\{1,...,\nu(X)\}$, for which we apply the identity given in \Cref{eq:swap-projections}, i.e.
    			\begin{equation*}
    				\begin{aligned}
    					P_i^{(M)}(\ad_i)^ka_i^\ell&=P_i^{(M)}(\ad_i)^ka_i^\ell P_i^{(M+d)} \qquad&\text{and}\qquad (\ad_i)^ka_i^\ell P_i^{(M)}&=P_i^{(M+d)}(\ad_i)^ka_i^\ell P_i^{(M)}\,,\\
    					\overline{P}_i^{(M)}(\ad_i)^ka_i^\ell&=\overline{P}_i^{(M)}(\ad_i)^ka_i^\ell \overline{P}_i^{(M-d)} \qquad&\text{and}\qquad (\ad_i)^ka_i^\ell \overline{P}_i^{(M)}&=\overline{P}_i^{(M-d)}(\ad_i)^ka_i^\ell \overline{P}_i^{(M)}\,,
    				\end{aligned}
    			\end{equation*}
    			combined with \Cref{eq:comp-projection-union-bound}
    			\begin{equation}\label{eq:expand-product-complementary-cut}
    				\overline{P}_X^{(M)}=I-\prod_{i=1}^{|X|}P_i^{(M)}=I-P_1+P_1\Bigl(I-\prod_{i=2}^{|X|}P_i^{(M)}\Bigr)=\sum_{i=1}^{|X|}\prod_{j=1}^{i-1} P_j^{(M)}\overline{P}_i^{(M)}\,,
    			\end{equation}
    			where we take the product over all elements in $X$ with labeling $i \in\{1,...,|X|\}$ and use the convention that an empty product is the identity. This shows for $\rho_{st}\coloneqq e^{st\widetilde{\cL}}(\rho)$
    			\begin{equation*}
    				\begin{aligned}
    					\cH[&\overline{\cP}_X(H_X)](\rho_{st})+L_{X,j}\overline{\cP}_X(\rho_{st})L_{X,j}^\dagger-\frac{1}{2}\{\overline{\cP}_X(L_{X,j}^\dagger L_{X,j}),\rho_{st}\}\\
    					&=-i\sum_{i=1}^{|X|}\Bigl(H\prod_{j=1}^{i-1} P_j^{(M)}\overline{P}_i^{(M)} + \prod_{j=1}^{i-1} P_j^{(M)}\overline{P}_i^{(M)}HP_X\overline{P}_i^{(M-d)}\Bigr)\rho_{st}+s.c.\\
    					&\qquad+\sum_{i=1}^{|X|}L_{X,j}\Bigl(\rho_{st}\overline{P}_i^{(M)}\prod_{j=1}^{i-1} P_j^{(M)}+\prod_{j=1}^{i-1} P_j^{(M)}\overline{P}_i^{(M)}\rho_{st}{P}_X\Bigr)L_{X,j}^\dagger\\
    					&\qquad-\frac{1}{2}\sum_{i=1}^{|X|}\Bigl(L_{X,j}^\dagger L_{X,j}\prod_{j=1}^{i-1} P_j^{(M)}\overline{P}_i^{(M)} + \prod_{j=1}^{i-1} P_j^{(M)}\overline{P}_i^{(M)}L_{X,j}^\dagger L_{X,j}P_X\overline{P}_i^{(M-d)}\Bigr)\rho_{st}+s.c.
    				\end{aligned}
    			\end{equation*}
    			where $s.c.$ stands for the symmetric conjugate, but in this context, it simply involves reordering the operators and taking the symmetric conjugate of the complex number. Then, we use the relative boundedness properties of polynomials of annihilation and creation operators proven in \Cref{lem:relative-bounds-number-op}, which proves 
    			\begin{equation*}
    				\begin{aligned}
    					\Bigl\|&\Bigl(\cH[\overline{\cP}_X(H_X)]+ \sum_{j=1}^{\nu(X)}L_{X,j}\overline{\cP}_X(\cdot)L_{X,j}^\dagger-\frac{1}{2}\{\overline{\cP}_X(L_{X,j}^\dagger L_{X,j}),\cdot\}\Bigr)(\rho_{st})\bigr\|_1\\
    					&\leq 2\Lambda|X|(d+1)^{2r_{\exp}}d!^{r_{\exp}}\max_{i\in X}\bigl(\|(N+I)_X^{\dd}\overline{P}_i^{(M-d)}\rho_{st}\|_1+\|\rho_{st}\overline{P}_i^{(M-d)}(N+I)_X^{\dd}\|_1\bigr)\\
    					&\qquad+\nu(X)\Lambda|X|(d/2+1)^{4r_{\exp}}(d/2)!^{2r_{\exp}}\max_{i\in X}\Bigl(\|(N+I)^{\dd/2}_X\rho_{st}\overline{P}_i(N+I)^{\dd/2}_X\|_{1}\\
    					&\qquad\qquad\qquad\qquad\qquad\qquad\qquad\qquad\qquad\qquad+\|(N+I)^{\dd/2}_X\overline{P}_i\rho_{st}(N+I)^{\dd/2}_X\|_{1}\Bigr)\\
    					&\qquad+\nu(X)\Lambda|X|(d/2+1)^{4r_{\exp}}d!^{r_{\exp}}\max_{i\in X}\Bigl(\|(N+I)_X^{\dd}\overline{P}_i^{(M-d)}\rho_{st}\|_1+\|\rho_{st}\overline{P}_i^{(M-d)}(N+I)_X^{\dd}\|_1\Bigr)\\
    					&\leq r_{\exp}\Lambda(d/2+1)^{4r_{\exp}}d!^{r_{\exp}}\biggl((2+\nu(X))\max_{i\in X}\Bigl(\|(N+I)_X^{\dd}\overline{P}_i^{(M-d)}\rho_{st}\|_1\\
						&\qquad\qquad\qquad\qquad\qquad\qquad\qquad\qquad\qquad\qquad+\|\rho_{st}\overline{P}_i^{(M-d)}(N+I)_X^{\dd}\|_1\Bigr)\\
    					&\qquad\qquad\qquad\qquad\qquad\qquad\qquad+\nu(X)\max_{i\in X}\Bigl(\|(N+I)^{\dd/2}_X\rho_{st}\overline{P}_i(N+I)^{\dd/2}_X\|_{1}\\
						&\qquad\qquad\qquad\qquad\qquad\qquad\qquad\qquad\qquad\qquad+\|(N+I)^{\dd/2}_X\overline{P}_i\rho_{st}(N+I)^{\dd/2}_X\|_{1}\Bigr)\biggr)
    				\end{aligned}
    			\end{equation*}
    			where we used $(d+1)\leq(d/2+1)^2$ and $(d/2)!^2\leq d!$. Next, we use the Cauchy–Schwarz inequality to show
    			\begin{equation*}
    				\begin{aligned}
    					\|(N+I)_X^{\dd}\overline{P}_i^{(M-d)}\rho_{st}\|_1,\,\|\rho_{st}\overline{P}_i^{(M-d)}(N+I)_X^{\dd}\|_1&\leq\sqrt{\tr[\overline{P}_i^{(M-d)}\rho_{st}(N+I)_X^{2\dd}]}\\
    					&\leq\sqrt[4]{\tr[\overline{P}_i^{(M-d)}\rho_{st}]\tr[\rho_{st}(N+I)_X^{4\dd}]}
    				\end{aligned}
    			\end{equation*}
    			and
    			\begin{equation*}
    				\begin{aligned}
    					\|(N+I)^{\dd/2}_X\rho_{st}\overline{P}_i(N+I)^{\dd/2}_X\|_{1},&\,\|(N+I)^{\dd/2}_X\overline{P}_i\rho_{st}(N+I)^{\dd/2}_X\|_{1}\\
    					&\leq\sqrt[4]{\tr[\overline{P}_i\rho_{st}]\tr[(N+I)^{2\dd}_X\rho_{st}]\tr[(N+I)^{\dd}_X\rho_{st}]}\\
    					&\leq\sqrt[4]{\tr[\overline{P}_i\rho_{st}]\tr[(N+I)^{2\dd}_X\rho_{st}]^2}\\
    					&\leq\sqrt[4]{\tr[\overline{P}_i\rho_{st}]\tr[(N+I)^{4\dd}_X\rho_{st}]}
    				\end{aligned}
    			\end{equation*}
    			so that
    			\begin{equation}\label{eq:approximation-lem-step2}
    				\begin{aligned}
    					\Bigl\|&\Bigl(\cH[\overline{\cP}_X(H_X)]+ \sum_{j=1}^{\nu(X)}L_{X,j}\overline{\cP}_X(\cdot)L_{X,j}^\dagger-\frac{1}{2}\{\overline{\cP}_X(L_{X,j}^\dagger L_{X,j}),\cdot\}\Bigr)(\rho_{st})\Bigr\|_1\\
    					&\qquad\leq 4(1+\nu(X))r_{\exp}\Lambda(d/2+1)^{4r_{\exp}}d!^{r_{\exp}}\max_{i\in X}\sqrt[4]{\tr[\overline{P}_i^{(M-d)}\rho_{st}]\tr[(N+I)^{4\dd}_X\rho_{st}]}\,.
    				\end{aligned}
    			\end{equation}
    			Next, we apply Young's inequality to 
    			\begin{equation*}
    				\tr[(N+I)^{4\dd}_X\rho_{st}]=\tr[\sqrt{\rho_{st}}(N+I)^{4\dd}_X\sqrt{\rho_{st}}]\leq\max_{i\in X}\tr[(N+I)^{4d|X|}_i\rho_{st}]
    			\end{equation*}
    			and \Cref{eq:markov-union-bound} combined with \Cref{lem:moment-prop-bound} to \Cref{eq:approximation-lem-step2}:
    			\begin{equation}\label{eq:approximation-simu-markov}
    				\begin{aligned}
    					\Bigl\|&\Bigl(\cH[\overline{\cP}_X(H_X)]+ \sum_{j=1}^{\nu(X)}L_{X,j}\overline{\cP}_X(\cdot)L_{X,j}^\dagger-\frac{1}{2}\{\overline{\cP}_X(L_{X,j}^\dagger L_{X,j}),\cdot\}\Bigr)(\rho_{st})\Bigr\|_1\\
    					&\qquad\qquad\leq 4(1+\nu(X))r_{\exp}\Lambda(d/2+1)^{4r_{\exp}}d!^{r_{\exp}}\frac{\sqrt{\max_{i\in V}M_i^{(\hat{k})}(st)}}{(M-d+1)^{k/4}}\\
    					&\qquad\qquad\leq 4(1+\nu(X))r_{\exp}\Lambda(d/2+1)^{4r_{\exp}}d!^{r_{\exp}}\frac{\sqrt{p4^{p+1} e^{\hat{k} \Gamma st} \max_{i \in V} M_i^{(\hat{k})}(0) + \Upsilon_{\hat{k}}(st)}}{(M-d+1)^{\hat{k}/4}}
    				\end{aligned}
    			\end{equation}
    			where $\hat{k}=\max_{X\in\cV^{(r)}}\{k,4dr_{\exp},K\}$ and 
    			\begin{equation*}
    				\Upsilon_{\hat{k}}(st)=4^{p+1} \Big( e^{\hat{k} \Gamma st} \frac{\hat{k}^{\xi \hat{k}}\Gamma'''}{\hat{k}\Gamma'} + \frac{\hat{k}^{\hat{k}}}{3}\Big)
    			\end{equation*}
    			given in \Cref{lem:moment-prop-bound}. Therefore, there is constant $c\geq0$ so that
    			\begin{equation}\label{eq:lem-approx-constant}
    				\begin{aligned}
    					\Bigl\|&\Bigl(\cH[\overline{\cP}_X(H_X)]+ \sum_{j=1}^{\nu(X)}L_{X,j}\overline{\cP}_X(\cdot)L_{X,j}^\dagger-\frac{1}{2}\{\overline{\cP}_X(L_{X,j}^\dagger L_{X,j}),\cdot\}\Bigr)(\rho_{st})\Bigr\|_1\\
					&\qquad\qquad\qquad\leq c (1+\nu(X))\hat{k}^{\xi\hat{k}}e^{\hat{k} \Gamma st/2}\frac{\sqrt{\max_{i \in V}\|\rho\|_{W^{2\hat{k},1}_i}}}{(M-d+1)^{\hat{k}/4}}\,.
    				\end{aligned}
    			\end{equation}
    			Summing over all $X\in\cV^{(r)}$ as in \Cref{eq:approximation-lem-step2} proves 
    			\begin{equation*}
    				\begin{aligned}
    					\|e^{t\widetilde{\cL}}(\rho)-e^{t\widetilde{\cL}^{(M)}}(\rho)\|_1&\leq t\mathcal{K}c \hat{k}^{\xi\hat{k}}e^{\hat{k} \Gamma t/2}\frac{\sqrt{\max_{i \in V}\|\rho\|_{W^{2\hat{k},1}_i}}}{(M-d+1)^{\hat{k}/4}}\,,
    				\end{aligned}
    			\end{equation*}
    			finishes the proof of the statement .
    		\end{proof}
            For special cases, the above approximation scheme can be improved. One seminal example is the Bose--Hubbard model, for which an efficient simulation scheme was demonstrated by providing a weak Lieb--Robinson bound depending polynomially on time \cite{Kuwahara.2024digital}, following the simulation scheme for Hamiltonians constructed in \cite{Haah.2021}. 
    		Before introducing our simulation algorithms, we present a brief and straightforward tool for approximating an input state:
    		\begin{cor}\label{cor:state-approximation}
    			Let $\rho\in\cT_1$ be a state and $k\in\N$ such that $\max_{i\in R}\|\rho\|_{W^{2k,1}_i}<\infty$, then
    			\begin{equation*}
    				\|\rho-\cP_R(\rho)\|_1\leq2|R|\frac{\sqrt{\max_{i\in R}\|\rho\|_{W^{2k,1}_i}}}{(M+1)^{k/2}}
    			\end{equation*}
    			where $\cP_R=\cP_R^{(M)}=P_R\cdot P_R$ with $P_R=P_R^{(M)}=\prod_{j\in R}P_j^{(M)}$ and $P_j^{(M)}\coloneqq\sum_{n\leq M}\ketbra{n}{n}_j$.
    		\end{cor}
    		\begin{proof}
    			The approximation result follows directly from the proof of \Cref{lem:moment-prop-bound}. Here, we consider
    			\begin{equation*}
    				\begin{aligned}
    					\|\rho-\cP_R(\rho)\|_1&= \|\rho \overline{P}_R+\overline{P}_R\rho P_R\|_1\\
    					&\overset{(1)}{\leq}\sum_{i=1}^{|R|}\|\rho \overline{P}_i^{(M)}\prod_{j=1}^{i-1} P_j^{(M)}\|_1+\|\prod_{j=1}^{i-1} P_j^{(M)}\overline{P}_i^{(M)}\rho\|_1\\
    					&\leq|R|\max_{i\in R}\Bigl(\|\rho \overline{P}_i^{(M)}\|_1+\|\overline{P}_i^{(M)}\rho\|_1\Bigr)\\
    					&\overset{(2)}{\leq}2|R|\sqrt{\max_{i\in R}\tr[\overline{P}_i^{(M)}\rho]}\\
    					&\overset{(3)}{\leq}2|R|\frac{\sqrt{\max_{i\in R}M_i^{(k)}(0)}}{(M+1)^{k/2}}
    				\end{aligned}
    			\end{equation*}
				where we used \Cref{eq:comp-projection-union-bound} in $(1)$, the Cauchy--Schwarz inequality in $(2)$ and \Cref{eq:markov-union-bound} in $(3)$, which finishes the proof.
    		\end{proof}
    		
    	\subsubsection{Qubit-bosonic approximation}
    		With the tools to approximate a quantum Markov semigroup and a quantum state in a finite energy Fock space, we start with the first simulation algorithm based on the scheme discussed in \cite[Cor.~2]{Cleve.2017} to approximate the projected Lindbladian above by $r$-local Pauli strings. Here, the gate complexity refers to the number of used Pauli strings.
    		\begin{thm}[Pauli-approximation]\label{thm:pauli-approx}
			Let $(\widetilde{\cL},\cT_f)$ be an operator satisfying \Cref{assum:local-GKSL,assum-lrb} (with constants $K$, locality $r_{\exp}$ and degree $d$), and set $\hat{k}=\max\{k,4dr_{\exp},K\}$ with $k\geq 8r_{\exp}(d+8)$. Then, the quantum Markov semigroup $e^{t\widetilde{\cL}}$ with input $\rho$ satisfying $\max_{i\in V}\|\rho\|_{W^{2\hat{k},1}_i}<\infty$ can be approximated up to error $\varepsilon$ by a Pauli circuit with gate complexity
    			\begin{equation*}
				\widetilde{\cO}\Biggl(te^{\cO(\hat{k}t)}\cK^{3}\sqrt{\frac{(\cK+|V|)\max_{i\in V}\big\{\|\rho\|_{W^{2\hat{k},1}_i}\}}{\varepsilon}}\Biggr),
    			\end{equation*}
			using the scheme constructed in \cite{Cleve.2017}, where $t$ denotes the evolution time and $\mathcal{K}=\sum_{X\in\cV^{(r)}}(1+\nu(X))$\,.
    		\end{thm}
            Due to the exponential time-dependence of the bound, the displayed complexity is polynomial for times up to logarithmic order in the relevant system-size parameters. For certain systems, such as the Bose--Hubbard model \cite{Kuwahara.2024digital}, the specific structure allows for an improved gate complexity.
    		\begin{proof}
    			First, the bosonic generator is approximated by \Cref{lem:approximation}, achieving the following bound:
    			\begin{equation*}
    				\begin{aligned}
    					\|e^{t\widetilde{\cL}}(\rho)-e^{t\widetilde{\cL}^{(M)}}(\rho)\|_1&\leq \mathcal{K}c \hat{k}^{\xi\hat{k}}t e^{\hat{k} \Gamma t/2}\frac{\sqrt{\max_{i \in V}\|\rho\|_{W^{2\hat{k},1}_i}}}{(M-d+1)^{\hat{k}/4}}\,,
    				\end{aligned}
    			\end{equation*}
    			and  the input state by \Cref{cor:state-approximation}, i.e.~for $\rho^{(M)}\coloneqq \cP_V^{(M)}(\rho)$
    			\begin{equation*}
    				\|\rho-\rho^{(M)}\|_1\leq2|V|\frac{\sqrt{\max_{i\in V}\|\rho\|_{W^{2k,1}_i}}}{(M+1)^{k/2}}\,,
    			\end{equation*}
    			so that
    			\begin{equation}\label{eq:finite-state-semigroup-approximation}
    				\begin{aligned}
    					\|e^{t\widetilde{\cL}}(\rho)-e^{t\widetilde{\cL}^{(M)}}(\rho^{(M)})\|_1&\leq \bigl(\mathcal{K}c \hat{k}^{\xi\hat{k}}t e^{\hat{k} \Gamma t/2}+2|V|\bigr)\frac{\sqrt{\max_{i \in V}\|\rho\|_{W^{2\hat{k},1}_i}}}{(M-d+1)^{\hat{k}/4}}\,.
    				\end{aligned}
    			\end{equation}
    			If we choose $M$ to be
    			\begin{equation}\label{eq:pauli-approx-assum1-M}
    				M \geq \Biggl(\frac{2\bigl(\mathcal{K} c \hat{k}^{\xi\hat{k}}t e^{\hat{k} \Gamma t/2} + 2|V|\bigr)\sqrt{\max_{i \in V}\|\rho\|_{W^{2\hat{k},1}_i}}}{\varepsilon}\Biggr)^{4/\hat{k}}+d-1,
    			\end{equation}
    			for any $ \varepsilon \in (0,1) $, then
    			\begin{equation*}
    				\|e^{t\widetilde{\cL}}(\rho) - e^{t\widetilde{\cL}^{(M)}}(\rho^{(M)})\|_1 \leq \frac{\varepsilon}{2}\,.
    			\end{equation*}
    			Moreover, we choose $M$ such that there is an $M_{\mathrm{Pauli}}\in\N$ with $M+d/2+1=2^{M_{\mathrm{Pauli}}}$. This choice is motivated by the fact that $e^{t\widetilde{\cL}^{(M)}}\circ\cP_V^{(M+d/2)}$ can be identified with a finite-dimensional semigroup because $\widetilde{\cL}^{(M)}$ is a bounded generator, so that $e^{t\widetilde{\cL}^{(M)}}$ can be expressed by its series expansion and 
    			\begin{equation*}
    				\begin{aligned}
    					\widetilde{\cL}^{(M)}&\circ\cP_V^{(M+d/2)}\\
    					&=\cP_V^{(M+d/2)}\circ\left(\sum_{\substack{X\in\cV^{(r)}}}\left(\cH[P_X^{(M)}H_XP_X^{(M)}] + \sum_{j=1}^{\nu(X)}\cL[P_X^{(M+d/2)}L_{X,j}P_X^{(M)}]\right)\right)\circ\cP_V^{(M+d/2)}\,.
    				\end{aligned}
    			\end{equation*}
    			Therefore, we can restrict the semigroup to the finite-dimensional Banach space of bounded operators $\cB(P_{V}^{(M+d/2)}\cH)$, which is by definition isomorphic to 
    			\begin{equation*}
    				\cB(P_{V}^{(M+d/2)}\cH)\simeq\C^{(M+d/2+1)^{|V|}\times (M+d/2+1)^{|V|}}=\C^{2^{|V|M_{\mathrm{Pauli}}}\times 2^{|V|M_{\mathrm{Pauli}}}}\,.
    			\end{equation*}
    			Since the generalized Pauli operators, i.e., tensor products of $\{I,\sigma_{x},\sigma_{y},\sigma_{z}\}$, form a basis of $\C^{2^{|V|M_{\mathrm{Pauli}}}\times 2^{|V|M_{\mathrm{Pauli}}}}$, operators of the form $P_X^{(M)}H_XP_X^{(M)}$, $L_{X,j}P_X^{(M)}=P_X^{(M+d/2)}L_{X,j}P_X^{(M)}\in\cB(P_X^{(M+d/2)}\cH_X)\otimes I_{V\backslash X}$ can be expressed in terms of a $2^{2|X|\,M_{\mathrm{Pauli}}}$-dimensional orthogonal system with non-trivial elements on $X$. We denote the restriction of $\widetilde{\cL}^{(M)}\circ\cP_V^{(M+d/2)}$ onto the space of Pauli operators $\C^{2^{|V|M_{\mathrm{Pauli}}}\times 2^{|V|M_{\mathrm{Pauli}}}}$ by $\widetilde{\cL}^{(M,\mathrm{Pauli})}$ defined by 
    			\begin{equation*}
    				P_X^{(M)}H_XP_X^{(M)},\qquad\text{and}\qquad P_X^{(M+d/2)}L_{X,j}P_X^{(M)}
    			\end{equation*}
    			for all $X\in\cV^{(r)}$ and $j\in\{1,...,\nu(X)\}$. Then, the above Hamiltonians and Lindblad operators can be expressed by the generalized Pauli operators $V_{X,i}$
    			\begin{equation*}
    				P_X^{(M)}H_XP_X^{(M)}=\sum_{i=0}^{2^{2|X|M_{\mathrm{Pauli}}}}\alpha_{X,i}V_{X,i}\qquad\text{and}\qquad P_X^{(M+d/2)}L_{X,j}P_X^{(M)}=\sum_{i=0}^{2^{2|X|M_{\mathrm{Pauli}}}}\beta_{X,j,i}V_{X,i}
    			\end{equation*}
    			In the above representation the Pauli norm defined in \cite[Eq.~7]{Cleve.2017} is given by 
    			\begin{equation}\label{eq:pauli-norm}
    				\|\widetilde{\cL}^{(M,\mathrm{Pauli})}\|_{\mathrm{Pauli}}=\sum_{X\in\cV^{(r)}}\Biggl(\sum_{i=0}^{2^{2|X|M_{\mathrm{Pauli}}}}|\alpha_{X,i}|+\sum_{j=1}^{\nu(X)}\Bigl(\sum_{i=0}^{2^{2|X|M_{\mathrm{Pauli}}}}|\beta_{X,j,i}|\Bigr)^2\Biggr)\,.
    			\end{equation}
    			Taking into account that the orthogonality of the generalized Pauli basis, Hölder's inequality shows that for example
    			\begin{equation*}
    				\begin{aligned}
    					|\alpha_{X,i}|&=\frac{1}{2^{|X|M_{\mathrm{Pauli}}}}\left|\tr[P_X^{(M)}H_XP_X^{(M)} V_{X,i}]\right|\\
    					&\leq\frac{1}{2^{|X|M_{\mathrm{Pauli}}}}\|P_X^{(M)}H_XP_X^{(M)}\|_\infty\|V_{X,i}\|_1\\
    					&=\|P_X^{(M)}H_XP_X^{(M)}\|_\infty
    				\end{aligned}
    			\end{equation*}
    			so that \Cref{lem:boundedness-projective-generator}, i.e.
    			\begin{equation*}
    				\begin{aligned}
    					\|P_{X}^{(M)}H_XP_{X}^{(M)}\|_{\infty}&\leq \Lambda (d+1)^{2r_{\exp}}d!^{r_{\exp}}(M+1)^{r_{\exp}d}\\
    					\|P_X^{(M+d/2)}L_{X,j}P_X^{(M)}\|_{\infty}&\leq\sqrt{\Lambda(d/2+1)^{4r_{\exp}}(d/2)!^{2r_{\exp}}(M+1)^{r_{\exp}d}}\,,
    				\end{aligned}
    			\end{equation*}
    			for all $X\in\cV^{(r)}$ and $j\in\{1,...,\nu(X)\}$, shows
    			\begin{equation*}
    				\begin{aligned}
    					\|&\widetilde{\cL}^{(M,\mathrm{Pauli})}\|_{\mathrm{Pauli}}\\
    					&\leq\sum_{X\in\cV^{(r)}}\Biggl(2^{2|X|M_{\mathrm{Pauli}}}\|P_X^{(M)}H_XP_X^{(M)}\|_\infty+\sum_{j=1}^{\nu(X)}\Bigl(2^{2|X|M_{\mathrm{Pauli}}}\|P_X^{(M+d/2)}L_{X,j}P_X^{(M)}\|_{\infty}\Bigr)^2\Biggr)\\
    					&\leq\cK2^{4r_{\exp}M_{\mathrm{Pauli}}}\Lambda(d/2+1)^{4r_{\exp}}d!^{r_{\exp}}(M+1)^{r_{\exp}d}\\
    					&\leq\cK\Lambda(d/2+1)^{4r_{\exp}}d!^{r_{\exp}}(M+d/2+1)^{r_{\exp}(d+4)}\\
    				\end{aligned}
    			\end{equation*}
    			Therefore, Theorem 1 in \cite{Cleve.2017} presents an algorithm with gate complexity (number of Pauli gates):
    			\begin{equation*}
    				\cO\Bigl(\widetilde{m}^2\widetilde{q}^2\widetilde{\tau}\frac{\bigl(\log(\widetilde{m}\widetilde{q}\widetilde{\tau}/\varepsilon)+\widetilde{n}\bigr)\log(\widetilde{\tau}/{\varepsilon})}{\log(\log(\widetilde{\tau}/\varepsilon))}\Bigr),
    			\end{equation*}
    			where $\widetilde{m}\leq\cK$, $\widetilde{q}\leq2^{2r_{\exp}M_{\mathrm{Pauli}}}$, $\widetilde{n}=|V|M_{\mathrm{Pauli}}$ and
    			\begin{equation*}
    				\|\widetilde{\cL}^{(M,\mathrm{Pauli})}\|_{\mathrm{Pauli}}\leq\cK\Lambda(d/2+1)^{4r_{\exp}}d!^{r_{\exp}}(M+d/2+1)^{r_{\exp}(d+4)}\,.
    			\end{equation*}
    			Inserting the above calculated upper bounds, we achieve a gate complexity of 
    			\begin{equation*}
					\begin{aligned}
						\cO\Bigl(\cK^3t&(M+d/2+1)^{r_{\exp}(d+8)}\\
						&\cdot \frac{\bigl(\log(\cK^2t(M+d/2+1)^{r_{\exp}(d+6)}/\varepsilon)+|V|M_{\mathrm{Pauli}}\bigr)\log(t\cK(M+d/2+1)^{r_{\exp}(d+4)}/{\varepsilon})}{\log(\log(t\cK(M+d/2+1)^{r_{\exp}(d+4)}/{\varepsilon}))}\Bigr).
					\end{aligned}
    			\end{equation*}
    			Finally, we choose a fixed $k\geq 4\widetilde{k}r_{\exp}(d+8)$ so that $\hat{k}/4\geq \widetilde{k}r_{\exp}(d+8)$ for $\widetilde{k}\geq1$ and the choice of $M$ in \Cref{eq:pauli-approx-assum1-M}
    			\begin{equation*}
    				(M+d/2+1)^{\widetilde{k}r_{\exp}(d+8)}=\cO((M-d+1)^{\hat{k}/4})=\cO\biggl(\frac{(t e^{\hat{k} \Gamma t/2}\cK+|V|)\max_{i\in V}\big\{\|\rho\|_{W^{\hat{k},1}_i}\big\}}{\varepsilon}\biggr),
    			\end{equation*}
    			which implies a gate complexity of order
    			\begin{equation*}
				\widetilde{\cO}\Bigl(te^{\hat{k}/\widetilde{k} \Gamma t/2}\cK^{3}(\cK+|V|)^{1/\widetilde{k}}\varepsilon^{-1/\widetilde{k}}(\max_{i\in V}\big\{\|\rho\|_{W^{2\hat{k},1}_i}\big\})^{1/\widetilde{k}}\Bigr),
    			\end{equation*}
    			where we suppressed logarithmic terms which completes the proof for $\widetilde{k}=2$.
    		\end{proof}
    		
    		Depending on the available resources, i.e., the possible quantum gates to build a quantum circuit that approximates the bosonic quantum Markov semigroups, we next present a scheme that reduces the simulation of a bosonic quantum Markov semigroup to an application of Lie's product formula, similar to the result in \cite{Kliesch.2011}. Specifically, we approximate the bosonic quantum Markov semigroup by a circuit of $r$-local, finite-dimensional quantum Markov semigroups. For that, we first provide an error bound for Lie's product formula.
    
    		\begin{lem}\label{lem:multi-trotter}
    			Let $\cL_1,...,\cL_K$ and $\sum_{j=1}^K\cL_j$ for $K\in\N$ be bounded generators on $(\cT_1,\|\cdot\|_1)$ of contractive uniformly continuous semigroups. Then,
    			\begin{equation*}
    				\Bigl\|e^{t\sum_{j=1}^\cK\cL_j}-\Bigl(\prod_{j=1}^\cK e^{t\frac{\cL_j}{n}}\Bigr)^{n}\Bigr\|_{1\rightarrow1}\leq\frac{t^2\cK(\cK-1)}{2n}\max_{j\in\{1,...,\cK\}}\{\|\cL_j\|_{1\rightarrow1}^2\}.
    			\end{equation*}
    		\end{lem}
    		\begin{proof}
    			In the first step of the proof, we follow the standard proof approach using the telescopic sum presented, for example, in Lemma 4.1 of \cite{Moebus.2024}, and the contractivity of the involved semigroups to show
    			\begin{equation*}
    				\begin{aligned}
    					\Bigl\|e^{t\sum_{j=1}^\cK\cL_j}-\Bigl(\prod_{j=1}^\cK e^{t\frac{\cL_j}{n}}\Bigr)^{n}\Bigr\|_{1\rightarrow1}&\leq n\Bigl\|e^{\frac{t}{n}\sum_{j=1}^\cK\cL_j}-\prod_{j=1}^\cK e^{t\frac{\cL_j}{n}}\Bigr\|_{1\rightarrow1}.
    				\end{aligned}
    			\end{equation*}
    			In the next step, we use the fundamental theorem of calculus twice to show:
    			\begin{equation*}
    				\begin{aligned}
    					\Bigl\|&e^{\frac{t}{n}\sum_{j=1}^\cK\cL_j}-\prod_{j=1}^\cK e^{t\frac{\cL_j}{n}}\Bigr\|_{1\rightarrow1}\\
    					&=\Bigl\|\int_{0}^{1}\frac{d}{ds_1}e^{s_1\frac{t}{n}\sum_{j=1}^\cK\cL_j}\prod_{j=1}^\cK e^{(1-s_1)t\frac{\cL_j}{n}}ds_1\Bigr\|_{1\rightarrow1}\\
    					&\leq\frac{t}{n}\int_{0}^{1}\Bigl\|e^{s_1\frac{t}{n}\sum_{j=1}^\cK\cL_j}\sum_{i=2}^\cK\Bigl[\cL_i,\prod_{j=1}^{i-1}e^{(1-s_1)t\frac{\cL_j}{n}}\Bigr]\prod_{j=i}^\cK e^{(1-s_1)t\frac{\cL_j}{n}}\Bigr\|_{1\rightarrow1}ds_1\\
    					&=\frac{t}{n}\int_{0}^{1}\Bigl\|e^{s_1\frac{t}{n}\sum_{j=1}^\cK\cL_j}\sum_{i=2}^\cK\Bigl[\cL_i,\prod_{j=1}^{i-1}e^{(1-s_1)t\frac{\cL_j}{n}}-I\Bigr]\prod_{j=i}^\cK e^{(1-s_1)t\frac{\cL_j}{n}}\Bigr\|_{1\rightarrow1}ds_1\\
    					&\overset{(1)}{\leq}\frac{t^2}{n^2}\int_{0}^{1}(1-s_1)\sum_{i=2}^\cK 2\|\cL_i\|_{1\rightarrow1}\sum_{j=1}^{i-1}\|\cL_j\|_{1\rightarrow 1}ds_1\\
    					&\leq\frac{t^2}{n^2}\frac{\cK(\cK-1)}{2}\max_{j\in\{1,...,\cK\}}\{\|\cL_j\|_{1\rightarrow1}^2\}\,,
    				\end{aligned}
    			\end{equation*}
    			where, we used the following identity in step (1):
    			\begin{equation*}
    				\biggl(\prod_{j=1}^\cK A_j\biggr)-I=\sum_{j=1}^\cK \prod_{i=1}^{j-1}A_i(A_j-I).
    			\end{equation*}
    			This finishes the proof of the statement.
    		\end{proof}
    		
    		Then, we combine the bounds found in \Cref{lem:approximation} and \Cref{lem:multi-trotter}. This combination yields the following simulation technique, which approximates the quantum Markov semigroup by a local uniformly continuous quantum Markov semigroup:
    		
    		\begin{prop}\label{prop:church-turing-finite2}
    			Let $(\widetilde{\cL},\cT_f)$ be an operator satisfying \Cref{assum:local-GKSL} and \Cref{assum-lrb} (with constants $K$, $d$ and locality $r_{\exp}$) and define the finite cutoff in terms of the Fock basis via $P_{X} = P^{(M)}_{X} \coloneqq \prod_{j\in X}P_j^{(M)}$ with $P_j^{(M)} \coloneqq \sum_{n\leq M}\ketbra{n}{n}_j$ by
    			\begin{equation*}
    				\widetilde{\cL}^{(M)} = \sum_{\substack{X\in\cV^{(r)}}}\left(\cH[P_XH_XP_X] + \sum_{j=1}^{\nu(X)}\cL[L_{X,j}P_X]\right)\,.
    			\end{equation*}
    			For $\hat{k}=\max\{k,4dr_{\exp},K\}$, there exist constants $c \geq 0$ and $\kappa \geq 0$ such that for every state $\rho\in\cT_1$ with $\max_{i\in V}\|\rho\|_{W^{2\hat{k},1}_i}<\infty$,
			\begin{equation*}
				\begin{aligned}
					\|e^{t\widetilde{\cL}}(\rho) - \Bigl(\prod_{X\in\cV^{(r)}} e^{t/n\widetilde{\cL}^{(M)}_X}\Bigr)^n(\rho^{(M)})\|_1
					&\leq \max\{\cK^2,|V|\}ct(1+t)e^{\kappa t}\sqrt{\max_{i \in V}\|\rho\|_{W^{2\hat{k},1}_i}} \frac{1}{n^{\frac{\hat{k}}{8r_{\exp}d+\hat{k}}}},
    				\end{aligned}
    			\end{equation*}
    			where $\mathcal{K}=\sum_{X\in\cV^{(r)}}(1+\nu(X))$. If \Cref{assum-lrb} satisfies $k=8r_{\exp}d \geq K$, we achieve the following bound:
    			\begin{equation*}
    				\begin{aligned}
    					\|e^{t\widetilde{\cL}}(\rho) - \Bigl(\prod_{X\in\cV^{(r)}} e^{t/n\widetilde{\cL}^{(M)}_X}\Bigr)^n(\rho^{(M)})\|_1 
					&\leq \max\{\cK^2,|V|\}ct(1+t)e^{\kappa t}\sqrt{\max_{i \in V}\|\rho\|_{W^{16d r_{\exp},1}_i}} \frac{1}{\sqrt{n}}\,.
    				\end{aligned}
    			\end{equation*}
    		\end{prop}
    		
    		\begin{proof}
    			By combining \Cref{lem:approximation,cor:state-approximation,lem:multi-trotter}, we achieve with $\rho^{(M)}\coloneqq \cP_V(\rho)$
    			\begin{equation*}
    				\begin{aligned}
    					\|&e^{t\widetilde{\cL}}(\rho)-\Bigl(\prod_{X\in\cV^{(r)}} e^{t/n\widetilde{\cL}^{(M)}_X}\Bigr)^n(\rho^{(M)})\|_1\\
    					&\leq\|e^{t\widetilde{\cL}}(\rho)-e^{t\widetilde{\cL}^{(M)}}(\rho)\|_1+\|\rho-\rho^{(M)}\|_1+\|e^{t\widetilde{\cL}^{(M)}}(\rho^{(M)})-\Bigl(\prod_{X\in\cV^{(r)}} e^{t/n\widetilde{\cL}^{(M)}_X}\Bigr)^n(\rho^{(M)})\|_1\\
    					&\leq \mathcal{K}\widetilde{c}_1te^{\hat{k} \Gamma t/2}\frac{\sqrt{\max_{i \in V}\|\rho\|_{W^{2\hat{k},1}_i}}}{(M-d+1)^{\hat{k}/4}}+2|V|\frac{\sqrt{\max_{i\in V}\|\rho\|_{W^{\hat{k},1}_i}}}{(M+1)^{\hat{k}/4}}+\frac{t^2\cK(\cK-1)}{2n}\max_{X\in\cV^{(r)}}\{\|\widetilde{\cL}_X^{(M)}\|_{1\rightarrow1}^2\}\\
    					&\leq (\mathcal{K}+|V|){c}_1te^{\hat{k} \Gamma t/2}\frac{\sqrt{\max_{i \in V}\|\rho\|_{W^{2\hat{k},1}_i}}}{(M-d+1)^{\hat{k}/4}}+\frac{t^2\cK(\cK-1)}{2n}\max_{X\in\cV^{(r)}}\{\|\widetilde{\cL}_X^{(M)}\|_{1\rightarrow1}^2\}\,,
    				\end{aligned}
    			\end{equation*}
    			where $\hat{k}^{\xi\hat{k}}$ is bounded, because $k$ is fixed, and hidden in $c_1$ (resp.~$\widetilde{c}_1$). Moreover, the norm $\|\widetilde{\cL}_X^{(M)}\|_{1\rightarrow1}$ can be upper bounded for all $X\in\cV^{(r)}$ by \Cref{lem:boundedness-projective-generator}:
    			\begin{equation*}
    				\begin{aligned}
    					\|\widetilde{\cL}_X^{(M)}(Y)\|_{1}&\leq\Bigl(2\|P_XH_XP_X\|_{\infty} + \sum_{j=1}^{\nu(X)}\|P_XL_{X,j}^\dagger\|_\infty\|L_{X,j}P_X\|_\infty+\|P_XL_{X,j}^\dagger L_{X,j}P_X\|_{\infty}\Bigr)\|Y\|_1\\
    					&\leq 2(1+\nu)\Lambda(d/2+1)^{4r_{\exp}}d!^{r_{\exp}}(M+1)^{r_{\exp}d}\\
    					&\leq 2(1+\nu)\Lambda(d/2+1)^{4r_{\exp}}d!^{r_{\exp}}(d+1)^{r_{\exp}d}(M-d+1)^{r_{\exp}d}
    				\end{aligned}
    			\end{equation*}
    			so that for $c_2=2(1+\nu)^2\Lambda^2(d/2+1)^{8r_{\exp}}d!^{2r_{\exp}}(d+1)^{2r_{\exp}d}$, we achieve
    			\begin{equation*}
    				\begin{aligned}
    					\|e^{t\widetilde{\cL}}(\rho)-\Bigl(\prod_{X\in\cV^{(r)}} &e^{t/n\widetilde{\cL}^{(M)}_X}\Bigr)^n(\rho^{(M)})\|_1\\
    					&\leq (\mathcal{K}+|V|)c_1te^{\hat{k} \Gamma t/2}\frac{\sqrt{\max_{i \in V}\|\rho\|_{W^{2\hat{k},1}_i}}}{(M-d+1)^{\hat{k}/4}}+\frac{t^2\cK^2}{n}c_2(M-d+1)^{2r_{\exp}d}\,.
    				\end{aligned}
    			\end{equation*}
    			Next, we choose $(M-d+1)=n^{\frac{4}{\hat{k}+8dr_{\exp}}}$ so that
    			\begin{equation*}
    				\begin{aligned}
    					\|e^{t\widetilde{\cL}}(\rho)&-\Bigl(\prod_{X\in\cV^{(r)}} e^{t/n\widetilde{\cL}^{(M)}_X}\Bigr)^n(\rho^{(M)})\|_1\\
    					&\leq \Bigl((\mathcal{K}+|V|)tc_1e^{\hat{k} \Gamma t/2}\sqrt{\max_{i \in V}\|\rho\|_{W^{2\hat{k},1}_i}}+t^2\cK^2 c_2\Bigr)\frac{1}{n^{\frac{\hat{k}}{8r_{\exp}d+\hat{k}}}}\,.
    				\end{aligned}
    			\end{equation*}
    			If $k=8dr_{\exp}\geq K$, then
    			\begin{equation*}
    				\begin{aligned}
    					\|e^{t\widetilde{\cL}}(\rho)&-\Bigl(\prod_{X\in\cV^{(r)}} e^{t/n\widetilde{\cL}^{(M)}_X}\Bigr)^n(\rho^{(M)})\|_1\\
						&\leq \Bigl((\mathcal{K}+|V|)tc_1e^{4dr_{\exp} \Gamma t}\sqrt{\max_{i \in V}\|\rho\|_{W^{16d r_{\exp},1}_i}}+t^2\cK^2 c_2\Bigr)\frac{1}{n^{\frac{1}{2}}}
    				\end{aligned}
    			\end{equation*}
    			finishes the proof.
    		\end{proof}
    		
    		This Trotter-approximation scheme directly implies the following simulation algorithm. Here, the gate complexity refers to the number of $r$-local quantum Markov semigroups.
    		\begin{cor}
    			Let $(\widetilde{\cL},\cT_f)$ be an operator satisfying the same assumptions as in \Cref{prop:church-turing-finite2}, in particular with $k=8r_{\exp}d\geq K$, where $K$ is given in \Cref{assum-lrb}. Then, every evolved state $\rho_t=e^{t\widetilde{\cL}}(\rho)$ with $\rho\in\cT_1$ satisfying $\max_{i\in V}\|\rho\|_{W^{16r_{\exp}d,1}_i}<\infty$ can be approximated by a concatenation of quantum Markov semigroups non-trivially defined on subsets of diameter $r$ and generated by
    			\begin{equation*}
    				\widetilde{\cL}^{(M)}_X = \cH[P_XH_XP_X] + \sum_{j=1}^{\nu(X)}\cL[L_{X,j}P_X],
    			\end{equation*}
			for all $X\in\cV^{(r)}$ using a gate complexity $\cO(\max\{\cK^4,|V|^2\}\varepsilon^{-2}te^{\cO(t)}\max_{i \in V}\|\rho\|_{W^{16d r_{\exp},1}_i})$. The projections are with respect to the Fock space defined in \Cref{prop:church-turing-finite2}.
    		\end{cor}

    	\subsubsection{Bosonic-bosonic approximation}
            Similar to the case mentioned above, we consider the approximation of a bosonic quantum Markov semigroup using a Trotter product formula. The key difference here is that we assume access to local building blocks of the quantum Markov semigroup to be simulated, so the simulation protocol does not aim to approximate a finite cutoff, but rather a global bosonic quantum Markov semigroup by means of local bosonic quantum Markov semigroups. In this context, we follow the ideas of~\cite[Thm.~4.2]{Moebus.2024}.    	
        	\begin{prop}\label{prop:church-turing-infinite}
        		Let $(\widetilde{\cL},\cT_f)$ be an operator satisfying \Cref{assum:local-GKSL,assum-lrb} (with constants $\xi$, $\Gamma$, $K$, degree $d$ and locality $r_{\exp}$), and assume that the closure of every local operator $(\widetilde{\cL}_X,\cT_f)$ generates a quantum Markov semigroup. Then for $\hat{k}=\max\{k,4dr_{\exp},K\}$, $\cK=\sum_{X\in\cV^{(r)}}(1+\nu(X))$, any state $\rho\in\cT_1$ with $\max_{i\in V}\|\rho\|_{W^{\hat{k},1}_i}<\infty$ and $t\geq0$
    	       	\begin{equation*}
                    \begin{aligned}
                        \biggl\|e^{t\widetilde{\cL}}(\rho)-\Bigl(\prod_{X\in\cV^{(r)}}e^{\frac{t}{n}\widetilde{\cL}_X}\Bigr)^n(\rho)\biggr\|_{1}\leq\frac{t^2}{n}ce^{\hat{k} \Gamma t}\hat{k}^{\xi \hat{k}}\cK^2\max_{i \in V} \|\rho\|_{W^{\hat{k},1}_i}\,.
                    \end{aligned}
    		      \end{equation*}
    	    \end{prop}
    		\begin{proof}
                Here, we follow the proof of Theorem 4.2 in \cite{Moebus.2024}. First, we apply the telescopic sum presented in Lemma 4.1 of \cite{Moebus.2024} and the contractivity of the involved semigroups to show
                \begin{equation*}
                    \begin{aligned}
                        \Bigl\|e^{t\widetilde{\cL}}(\rho)-\Bigl(\prod_{X\in\cV^{(r)}}e^{\frac{t}{n}\widetilde{\cL}_X}\Bigr)^{n}(\rho)\Bigr\|_{1}&\leq n\max_{j\in\{0,...,n-1\}}\Bigl\|\Bigl(e^{\frac{t}{n}\widetilde{\cL}}-\prod_{X\in\cV^{(r)}}e^{\frac{t}{n}\widetilde{\cL}_X}\Bigr)e^{\frac{j}{n}t\widetilde{\cL}}(\rho)\Bigr\|_{1}.
                    \end{aligned}
                \end{equation*}
                In the next step, we use the fundamental theorem of calculus and the identity:
                \begin{equation*}
                    \biggl(\prod_{j=1}^J A_j\biggr)-I=\sum_{j=1}^J \prod_{i=1}^{j-1}A_i(A_j-I),
                \end{equation*}
                twice to show:
                \begin{equation*}
                    \begin{aligned}
                        \Bigl(e^{\frac{t}{n}\widetilde{\cL}}&-\prod_{X\in\cV^{(r)}}e^{\frac{t}{n}\widetilde{\cL}_X}\Bigr)(\rho_{s})\\
                        &\quad=\Bigl(e^{\frac{t}{n}\widetilde{\cL}}-I-\sum_{j=1}^{|\cV^{(r)}|}\prod_{i=1}^{j-1}e^{\frac{t}{n}\widetilde{\cL}_{X_i}}\bigl(e^{\frac{t}{n}\widetilde{\cL}_{X_j}}-I\bigr)\Bigr)(\rho_{s})\\
                        &\quad=\frac{t}{n}\int_0^1\Bigl(e^{\tau_1\frac{t}{n}\widetilde{\cL}}\widetilde{\cL}-\sum_{j=1}^{|\cV^{(r)}|}\prod_{i=1}^{j-1}e^{\frac{t}{n}\widetilde{\cL}_{X_i}} e^{\tau_1\frac{t}{n}\widetilde{\cL}_{X_j}}\widetilde{\cL}_{X_j}\Bigr)(\rho_{s})d\tau_1\\
                        &\quad=\frac{t^2}{n^2}\int_0^1\int_0^1\tau_1\biggl(e^{\tau_1\tau_2\frac{t}{n}\widetilde{\cL}}\widetilde{\cL}^2-\sum_{j=1}^{|\cV^{(r)}|}\sum_{k=1}^{j}\prod_{\ell=1}^{k-1}e^{\frac{t}{n}\widetilde{\cL}_{X_\ell}} e^{\tau_1\tau_2\frac{t}{n}\widetilde{\cL}_{X_k}}\widetilde{\cL}_{X_k}\widetilde{\cL}_{X_j}\biggr)(\rho_{s})d\tau_2d\tau_1,
                    \end{aligned}
                \end{equation*}
                where $\rho_{s}\coloneqq e^{\frac{j}{n}t\widetilde{\cL}}(\rho)$ for any $j\in\{0,...,n-1\}$. Next, we apply the contractivity of the involved semigroups and the triangle inequality multiple times to obtain:
                \begin{equation*}
                    \begin{aligned}
                        \biggl\|\Bigl(e^{\frac{t}{n}\widetilde{\cL}}-\prod_{X\in\cV^{(r)}}e^{\frac{t}{n}\widetilde{\cL}_X}\Bigr)(\rho_{s})\biggr\|_{1}&\leq\frac{t^2}{2n}\Bigl(\|\widetilde{\cL}^2(\rho_{s})\|_1+\sum_{X,Y\in\cV^{(r)}}\|\widetilde{\cL}_{X}\widetilde{\cL}_{Y}(\rho_{s})\|_1\Bigr)\\
                        &\leq\frac{t^2}{n}\sum_{X,Y\in\cV^{(r)}}\|\widetilde{\cL}_{X}\widetilde{\cL}_{Y}(\rho_{s})\|_1\,.
                    \end{aligned}
                \end{equation*}
                Next, we apply the relative bounds given in \Cref{lem:relative-bounds-number-op}. First note that the concatenation of two operators $\widetilde{\cL}_X\circ\widetilde{\cL}_Y$ is a sum over $4(1+\nu(X))(1+\nu(Y))$ terms given by left and right multiplication of products of the operators
                \begin{equation*}
                    I, H_X,\,L_{X,j},\,L_{X,j}^\dagger,\,L_{X,j}^\dagger L_{X,j},\,H_Y,\,L_{Y,j},\,L_{Y,j}^\dagger,\,L_{Y,j}^\dagger L_{Y,j}\,.\,
                \end{equation*}
                Since all of these operators are built from the monomials $(\ad_i)^{k}a_i^{\ell}$ for some $k,\ell\in\{0,...,d\}$, we need to control terms like $(\ad_i)^{k_1}a_i^{\ell_1}(\ad_i)^{k_2}a_i^{\ell_2}$ for $k_1,\ell_1,k_2,\ell_2\in\{0,...,d\}$. Due to \Cref{lem:reordering-product-monomials}
    		      \begin{equation*}
                    a_i^{\ell}(\ad_i)^{k}=\sum_{j=0}^{\min\{\ell,k\}}j!\binom{\ell}{j}\binom{k}{j}(\ad_i)^{k-j}a_i^{\ell-j}\,,
    		      \end{equation*}
                we have
                \begin{equation*}
                    (\ad_i)^{k_1}a_i^{\ell_1}(\ad_i)^{k_2}a_i^{\ell_2}=\sum_{j=0}^{\min\{\ell_1,k_2\}}j!\binom{\ell_1}{j}\binom{k_2}{j}(\ad_i)^{k_1+k_2-j}a_i^{\ell_1+\ell_2-j}\,.
                \end{equation*}
                This shows that the above product operators are defined on the region $X\cup Y$, are $2r_{\exp}$-local, have maximal degree $2d$, and have coefficients bounded by the loose upper bound $(d!)^2\Lambda$. With that, \Cref{lem:relative-bounds-number-op} shows
        		\begin{equation*}
        			\begin{aligned}
    	       			\biggl\|\Bigl(e^{\frac{t}{n}\widetilde{\cL}}-\prod_{X\in\cV^{(r)}}e^{\frac{t}{n}\widetilde{\cL}_X}\Bigr)&(\rho_{s})\biggr\|_{1}\\
    			     	&\leq\frac{t^2}{n}\Lambda (d+1)^{8r_{\exp}}(2d)!^{2r_{\exp}+2}\cK^2\max_{X,Y\in\cV^{(r)}}\|(N+I)^{2\dd}_{X\cup Y}\rho_s(N+I)^{2\dd}_{X\cup Y}\|_{1}\,.
                    \end{aligned}
    		      \end{equation*}
                Next, we use Young's inequality and $|X\cup Y|\leq 2r_{\exp}$ to show 
    		      \begin{equation*}
                    \|(N+I)^{2\dd}_{X\cup Y}\rho_s(N+I)^{2\dd}_{X\cup Y}\|_{1}\leq\max_{i\in X\cup Y}\tr[(N+I)^{8dr_{\exp}}_{i}\rho_s]\,.
                \end{equation*}
                Finally, the moment propagation bound in \Cref{lem:moment-prop-bound} shows
                \begin{equation*}
                    \begin{aligned}
                        \biggl\|\Bigl(e^{\frac{t}{n}\widetilde{\cL}}&-\prod_{X\in\cV^{(r)}}e^{\frac{t}{n}\widetilde{\cL}_X}\Bigr)(\rho_{s})\biggr\|_{1}\\
                        &\leq\frac{t^2}{n}\Lambda (d+1)^{8r_{\exp}}(2d)!^{2r_{\exp}+2}\cK^2\max_{i\in V}\tr[(N+I)^{8dr_{\exp}}_{i}\rho_s]\\
                        &\leq\frac{t^2}{n}\Lambda (d+1)^{8r_{\exp}}(2d)!^{2r_{\exp}+2}\cK^2\Bigl(p4^{p+1} e^{k \Gamma t} \max_{i \in V} M_i^{(k)}(0) + \Upsilon_k(t)\Bigr)\\
                        &\leq\frac{t^2}{n}ce^{k \Gamma t}k^{\xi k}\cK^2\max_{i \in V} \|\rho\|_{W^{k,1}_i}\\
                    \end{aligned}
                \end{equation*}
                for $k\geq 8dr_{\exp}$ satisfying \Cref{assum:local-GKSL}, which finishes the proof.
            \end{proof}
            This Trotter-approximation scheme directly shows the following simulation algorithm. Here, the gate complexity refers to the number of $r$-local quantum Markov semigroups.
    	    \begin{cor}
                Let $(\widetilde{\cL},\cT_f)$ be an operator satisfying the same assumption given in \Cref{prop:church-turing-infinite}, in particular with $k=8r_{\exp}d\geq K$, where $K$ is given in \Cref{assum-lrb}. Then, every evolved state $\rho_t=e^{t\widetilde{\cL}}(\rho)$ with $\rho\in\cT_1$ satisfying $\max_{i\in V}\|\rho\|_{W^{16r_{\exp}d,1}_i}<\infty$ can be approximated by a concatenation of $r$-local semigroups defined by the generators $\{\cL_X\}_{X\in\cV^{(r)}}$ using a gate complexity $\cO(\cK^2\varepsilon^{-1}t^2e^{\cO(t)}\max_{i \in V}\|\rho\|_{W^{16d r_{\exp},1}_i})$.
            \end{cor}

            In the above results, we used the following identity:
    		\begin{lem}\label{lem:reordering-product-monomials}
    			Let $k,\ell\in\N$, then
    			\begin{equation*}
    				a^{\ell}(\ad)^{k}=\sum_{j=0}^{\min\{\ell,k\}}j!\binom{\ell}{j}\binom{k}{j}(\ad)^{k-j}a^{\ell-j}\,,
    			\end{equation*}
    		\end{lem}
    		\begin{proof}
    			We prove the statement by induction over $k$ and $\ell$. For that, we denote
    			\begin{equation*}
    				A(\ell,k)\coloneqq \sum_{j=0}^{\min\{\ell,k\}}j!\binom{\ell}{j}\binom{k}{j}(\ad)^{k-j}a^{\ell-j}\,.
    			\end{equation*}
    			First, we state with the induction start, which holds true because of 
    			\begin{equation*}
    				\begin{aligned}
    					A(\ell,k)=\begin{cases}
    						I & \text{if}\quad k=\ell=0\\
    						a^\dagger & \text{if}\quad k=0,\,\ell=1\\
    						a & \text{if}\quad k=1,\,\ell=0\\
    						a^\dagger a+I &\text{if}\quad k=\ell=1
    					\end{cases}
    				\end{aligned}
    			\end{equation*}
    			Then, we continue with the induction step. For that, we order the set $(\ell,k)\in\N^2$ satisfying $k\leq \ell$ by the lexicographical order so that 
    			\begin{equation*}
    				(\ell,k)+1=\begin{cases}
    					(\ell+1,0) & \text{if}\qquad \ell=k\\
    					(\ell,k+1) & \text{if}\qquad k<\ell\,.
    				\end{cases}
    			\end{equation*}
    			Therefore, we consider the following two cases assuming that the statement holds true for a $(\widetilde{\ell},\widetilde{k})\in\N^2$ satisfying $\widetilde{k}\leq \widetilde{\ell}$ and $(\widetilde{\ell},\widetilde{k})\leq(\ell,k)$ in the sense of the lexicographic order:
    			
    			\textit{Case I:} $\ell=k$
    				\begin{equation*}
    					(\ad)^{\ell+1}=A(\ell+1,0)
    				\end{equation*}
    				
    			\textit{Case II:} $k<\ell$\\
    				In the first step, we apply the commutation relation iteratively to commute one annihilation operator through $a^{\ell}$. Then, we apply the
    				\begin{equation*}
    					\begin{aligned}
    						a^{\ell}(\ad)^{k+1}&=\ad a^{\ell}(\ad)^{k}+\ell a^{\ell-1}(\ad)^{k}\\
    						&=\ad A(\ell,k)+\ell A(\ell-1,k)\\
    						&=\sum_{j=0}^{k}j!\binom{\ell}{j}\binom{k}{j}(\ad)^{k+1-j}a^{\ell-j}+\ell\sum_{j=0}^{k}j!\binom{\ell-1}{j}\binom{k}{j}(\ad)^{k-j}a^{\ell-1-j}\\
    						&=(\ad)^{k+1}a^{\ell}+ \sum_{j=1}^{k}j!\binom{\ell}{j}\binom{k}{j}(\ad)^{k-j+1}a^{\ell-j}\\
    						&\qquad+\ell\sum_{j=1}^{k}(j-1)!\binom{\ell-1}{j-1}\binom{k}{j-1}(\ad)^{k-j+1}a^{\ell-j}+\ell k!\binom{\ell-1}{k}a^{\ell-k-1}\\
    						&=(\ad)^{k+1}a^{\ell}+ \sum_{j=1}^{k}j!\binom{\ell}{j}\Biggl(\binom{k}{j}+\binom{k}{j-1}\Biggr)(\ad)^{k-j+1}a^{\ell-j}\\
    						&\qquad+(k+1)!\binom{\ell}{k+1}a^{\ell-k-1}\\
    						&=(\ad)^{k+1}a^{\ell}+ \sum_{j=1}^{k}j!\binom{\ell}{j}\binom{k+1}{j}(\ad)^{k-j+1}a^{\ell-j}+(k+1)!\binom{\ell}{k+1}a^{\ell-k-1}\\
    						&=A(\ell,k+1)
    					\end{aligned}
    				\end{equation*}
    				This shows that the statement holds true for $(\ell,k)+1$ if $k\leq\ell$. The case $\ell\leq k$ can be shown by swapping the roles of $\ell$ and $k$, which finishes the proof by induction.
    		\end{proof}

    \subsection{Local simulation of bosonic Lindbladian}
        Next, we consider locally defined expectations and show that every bosonic quantum Markov semigroup satisfying \Cref{assum:local-GKSL,assum-lrb} can be approximated by a finite-dimensional quantum Markov semigroup defined on a neighborhood of the support of the observable in question, independently of the system size (i.e., the number of modes). Following, for example, a Trotter product formula scheme, this local quantum Markov semigroup can in turn be approximated by local finite-dimensional building blocks:
		\begin{prop}\label{prop:simulation-local-observables}
			Let $(\widetilde{\cL},\cT_f)$ be an operator defined on a bosonic $D$-dimensional lattice satisfying \Cref{assum:local-GKSL,assum-lrb}. Given $M\in\N$, we denote the finite-rank projection on a region $R\subset V$ as $P\coloneqq \prod_{j\in R}P^{(M)}_{j}$ where $P^{(M)}_{i}=\sum_{n\le M}|n\rangle\langle n|_i$ and
			\begin{equation*}
				\widetilde{\cL}^{(M)}_{R} = \sum_{X\in\cV^{(r)}_R}\left(\cH[PH_XP] + \sum_{j=1}^{\nu(X)}\cL[L_{X,j}P]\right)\,.
			\end{equation*}			
			Then, for $t\geq0$, $k\in\N$ satisfying \Cref{assum-lrb} with $\hat{k}\coloneqq \max\{K,4dr_{\exp}\}$, $(M-d+1)=\cO(n^{\frac{4}{\hat{k}+6dr_{\exp}}})$, $\max_{j\in V}\|\rho\|_{W^{2\hat{k},1}_j}<\infty$, $|R|=\cO(|T|n^{\frac{D4r_{\exp}d}{\hat{k}+6dr_{\exp}}})$ such that $\dist(\partial_r R,T)=\cO((M-d+1)^{r_{\exp}d})$ and any bounded observable $O_T$ supported on a region $T\subset R[-r]$, there is a constant $h_1\ge 0$ independent of $n$, $|T|$, $|R|$ and $|V|$ such that, whenever the exponent of $n$ below is positive,
			\begin{equation*}
				\begin{aligned}
					&\frac{\left|\tr[e^{t\widetilde{\cL}}(\rho)O_T]-\tr[\Bigl(\prod_{X\in\cV^{(r)}_R}e^{\frac{t}{n}\widetilde{\cL}_R^{(M,X)}}\Bigr)^{n}(P\rho P)O_T]\right|}{\|O_T\|_\infty}\\
					&\,\leq h_1|T|^{r+1}\biggl(\hat{k}^{\xi \hat{k}}e^{\hat{k} \Gamma t}\max_{j\in V}\|\rho\|_{W^{2\hat{k},1}_j}+t^2K_R^2\|\rho\|_1\biggr)\frac{1}{n^{\frac{\hat{k}-2(1+2(r+1)D)dr_{\exp}}{6dr_{\exp}+\hat{k}}}}
				\end{aligned}
			\end{equation*}
			where $K_R=|\{X\,|\,X\in\cV^{(r)}\wedge X\subset R\}|$.
		\end{prop}
		\begin{proof}
			First, we approximate the quantum Markov semigroup under consideration by a locally defined and finite-dimensional quantum Markov semigroup via the weak Lieb--Robinson information propagation bounds proven in \Cref{thm:bosonicLRB}. Then, the finite cutoff is approximated by the Trotter-product formula presented in \Cref{lem:multi-trotter,cor:state-approximation}: For $t\geq0$, $k$ satisfying \Cref{assum-lrb}, $M\in\N$ and any bounded observable $O_T$ supported on a region $T\subset R[-r]\coloneqq \{v\in V\,|\,\dist(v,R^c)> r\}$, there are constants $c_1,c_2,\Gamma,\xi,\xi_2\ge 0$ independent of $|T|$, $|R|$ and $|V|$ such that
			\begin{equation*}
				\begin{aligned}
                    &\frac{\left|\tr[e^{t\widetilde{\cL}}(\rho)O_T]-\tr\Bigl[\Bigl(\prod_{X\in\cV^{(r)}_R}e^{\frac{t}{n}\widetilde{\cL}_R^{(M,X)}}\Bigr)^{n}(P \rho P)O_T\Bigr]\right|}{\|O_T\|_\infty}\\
					&\qquad\qquad\leq\left|\tr\Bigl[\Bigl(e^{t\widetilde{\cL}}-e^{t\widetilde{\cL}_R^{(M)}}\Bigr)(\rho)\frac{O_T}{\|O_T\|_\infty}\Bigr]\right|\\
                    &\qquad\qquad\qquad+\left|\tr\Bigl[\biggl(e^{t\widetilde{\cL}_R^{(M)}}-\Bigl(\prod_{X\in\cV^{(r)}_R}e^{\frac{t}{n}\widetilde{\cL}_R^{(M,X)}}\Bigr)^{n}\biggr)(\rho)\frac{O_T}{\|O_T\|_\infty}\Bigr]\right|\\
                    &\qquad\qquad\qquad+\left|\tr\Bigl[\Bigl(\prod_{X\in\cV^{(r)}_R}e^{\frac{t}{n}\widetilde{\cL}_R^{(M,X)}}\Bigr)^{n}(P\rho P-\rho)\frac{O_T}{\|O_T\|_\infty}\Bigr]\right|\\
					&\qquad\qquad\leq c_1 |R|^{r+1}e^{\hat{k} \Gamma t}\frac{\sqrt{\max_{j\in V}\tr[\rho (N_j+I)^{\hat{k}}] + \hat{k}^{\xi  \hat{k}}}}{(M-d+1)^{\hat{k}/4-r_{\exp}d/2}}\\
                    &\qquad\qquad\qquad+c_2|R|^r\frac{F^\downarrow_\mu(r)}{C_\mu^2}\,e^{t\xi_2(M+1)^{r_{\exp}d}\frac{C_\mu}{F^\downarrow_\mu(r)}}\sum_{x\in \partial_rR}\sum_{y\in T}F^\downarrow_{\mu}(\operatorname{dist}(x,y))\\
                    &\qquad\qquad\qquad+\frac{t^2K_R(K_R-1)}{2n} (4\Lambda(d/2+1)^{4r_{\exp}}d!^{r_{\exp}}(M+1)^{r_{\exp}d})^2\|\rho\|_1\\
                    &\qquad\qquad\qquad+2|R|\frac{\sqrt{\max_{i\in V}\|\rho\|_{W^{2\hat{k},1}_i}}}{(M+1)^{\hat{k}/2}}
				\end{aligned}
			\end{equation*}
            Beyond \Cref{thm:bosonicLRB} and \Cref{lem:multi-trotter}, we used \Cref{lem:boundedness-projective-generator} to upper bound the generator norms appearing in the Trotter product-formula bound \Cref{lem:multi-trotter} and defined $K_R=|\cV^{(r)}_R|$. Next, we chose again $(M-d+1)=n^{\frac{4}{\hat{k}+6dr_{\exp}}}$ (see proof of \Cref{prop:church-turing-finite2}) so that
			\begin{equation*}
				\begin{aligned}
					&\frac{\left|\tr[e^{t\widetilde{\cL}}(\rho)O_T]-\tr[\Bigl(\prod_{X\in\cV^{(r)}_R}e^{\frac{t}{n}\widetilde{\cL}_R^{(M,X)}}\Bigr)^{n}(P\rho P)O_T]\right|}{\|O_T\|_\infty}\\
					&\,\leq \biggl(c_1 |R|^{r+1}e^{\hat{k} \Gamma t}\hat{k}^{\xi  \hat{k}}\sqrt{\max_{j\in V}\|\rho\|_{W^{2\hat{k},1}_j}}+t^2K_R^28(\Lambda(d/2+1)^{4r_{\exp}}d!^{r_{\exp}}(d+1)^{r_{\exp}d})^2\|\rho\|_1\biggr)\frac{1}{n^{\frac{\hat{k}-2dr_{\exp}}{6dr_{\exp}+\hat{k}}}}\\
					&\quad\quad+c_2|R|^r\frac{F^\downarrow_\mu(r)}{C_\mu^2}\,\|F^\downarrow\|\,\min\{|T|,|\partial_rR|\}\,e^{-\mu \bigl(d(T,\partial_rR)-t\frac{\xi_2 C_\mu}{\mu F_\mu^\downarrow(r)}2^{r_{\exp}d/2}(M-d+1)^{r_{\exp}d}\bigr)}\|\rho\|_1\,,
				\end{aligned}
			\end{equation*}
            where we followed \Cref{eq:thm-exponential-proof-step} in the proof of \Cref{thm:bosonicLRB} to achieve the last line. Finally,
			\begin{equation}
				\dist(\partial_r R,T)\geq 2t\frac{\xi_2 C_\mu}{\mu F_\mu^\downarrow(r)}2^{r_{\exp}d/2}(M-d+1)^{r_{\exp}d}
			\end{equation}
			implies $\dist(\partial_r R,T)=\cO(n^{\frac{4r_{\exp}d}{\hat{k}+6dr_{\exp}}})$ because of $(M-d+1)=n^{\frac{4}{\hat{k}+6dr_{\exp}}}$. On a $D$-dimensional lattice, we can choose $R$ such that $|R|=\cO(|T|n^{\frac{D4r_{\exp}d}{\hat{k}+6dr_{\exp}}})$. Therefore, there is a constant $h\geq 0$ such that
			\begin{equation*}
				\begin{aligned}
					&\frac{\left|\tr[e^{t\widetilde{\cL}}(\rho)O_T]-\tr[\Bigl(\prod_{X\in\cV^{(r)}_R}e^{\frac{t}{n}\widetilde{\cL}_R^{(M,X)}}\Bigr)^{n}(P\rho P)O_T]\right|}{\|O_T\|_\infty}\\
					&\,\leq h_1|T|^{r+1}\biggl(\hat{k}^{\xi \hat{k}}e^{\hat{k} \Gamma t}\max_{j\in V}\|\rho\|_{W^{2\hat{k},1}_j}+t^2K_R^2\|\rho\|_1\biggr)\frac{1}{n^{\frac{\hat{k}-2(1+2(r+1)D)dr_{\exp}}{6dr_{\exp}+\hat{k}}}}
				\end{aligned}
			\end{equation*}
			which finishes the proof of the result.
		\end{proof}
		Note that higher-order Trotter-Suzuki formulas can be used to improve the above result in terms of the convergence rate in $n$ at the cost of a larger $k$ --- higher regularity on the input state. An intriguing direction for future work are multi-product Trotter formulas achieving higher order simulation rates (see for instance \cite{Childs.2021,Zhuk.2024}). Another interesting question is how to simulate higher-order bosonic Hamiltonians using fundamental bosonic operations, as presented in \cite{Kraus.2007}. It is worth noting that the fundamental question of whether the target operator generates a quantum dynamical semigroup remains a non-trivial open problem for Hamiltonians defined by higher-order polynomials in annihilation and creation operators.
        
\section{Learning Bosonic Hamiltonians}\label{sec:learning}
	In this section, we show how the results of \Cref{sec:lrb} can be used to derive learning procedures for the Bose--Hubbard model as well as the regularized Hamiltonian. For this, we will use the following two lemmas:
	\begin{lem}[Hoeffding's inequality \cite{Hoeffding.1963propability}]\label{lem:hoeffy}
		Let $N_{\mathrm{samp}}\in\N$, $X_1,\dots, X_{N_{\mathrm{samp}}}$ be independent random variables such that $a\le X_j\le b$, $a\le b$, almost surely. Consider the sum $S_{N_{\mathrm{samp}}}\coloneqq\sum_{j=1}^{N_{\mathrm{samp}}} X_j$. Then,
		\begin{align*}
			\mathbb{P}\Big(|S_{N_{\mathrm{samp}}}-\mathbb{E}[S_{N_{\mathrm{samp}}}]|\ge\epsilon\Big)\le 2\operatorname{exp}\left(-\frac{2\epsilon^2}{{N_{\mathrm{samp}}}(b-a)^2}\right)\,.
		\end{align*}
	\end{lem}

	\begin{lem}[Multivariate Markov brothers' inequality, see \cite{Harris2008,Harris.2002markov}] \label{Markovbroth}
		Given a real-valued polynomial $P$ of $K$ variables, and of maximum degree $D$, for all $k\le D$ 
		\begin{align*}
			\sup\big\{\max_{x_1,\dots,x_k}|\partial_{x_1}\dots \partial_{x_k}P(x)|\,:\,\|x\|_2\le 1\big\}\le 2^{2k-1}T^{(k)}_D(1)\,\sup\big\{|P(x)|\,:\,\|x\|_2\le 1\big\}\,.
		\end{align*}
		where $\|x\|_2$ denotes the Euclidean norm of $x\in \mathbb{R}^K$, and 
		\begin{align*}
			T^{(k)}_D(1)\coloneqq \frac{D^2(D^2-1^2)(D^2-2^2)\dots (D^2-(k-1)^2)}{1\cdot 3\cdot 5\dots (2k-1)}\,.
		\end{align*}
	\end{lem} 

\subsection{Learning from moment bounds}\label{vanillaalgobound}
	We start with a relatively simple finite-difference scheme for the Bose--Hubbard model, which already entails difficulties caused by the unboundedness of the Hamiltonian.
	
\subsubsection{The Bose--Hubbard model}
\begin{figure}[H]
		\begin{center}
			\includegraphics[width=0.8\textwidth]{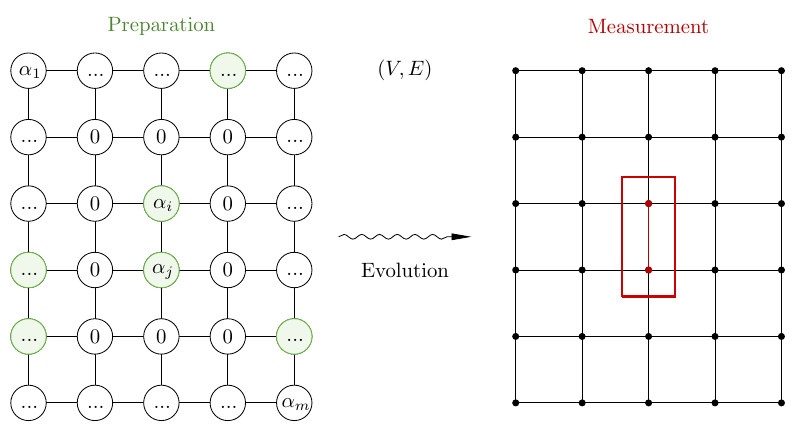}
		\end{center}	
		\caption{The coherent input states and the heterodyne measurement of the evolved state admit a local structure with respect to the underlying graph $(V,E)$. Here, the coherent input state is specified by the values $\alpha\in\C^m$ on isolated edges surrounded by vacuum states.}
		\label{fig2}
	\end{figure}

	First, we briefly recall the definition of the Bose--Hubbard model \Cref{ex:Bose-Hubbard} defined on a graph $(V,E)$:
	\begin{equation*}
		H\coloneqq \sum_{(i,j)\in E}\lambda^{(i,j)}\,a_i^\dagger\,a_j+\sum_{i\in V}u^{(i)}N_i(N_i-I)+\sum_{i\in V}\mu^{(i)}N_i
	\end{equation*}
	with complex coefficients $\lambda^{(i,j)}$ and real coefficients $u^{(i)}$ and $\mu^{(i)}$ whose absolute values are upper bounded by $\Lambda\geq0$. As discussed, the local structure can be exploited by the infinitesimal description of the quantum Markov time evolution. For that, we prepare coherent states $\ketbra{\alpha}{\alpha}$, let them evolve and measure after short time steps with heterodyne measurements $P_\beta^{(e)}\coloneqq\frac{1}{\pi^{|e|}}\ketbra{\beta}{\beta}_e\otimes I$ where we allow $e=(i,i)=i\in V$ by slight abuse of notation indicated by $e\in V\cup E$ and use the convention $|e|=|(i,j)|=2$ for $i\neq j$ and $1$ in the case of equality. Similarly, a coherent state $\ketbra{\alpha}{\alpha}$ with $\alpha\in\C^m$ is prepared satisfying $\alpha_i=0$ for all $i\in\{j\in V\,|\,1\leq\dist(e,j)\leq2\}$. As illustrated in \Cref{fig2}, the coherent states are non-trivial only on isolated vertices or edges, which allows us to exploit the locality of the measurement and parallelize at the same time.
    
    The beauty of the choice of the coherent input state and heterodyne measurement can be seen by calculating the first Taylor term of the probability density defined via Born's rule and the considered dynamics. A short calculation reveals
	\begin{equation}\label{alg1:polynomial}
		\begin{aligned}
			-\frac{\pi}{\abs{\braket{\alpha,\beta}}^2}i\tr[P_\beta^{(i)}[H,\ketbra{\alpha}{\alpha}]]&=-i\,u^{(i)}(\overline{\beta}_i^2\alpha_i^2-\overline{\alpha}_i^2\beta_i^2)-i\,\mu^{(i)}(\overline{\beta}_i\alpha_i-\overline{\alpha}_i\beta_i)\eqqcolon g_i(\alpha_i,\beta_i)\\
			-\frac{\pi^2}{\abs{\braket{\alpha,\beta}}^2}i\tr[P_\beta^{(i,j)}[H,\ketbra{\alpha}{\alpha}]]&=-i\,\lambda^{(e)}\left(\overline{\beta_i}\alpha_j-\overline{\alpha}_i\beta_j\right)+h.c.+g_i(\alpha_i,\beta_i)+g_j(\alpha_j,\beta_j)\\
            &\eqqcolon g_{ij}(\alpha_i,\alpha_j,\beta_i,\beta_j)\,,
		\end{aligned}
	\end{equation}
	where $i\in V$, $(i,j)\in E$ and $h.c.$ stands for the Hermitian conjugate of the term before. It is important to note that the coefficients of the polynomials $g_i$ and $g_{ij}$ are the defining parameters of the Hamiltonian. Moreover, $\tr[P_\beta^{(e)}[H,\ketbra{\alpha}{\alpha}]]$ depends only on the local parameters defining the Hamiltonian on the support of $P_\beta^{(e)}$ for $e\in V\cup E$. For this reason, we use the convention of reducing $\alpha_e=(\alpha_i,\alpha_j)\in\C^{|e|}$, or simply $\alpha$ if the reduction to $e$ is clear from the context. To extract the coefficients of the above polynomials, we first need to approximate the values on the left-hand side by a finite-difference scheme. For that, we need to control the error given by Taylor's theorem with Lagrange remainder:
	\begin{equation}\label{eq:bh-finite-diff-error}
		\begin{aligned}
			\Big| \frac{|\langle{\alpha}|{\beta}\rangle|^2}{\pi^{|e|}}\,g_{e}(\alpha,\beta)-t^{-1}\tr\big[(\rho_t-\rho)P_\beta^{(e)}\big]\Big|&\leq\frac{t}{2}\sup_{s\in[0,1]}\abs{\tr\Bigl[[[P_\beta^{(e)},H_{e[1]}],H_{e[2]}]\rho_s\Bigr]}\\
			&\le \frac{t}{2}\sup_{s\in[0,1]}\big\|[H_{e[1]},[H_{e[2]},\rho_s]]\big\|_1\\
			&\leq \Lambda^2\,9\gamma t \sup_{s\in[0,1]} M^{(4)}_{e[2]}(s)\,,
		\end{aligned}
	\end{equation}
    where we applied the relative bounds proven in \Cref{lem:relative-bounds-number-op}, in particular \Cref{eq:rel-boundedness-single-mode-term}, for any $e\in E\cup V$, $\alpha,\beta\in\C^{|e|}$ and $M^{(k)}_{e[2]}(s)\coloneqq \tr\big[\rho_s \prod_{i\in e[2]}(I+N_i)^k\big]$ using the notation introduced in \Cref{sec:lrb}, i.e.~$e[r]\coloneqq\{v\in V\,|\,\dist(v,e)\leq r\}$ for $r\in\{1,2\}$. In the last line, we follow the proof of \Cref{lem:rel-bounds}. Note that the upper bound above, in particular $M^{(4)}_{e[2]}(s)$, does not depend only on the input moments of the sites in $e$, because the time evolution spreads information. However, the Bose--Hubbard model satisfies the moment propagation bound in \Cref{lem:moment-prop-bound} by the proof of \Cref{prop:LRB-bose-hubbard}, i.e.~for all $t\ge 0$, $j\in V$ and $\kappa\leq k$ the following bound holds
	\begin{equation*}
		M_{e[2]}^{(k)}(t)\le 16e^{k|{e[2]}|\Gamma t}\,\max_{i\in V}\big\{\|\rho\|_{W^{|{e[2]}|k,1}_i}\big\}+\Upsilon_k(t)\,,
	\end{equation*}
	where $\Gamma=\Gamma' + \gamma \Gamma''$ is defined in \Cref{prop:LRB-bose-hubbard} and $\Upsilon_k$ is defined in \Cref{eq:dgl-linear-upper-bound} satisfying $\Upsilon_k=k^{\mathcal{O}(k)}$. Together this results in 
	\begin{equation}\label{eqboundmomentfinitediff1}
        \begin{aligned}
            \Big| \frac{1}{\pi^{|e|}}\,g_{e}(\alpha,\beta)|\langle{\alpha}|{\beta}\rangle|^2-t^{-1}\tr\big[(\rho_t-\rho)P_\beta^{(e)}\big]\Big|&=\cO(t\max_{i\in V}\big\{\|\rho\|_{W^{|{e[2]}|k,1}_i}\big\})\\
            &=\cO(t)
        \end{aligned}
	\end{equation}
	which validates the finite-difference scheme. Here, we used that $\|\alpha\|_{\infty}=\eta$ is a predefined constant (see \eqref{extract-coeff-bose-hubbard}) so that \Cref{lem:coherent-state-sobolev-norm} shows that $\max_{i\in V}\big\{\|\rho\|_{W^{|{e[2]}|k,1}_i}\big\}$ is just a constant. Keeping this bound in mind, we discretize the continuous measurement on $R_{\beta}\coloneqq [0,\beta_1]\times[0,\beta_2]$ for $\beta=\beta_1+i\beta_2\in\C$ if the measurement is locally defined on a vertex $i$ (or on $R_{\beta}=[0,\beta_1]\times[0,\beta_2]\times[0,\beta_3]\times[0,\beta_4]$ for $\beta=(\beta_1+i\beta_2,\beta_3+i\beta_4)\in\C^2$ if it is defined on an edge $e=(i,j)$). Then,
	\begin{equation}\label{extract-coeff-bose-hubbard}
		\begin{aligned}
			u^{(i)}&=-\frac{1}{2}\int_{R_{(1+i)}}\left(g_i(1,\beta_i)+g_i(-1,\beta_i)\right)d\beta_i\\
			\mu^{(i)}&=-\frac{1}{2}\int_{R_{(1+i)}}\left(g_i(1,\beta_i)-g_i(-1,\beta_i)\right)d\beta_i\\
			\RE(\lambda^{(i,j)})&=-\frac{1}{2}\int_{R_{(1+i)}}\int_{R_{(1+i)}} g_{ij}(1,1,\beta_i,\beta_j)-g_i(1,\beta_i)-g_j(1,\beta_j)d\beta_id\beta_j\\
			\IM(\lambda^{(i,j)})&=-\frac{1}{2}\int_{R_{(1+i)}}\int_{R_{(1+i)}} g_{ij}(-i,i,\beta_i,\beta_j)-g_i(-i,\beta_i)-g_j(i,\beta_j)d\beta_id\beta_j
		\end{aligned}
	\end{equation}
	is a possible choice to extract the coefficients of the underlying evolution. Therefore, the parameters can be extracted by choosing the coherent-state amplitudes and the output region of the heterodyne measurement appropriately. However, estimating the left-hand side in \eqref{extract-coeff-bose-hubbard} requires a finite-difference scheme on the regions $R_{\beta}$.
	
	Next, we turn the bound found in \Cref{eq:bh-finite-diff-error} as well as the exact scheme in \Cref{extract-coeff-bose-hubbard} into a guarantee for learning the coefficients of the Bose--Hubbard Hamiltonian. Our method is detailed in \Cref{alg:bose-hubbard-finite-diff} below, where we denote the indicator function of the set $R_\beta$ by $1_{R_\beta}$. Moreover, we define
    \begin{equation}\label{eq:bose-hubbard-alg-input}
        \begin{aligned}
            \cN_{2}^{\cS}&=\{-1,1\}\qquad&\qquad\cN_{2}^{\cM}&=\{1+i\}\subset\C\\
            \cN_{4}^{\cS}&=\{(1,1),(-i,i)\}\qquad&\qquad\cN_{4}^{\cM}&=\{(1+i,1+i)\}\subset\C^2
        \end{aligned}
    \end{equation}   
    
    \begin{algorithm}[H]
        \caption{Finite-difference scheme for the Bose--Hubbard model}\label{alg:bose-hubbard-finite-diff}
		
        \KwIn{$N_{\operatorname{samp}}\in\mathbb{N}$ and $\cN_{2}^{\cS}$, $\cN_{2}^{\cM}\subset\C$, $\cN_{4}^{\cS}$, $\cN_{4}^{\cM}\subset\C^2$ defined in \Cref{eq:bose-hubbard-alg-input}.}
		\KwOut{$\hat{Q}_{\alpha,\beta}^{(e,N_{\operatorname{samp}})}(t)\in\mathbb{R}$ for all $e\in E\cup V$ with $\alpha\in\cN_{2|e|}^{\cS}$, $\beta\in \cN_{2|e|}^{\cM}$.}
		\BlankLine
		Partition $V$ in $\nu_1=\mathcal{O}(1)$ sets $B_1^{(1)},...,B_{\nu_1}^{(1)}$ and $E$ in $\nu_2=\mathcal{O}(1)$ sets $B_1^{(2)},...,B_{\nu_2}^{(2)}$ satisfying $\forall k\in\{1,2\}, \forall j\in\{1,...,\nu_k\}, \forall b_1,b_2\in B_j^{(k)}$ : $\dist(b_1,b_2)\geq2$;\\
		\For{$k\in\{1,2\}$}{
			\For{$j \in\{1,...,\nu_k\}$, $\alpha\in\cN_{2|e|}^{\cS}$}{
				\For{$i_\alpha\in\{1,...,N_{\operatorname{samp}}\}$}{
					Prepare state $\ket{\alpha}\coloneqq \otimes_{e\in B_j^{(k)}}|\alpha\rangle_{e}\otimes |0\rangle_{(B_j^{(k)})^c}$;\\
					Run evolution generated by $-i[H,\cdot]$ on $\ket{\alpha}$ up to time $t=\cO(\varepsilon)$;\\
                    Perform heterodyne measurements on $B_j^{(k)}$\\
                    $\rightarrow$ get outcomes $\beta^{(i_\alpha)}_e\in \mathbb{C}^{|e|}$ $\forall e\in B_j^{(k)}$;
	   			}
			}
		}
		\Return{$\hat{Q}^{(e,N_{\operatorname{samp}})}_{(\alpha,\beta)}(t)\coloneqq\frac{1}{N_{\operatorname{samp}}t}\sum_{i_\alpha=1}^{N_{\operatorname{samp}}}\, \Big(\frac{1_{R_\beta}(\beta^{(i_\alpha)}_e)}{|\langle \alpha|\beta^{(i_\alpha)}_e\rangle|^2}-\frac{|R_\beta|}{\pi^{|e|}}\Big).$}
	\end{algorithm}
    
    The estimators $\hat{Q}^{(e,N_{\operatorname{samp}})}_{(\alpha,\beta)}(t)$ constructed above are sums of i.i.d.~random variables with
	\begin{align}\label{eqexpectationestimatorintro}
		\mathbb{E}\Big[\hat{Q}^{(e,N_{\operatorname{samp}})}_{(\alpha,\beta)}(t)\Big]= t^{-1}\int_{R_\beta} \frac{\tr\big[(\rho_t-\rho)P_{\beta'}^{(e)}\big]}{|\langle \alpha|\beta'\rangle|^2}\,d^{2|e|}\beta'\,,
	\end{align}
	where $\rho\coloneqq \ketbra{\alpha}{\alpha}$ and $\rho_t=e^{-it[H,\cdot]}(\rho)$. Moreover, a joint use of Hoeffding's concentration inequality (see \Cref{lem:hoeffy}) and the union bound show with probability $1-\delta\in(0,1)$ and for any $\alpha\in\cN_{2}^{\cS}$, $\beta\in\cN_{2}^{\cM}$ or $\alpha\in\cN_{4}^{\cS}$, $\beta\in\cN_{4}^{\cM}$ (see \Cref{eq:bose-hubbard-alg-input}) and any $e \in E\cup V$ that the random variables $\hat{Q}^{(e,N_{\operatorname{samp}})}_{(\alpha,\beta)}(t)$ satisfy
	\begin{align}\label{eqexpectationestimator2}
		\Big|\hat{Q}^{(e,N_{\operatorname{samp}})}_{(\alpha,\beta)}(t)-\mathbb{E}\big[\hat{Q}^{(e,N_{\operatorname{samp}})}_{(\alpha,\beta)}(t)\big]\Big|\le \epsilon_{\operatorname{stat}}
	\end{align}
	as long as the number of measurements satisfies:
    \begin{align}\label{eqT11}
		N_{\operatorname{samp}}= \, \Omega\left(\frac{1}{(t\cdot {\epsilon_{\operatorname{stat}}})^2}\ln\left(\frac{|V|}{\delta}\right)\right)\,.
	\end{align}
    where we used the finite connectivity of the graph showing that $|E|\leq\gamma|V|$. Here, we applied the Hoeffding's bound given in \Cref{lem:hoeffy}, which requires a lower and an upper bound on the estimator $\hat{Q}^{(e,N_{\operatorname{samp}})}_{(\alpha,\beta)}(t)$, which is given because $\min_{\beta' \in R_\beta} |\braket{\alpha|\beta'}|^2>0$ is a constant by construction. Next, we denote
	\begin{align}\label{eqQe}
		Q^{(e)}_\alpha(\beta)&\coloneqq \frac{1}{\pi^{|e|}}\,\int_{R_\beta}\,g_e(\alpha,\beta')\,d^{2|e|}\beta'
	\end{align}
	for which \eqref{eqboundmomentfinitediff1} shows
	\begin{align}\label{eqboundmomentfinitediff23}
		\Big|Q^{(e)}_{\alpha}(\beta)-\mathbb{E}\big[\hat{Q}^{(e,N_{\operatorname{samp}})}_{(\alpha,\beta)}(t)\big]\Big|=\mathcal{O}(t)\,.
	\end{align}
	In summary, combining \eqref{eqboundmomentfinitediff23} with \eqref{eqexpectationestimator2}, we see that
	\begin{align}\label{eqQQhatbound}
		\Big| Q^{(e)}_{\alpha}(\beta)- \hat{Q}^{(e,N_{\operatorname{samp}})}_{(\alpha,\beta)}(t)\Big|=\mathcal{O}({\epsilon}_{\operatorname{stat}}+ t)\,.
	\end{align}
	For very small time steps $t=\cO(\epsilon)$ (see \Cref{alg:bose-hubbard-finite-diff}), this shows that our estimator approximates the finite-difference expression well and thereby the left-hand side in \Cref{alg1:polynomial}. Therefore, we can learn the parameters by the scheme presented in \Cref{extract-coeff-bose-hubbard} with probability $1-\delta$ and up to precision $\epsilon$ as long as the evolution time for each run scales as $t=\mathcal{O}(\epsilon)$ and the total number of samples is at least $N_{\operatorname{samp}}=\mathcal{O}\left(\frac{1}{{\epsilon}^4}\ln\left(\frac{|V|}{\delta}\right)\right)$. Overall, our scheme admits a total evolution time bounded as 
	\begin{equation*}
		T_{\operatorname{evo}}={\mathcal{O}}\left(\epsilon^{-3}\ln\left(\frac{|V|}{\delta}\right)\right)\, ,
	\end{equation*}
    which is the main figure of merit in this work. We summarize this result in the following Proposition.
    
	\begin{prop}[Vanilla strategy for the Bose--Hubbard model]\label{bh-Vanillastrategythm}
		Combining \Cref{alg:bose-hubbard-finite-diff} with the direct method to extract the coefficients discussed in \Cref{extract-coeff-bose-hubbard}, we get {estimators} $\hat{\lambda}^{(e)}, \hat{u}^{(j)}$ and $\hat{\mu}^{(j)}$ of ${\lambda}^{(e)}, {u}^{(j)}$ and $\mu^{(i)}$,  such that for all $e\in E$ and $j\in V$, $|\hat{\lambda}^{(e)}-\lambda^{(e)}|,\,|\hat{u}^{(j)}-u^{(j)}|,\,|\hat{\mu}^{(j)}-\mu^{(j)}|\le \epsilon$ with probability $1-\delta$ as long as the evolution time for each run scales as $t=\mathcal{O}(\epsilon)$ and the total number of samples is at least $N_{\operatorname{samp}}=\mathcal{O}\left(\frac{1}{{\epsilon}^4}\ln\left(\frac{|V|}{\delta}\right)\right)$ so that our scheme uses a total evolution time of
		\begin{equation*}
			T_{\operatorname{evo}}=\mathcal{O}\big(\epsilon^{-3}\ln\big(\frac{|V|}{\delta}\big)\big)\,.
		\end{equation*}
	\end{prop}
    
\subsubsection{Polynomial two-body interaction}
	In the next step, we generalize the above scheme to polynomial two-body interactions (\ref{ex:general-hamiltonian}) defined by  $\cH^{(d)}=\cH[H_E]$ with 
	\begin{align*}
		H_E=\sum_{e\in E}H_e\qquad\text{ with }\qquad H_e\coloneqq \sum_{k,\ell,k',\ell'=0}^d\,\lambda^{(e)}_{k\ell k'\ell'}\,(a_i^\dagger)^k\,a_i^\ell\,(a_j^\dagger)^{k'}\,a_j^{\ell'}
	\end{align*}
	for complex coefficients $\lambda^{(e)}_{k\ell k'\ell'}$ upper bounded in absolute value by $\Lambda$ and $\lambda^{(e)}_{k\ell k'\ell'}=\overline{\lambda}^{(e)}_{\ell k \ell' k'}$. Note that we can fix $\lambda_{0000}=0$, due to gauge freedom.  To avoid superfluous technicalities, we further assume that $\lambda^{(e)}_{00k'\ell'}=\lambda^{(e)}_{k\ell 00}=0$. As we will see later on, this assumption will allow us to learn the coefficients exclusively from two-mode heterodyne measurements, but one can remove it by further assuming access to single-mode measurements as in the Bose--Hubbard model. Then, we directly include the $p$-photon dissipation, i.e.~for $\alpha\in\mathbb{C}^{m}$ and $p\in\mathbb{N}$
	\begin{align*}
		\cL^{(\alpha,p)}\coloneqq \sum_{j\in V}\cL[L_j^{(\alpha_j,p)}],\qquad \text{ where }\quad L_j^{(\alpha_j,p)}\coloneqq a_j^p-\alpha_j^p I\,,
	\end{align*}
	in our analysis. Therefore, we consider the regularized Hamiltonian (\ref{ex:reg-hamiltonian}) 
	\begin{equation*}
		\widetilde{\cL}^{(\alpha)}=\cH^{(d)}+\cL^{(\alpha,p)}\,.
	\end{equation*}
    As in the case of the Bose--Hubbard model, we define for $e\in E$ the heterodyne measurement by $P_\beta^{(e)}\coloneqq\frac{1}{\pi^{2}}\ketbra{\beta}{\beta}_e\otimes I$ with $\beta=(\beta_{i,R}+i\beta_{i,I},\beta_{j,R}+i\beta_{j,I})\cong(\beta_{i,R}, \beta_{i,I},\beta_{j,R}, \beta_{j,I})\in\mathbb{R}^4\cong\mathbb{C}^2$ and a coherent state $\ketbra{\alpha}{\alpha}$ with $\alpha\in\C^m$ satisfying $\alpha_i=0$ for all $i\in\partial e\coloneqq\{j\in V\,|\,\dist(e,j)=1\}=e[1]\backslash e$. Moreover, we assume that the above $\alpha$ defines both the input coherent state and the dissipation $\cL^{\alpha,p}$. Then, the defining property of the coherent state, $a_j\ket{\alpha}=\alpha_j\ket{\alpha}$, is used similarly to \Cref{alg1:polynomial} so that:
	\begin{equation*}
		\begin{aligned}
			\frac{\pi^2}{\abs{\braket{\alpha,\beta}}^2}\tr[P_\beta^{(e)}\widetilde{\cL}^{(\alpha)}(\ketbra{\alpha}{\alpha})]&\eqqcolon g_e(\alpha, \beta)
		\end{aligned}
	\end{equation*}
	with the polynomial
    \begin{equation}\label{alg2:polynomial}
        g_e(\alpha, \beta)=i\sum_{k\ell k'\ell'}\lambda^{(e)}_{k\ell k'\ell'}\left(\overline{\alpha}_i^{k}\overline{\alpha}_j^{k'}\beta_i^{\ell}\beta_j^{\ell'}-\overline{\beta}_i^k\overline{\beta}_j^{k'}\alpha_i^\ell\alpha_j^{\ell'}  \right)\,.
    \end{equation}
    Here, we used that for a given edge $e=(i,j)\in E$, $\ketbra{\alpha}{\alpha}=\ketbra{0}{0}_{\partial e}\otimes\ketbra{\alpha}{\alpha}_e\otimes\ketbra{\alpha}{\alpha}_{e[1]^c}$ and $P_\beta^{(e)}\ketbra{\alpha}{\alpha}=\ketbra{0}{0}_{\partial e}P_\beta^{(e)}\ketbra{\alpha}{\alpha}$ showing that
    \begin{equation*}
        \tr\Bigl[P_\beta^{(e)}[H_{e'},\ketbra{\alpha}{\alpha}]\Bigr]=0
    \end{equation*}
    for all $e'\neq e$ because $a_{j'}\ket{\alpha_{j'}}=0$ for $j'\in\partial e$ and $[H_{e'},\ketbra{\alpha}{\alpha}_{e[1]^{c}}]=0$. Here, we used the notation $\ketbra{\alpha}{\alpha}_{e[1]^{c}}=\bigotimes_{j\in e[1]^{c}}\ketbra{\alpha_j}{\alpha_j}$. Before solving the above polynomial for its coefficients, we discuss why a finite-difference scheme can approximate the left-hand side in \Cref{alg2:polynomial} well. As for the Bose--Hubbard model, we need to assume a moment propagation bound (see \Cref{assum-lrb}). If this assumption is satisfied, we can again apply \Cref{lem:moment-prop-bound} to show
	\begin{align*}
		\Big| \frac{1}{\pi^2}\,g_e(\alpha,\beta)|\langle{\alpha}|{\beta}\rangle|^2-t^{-1}\tr\big[(\rho_t-\rho)P_\beta^{(e)}\big]\Big|= \mathcal{O}(t)\,.
	\end{align*}
	for all fixed $\alpha,\beta\in \mathbb{C}^2$. Note that the input $\alpha$ relates to the defining values $\alpha_i,\alpha_j$ of the coherent input state with $e=(i,j)$. In \Cref{prop:LRB-regularized-hamiltonian}, we proved that for $p\geq 2(d+1)$ and $k-2+p\geq\frac{k(k+d)}{k-d}$, where $k$ stands for the power of the considered moment, the regularized Hamiltonian always satisfies the moment propagation bound. In the next lemma, we exploit the structure of $g_e$ to extract the coefficients from multiple differentiations of it.
	\begin{lem}\label{Lemmacoeffpoly}
		Let $g_e(\tilde \alpha, \tilde \beta)$ be as in Equation \eqref{alg2:polynomial} with $\tilde \alpha$, $\tilde \beta \in \mathbb C^2$. Let $x$, $y \in \mathbb R^2$. Let us define the real polynomials
        \begin{equation*}
            g_e^{(\mathrm{Im})}(x,y) := g_e(x, y)\, , \qquad  g_e^{(\mathrm{Re})}(x,y) := g_e(x, e^{-i\frac{\pi}{2(\ell+\ell')} }y) \, .
        \end{equation*}
        Let now $\alpha$, $\beta \in \mathbb R^2$. Assuming that $\lambda^{(e)}_{k\ell 00}=\lambda^{(e)}_{00k'\ell'}=0$ and $\ell+\ell'>0$, we get
		\begin{align*}
    		\operatorname{Im}\big(\lambda^{(e)}_{k\ell k'\ell'}\big)&=-\frac{\partial^k_{{\alpha}_i}\partial^{k'}_{{\alpha}_j}\partial_{\beta_i}^{\ell}\partial_{\beta_j}^{\ell'}\, g_e^{(\mathrm{Im})}(\alpha, \beta)\big|_{\alpha,\beta=0}}{2k'! \ell'! k!\ell!}\\
    		\operatorname{Re}\big(\lambda^{(e)}_{k\ell k'\ell'}\big)&=\frac{\partial^k_{{\alpha}_i}\partial^{k'}_{{\alpha}_j}\partial_{\beta_i}^{\ell}\partial_{\beta_j}^{\ell'}\, g_e^{(\mathrm{Re})}(\alpha, \beta)\big|_{\alpha,\beta=0}}{2k'!\ell'! k!\ell!}\,.
 		\end{align*}
 	\end{lem}
	\begin{proof}
	We take $e=(i,j)$. Using that $\overline{\lambda}^{(e)}_{k\ell k'\ell'}={\lambda}^{(e)}_{\ell k\ell'k'}$, we see that the last sum above simplifies to
	\begin{align*}
		g^{(\mathrm{Im})}_e(\alpha,\beta)=i\sum_{k\ell k'\ell'}\big(\lambda^{(e)}_{k\ell k'\ell'}-\overline{\lambda}^{(e)}_{k\ell k'\ell'}\big){\alpha}_i^{k}{\alpha}_j^{k'}\beta_i^{\ell}\beta_j^{\ell'}=-2\sum_{k\ell k'\ell'}\operatorname{Im}\big(\lambda^{(e)}_{k\ell k'\ell'}\big)\,{\alpha}_i^{k}{\alpha}_j^{k'}\beta_i^{\ell}\beta_j^{\ell'}\,.
	\end{align*}
	Therefore, we get that
	\begin{align*}
		\partial^k_{{\alpha}_i}\partial^{k'}_{{\alpha}_j}\partial_{\beta_i}^{\ell}\partial_{\beta_j}^{\ell'}\,g_e^{(\mathrm{Im})}(\alpha, \beta)\big|_{\alpha,\beta=0}=-2k!k'!\ell !\ell'!\operatorname{Im}\big(\lambda^{(e)}_{k\ell k'\ell'}\big)\,.
	\end{align*}
	Similarly, we can then show that we get
	\begin{align*}
		\partial^k_{{\alpha}_i}\partial^{k'}_{{\alpha}_j}\partial_{\beta_i}^{\ell}\partial_{\beta_j}^{\ell'}\,g_e^{(\mathrm{Re})}(\alpha, \beta)\big|_{\alpha,\beta=0}=2 k!k'!\ell !\ell'!\operatorname{Re}\big(\lambda^{(e)}_{k\ell k'\ell'}\big)\,.
	\end{align*}
	The remaining case $\ell+\ell'=0$ is recovered from the coefficient with exchanged creation and annihilation multi-indices using the Hermiticity relation $\lambda^{(e)}_{k\ell k'\ell'}=\overline{\lambda}^{(e)}_{\ell k\ell' k'}$.
	\end{proof} 

Therefore, successive derivatives of $g_e^{(\mathrm{Re})}$ and $g_e^{(\mathrm{Im})}$ at $0$ are associated with the coefficients $\lambda^{(e)}_{k\ell k'\ell '}$. Next, we use the same method as in \Cref{alg:bose-hubbard-finite-diff} to learn the polynomials $g_e$ and their derivatives. First, we learn another polynomial associated with $g_e$, namely its integral in $\beta$ over the set $R_\beta\coloneqq [0,\beta_1]\times \dots \times [0,\beta_4]$, for $\alpha$ and $\beta$ belonging to amplitude and measurement grids denoted by $\cN_{\cS}\subset\mathbb R^2$ and $\cN_{\cM}\subset\mathbb R^4$, with maximal norm bounded by $1$ for simplicity. These sets are chosen below according to \Cref{eq:chebyshev-nodes-finite-diff}, and the protocol is detailed in \Cref{alg:poly-interaction-finite-diff}, where we denote the indicator function of the set $R_\beta$ as $1_{R_\beta}$.

\medskip

\begin{algorithm}[H]
    \caption{Finite-difference scheme for the polynomial interaction Hamiltonian}\label{alg:poly-interaction-finite-diff}
	\KwIn{$N_{\mathrm{samp}}\in\mathbb{N}$ and sets $\cN_{\cS}\subset\R^2$, $\cN_{\cM}\subset\R^4\cong\C^2$.}
    \KwOut{$\hat{Q}_{\alpha,\beta}^{(e,N_{\operatorname{samp}})}(t)\in\mathbb{R}$ for all $e\in E$ with $\alpha\in\cN_{\cS}$, $\beta\in \cN_{\cM}$.}
    \BlankLine
	Partition $E$ in $\nu=\mathcal{O}(1)$ sets $B_1,...,B_{\nu}$ satisfying $\forall j\in\{1,...,\nu\}, \forall b_1,b_2\in B_j$ : $\dist(b_1,b_2)\geq2$;\\
	\For{$j \in\{1,...,\nu\}$, $\alpha\in\cN_{\cS}$}{
		\For{$i_\alpha\in\{1,...,N_{\operatorname{samp}}\}$}{
			Prepare state $\ket{\alpha}\coloneqq \otimes_{e\in B_j}|\alpha\rangle_{e}\otimes |0\rangle_{B_j^c}$;\\
			Run evolution generated by $\widetilde{\cL}^{(\alpha)}$ on $\ket{\alpha}$ up to time $t=\cO(\varepsilon)$;\\
            Perform heterodyne measurements on $B_j$\\
            $\rightarrow$ get outcomes $\beta^{(i_\alpha)}_e\in\R^4\cong \C^2$ $\forall e\in B_j$;
	   	}
	}
    \Return{$\hat{Q}^{(e,N_{\operatorname{samp}})}_{(\alpha,\beta)}(t)\coloneqq\frac{1}{N_{\operatorname{samp}}t}\sum_{i_\alpha=1}^{N_{\operatorname{samp}}}\, \Big(\frac{1_{R_\beta}(\beta^{(i_\alpha)})}{|\langle \alpha|\beta^{(i_\alpha)}\rangle|^2}-\frac{|R_\beta|}{\pi^2}\Big).$}
\end{algorithm}

\medskip
Then, we follow the same steps as for the Bose--Hubbard model. First, we calculate the expectation and apply Hoeffding's concentration bound to approximate the estimator by its expectation (see \Cref{eqexpectationestimatorintro}, (\ref{eqexpectationestimator2}) and (\ref{eqT11})). Next, we apply the error bound for the finite-difference scheme to the expectation of the estimator (see \Cref{eqboundmomentfinitediff1} and (\ref{eqQe})). In summary, for $(\alpha,\beta)\in\cN_{\cS}\times \cN_{\cM}$
\begin{align}\label{eqQQhatbound-poly}
    \Big| Q^{(e)}_{\alpha}(\beta)- \hat{Q}^{(e,N_{\operatorname{samp}})}_{(\alpha,\beta)}(t)\Big|=\mathcal{O}({\epsilon}_{\operatorname{stat}}+ t)\,.
\end{align}
The connection to the parameters $\lambda^{(e)}_{k\ell k'\ell'}$ can then be made through the use of \Cref{Lemmacoeffpoly}. By the definition $Q^{(e)}_\alpha(\beta)\coloneqq \frac{1}{\pi^2}\,\int_{R_\beta}\,g_e(\alpha,\beta')\,d^4\beta'$, we see that
\begin{align} \label{eq:g-from-Q}
    g_e(\alpha, \beta)
    =\pi^2 \partial_{\beta_{i,R}}\partial_{\beta_{i,I}}\partial_{\beta_{j,R}}\partial_{\beta_{j,I}}Q^{(e)}_\alpha( \beta)\big|_{\beta}\,.
\end{align}
Together with \Cref{Lemmacoeffpoly}, we can then make a change of variables in $g_e(\alpha, \beta)$ to define $g^{(\mathrm{Re})}_e(\alpha, \beta)$ and $g^{(\mathrm{Im})}_e(\alpha, \beta)$ and take partial derivatives in order to recover $\operatorname{Re}\big(\lambda^{(e)}_{k\ell k'\ell'}\big)$ and $\operatorname{Im}\big(\lambda^{(e)}_{k\ell k'\ell'}\big)$. Next, we perform exact polynomial interpolation of the data $(\alpha, \beta, \hat{Q}_{(\alpha,\beta)}^{(e,T)}(t))$ generated by \Cref{alg:poly-interaction-finite-diff} using outlier-robust multivariate polynomial interpolation \cite{Arora.2024}, treating $Q_\alpha^{(e)}(\beta)$ as a polynomial in $6$ real variables, i.e.~$\alpha\in\R^2$ and $\beta\in\C^2\cong\R^4$. This results in a polynomial $\widetilde{Q}^{(e)}:(\alpha,\beta)\mapsto \widetilde{Q}_\alpha^{(e)}(\beta)$ of degree at most $d$ in each real variable. It follows from \Cref{lem:det-multivar-poly-int} with $n=6$ and $M=d+1$ in combination with \Cref{eqQQhatbound-poly} that for any $(\alpha,\beta)\in [-1,1]^{6}$,
\begin{equation}\label{eq:chebyshev-nodes-finite-diff}
    \begin{aligned}
        \big|Q_\alpha^{(e)}(\beta)-\widetilde{Q}^{(e)}_\alpha(\beta)\big|&\le 3 \max_{(\alpha,\beta)\in\cN_{\cS}\times \cN_{\cM}}\, |Q_\alpha^{(e)}(\beta)-\hat{Q}_{(\alpha,\beta)}^{(e,N_{\operatorname{samp}})}(t)|\\
        &=\mathcal{O}( \epsilon_{\operatorname{stat}}+t)\,.
    \end{aligned}
\end{equation}
For the polynomial interpolation to work, we choose the sets of points $(\alpha,\beta)\in\cN_{\cS}\times \cN_{\cM}$ on which we measure according to the Chebyshev partition in Definition \ref{defi:chebyshev-partition}, such that there is at least one point in each box $C_j$ defined in \Cref{eq:chebyshev-box}. Since each coordinate can be chosen independently according to a $(m,1)$-Chebyshev partition, we can obtain such a choice of points with $|\cN_{\cS}|=\mathcal O(d^2)$ and $|\cN_{\cM}| = \mathcal O(d^4)$. In the presence of outliers, one can instead choose the points $(\alpha, \beta)$ according to the multivariate Chebyshev distribution and use the algorithm in \cite{Arora.2024} (see \Cref{lem:det-multivar-poly-int}). 

Therefore, combining \Cref{eq:g-from-Q} with a multivariate Markov brothers' inequality (see \Cref{Markovbroth}), we have obtained an estimator 
\begin{equation*}
    \widetilde g_e(\alpha, \beta)
    :=\pi^2 \partial_{\beta_{i,R}}\partial_{\beta_{i,I}}\partial_{\beta_{j,R}}\partial_{\beta_{j,I}}\widetilde Q^{(e)}_\alpha( \beta)\big|_{\beta}
\end{equation*}
such that, assuming $d = \mathcal O(1)$,
\begin{equation*}
   \sup_{(\alpha,\beta)\in [-1,1]^{6}} |g_e(\alpha, \beta) - \widetilde g_e(\alpha, \beta)| =\mathcal{O}\big({\epsilon}_{\operatorname{stat}}+t\big)\,.
\end{equation*}
Further uses of a multivariate Markov brothers' inequality then imply that for any coefficient $\lambda^{(e)}_{k\ell k'\ell'}$, defining $\hat{\lambda}^{(e)}_{k\ell k'\ell'}$ by replacing the $g$'s by $\widetilde{g}$'s in \Cref{Lemmacoeffpoly}, we finally get for all $k,\ell,k',\ell'$ that
\begin{align*}
    |{\lambda}^{(e)}_{k\ell k'\ell'}-\hat{\lambda}^{(e)}_{k\ell k'\ell'}|&=\mathcal{O}\big({\epsilon}_{\operatorname{stat}}+t\big)\,,
\end{align*}
where we assumed that $d,\Lambda=\mathcal{O}(1)$. Imposing each of the variables in the above brackets to be of order $\epsilon$, we arrive at the main result of this section:

\begin{prop}[Vanilla strategy for poly.~two-body interaction]\label{Vanillastrategythm}
   Combining \Cref{alg:poly-interaction-finite-diff} with exact polynomial interpolation of its output, we get estimators $\hat{\lambda}^{(e)}_{k\ell k'\ell'}$ of $\lambda^{(e)}_{k\ell k'\ell'}$ such that for all $e\in E$ and all $k,\ell,k',\ell'$, $|\hat{\lambda}^{(e)}_{k\ell k'\ell'}-\lambda^{(e)}_{k\ell k'\ell'}|\le \epsilon$ with probability $1-\delta$ as long as the evolution time for each run scales as $t=\mathcal{O}(\epsilon)$ and the total number of samples is at least $N_{\operatorname{samp}}=\mathcal{O}\left(\frac{1}{{\epsilon}^4}\ln\left(\frac{|E|}{\delta}\right)\right)$ so that our scheme uses a total evolution time of
   \begin{equation*}
   		T_{\operatorname{evo}}=\mathcal{O}\left(\epsilon^{-3}\ln\left(\frac{|E|}{\delta}\right)\right)\,.
   \end{equation*}
\end{prop}

\begin{proof}
The total number of samples needed in \Cref{alg:poly-interaction-finite-diff} is equal to $N_{\mathrm{samp}}\cdot |\cN_{\cS}|\cdot \nu$, where $|\cN_{\cS}|\cdot \nu$ corresponds to the total number of preparation setups. Here, $\nu$ is the number of colour classes for the distance-two conflict graph on the edge set $E$; for bounded-degree interaction graphs this is a constant (for instance, a greedy colouring gives $\nu=\mathcal{O}(\gamma^2)$ when the graph degree is bounded by $\gamma$). The result follows from \eqref{eqT11} with $\epsilon_{\operatorname{stat}}=\cO(\epsilon)$.
\end{proof}

\subsection{Improved sample complexity via weak Lieb--Robinson bounds}\label{LRsamplecomplexity}

We now aim at improving the $\epsilon$-dependence of the sample complexity found in \Cref{bh-Vanillastrategythm,Vanillastrategythm} using the weak Lieb--Robinson estimates from \Cref{thm:bosonicLRB}, which hold for all generators satisfying \Cref{assum:local-GKSL,assum-lrb}. These two assumptions are, for instance, covered by the Bose--Hubbard model, any quadratic Hamiltonian, and the regularized polynomial two-body interaction Hamiltonian (see \Cref{eq:regularized-hamiltonian}). 
Here, it is important to mention that the photon dissipation does not depend on the input state but must be known beforehand. In contrast, in the finite-difference scheme (see \Cref{Vanillastrategythm}), the photon dissipation is defined by the same $\alpha \in \mathbb{R}^{|V|}$ as the input state. Since this is no longer necessary, the following scheme can also learn models such as
\begin{equation*}
	\widetilde{\cL}^{(0)} = \cH^{(d)} + \cL[a^p]\,.
\end{equation*}
This can be generalized to any dissipation, as long as it is known beforehand and the whole operator generates a quantum Markov semigroup. Therefore, we denote the generator by $\widetilde{\cL}$, which stands for the Bose--Hubbard model \Cref{ex:Bose-Hubbard}, any quadratic Hamiltonian \Cref{ex:Gaussian}, or a Hamiltonian regularized by a fixed power and shift of the photon number operator \Cref{ex:reg-hamiltonian}.

For simplicity, we present one scheme that covers all examples. We first write it for the regular $D$-dimensional lattice $V=\{-n,...,n\}^D$. The same proof applies, without changing the stated scaling except for constants, to any family of interaction graphs with uniformly bounded growth, meaning that there are constants $C_{\operatorname{vol}},D_{\operatorname{vol}}$ independent of $|V|$ such that every graph ball satisfies $|B(x,R)|\leq C_{\operatorname{vol}}(1+R)^{D_{\operatorname{vol}}}$. Indeed, the only place where the lattice structure is used is the partition of vertices and edges into classes whose enlarged neighborhoods are disjoint. For a neighborhood radius $\ell_{\mathrm{LR}}=\mathcal{O}(\operatorname{polylog}(\varepsilon_{\mathrm{LR}}^{-1}))$, form the conflict graph on $V\cup E$ by connecting two vertices or edges when their enlarged neighborhoods intersect. Under the bounded-growth condition this conflict graph has degree $\mathcal{O}(\ell_{\mathrm{LR}}^{D_{\operatorname{vol}}})$, so a greedy coloring gives $\mathcal{O}(\ell_{\mathrm{LR}}^{D_{\operatorname{vol}}})=\operatorname{polylog}(\varepsilon_{\mathrm{LR}}^{-1})$ color classes. On an arbitrary bounded-degree graph without bounded growth the same construction remains valid, but the number of color classes is the corresponding packing number (bounded by an exponential in $\ell_{\mathrm{LR}}$ in the worst case); in that case the sample and runtime bounds acquire this geometry-dependent factor rather than a purely polylogarithmic overhead.

We begin by describing an estimation algorithm that serves as a central subroutine in our learning procedure. Before presenting the algorithm, we fix some notation. For an edge or vertex $e\in E\cup V$, let $\ell_{\mathrm{LR}}=\mathcal{O}(\operatorname{polylog}(\varepsilon_{\mathrm{LR}}^{-1}))$ be the radius supplied by the weak Lieb--Robinson bound and set $R_e=e[\ell_{\mathrm{LR}}]\subset V$. On the lattice, one may choose $R_e$ inside a rectangle centered around $e$ with side length $\mathcal{O}(\ell_{\mathrm{LR}})$; on a bounded-growth graph, $R_e$ denotes the corresponding graph-metric neighborhood. With this, $P=P_{R_e}^{(M)}$ denotes the finite-rank projection on the region $R_e$ introduced in \Cref{thm:bosonicLRB}. As before, the preparation and measurement are defined by $\alpha$ and $\beta$ belonging to a given set of points denoted by $\cN_{K}^{\cS}\subset \R^K$ for the input states and $\cN_{K}^{\cM}\subset \R^K$ for the heterodyne measurement, with maximal norm bounded by $1$ for simplicity. These sets are specified later in \Cref{eq:learning-algorithm-chebpoly} and the protocol is detailed in \Cref{protocollearningLRprojected} below:

\begin{algorithm}[H]
    \caption{Refined scheme for bosonic Hamiltonian learning}\label{protocollearningLRprojected}
    \KwIn{$N_{\mathrm{samp}}\in\mathbb{N}$, time grid $\cN_t\subset [b_1,b_2]$ with $0<b_1<b_2$, and finite sets $\cN_{1}^{\cS}$, $\cN_{2}^{\cM}$, $\cN_{2}^{\cS}$, $\cN_{4}^{\cM}$.}
    \KwOut{$\hat{L}_{\alpha,\beta}^{(e,N_{\mathrm{samp}})}(t)\in\mathbb{R}$ for all $e\in E\cup V$, $\alpha\in\cN_{|e|}^{\cS}$, $\beta\in \cN_{2|e|}^{\cM}$ and $t\in\cN_t$.}
	\BlankLine
	Partition $V$ in $\nu_1=\mathcal{O}(\operatorname{polylog}(\varepsilon_{\mathrm{LR}}^{-1}))$ sets $B_1^{(1)},...,B_{\nu_1}^{(1)}$ and $E$ in $\nu_2=\mathcal{O}(\operatorname{polylog}(\varepsilon_{\mathrm{LR}}^{-1}))$ sets $B_1^{(2)},...,B_{\nu_2}^{(2)}$ satisfying $\forall k\in\{1,2\}, \forall j\in\{1,...,\nu_k\}, \forall b_1\neq b_2\in B_j^{(k)}$ : $R_{b_1}\cap R_{b_2}=\emptyset$;\\
	\For{$k\in\{1,2\}$}{
		\For{$j \in\{1,...,\nu_k\}$, $\alpha\in\cN_{|e|}^{\cS}$, $t\in\cN_t$}{
			\For{$i_\alpha\in\{1,...,N_{\operatorname{samp}}\}$}{
				Prepare state $\ket{\alpha}\coloneqq \otimes_{e\in B_j^{(k)}}|\alpha\rangle_{e}\otimes |0\rangle_{(B_j^{(k)})^c}$;\\
				Run evolution generated by $\widetilde{\cL}$ on $\ket{\alpha}$ up to time $t$;\\
                Perform heterodyne measurements on $B_j^{(k)}$\\
                $\rightarrow$ get outcomes $\beta^{(i_\alpha)}_e\in \mathbb{C}^{|e|}$ for all $e\in B_j^{(k)}$;
	   		}
		}
	}
	\Return{$\hat{L}^{(e,N_{\mathrm{samp}})}_{(\alpha,\beta)}(t)\coloneqq  \frac{1}{N_{\mathrm{samp}}}\sum_{i_{\alpha,t}=1}^{N_{\mathrm{samp}}}\, e^{|\beta_e^{(i_{\alpha,t})}|^2+|\alpha|^2}\,1_{R_\beta}(\beta_e^{(i_{\alpha,t})})$.}
\end{algorithm}
\medskip

The claimed partitions in \Cref{protocollearningLRprojected} can be constructed as follows. First, we cover $V$ with non-overlapping $\ell^1$-balls of radius $\ell_{\mathrm{LR}}$. Then, we define $B_1^{(1)}$ to be the set of center points of these $\ell^1$-balls. The sets $B_j^{(1)}$ are then defined by shifts of the initial set $B_1^{(1)}$, which implies that the number of such sets scales with the volume of the $\ell^1$-balls: $\mathcal{O}(\ell_{\mathrm{LR}}^D)=\mathcal{O}(\operatorname{polylog}(\varepsilon_{\mathrm{LR}}^{-1}))=\nu_1$. The partition of the edges is defined analogously, with non-overlapping enlargements $e[\ell_{\mathrm{LR}}]$ in place of balls around vertices. Similarly, the center edges define the set $B_1^{(2)}$, and the sets $B_j^{(2)}$ are obtained via shifts, so that the overall number of partition elements again scales with $\nu_2$.

\medskip

We now explain what quantities are being estimated in \Cref{protocollearningLRprojected} and how we can use those estimates for learning the Hamiltonian coefficients. Recall that we use the convention that $e=(i,i)=i\in V$ denotes edges and vertices at the same time indicated by $e\in E\cup V$. Moreover, $|e|=1$ if $e\in V$ and $|e|=2$ if $e\in E$.
First, observe that the constructed estimators by definition have the following expectation values:
\begin{align*}
    \mathbb{E}\left[\hat{L}^{(e,N_{\mathrm{samp}})}_{(\alpha,\beta)}(t)\right] 
    = \int_{R_\beta}e^{|\beta'|^2+|\alpha|^2} \operatorname{tr}\left[P_{\beta'}^{(e)}\, e^{t\widetilde{\mathcal{L}}}(\ketbra{\alpha}{\alpha})\right]d^{2|e|}\beta'\,,
\end{align*}
where $\alpha\in\R^{|e|}$ defines $\ket{\alpha}$ by extending $\alpha$ (or $\alpha_e$ if not clear from context) by vacuum vectors to a coherent state in a Fock space with $|R_e|$ modes (see definition in \Cref{protocollearningLRprojected}). A joint use of Hoeffding's concentration inequality (see \Cref{lem:hoeffy}) and the union bound yields
\begin{equation}\label{eq:learning-hoeffding}
    \Big|\hat{L}^{(e,N_{\mathrm{samp}})}_{(\alpha,\beta)}(t)-\mathbb{E}\big[\hat{L}^{(e,N_{\mathrm{samp}})}_{(\alpha,\beta)}(t)\big]\Big|\le \epsilon_{\operatorname{stat}}
\end{equation}
holds simultaneously for all $e\in E\cup V$, $\alpha\in\cN_{|e|}^{\cS},\,\beta\in \cN_{2|e|}^{\cM}$, $t\in\cN_t$ with probability $1-\delta$ as long as the number of measurements for each $\alpha, j\in\{1,...,\nu_{|e|}\}$ satisfies: 
\begin{align*}
    N_{\mathrm{samp}}= \, \mathcal{O}\left(\frac{1}{{\epsilon^2_{\operatorname{stat}}}}\ln\left(\frac{(|E|\cdot |\cN_{2}^{\cS}|\cdot |\cN_{4}^{\cM}|+|V|\cdot |\cN_{1}^{\cS}|\cdot |\cN_{2}^{\cM}|)\cdot |\cN_t|}{\delta}\right)\right)\,.
\end{align*}
That is, the estimates $\hat{L}^{(e,N_{\mathrm{samp}})}_{(\alpha,\beta)}(t)$ are reliable approximations to their expectation values, with high probability. Notice that, compared with the bound obtained using the finite-difference scheme in \eqref{eqT11}, the bound on $N_{\mathrm{samp}}$ here does not scale inverse polynomially with $t$.

We now consider these expectation values in more detail and relate them to our quantities of interest. To this end, we use the weak Lieb--Robinson-type bound provided in \Cref{thm:bosonicLRB,cor:lrb-projective-input}, as well as \Cref{prop:propagation-quadratic-Hamiltonian,prop:LRB-bose-hubbard,prop:LRB-regularized-hamiltonian}, which verify the assumptions. Then, for constant time $t=\cO(1)$ and a given coherent input state $\ketbra{\alpha}{\alpha}$ given by $\alpha\in \C^{|V|}\cong\mathbb{R}^{2|V|}$ with $\|\alpha\|_\infty=\mathcal{O}(1)$, \Cref{cor:lrb-projective-input} shows that there exist constants $a,b,c>0$ such that, for any $T\subset R\subset V$ with $|T|=\mathcal{O}(1)$, $\operatorname{dist}(\partial E_R,T)$ large enough (see \Cref{thm:bosonicLRB}), $M= \mathcal{O}\big( \operatorname{dist}(\partial E_R,T)^{\frac{1}{r_{\exp}d}} \big)$ and any observable $O_T$ defined on $T$ 
\begin{equation}\label{LRBsimpleprime}
    \frac{\left|\tr[O_T(\rho_t-\rho_t^{(M)})]\right|}{\|O_T\|_\infty}\le a\,  |R|^{3}\,e^{-b
     \operatorname{dist}(\partial E_R,T)^{c}}\,.
\end{equation}
Here, $r=2=r_{\exp}$, $d$ is the degree of the generator (see \Cref{assum:local-GKSL}) and we denote $\rho_t=e^{t\widetilde{\mathcal{L}}}(\ketbra{\alpha}{\alpha})$, $\rho^{(M)}_t=e^{t\widetilde{\mathcal{L}}^{(M)}_{R}}(\ketbra{\psi_\alpha}{\psi_\alpha})=e^{t\widetilde{\mathcal{L}}^{(M)}_{R}}\circ\cP(\ketbra{\psi_\alpha}{\psi_\alpha})$ with $\ket{\psi_\alpha}=P\ket{\alpha}$, $P\coloneqq \prod_{j\in R}P^{(M)}_{j}$ and $\cP=P\cdot P$. Using \Cref{LRBsimpleprime}, we can adapt the method developed in \cite{Stilckfranca.2024} to the present bosonic setting. 
For our purposes, we fix $T=e$ for any edge or vertex $e\in E\cup V$ and choose $R=R_e=e[\ell_{\mathrm{LR}}]$ as in \Cref{protocollearningLRprojected}. Then, the effect operator $P_\beta^{(e)}= \frac{1}{\pi^{|e|}}\ketbra{\beta}{\beta}_e\otimes I_{e^c}$ satisfies
\begin{align*}
    \Big|\tr\big[(\rho_t-\rho^{(M)}_t){P_\beta^{(e)}}\big]\Big|\le \epsilon_{\operatorname{LR}}
\end{align*}
because $\operatorname{dist}(\partial E_{R_e},e)=\mathcal{O}(\operatorname{polylog}(\epsilon_{\operatorname{LR}}^{-1}))$ ($\partial E_{R}\coloneqq\{X\in\cV^{(r)}\,|\,X\cap R\neq\emptyset\neq X\cap R^c\}$ denotes boundary interaction supports) for any $\epsilon_{\operatorname{LR}}>0$ by the definition of $R_e$. Next, we approximate the finite-dimensional superoperator exponential $e^{t\widetilde{\cL}^{(M)}_{R}}$ by its Taylor expansion of order $K$:
\begin{align*}
    \left|\tr\big[\rho_t^{(M)}{P_\beta^{(e)}}\big]-T_{R_e,K}(t)\right|\le \frac{(t\|\widetilde{\cL}^{(M)}_{R_e}\circ\cP\|_{1\rightarrow 1})^{K+1}}{(K+1)!}\, , 
\end{align*}
where
\begin{equation*}
	T_{R_e,K}(t)\coloneqq \sum_{j=0}^K\,\frac{t^j}{j!}\,\tr\Big[(\widetilde{\cL}_{R_e}^{(M)})^j\left(|\psi_\alpha\rangle\langle \psi_\alpha|\right){P_\beta^{(e)}}\Big]\,.
\end{equation*}
The $1\rightarrow1$ operator norm above can be further controlled using one of our proven statements for the Bose--Hubbard model \Cref{ex:Bose-Hubbard}, any two-body Hamiltonian defined by a polynomial in annihilation and creation operators on at most two modes \Cref{ex:general-hamiltonian}, which satisfies \Cref{assum:local-GKSL,assum-lrb}, or our Hamiltonian regularized by photon dissipation \Cref{ex:reg-hamiltonian}. For all cases, we provide upper bounds on the following operator norm in \Cref{cor:boundedness-projective-generator-bose-hubbard,cor:boundedness-projective-generator-quadratic-generalized,cor:boundedness-projective-generator-reg-hamiltonian}, but focus on the latter since this bound is an upper bound for all examples. To apply the bound, first recall that for $\cP=P\cdot P$
\begin{equation}\label{eq:proof-redefine-projected-generators}
    \cH[PHP]\circ\cP=\cP\cH[PHP]\cP\qquad\text{and}\qquad \cL[(a^p-\alpha^p I)P]\cP=\cP\cL[a^p-\alpha^p I]\cP
\end{equation}
because of $a^pP=Pa^pP$ (see \Cref{eq:swap-projections}). This shows by definition of $\cP$
\begin{equation*}
    \|\widetilde{\cL}^{(M)}_{R_e}\circ\cP\|_{1\rightarrow 1}\leq\|\widetilde{\cL}^{(\alpha)}_{R_e}\circ\cP\|_{1\rightarrow 1}
\end{equation*}
so that \Cref{cor:boundedness-projective-generator-reg-hamiltonian} provides the bound
\begin{equation*}
    \begin{aligned}
		\|\widetilde{\cL}^{(M)}_{R_e}\circ\cP\|_{1\rightarrow 1}&\leq|R_e|(1+2\eta)(p!+1)(M+1)^{p}+2|E_{R_e}|\Lambda \widetilde{d}^{4}(\widetilde{d}!)^{2}(M+1)^{2\widetilde{d}}
	\end{aligned}
\end{equation*}
with $\eta=\|\alpha\|_\infty$, Hamiltonian degree $\widetilde{d}$ and photon-loss order $p$ (see \Cref{ex:reg-hamiltonian}). By the choice $p\geq 2(\widetilde{d}+1)$ (see \Cref{prop:LRB-regularized-hamiltonian}), the dissipative contribution dominates the cutoff scaling of the displayed upper bound, so that
\begin{equation*}
    \begin{aligned}
		\|\widetilde{\cL}^{(M)}_{R_e}\circ\cP\|_{1\rightarrow 1}&\leq|R_e|(1+2\eta)(p!+1)(M+1)^{p}+2|E_{R_e}|\Lambda \widetilde{d}^{4}(\widetilde{d}!)^{2}(M+1)^{2\widetilde{d}}\\
        &\leq \kappa |R_e|^2(M+1)^p
	\end{aligned}
\end{equation*}
for a constant $\kappa>0$. Next, recall that bounded growth gives $|R_e|=\mathcal{O}(\operatorname{polylog}(\epsilon_{\operatorname{LR}}^{-1}))$ and $M=\mathcal{O}\big( \operatorname{dist}(\partial E_R,T)^{\frac{1}{r_{\exp}d}} \big)=\mathcal{O}(\operatorname{polylog}(\epsilon_{\operatorname{LR}}^{-1}))$. Combining the above bounds shows
\begin{align*}
    \left|\tr\big[\rho_t^{(M)}{P_\beta^{(e)}}\big]-T_{R_e,K}(t)\right|\le \frac{(t\kappa |R_e|^2(M+1)^p)^{K+1}}{(K+1)!}\,.
\end{align*}
With $(K+1)!=(K+1)\cdots \bigl\lfloor\frac{K+1}{2}\bigr\rfloor\cdots 1\geq \bigl(\lfloor (K+1)/2\rfloor\bigr)^{\lceil (K+1)/2\rceil}\geq \bigl( K/2\bigr)^{(K+1)/2}$,
\begin{equation*}
    \begin{aligned}
        \frac{(t\kappa |R_e|^2(M+1)^p)^{K+1}}{(K+1)!}\leq\Bigl(\frac{2t^2\kappa^2|R_e|^4(M+1)^{2p}}{ K}\Bigr)^{(K+1)/2}\leq \varepsilon_{\operatorname{poly}}
    \end{aligned}
\end{equation*}
if $ K  \geq 2t^2\kappa^2|R_e|^4(M+1)^{2p} e^2$, which shows
\begin{equation*}
    \begin{aligned}
        \Bigl(\frac{2t^2\kappa^2|R_e|^4(M+1)^{2p}}{K}\Bigr)^{(K+1) /2}\leq\Bigl(\frac{1}{e}\Bigr)^{K+1}=e^{-(K+1)}
    \end{aligned}
\end{equation*}
and $K\geq-\ln(\varepsilon_{\operatorname{poly}})$. Together this shows
\begin{align*}
    \left|\tr\big[\rho_t^{(M)} {P_\beta^{(e)}}\big]-T_{R_e,K}(t)\right|\le \epsilon_{\operatorname{poly}}\qquad\text{ for }\qquad K\geq \max\{-\ln(\epsilon_{\operatorname{poly}}), 2t^2\kappa^2|R_e|^4(M+1)^{2p}e^2\}\,.
\end{align*}
Next, the functions $T_{R_e,K}(t)$ can be expressed as
\begin{align*}
    T_{R_e,K}(t)
    &=\sum_{j=0}^K\,\frac{t^j}{j!}\,\tr\Big[(\widetilde{\cL}_{R_e}^{(M)})^j\left(|\psi_\alpha\rangle\langle \psi_\alpha|\right){P_\beta^{(e)}}\Big]\\
    &= \sum_{j=0}^K\,\frac{t^j}{j!}\,\tr\Big[(\widetilde{\cL}_{R_e}^{(M)})^j\left(|\psi_\alpha\rangle\langle \psi_\alpha|\right) P{P_\beta^{(e)}} P\Big]\, .
\end{align*}
Here, we used the definition of $\ket{\psi_\alpha}=P\ket{\alpha}$, \Cref{eq:proof-redefine-projected-generators} and the commutativity of the trace. Extracting the normalization factors, we can write this as 
\begin{align*}
    T_{R_e,K}(t) = \frac{e^{-|\alpha|^2-|\beta|^2}}{\pi^{|e|}} \,g_e (\alpha, \beta, t)\, ,
\end{align*}
where $g_e (\alpha, \beta, t)$ is now a polynomial in all its variables and has degree at most 
\begin{equation*}
    M'=\mathcal{O}(\operatorname{polylog}(\varepsilon_{\operatorname{LR}}^{-1},\varepsilon_{\operatorname{poly}}^{-1}))
\end{equation*}
in each of its variables. Additionally, we choose $\alpha = \alpha_e$ with $\alpha_i = 0$ for all $i \in R_e \setminus \{e\}$. Using the same reasoning as in \Cref{alg2:polynomial}, we see that $T_{R_e,1}(t)$ is determined solely by the coefficients of the generator restricted to $e$, and not the full region $R_e$. Up to this point, we have argued that
\begin{equation}\label{eq:taylor-approximation}
    \max_{\substack{
        \alpha \in [-1,1]^{|e|}\\
        \beta \in [-1,1]^{2|e|}\\
        t \in [b_1, b_2]
    }}
    \left| 
        \operatorname{tr} \left[ P^{(e)}_\beta e^{t\widetilde{\mathcal{L}}} \left( \ketbra{\alpha}{\alpha} \right) \right] 
        - \frac{e^{-|\beta|^2 - |\alpha_e|^2}}{\pi^{|e|}} \, g_e(\alpha, \beta, t)
    \right| 
    \leq \varepsilon_{\mathrm{poly}} + \varepsilon_{\mathrm{LR}}\,.
\end{equation}
Here, we choose $\alpha$ and $\beta$ to lie in the $\ell^\infty$ ball of radius 1, and $t$ is taken as described in \Cref{protocollearningLRprojected}. Note that this can be extended as long as the intervals are uniformly bounded for the whole learning process. Combining this approximation guarantee with the observations that we made about our estimators above, if we define
\begin{align*}
    L^{(e)}_\alpha(\beta,t)\coloneqq {\frac{1}{\pi^{|e|}}}\int_{R_\beta}\,g_e(\alpha_e,\beta',t)\,d^{2|e|}\beta'\,,
\end{align*}
we see that (see \Cref{eq:taylor-approximation}), for $|\alpha|,|\beta|=\mathcal{O}(1)$ as in our algorithm,
\begin{align*}
    \left\lvert\mathbb{E}\left[\hat{L}^{(e,N_{\mathrm{samp}})}_{(\alpha,\beta)}(t)\right] - L^{(e)}_\alpha(\beta,t) \right\rvert = \mathcal{O}(\varepsilon_{\operatorname{poly}}+\varepsilon_{\operatorname{LR}})\, ,
\end{align*}
and therefore (see \Cref{eq:learning-hoeffding})
\begin{align*}
    \left\lvert \hat{L}^{(e,N_{\mathrm{samp}})}_{(\alpha,\beta)}(t) - L^{(e)}_\alpha(\beta,t) \right\rvert = \mathcal{O}(\varepsilon_{\operatorname{poly}}+\varepsilon_{\operatorname{LR}} + \epsilon_{\operatorname{stat}})\, ,
\end{align*}
holds simultaneously for all $e\in E\cup V$, $\alpha\in\cN_{|e|}^{\cS},\,\beta\in \cN_{2|e|}^{\cM}$, $t\in\cN_t$ with probability $1-\delta$ as long as the number of samples $N_{\mathrm{samp}}$ scales as chosen in detail above.

Next, for each vertex and edge $e\in E\cup V$ we choose the sets $\cN_t$, $\cN_{|e|}^{\cS}$ and $\cN_{2|e|}^{\cM}$ of cardinality $|\cN_t|=\mathcal{O}(M')$, $|\cN_{|e|}^{\cS}| =\mathcal O((M')^{|e|})$ and $|\cN_{2|e|}^{\cM}| =\mathcal O((M')^{2|e|}) $. They are chosen such that we have at least one point in each $C_j$ (see \Cref{eq:shifted-chebyshev-box}) of a $(b_1, b_2, m, 3|e|)$-Chebyshev partition as in \Cref{defi:shifted-chebyshev-partition} with $m=\mathcal O(M')$. Here, we choose for the time interval $b_1 = (M')^{-2}$ and $b_2 = b_1 + 2$.

Using outlier-robust multivariate polynomial interpolation as described in~\cite{Arora.2024} on the data $\hat{L}^{(e,N_{\mathrm{samp}})}_{\alpha,\beta}(t)$ for $(\alpha,\beta,t) \in \cN_{|e|}^{\cS} \times \cN_{2|e|}^{\cM} \times \cN_t$, we obtain a polynomial $\widetilde{L}^{(e)}_\alpha(\beta, t)$ of degree at most $M'$ in each of its variables, satisfying
\begin{equation}\label{eq:learning-algorithm-chebpoly}
    \sup_{\substack{t \in [b_1, b_2] \\ (\alpha, \beta) \in [-1,1]^{3|e|}  }}|\widetilde L^{(e)}_\alpha(\beta,t) - L^{(e)}_\alpha(\beta,t)| = \mathcal{O}(\varepsilon_{\operatorname{poly}}+\varepsilon_{\operatorname{LR}} + \epsilon_{\operatorname{stat}}) \,.
\end{equation}
We can use a multivariate Markov brothers' inequality (\Cref{Markovbroth}) to infer that
\begin{equation*}
     \sup_{\substack{t \in [b_1, b_2] \\ (\alpha, \beta) \in [-1,1]^{3|e|}  }}|\widetilde g_e(\alpha, \beta, t) - g_e(\alpha, \beta, t)| = \mathcal{O}(\varepsilon_{\operatorname{poly}}+\varepsilon_{\operatorname{LR}} + \epsilon_{\operatorname{stat}}) \,,
\end{equation*}
with $\widetilde g_e(\alpha, \beta, t) = \pi^{|e|} \partial_{\beta_e} \widetilde L^{(e)}_\alpha(\beta,t)\big|_{\beta}$ and $\partial_{\beta_e} = \partial_{\beta_{i,R}}\partial_{\beta_{i,I}}$ if $|e|=1$ and $\partial_{\beta_e} = \partial_{\beta_{i,R}}\partial_{\beta_{i,I}}\partial_{\beta_{j,R}}\partial_{\beta_{j,I}}$ if $|e|=2$. Next, let $x \in \mathbb R^{|e|}$ and $y \in \mathbb R^{|e|}$. Then, we define $g^{(\mathrm{Im})}_e(x, y, t):= g_e(x, y, t)$, $\widetilde g^{(\mathrm{Im})}_e(x, y, t):= \widetilde g_e(x, y, t)$, $g^{(\mathrm{Re})}_e(\alpha, \beta, t)= g_e(x, e^{i\varphi_e}y, t)$ and $\widetilde g^{(\mathrm{Re})}_e(\alpha, \beta, t)= \widetilde g_e(x, e^{i\varphi_e}y, t)$ for suitably chosen phases $\varphi_e \in [0,2\pi)$. For $i_\alpha \in \N^{|e|}$ and $i_\beta \in \N^{|e|}$, let $\partial^{i_\zeta}_{\zeta_e} := \partial^{i_{\zeta}}_{\zeta_{i}}$ if $|e|=1$ and $\partial^{i_\zeta}_{\zeta_e} := \partial^{i_{\zeta,1}}_{\zeta_{i}}\partial^{i_{\zeta,2}}_{\zeta_{j}}$ if $|e|=2$ for $\zeta \in \{\alpha, \beta\}$. Then, a multivariate Markov brothers' inequality (\Cref{Markovbroth}) together with \Cref{lem:time-at-zero} ensures that for $\# \in \{\mathrm{Re}, \mathrm{Im}\}$
\begin{align*}
    &\left|\partial_t \partial^{i_\alpha}_{\alpha_e} \partial^{i_\beta}_{\beta_e} g^{(\#)}_e(x, y, t)|_{\alpha,\beta,t=0} - \partial_t \partial^{i_\alpha}_{\alpha_e} \partial^{i_\beta}_{\beta_e} \widetilde g^{(\#)}_e(x, y, t)|_{\alpha,\beta,t=0}\right| \\ 
    &\hspace{2cm}=\mathcal{O}\Big((M')^{4|e|+2}(\epsilon_{\operatorname{stat}}+\epsilon_{\operatorname{LR}}+\epsilon_{\operatorname{poly}})\Big) \\
    &\hspace{2cm}=\mathcal{O}\Big(\operatorname{polylog}\big(\epsilon_{\operatorname{LR}}^{-1},\epsilon_{\operatorname{poly}}^{-1} \big)\,(\epsilon_{\operatorname{stat}}+\epsilon_{\operatorname{LR}}+\epsilon_{\operatorname{poly}})\Big)\,.
\end{align*}
It remains  to verify that the coefficients $\lambda_{k\ell}^{(e)}$ for $e\in V$ and $\lambda_{k\ell k'\ell'}^{(e)}$ for $e\in E$ of $H$ can be reconstructed from these very coefficients.

In the following, we focus on the case $e \in E$, assuming that there are no on-site terms, i.e. $\lambda^{(e')}_{00k'\ell'}=\lambda^{(e')}_{k\ell 00}=0$. Later, the same procedure can be used to exclusively learn the on-site terms, and in a next step, the coefficients on the edges can be determined by subtracting the on-site terms for $e \in V$. First, we have
\begin{equation*}
    \partial_t {\frac{1}{\pi^2}}g^{(\mathrm{Im})}_e(\alpha_e,\beta,t)|_{t=0} =e^{|\beta|^2+|\alpha|^2}\langle \alpha|P(\widetilde{\mathcal{L}}_{R_e}^{(M)\dagger})(P{P_{\beta}^{(e)}} P\otimes I_{e^c})P|\alpha\rangle\,,
\end{equation*}
where $P=P_{R_e}^{(M)}$. Using the structure of the coherent input state $\ket{\alpha}=\ket{\alpha_i,\alpha_j}\otimes\ket{0}_{R_e\backslash\{e\}}$ and $P_{\beta}^{(e)}=\ketbra{\beta}{\beta}_e\otimes I_{R_e\backslash\{e\}}$, the same reasoning given in \Cref{alg2:polynomial} shows that the contributing coefficients of $H$ in $\partial_t {\frac{1}{\pi^2}}g^{(\mathrm{Im})}_e(\alpha_e,\beta,t)|_{t=0}$ reduce to the coefficients of $H$ defining the local terms on $e=(i,j)$ and the photon-dissipation $\cL^{(\alpha,p)}_e$ (see \Cref{ex:reg-hamiltonian}) on $e$. To track the contribution of the photon-dissipation, we define the function
\begin{equation}\label{eq:contribution-dissipation}
    c_{\operatorname{diss}}(\alpha, \beta) \coloneqq -e^{|\alpha_{e}|^2 + |\beta|^2} \, \langle \alpha_e | \cL^{(\alpha,p) \dagger}_e (|\beta\rangle\langle \beta|_e) | \alpha_e \rangle\,,
\end{equation}
which appears in the case of the regularized Hamiltonian~\Cref{ex:reg-hamiltonian}, but is known and can therefore be subtracted in the following. Next, we decompose the coherent states in the Fock basis as in \Cref{eq:coherent-state} to connect the constructed polynomial \eqref{eq:learning-algorithm-chebpoly} to the defining parameter of the Hamiltonian: Denoting $P\equiv P_e^{(M)}$,
\begin{equation}\label{eq:refined-learning-use-locallity}
    \begin{aligned}
        h_e(\alpha_e,\beta)&\coloneqq -e^{|\beta|^2+|\alpha|^2}\langle \alpha|P(\widetilde{\mathcal{L}}_{R_e}^{(M)\dagger})(P{P_{\beta}^{(e)}} P\otimes I_{e^c})P|\alpha\rangle\,\\
        &=-e^{|\alpha_{e}|^2+|\beta|^2}\,\Bigl(i\langle \alpha_e| P\big[PH_{e}P,P|\beta\rangle\langle \beta|_e P\big] P|\alpha_e\rangle+\langle \alpha_e|\cL^{(\alpha,p) \dagger}_e(|\beta\rangle\langle \beta|_e)|\alpha_e\rangle\Bigr)\\
        &=-i \sum_{u,v,u',v'=0}^M\,\frac{\alpha_i^u\alpha_j^{u'}\beta_i^{v}\beta_j^{v'}}{\sqrt{u!v!u'!v'!}}\,  \big(\langle uu'|H_e|vv'\rangle-\langle vv'|H_e|uu'\rangle\big)+c_{\operatorname{diss}}(\alpha,\beta)\\
        &=2\sum_{u,v,u',v'=0}^{M}\,\frac{\alpha_i^u\alpha_j^{u'}\beta_i^{v}\beta_j^{v'}}{\sqrt{u!v!u'!v'!}}\,  \operatorname{Im}\big(\langle uu'|H_e|vv'\rangle\big)+c_{\operatorname{diss}}(\alpha,\beta)\,,
    \end{aligned}
\end{equation}
where we used the assumption $\lambda^{(e')}_{00k'\ell'}=\lambda^{(e')}_{k\ell 00}=0$, so that $\langle \alpha|P\big[PH_{e'}P,P|\beta\rangle\langle\beta|_e\otimes I_{e^c}P\big]P|\alpha\rangle=0$ for $e\ne e'$. Therefore, one can access the imaginary parts of the matrix elements of the interaction $H_e$ in the Fock basis up to local Fock numbers $d$ by differentiating the polynomial $h_e$ up to $d$ times in $\alpha_i,\alpha_j,\beta_i$ and $\beta_j$ at $0$. In the case where the generator of the considered evolution is a regularized Hamiltonian (see \Cref{eq:regularized-hamiltonian}), it is important to note that $c_{\operatorname{diss}}(\alpha, \beta)$ is completely known beforehand, so that it contributes only an offset after differentiating and evaluating at $0$. This known additive term can be corrected for without worsening the accuracy, so that approximate access to the derivatives of $g^{(\mathrm{Im})}_e$ indeed provides approximate access to the derivatives of $h_e$.

A similar procedure to the one described up to now, by appending phases of the form $e^{-i\frac{\pi}{2(u+v')}}$ to $\beta$ in the definition of $g^{(\mathrm{Re})}_e$ similarly to what was done in \Cref{vanillaalgobound}, allows us to extract the real parts of these coefficients. It therefore remains to argue that the coefficients $\lambda^{(e)}_{k\ell k'\ell'}$ can be retrieved from $\langle uu'|H_e|vv'\rangle$. Indeed,
\begin{align*}
    \langle& uu'|H_e|vv'\rangle\\
    &=\sum_{k\ell k'\ell'}\lambda^{(e)}_{k\ell k'\ell'}\,\langle u|(a_i^{\dagger})^ka_i^{\ell}|v\rangle \,\langle u' |(a_j^\dagger)^{k'}a_j^{\ell'}|v'\rangle\\
    &=\sum_{k\ell k'\ell'} \lambda^{(e)}_{k\ell k'\ell'}\delta_{u-k,v-\ell}\,
    \delta_{u'-k',v'-\ell'}\\
    &\quad\cdot\delta_{v\geq \ell}\delta_{u\geq k}\delta_{v'\geq \ell'}\delta_{u'\geq k'}\sqrt{v\dots (v-\ell+1)\,u\dots (u-k+1) v'\dots (v'-\ell'+1)u'\dots (u'-k'+1)}\\
    &= \sum_{kk'}\lambda_{k(v+k-u)k' (v'+k'-u') }^{(e)}\\
    &\quad\delta_{v\geq \ell}\delta_{u\geq k}\delta_{v'\geq \ell'}\delta_{u'\geq k'}\cdot\sqrt{v\dots (u-k+1)\,u\dots (u-k+1) v'\dots (u'-k'+1)u'\dots (u'-k'+1)}\,.
\end{align*}
Then, we start with the lowest-level non-zero coefficients of the form $\lambda^{(e)}_{0101}$. For this, take $u=0=u'$ and $v=1=v'$. The sum above reduces to $\langle 00|H_e|11\rangle= \lambda_{0101}^{(e)}$. Higher-level coefficients are learned by increasing the values of $u,v,u',v'$ iteratively. 
Here, note that we only care about summands with $k,k'\leq d=\mathcal{O}(1)$, for which the factors under the square root are in turn at most $\mathcal{O}(1)$, and that the resulting relations can always only involve a number of coefficients that depends on $d=\mathcal{O}(1)$. The above explained method to relate $L_{\alpha}^{(e)}(\beta,t)$ to the coefficients of the terms of the two-body interactions directly reduces to on-site terms. Here it is important that existing two-body interactions do not appear in the calculation \eqref{eq:refined-learning-use-locallity} by the same reasoning as in \Cref{alg2:polynomial}. This allows us to learn the on-site terms. In a next step the two-body interaction coefficients are learned by subtracting the now-known on-site terms in \Cref{eq:refined-learning-use-locallity}. Therefore, we have proved the following result:

\begin{thm}[Refined strategy]\label{refinedstrategy}
	For the Bose--Hubbard model (\ref{ex:Bose-Hubbard}), the regularized Hamiltonian (\ref{ex:reg-hamiltonian}), or any polynomial two-body interaction Hamiltonian (\ref{ex:general-hamiltonian}) satisfying our weak Lieb--Robinson-type information propagation bound (\ref{thm:bosonicLRB}) on a fixed-dimensional lattice or, more generally, on an interaction graph with uniformly bounded growth as described above, the method described in the previous paragraph provides estimators for $\hat{\lambda}^{(e)}_{k\ell k'\ell'},\,\hat{u}^{(i)},\,\hat{\mu}^{(i)}$ of $\lambda^{(e)}_{k\ell k'\ell'},\,{u}^{(i)},\,{\mu}^{(i)}$ such that, with probability $1-\delta$, we have $|\hat{\lambda}^{(e)}_{k\ell k'\ell'}-\lambda^{(e)}_{k\ell k'\ell'}|,\,|\hat{u}^{(i)}-u^{(i)}|,\,|\hat{\mu}^{(i)}-\mu^{(i)}|\le \epsilon$ simultaneously for all $k,\ell, k',\ell'\in\{1,...,d\}$, $e\in E$ and $i\in V$, as long as the evolution time for each run scales as $t=\widetilde{\mathcal{O}}(1)$ and the total number of samples is at least $N_{\operatorname{samp}}=\widetilde{\mathcal{O}}\big({\epsilon}^{-2}\ln\Big(\frac{|V|+|E|}{\delta}\Big)\big)$. Our scheme uses a total evolution time of 
    \begin{equation*}
        T_{\operatorname{evo}}=\widetilde{\mathcal{O}}\left(\frac{1}{\epsilon^2}\ln\left(\frac{|V|+|E|}{\delta}\right)\right)\,,
    \end{equation*}
    where $\widetilde{\cO}$ suppresses factors polylogarithmic in $\varepsilon^{-1}$.
\end{thm}

\begin{rmk*}
    Note that the dissipation in the regularized Hamiltonian is independent of the input state but needs to be known beforehand. Another example of interest is
    \begin{equation*}
        \widetilde{\cL}^{(0)} = \cH^{(d)} + \cL[a^p]\,,
    \end{equation*}
    which, for example, also includes the Bose--Hubbard model with local particle loss. Following the same proof strategy, the scheme extends to any known dissipation which, together with the Hamiltonian, satisfies our weak Lieb--Robinson-type information propagation bounds (see \Cref{thm:bosonicLRB}) and boundedness of the projected generators.
\end{rmk*}

\subsection{Auxiliary result: Multivariate polynomial interpolation}\label{subsec:interpolation}

In this section, we state in a self-contained manner the results from \cite{Arora.2024} on outlier-robust polynomial interpolation that we use in our algorithms. The following can be found as Definition 2.3 in \cite{Arora.2024}:
\begin{defi} \label{defi:chebyshev-partition} Let $m$, $n \in \mathbb N$. Then, the $(m,n)$-Chebyshev partition of the cube $[-1,1]^n$ is a set of $m^n$ cells $C_j$ indexed by $j \in \{1, \ldots, m\}^n$ such that
\begin{equation} \label{eq:chebyshev-box}
    C_j = \left[\cos\left(\frac{\pi j_1}{m}\right), \cos\left(\frac{\pi (j_1-1)}{m}\right)\right] \times \ldots \times \left[\cos\left(\frac{\pi j_n}{m}\right), \cos\left(\frac{\pi (j_n-1)}{m}\right)\right] \, .
\end{equation}
\end{defi}

\begin{lem}\label{lem:det-multivar-poly-int}
    Let $n=\mathcal O(1)$, $f:[-1,1]^n \to \mathbb R$ be a polynomial of at most degree $M$ in each of the variables and $\sigma > 0$. Suppose that $m=\mathcal O(M)$ and we are given $m^n$ inputs $x_j \in [-1,1]^n$ with $x_j \in C_j$ as in \Cref{eq:chebyshev-box} for all $j \in \{1, \ldots, m\}^n$. Moreover, we are given $m^n$ noisy outputs such that $|y_j - f(x_j)| \leq \sigma$. Then, there exists an algorithm that returns a polynomial $\tilde f$ of at most degree $M$ in each variable such that 
    \begin{equation*}
        \|f-\tilde f\|_{L^\infty([-1,1]^n)} \leq 3 \sigma \, .
    \end{equation*}
    Moreover, the runtime of the algorithm is $\mathcal O(\log(\|f\|_{L^\infty([-1,1]^n)}/\sigma) \mathrm{poly}(M))$.
\end{lem}
\begin{proof}
    This follows from \cite[Theorem 3.4]{Arora.2024} with $\alpha =0$, $\epsilon = 1/2$ and $\eta = \min\{\sigma,1\}$.
\end{proof}

There also exists an algorithm which is robust with respect to outliers, at the cost of having to sample the points $x_j$ at random and a logarithmic overhead in sample complexity:
\begin{lem}\label{lem:random-multivar-poly-int}
    Let $\delta \in (0, \frac{1}{2}]$, $\sigma >0$, $n = \mathcal O(1)$ and $f:[-1,1]^n \to \mathbb R$ be a polynomial of at most degree $M$ in each of the variables. Let us assume we are given $N=\mathcal O(M^n \log(M/\delta))$ pairs $(x_j, y_j)$. Each $x_j \in [-1,1]^n$ is drawn independently from the multidimensional Chebyshev distribution and is an outlier with probability $1-\rho$, where $\rho=\mathcal O(1) < \frac{1}{2}$. After generating the $x_j$, an adversary picks for each non-outlier a number $y_j$ such that $|y_j-f(x_j)| \leq \sigma$ and for outliers any $y_j \in \mathbb R$. Then, there exists an algorithm that returns a polynomial $\tilde f$ of at most degree $M$ in each variable such that 
    \begin{equation*}
        \|f-\tilde f\|_{L^\infty([-1,1]^n)} \leq 3 \sigma \, .
    \end{equation*}
 Moreover, the runtime of the algorithm is $\mathcal O(\log(\|f\|_{L^\infty([-1,1]^n)}/\sigma) \mathrm{poly}(M, \log(\delta)))$.
\end{lem}
\begin{proof}
    This follows from \cite[Theorem 3.1]{Arora.2024} with $\epsilon = \frac{1}{2}$ and $\eta = \frac{1}{2}\sigma$.
\end{proof}

If the multivariate polynomial is defined on $[b_1,b_2] \times [-1,1]^n$ rather than on $[-1,1]^n$, with $0\leq b_1 < b_2$, we adapt the multivariate interpolation results by defining the new polynomial $\hat f(x,y) := f((b_1+b_2)/2 + x\cdot (b_2-b_1)/2,y)$ for all $x \in [-1, 1]$ and $y \in [-1,1]^n$, which is now a polynomial on $[-1,1]^n$. We only need to change the Chebyshev partition accordingly:
\begin{defi} \label{defi:shifted-chebyshev-partition} Let $m$, $n \in \mathbb N$ and $b_1$, $b_2 \in \mathbb R$ such that $0\leq b_1 < b_2$. Then, the $(b_1, b_2, m,n)$-shifted Chebyshev partition of the deformed cube $[b_1,b_2] \times [-1,1]^n$ is a set of $m^{n+1}$ cells $C_j$ indexed by $j \in \{1, \ldots, m\}^{n+1}$ such that
\begin{align} \label{eq:shifted-chebyshev-box}
    C_j =& \left[\frac{b_1+b_2}{2}+\cos\left(\frac{\pi j_0}{m}\right)\left(\frac{b_2-b_1}{2}\right),\frac{b_1+b_2}{2}+\cos\left(\frac{\pi (j_0-1)}{m}\right)\left(\frac{b_2-b_1}{2}\right)\right] \\ &\times \left[\cos\left(\frac{\pi j_1}{m}\right), \cos\left(\frac{\pi (j_1-1)}{m}\right)\right] \times \ldots \times \left[\cos\left(\frac{\pi j_n}{m}\right), \cos\left(\frac{\pi (j_n-1)}{m}\right)\right] \, . \nonumber
\end{align}
\end{defi}
In our algorithm, we estimate the difference between time derivatives on an interval $[0,b_2]$ from a polynomial interpolation on $[b_1, b_2]$. To do this, we use the following consequence of the Markov brothers' inequality, which follows the same proof as \cite[Corollary E.1]{Stilckfranca.2024}.

\begin{lem} \label{lem:time-at-zero}
    Let $f:[0, b_2] \times [-1,1]^n\to\mathbb{R}$ be a polynomial of degree at most $M$ in each of the variables.
    Let $\sigma>0$. Take $0<b_1= M^{-2}$ and $b_2 = 2+b_1$.
    Let $\tilde f: [0, b_2] \times [-1,1]^n\to\mathbb{R}$ be another polynomial such that 
    \begin{align*}
        \|f-\tilde f\|_{L^\infty([b_1, b_2] \times [-1,1]^n)}\leq\sigma
    \end{align*}
    Then, for any $x \in [-1,1]^n$,
    \begin{align*}
        \lvert \partial_t f(t, x)|_{t=0} - \partial_t\widetilde{f}(t,x)|_{t=0}\rvert 
        \leq eM^2\sigma\,.
    \end{align*}
\end{lem}

\newpage
\sloppy
\setlength{\bibitemsep}{0.5ex}
\printbibliography[heading=bibnumbered]

\end{document}